\documentclass[a4paper,11pt]{article}

\usepackage{jheppub_mod} 
\usepackage[T1]{fontenc} 

\usepackage{hyperref}
\usepackage{graphicx}
\usepackage{amsmath,amssymb,slashed}
\usepackage{booktabs,tabulary}
\usepackage{dsfont}
\usepackage{bm}
\usepackage{color}
\usepackage[dvipsnames]{xcolor}
\usepackage{multirow}
\usepackage{subcaption}
\usepackage{tikz-feynman}
\usepackage{multirow}
\usepackage{cancel}
\usepackage{xspace}   
\usepackage{xfrac}

\newcommand{\Fpi}{F_0}
\newcommand{\mpi}{M_{\pi}}
\newcommand{\mpii}{M_{\pi^0}}

\newcommand{\metap}{M_{\eta'}}

\newcommand{\MeV}{\,\text{MeV}}

\newcommand{\beq}{\begin{equation}}
\newcommand{\eeq}{\end{equation}}
\newcommand{\diff}{\text{d}}

\newcommand{\M}{\mathcal{M}}
\newcommand{\N}{\mathcal{N}}

\renewcommand{\O}{\mathcal{O}}
\renewcommand{\L}{\mathcal{L}}

\renewcommand{\Re}{\text{Re}\,}
\renewcommand{\Im}{\text{Im}\,}
\newcommand{\chpt}{$\chi$PT\xspace}

\newcommand{\etap}{\eta^{(\prime)}}
\newcommand{\LEW}{\Lambda_{\text{EW}}}
\newcommand{\munu}{{\mu\nu}}
\newcommand{\dabc}{d_{abc}}
\newcommand{\fabc}{f_{abc}}
\newcommand{\PL}{P_{\text{L}}}
\newcommand{\PR}{P_{\text{R}}}

\newcommand{\PRL}{P_{\text{R/L}}}
\newcommand{\hc}{\text{h.c.}}

\newcommand{\eqwith}{\, ,\hspace{.5cm} \mathrm{with} \hspace{.5cm}}
\newcommand{\eqand}{\hspace{.5cm} \mathrm{and} \hspace{.5cm}}

\definecolor{gray75}{gray}{0.75}
\def\vev#1{\big\langle #1 \big\rangle}

\allowdisplaybreaks[1]

\def\vecsign{\mathchar"017E}
\def\dvecsign{\smash{\stackon[-2.17pt]{\vecsign}{\rotatebox{180}{$\vecsign$}}}}
\def\dvec#1{\def\useanchorwidth{T}\stackon[-4.2pt]{#1}{\,\dvecsign}}
\usepackage{stackengine}
\stackMath

\def\cevsign{\smash{\stackon[5.pt]{\ }{\rotatebox{180}{$\vecsign$}}}}
\def\cev#1{\def\useanchorwidth{T}\stackon[-4.2pt]{#1}{\,\cevsign}}
\usepackage{stackengine}
\stackMath

\makeatletter
	\newenvironment{aligneq}
    {\begin{equation}\begin{alignedat}{20}}{\end{alignedat}\end{equation}}
    \usepackage{etoolbox}
    \newcommand*\NoIndentAfterEnv[1]{
  \AfterEndEnvironment{#1}{\par\@afterindentfalse\@afterheading}}
\makeatother
\NoIndentAfterEnv{aligneq}
\usepackage{extarrows} 

\allowdisplaybreaks[1]

\title{\boldmath{$C$ and $CP$ violation in effective field theories}}

\author[a]{Hakan Akdag,}
\author[a]{Bastian Kubis,}
\author[b]{and Andreas Wirzba}

\affiliation[a]{
Helmholtz-Institut f\"ur Strahlen- und Kernphysik (Theorie) and \\
Bethe Center for Theoretical Physics, Universit\"at Bonn, 53115 Bonn, Germany}

\affiliation[b]{Institut  f\"{u}r Kernphysik (Theorie), 
           Institute for Advanced Simulation, and \\
           J\"ulich Center for Hadron Physics,
           Forschungszentrum J\"ulich,  
           52425 J\"{u}lich, Germany}

\emailAdd{akdag@hiskp.uni-bonn.de}
\emailAdd{kubis@hiskp.uni-bonn.de}
\emailAdd{a.wirzba@fz-juelich.de}

\abstract{The quest for new sources of the simultaneous violation of $C$  and $CP$ symmetry was popular in the 1960s and has since been mostly neglected for more than half a century. In this work we revisit fundamental quark-level operators that break $C$ and $CP$ up to and including mass dimension 8 for flavor-conserving transitions, relying on the complete operator sets of the so-called Standard Model effective field theory and the low-energy effective field theory. 
With the formalism of chiral perturbation theory, we match these quark operators to 
light-meson physics, derive $C$- and $CP$-odd Lagrangians for several processes in the $\eta$, $\eta'$, and pion sectors, and furthermore, as a proof of principle, give estimates for the respective observables in explicit dependence of the underlying high-energy scale for new physics.}

\begin{document} 
\maketitle

%---------------------------------------------------------------
\section{Introduction}
\label{sec:intro}
%---------------------------------------------------------------
The efforts in search of new physics have reached a milestone with the observation of the Higgs boson at CERN's Large Hadron Collider~\cite{ATLAS:2012yve,CMS:2012zhx}, one of the most important building blocks of the Standard Model of particle physics (SM). Until now, there is no evidence for new particles other than the ones contained in the SM, at least up to an energy scale of the order $1\,\text{TeV}$. However, physics beyond the Standard Model (BSM) can arise from heavy particles with masses above some unknown high-energy scale $\Lambda$ exceeding the TeV range, which is out of experimental reach for the foreseeable future. 
It is commonly agreed on that---if such heavy degrees of freedom exist---the SM provides merely an effective description of the underlying beyond-Standard-Model theory. A convenient way to approach the effects of the latter below the scale $\Lambda$ concerns the construction of effective field theories, providing a fundamental, model-independent framework. According to the Appelquist--Carazzone theorem~\cite{Appelquist:1974tg} 
heavy particles decouple in a perturbative manner with (anomaly-free) interactions in the low-energy 
range. Hence the underlying BSM theory can be described by the \textit{Standard Model effective field theory} (SMEFT), whose perturbative expansion in powers of the small parameter $1/\Lambda$ yields
\beq\label{eq:SMEFT_expansion}
    \L_\text{SMEFT}= \L_\text{SM}
    +\frac{1}{\Lambda} \L_5
    +\frac{1}{\Lambda^2}\L_6
    +\frac{1}{\Lambda^3}\L_7+\ldots \eqwith \L_D\equiv\sum_i C_i^DQ_i^D\,.
\eeq
In this operator-product-type of expansion the quantity $\L_\text{SM}$ denotes the renormalizable Lagrangian of the SM with mass dimension 4, the non-renormalizable Lagrangians $\L_D$ include operators $Q^D_i$ with mass dimensions $D>4$ that are suppressed by powers of $1/\Lambda$, and the $C_i^D$ are the respective dimensionless coupling constants known as Wilson coefficients.

In general $\L_\text{SMEFT}$ is richer than $\L_\text{SM}$ itself, as 
the $\L_D$ are only restricted by Lorentz invariance, the same $SU(3)_C\times SU(2)_L\times U(1)_Y$ gauge group, and the same particle content as the SM.
In the recent past, much effort has been devoted to the systematic construction of complete sets of operators contributing to the $\L_D$, which includes the classification and counting of all possible $Q_i^D$ (cf.\ Ref.~\cite{Murphy:2020rsh} for a short overview) as well as eliminating redundant operators with the aid of equations of motion, partial integration, and Fierz identities.
The complete operator bases up to and including dimension 9 in SMEFT have been counted and/or computed in Refs.~\cite{Weinberg:1979sa, Buchmuller:1985jz, Grzadkowski:2010es, Lehman:2014jma, Henning:2015alf, Liao:2016hru,Li:2020gnx,Murphy:2020rsh, Liao:2020jmn, Li:2020xlh}.
If we exclude the existence of, yet unobserved, hypothetical light degrees of freedom that couple extremely weakly to the SM particle content, such as axions and sterile neutrinos, SMEFT reduces to the SM in the limit of small energies. An extension of SMEFT including new light particles can for instance be found in Refs.~\cite{delAguila:2008ir, Aparici:2009fh, Bhattacharya:2015vja,Liao:2016qyd,Dekens:2020ttz,Li:2021tsq, Galda:2021hbr}.

However, when going to even smaller energies, SMEFT may not be the most convenient theory to describe phenomena exclusively occurring below the electroweak scale $\LEW\lesssim v$, with $v$ denoting the Higgs vacuum expectation value (vev). In this scenario, particles of the SM with masses larger than $\LEW$ are no observable degrees of freedom anymore and the gauge group reduces to $SU(3)_C\times U(1)_{\rm em}$. This means that the top quark, the weak gauge bosons $W^\pm$ and $Z^0$, and the Higgs are integrated out and we are left with the dynamics of QCD and QED only, while effects of weak interactions are implicitly encoded in the constant Wilson coefficients, also giving rise to point-like interactions with neutrinos. Nevertheless, the construction of a consistent and complete basis of such a theory, without operator redundancies and terms of repeated flavors compositions, and its matching to SMEFT at larger energies is quite involved~\cite{Murphy:2020cly}. The appropriate theory to handle effects in this energy regime is known as \textit{low-energy effective field theory}~(LEFT), which is in principle a valid theory on its own, even without recourse to SMEFT. The continuation of LEFT to energies above $\LEW$ may for instance as well be given by the \textit{Higgs effective field theory}~\cite{Grinstein:2007iv, Alonso:2012px, Buchalla:2013rka, Gavela:2014uta, Brivio:2016fzo, Sun:2022ssa, Sun:2022snw, Sun:2022aag},
which, in comparison to SMEFT, does not rely on the conjecture that the Higgs belongs to an electroweak doublet.\footnote{See Ref.~\cite{Brivio:2017vri} for a comprehensive review.} In either way, the operator product expansion of LEFT proceeds in analogy to Eq.~\eqref{eq:SMEFT_expansion} with a small expansion parameter that can be chosen as $1/v$. With its own Wilson coefficients $\tilde C_i^d$ and operators $\tilde  Q_i^d$ of dimension $d$, the LEFT Lagrangian can be written as
\beq\label{eq:LEFT_expansion}
    \L_\text{LEFT}=v\tilde \L_3 + \L_{\nu\, \text{kin}} + \L_\text{QCD+QED}
    +\frac{1}{v} \tilde \L_5
    +\frac{1}{v^2}\tilde \L_6 +\ldots \eqwith \tilde \L_d\equiv\sum_i \tilde C_i^d\tilde Q_i^d\,,
\eeq
where the three-dimensional Lagrangian $\tilde \L_3$ refers to a mass term of a possible Majorana neutrino and the four-dimensional $\L_{\nu\text{kin}}$ includes the kinetic term for neutrinos.
All other dependencies on the Higgs vev are hidden in the respective Wilson coefficients and do not contribute to the dimensional analysis, so that the different power counting schemes in SMEFT and in LEFT have to be clearly distinguished. 
However, assuming SMEFT to be the underlying theory of LEFT, we can account for the additional suppression in $1/\Lambda$ by redefining the Wilson coefficients of LEFT as~\cite{Jenkins:2017jig}
\beq\label{eq:LEFT_suppression}
    \tilde C_i^d\rightarrow 
    \left(\frac{v}{\Lambda}\right)^{\sum_j(D_j-4)} \tilde C_i^d\,,
\eeq
where the $D_j$ denote the dimension of SMEFT operators that can be combined in one Feynman diagram to construct the desired LEFT operator. 

Hitherto, the investigation of the complete basis in LEFT extends up to dimension 9~\cite{Jenkins:2017dyc,Jenkins:2017jig, Dekens:2019ept, Li:2020tsi, Murphy:2020cly, Liao:2020zyx}, and the inclusion of axions or sterile neutrinos in LEFT can be found in Refs.~\cite{Li:2020lba, Chala:2020vqp,Li:2021tsq, Dekens:2022gha}.
A famous and maybe historically most important example of a LEFT operator is the Fermi theory of weak interactions~\cite{Fermi:1934sk,Fermi:1934hr}.

Our analysis aims at a rigorous derivation of $C$- and $CP$-violating sources in the mesonic sector arising from fundamental quark operators in LEFT.\footnote{An alternative path for the $C$- and $CP$-violating sources could be realized in SMEFT at the $W$-scale, cf.\ Refs.~\cite{Shi:2017ffh, Gardner:prep}, and should yield a set of operators that is equivalent to the one considered and presented in this work.} 
These, due to the $CPT$ theorem, $T$-odd and $P$-even (ToPe) contributions are of particular interest for cosmology, as they provide, according to the Sakharov conditions~\cite{Sakharov:1967dj}, one prerequisite (in addition to the $T$-odd, $P$-odd contributions) for the dynamical creation of the matter--antimatter asymmetry during the baryogenesis. On the other hand, theoretical work about this class of symmetry violation is severely lacking in contrast to the one for $C$-even and $P$-odd phenomena.
First thoughts on the structure of these BSM operators, independent of which effective theory they could possibly belong to, were already made in the 1990s~\cite{Khriplovich:1990ef} and extended throughout that decade~\cite{Conti:1992xn,Engel:1995vv,Ramsey-Musolf:1999cub,Kurylov:2000ub}. It was claimed that the first $C$- and $CP$-odd operators at low energies appear at dimension 7 and read\footnote{There are different formulations of the four-fermion operator in the cited literature, which are consistent with each other when using the Gordon identity.}
\beq\label{eq:ToPe_historic}
\begin{alignedat}{5}
    &\bar\psi \dvec D_\mu \gamma_5 \psi \bar \chi \gamma^\mu \gamma_5 \chi\,,\\
    &\bar\psi \sigma_\munu \lambda^a\psi F^{\mu\lambda}G^{a\nu}_\lambda \,,\\
    &\bar\psi \sigma_\munu \psi F^{\mu\lambda}Z^{\nu}_\lambda \,,
\end{alignedat}
\eeq
for up to two fermion fields $\psi$ and $\chi$.
An aspect withheld in these analyses is that the listed operators are chirality-violating and thus need to be equipped with an additional Higgs field. Therefore, according to a naive power counting, all these operators are of dimension 8 in SMEFT and of dimension 7 in LEFT. As the Higgs is integrated out in the latter theory, its vev, which is absorbed in the Wilson coefficients, does not contribute to the dimensional power.
However, this leads to another unpleasant inconsistency: the heavy $Z^0$-boson is not integrated out although we are exclusively dealing with interactions at low energies. 
We have not yet specified chirality-conserving $C$- and $CP$-odd operators of dimension 8 in LEFT. These are a priori not suppressed with respect to the chirality-violating ones at dimension 7 in LEFT, as both can originate from operators of dimension 8 in SMEFT and thus have the same suppression in $1/\Lambda$. In a similar way this point was observed in nucleon EDM analyses~\cite{Maekawa:2011vs,deVries:2012ab,Dekens:2013zca}, which found that $T$- and $P$-odd chirality-violating operators of dimension $5$ in LEFT can effectively be of the same order of magnitude as chirality-conserving ones at dimension $6$ in LEFT. 
Furthermore, at the time the operators in Eq.~\eqref{eq:ToPe_historic} had been proposed, the rigorously derived complete operator basis in LEFT was not available. 

Going down to even smaller energies, cf.\ Fig.~\ref{fig:BSM_scales}, some effort has been devoted to constructing chiral effective theories from underlying quark-level operators in SMEFT or LEFT, to access BSM phenomena including mesonic interactions below the hadronic scale of $\Lambda_\chi\approx 1.2 \,\text{GeV}$. Most of these works rely (implicitly) on the spurion analysis introduced in the work of Gasser and Leutwyler~\cite{Gasser:1983yg,Gasser:1984gg} and were for instance applied to neutrinoless double beta decays ~\cite{Prezeau:2003xn,Graesser:2016bpz, Cirigliano:2017ymo, Cirigliano:2018yza}, baryon- and lepton-number-violating interactions~\cite{Dekens:2018pbu,Liao:2019gex, Liao:2020roy,He:2020jly,  Liao:2021qfj, He:2021mrt}, neutron--antineutron oscillation~\cite{Bijnens:2017xrz}, $CP$ violation in axion interactions~\cite{Dekens:2022gha}, EDM analyses in the chiral $SU(2)$ case~\cite{deVries:2010ah, Maekawa:2011vs,deVries:2012ab, Bsaisou:2014oka}, or in Lorentz- and $CPT$-violating extensions of the SM~\cite{Kamand:2016xhv, Kamand:2017bzl, Altschul:2019beo}. However, probably due to the missing set of ToPe operators on the quark level, a rigorous and complete derivation of $C$- and $CP$-violating mesonic operators is still missing in the literature. 

For the application of these ToPe operators, the $\eta$ meson is of particular interest, because it is an eigenstate of $C$. It allows us to investigate ToPe forces in the absence of the weak interaction, and provides an ideal stage to probe $C$ and $CP$ violation outside the nuclear arena (see the review~\cite{Gan:2020aco} and references therein), which does not place rigorous bounds to constrain ToPe forces~\cite{Simonius:1975ve}. Furthermore, the new efforts of the  REDTOP~\cite{Gatto:2016rae,Gatto:2019dhj,REDTOP:2022slw} and JEF~\cite{Gan:2015nyc,JEF:2016,Gan:2017kfr} collaborations to search for rare $\eta$ decays underline the timeliness of model-independent $C$- and $CP$-violating operators in the $\eta$ sector. 
This complements renewed recent interest in the feasibility to probe $P$- and $CP$-violating operators in $\eta$ and $\eta'$ decays~\cite{Sanchez-Puertas:2018tnp,Escribano:2022wug,Zillinger:2022eva}, despite strong constraints from electric dipole moments.

At this point we would like to emphasize the importance of the effective field theory (EFT) approaches applied in this work. The breakdown of different energy scales in the EFT spirit is essential to incorporate the appropriate suppression of ToPe forces in terms of $\Lambda$, $v$, and $\Lambda_\chi$. Furthermore, given the unknown underlying mechanism above the scale $\Lambda$, the use of $\chi$PT in analogy to its usual application in the non-perturbative realm of QCD is inevitable.
We will confirm in the following that ToPe operators necessarily are of higher dimension than, e.g., the lowest-dimensional $T$-odd and $P$-odd operators that can generate EDMs; EFT naturalness arguments, both in SMEFT and LEFT, therefore suggest the former to be further suppressed compared to the latter, and hence more difficult to detect.  The motivation to undertake the present study at this point is clearly experiment-driven: the interpretation of new, improved limits on $C$- and $CP$-violating effects requires a dedicated theory framework, which we here provide.  Even if new measurements continue to be essentially null tests, they can be viewed as tests of the SMEFT or LEFT picture of BSM physics, which is especially important when compared to similar EFT approaches in EDM analyses mentioned above.

In this work we thoroughly revisit the $C$- and $CP$-violating operators up to and including dimension 8 in LEFT and provide the first complete set of these operators. Readers who are primarily interested in light-meson applications and implications for experimental analyses may proceed directly to Sect.~\ref{sec:applications}. With focus on the application to ToPe forces in $\eta$ decays, we restrict our analysis to flavor-conserving quark operators (with couplings to the gluon and photon fields). Additionally we quote the corresponding semi-leptonic operators. A generalization to flavor-changing processes can be carried out in complete analogy and is left for future analyses, if the phenomenological interest in these operators is given. 
To undertake this venture, we start with fundamental $C$- and $CP$-violating operators on the quark level from LEFT in Sect.~\ref{sec:LEFT_classification}, by first providing a concise overview of discrete space-time symmetries in Sect.~\ref{sec:discrete_symm_LEFT} and summarizing the full list of ToPe operators in Sect.~\ref{sec:LEFT_summary}, whose derivation from known LEFT bases is sketched in App.~\ref{app:LEFT}. Subsequently, in Sect.~\ref{sec:Matching}, we match these quark operators to mesonic physics. For this endeavor we rely on the well-established techniques from chiral perturbation theory (\chpt) and introduce the latter to the reader in Sect.~\ref{sec:Standard_ChPT}.
We summarize the discrete symmetries of each building block of \chpt in Sect.~\ref{sec:discrete_symm_ChPT}, and subsequently
explain the matching procedure between LEFT and \chpt in Sect.~\ref{sec:principles_of_matching}. Afterwards we illustrate the principles of matching in detail at the hand of the original dimension-7 operators in LEFT and furthermore translate the dimension-8 operators in Sects.~\ref{sec:Matching_LEFT_dim_7} and~\ref{sec:Matching_LEFT_dim_8}, respectively. As a short intermediate summary we provide an overview of the corresponding overall $C$- and $CP$-violating Lagrangian in Sect.~\ref{sec:intermediate_summary}. In Sect.~\ref{sec:Large_Nc} the ToPe chiral theory is taken to the large-$N_c$ limit, allowing for a consistent description of the $\eta'$. Finally, we sketch the application of our formalism to various flavor-conserving decays of $\eta$ and $\eta'$ mesons in Sect.~\ref{sec:applications} and close with a brief summary in Sect.~\ref{sec:summary}.

\begin{figure}[t!]
\centering
    \newcommand{\colora}{violet}
	\newcommand{\colorb}{MidnightBlue}
	\newcommand{\colorc}{PineGreen}
	\newcommand{\colord}{Mahogany}
	\newcommand{\colorx}{black}
	\newcommand{\linewidthboxex}{.5pt}
	\newcommand{\boxcorners}{0.4cm}
	\newcommand{\boxwidtha}{1.2}
	\newcommand{\boxwidthb}{3.}
	\newcommand{\boxwidthc}{5}
	\newcommand{\boxwidthd}{3.}
	\newcommand{\boxheight}{.5}
	\newcommand{\horizontaldistance}{1.5}
	\newcommand{\verticaldistance}{1}
	\newcommand{\scaleplot}[1]{\scalebox{1.}{#1}}
	\newcommand{\labledistance}{	0.7*\verticaldistance}
	%\scalebox{1}{
	\resizebox{\columnwidth}{!}{
	\hspace{-.3cm}
		\begin{tikzpicture}[baseline=0cm]
		%------BSM theory------
		% theory
		\draw[line width=\linewidthboxex, \colora, rounded corners=\boxcorners, align=center] (-\boxwidtha,-\boxheight) rectangle (\boxwidtha,\boxheight) 
		node[pos=0.5, black] (origin) {BSM physics}
		node[above=\labledistance of origin, black] {Theory};
		% energy
 		\draw[line width=\linewidthboxex,\colora, rounded corners=\boxcorners, align=center] (\boxwidtha+\verticaldistance,-\boxheight) rectangle (\boxwidtha+\verticaldistance+\boxwidthb,\boxheight) 
 		node[pos=0.5, black] (origin) {$p\geq\Lambda$}
		node[above=\labledistance of origin, black] {Scale};
 		% gauge group
  		\draw[line width=\linewidthboxex,\colora, rounded corners=\boxcorners, align=center] (\boxwidtha+2*\verticaldistance+\boxwidthb,-\boxheight) rectangle (\boxwidtha+2*\verticaldistance+\boxwidthb+\boxwidthc,\boxheight) 
  		node[pos=0.5, black] (origin) {\textbf{?}}
		node[above=\labledistance of origin, black] {Gauge group};
  		 % particle content
  		\draw[line width=\linewidthboxex,\colora, rounded corners=\boxcorners, align=center] (\boxwidtha+3*\verticaldistance+\boxwidthb+\boxwidthc,-\boxheight) rectangle (\boxwidtha+3*\verticaldistance+\boxwidthb+\boxwidthc+\boxwidthd,\boxheight) 
  		node[pos=0.5, black] (origin) {\textbf{?}}
		node[above=\labledistance of origin, black] {Particle content};
		%------SMEFT------
		% theory
		\draw[line width=\linewidthboxex, \colorb, rounded corners=\boxcorners, align=center] (-\boxwidtha,-\boxheight-\horizontaldistance) rectangle (\boxwidtha,\boxheight-\horizontaldistance) 
		node[pos=0.5, black] {SMEFT};
		% energy
 		\draw[line width=\linewidthboxex,\colorb, rounded corners=\boxcorners, align=center] (\boxwidtha+\verticaldistance,-\boxheight-\horizontaldistance) rectangle (\boxwidtha+\verticaldistance+\boxwidthb,\boxheight-\horizontaldistance) 
 		node[pos=0.5, black] {$\Lambda>p\geq\Lambda_{\text{EW}}$};
 		% gauge group
  		\draw[line width=\linewidthboxex,\colorb, rounded corners=\boxcorners, align=center] (\boxwidtha+2*\verticaldistance+\boxwidthb,-\boxheight-\horizontaldistance) rectangle (\boxwidtha+2*\verticaldistance+\boxwidthb+\boxwidthc,\boxheight-\horizontaldistance) 
  		node[pos=0.5, black] {$SU(3)_C\times SU(2)_L \times U(1)_Y$};
  		 % particle content
  		\draw[line width=\linewidthboxex,\colorb, rounded corners=\boxcorners, align=center] (\boxwidtha+3*\verticaldistance+\boxwidthb+\boxwidthc,-\boxheight-\horizontaldistance) rectangle (\boxwidtha+3*\verticaldistance+\boxwidthb+\boxwidthc+\boxwidthd,\boxheight-\horizontaldistance) 
  		node[pos=0.5, black] {$L_i,Q_i,l_i,u_i,d_i,$\\ $H, G, A, Z, W^\pm$};
		%------LEFT------
		% theory
		\draw[line width=\linewidthboxex, \colorc,rounded corners=\boxcorners, align=center] (-\boxwidtha,-\boxheight-2*\horizontaldistance) rectangle (\boxwidtha,\boxheight-2*\horizontaldistance) 
		node[pos=0.5, black] {LEFT};
		% energy
 		\draw[line width=\linewidthboxex,\colorc, rounded corners=\boxcorners, align=center] (\boxwidtha+\verticaldistance,-\boxheight-2*\horizontaldistance) rectangle (\boxwidtha+\verticaldistance+\boxwidthb,\boxheight-2*\horizontaldistance) 
 		node[pos=0.5, black] {$\Lambda_{\text{EW}}>p\geq\Lambda_\chi$};
 		% gauge group
  		\draw[line width=\linewidthboxex,\colorc, rounded corners=\boxcorners, align=center] (\boxwidtha+2*\verticaldistance+\boxwidthb,-\boxheight-2*\horizontaldistance) rectangle (\boxwidtha+2*\verticaldistance+\boxwidthb+\boxwidthc,\boxheight-2*\horizontaldistance) 
  		node[pos=0.5, black] {$SU(3)_C \times U(1)_Q$};
  		 % particle content
  		\draw[line width=\linewidthboxex,\colorc, rounded corners=\boxcorners, align=center] (\boxwidtha+3*\verticaldistance+\boxwidthb+\boxwidthc,-\boxheight-2*\horizontaldistance) rectangle (\boxwidtha+3*\verticaldistance+\boxwidthb+\boxwidthc+\boxwidthd,\boxheight-2*\horizontaldistance) 
  		node[pos=0.5, black] {$\psi, \ell, \nu, G, A$};	
		%------ChPT------
		% theory
		\draw[line width=\linewidthboxex, \colord, rounded corners=\boxcorners, align=center] (-\boxwidtha,-\boxheight-3*\horizontaldistance) rectangle (\boxwidtha,\boxheight-3*\horizontaldistance) 
		node[pos=0.5, black] {$\chi$PT};
		% energy
 		\draw[line width=\linewidthboxex,\colord, rounded corners=\boxcorners, align=center] (\boxwidtha+\verticaldistance,-\boxheight-3*\horizontaldistance) rectangle (\boxwidtha+\verticaldistance+\boxwidthb,\boxheight-3*\horizontaldistance) 
 		node[pos=0.5, black] {$\Lambda_\chi > p$};
 		% gauge group
  		\draw[line width=\linewidthboxex,\colord, rounded corners=\boxcorners, align=center] (\boxwidtha+2*\verticaldistance+\boxwidthb,-\boxheight-3*\horizontaldistance) rectangle (\boxwidtha+2*\verticaldistance+\boxwidthb+\boxwidthc,\boxheight-3*\horizontaldistance) 
  		node[pos=0.5, black] {$SU(3)_R\times SU(3)_L \times U(1)_Q$};
  		 % particle content
  		\draw[line width=\linewidthboxex,\colord, rounded corners=\boxcorners, align=center] (\boxwidtha+3*\verticaldistance+\boxwidthb+\boxwidthc,-\boxheight-3*\horizontaldistance) rectangle (\boxwidtha+3*\verticaldistance+\boxwidthb+\boxwidthc+\boxwidthd,\boxheight-3*\horizontaldistance) 
  		node[pos=0.5, black] {$\pi,A,\ell,\nu$};
		\end{tikzpicture}
	}
	\caption{Comparison of effective field theories at different energy scales $p$.}
	\label{fig:BSM_scales}
\end{figure}

\boldmath
\section{Effective beyond Standard Model theories: 
fundamental $C$- and $CP$-violating operators}
\label{sec:LEFT_classification}
\unboldmath
%---------------------------------------------------------------

In this section we shortly introduce the LEFT bases under consideration, as well as the notation and conventions used for our analysis.

Below the electroweak scale, the important degrees of freedom in the SM are two up-type and three down-type quarks, which we summarize by $\psi\in\{u,c,d,s,b\}$, three charged leptons $\ell\in\{e,\mu,\tau\}$, and the corresponding left-handed neutrinos $\nu_L\in\{\nu^e_L,\nu^\mu_L,\nu^\tau_L\}$. Except for the latter, these fermions are described by the QCD and QED Lagrangians by means of\footnote{One can in principle also include the QCD $\theta$-term.}
\begin{aligneq}\label{eq:QED+QCD}
    \L_\text{QED+QCD}=&-\frac{1}{4}F_\munu F^\munu-\frac{1}{4}G^a_\munu G^{a\munu}+\sum_{\psi} \bar\psi \big(i\slashed D -m_\psi \big) \psi +\sum_{\ell} \bar\ell \big(i\slashed D -m_\ell \big) \ell \,,
\end{aligneq}
with photonic and gluonic field-strength tensors $F_\munu,G^a_\munu$. We denote the representations of these tensors in dual space by
$\tilde{X}_\munu =\frac{1}{2}\epsilon_{\mu\nu\alpha\beta}X^{\alpha\beta}$ with $X_\munu\in \{F_\munu, G_\munu^a\}$.
The gauge covariant derivative acting on quarks is chosen to be 
\begin{aligneq}
    D_\mu \psi&=\big(\vec{\partial}_\mu+ieQ A_\mu+igT^aG^a_\mu\big)\psi\,,\\
    D_\mu \bar\psi&=\bar\psi\big(\cev{\partial}_\mu-ieQ A_\mu-igT^aG^a_\mu\big)\,,
\end{aligneq}
with $D_\mu \bar\psi\equiv(D_\mu\psi)^\dagger \gamma^0$. In this equation $Q$ and $T^a=\frac12\lambda^a$ are the generators of $U(1)_{\rm em}$ and ${SU}(3)_C$, respectively, where $\lambda^a$ denote the Gell-Mann matrices obeying the relations
\beq
    [T_a,T_b]=i\fabc T_c\,, \quad\quad  \{T_a,T_b\}=\frac{1}{3}\delta_{ab}+\dabc T_c\,.
\eeq
The quantities $\fabc$ and $\dabc$ denote the totally antisymmetric and symmetric structure constants of $SU(3)$. The gauge covariant derivative acting on $\ell$ is defined analogously, but without a coupling to gluons.
We will henceforth indicate the direction to which the derivative acts by arrows, i.e., $\bar\psi\vec{D}_\mu \psi= \bar\psi(D_\mu \psi)$ and  $\bar\psi\cev{D}_\mu \psi= (D_\mu\bar\psi) \psi$. It proves useful to introduce a hermitian version of the gauge covariant derivative by 
\beq
\bar \psi i\dvec{D}_\mu \psi\equiv \bar \psi i\vec{D}_\mu \psi - \bar \psi i\cev{D}_\mu \psi \,.
\eeq
The remaining Lagrangian terms in the LEFT Lagrangian from Eq.~\eqref{eq:LEFT_expansion} are built from the same degrees of freedom contained in Eq.~\eqref{eq:QED+QCD} obeying the gauge group $SU(3)_c \times U(1)_{\text{em}} $.
The choice of LEFT basis is not unique, we can for instance employ the Gordon identity, equations of motion, Fierz identities, as well as integration by parts to shift operators between the classes. In this work we consider the LEFT basis derived in Ref.~\cite{Jenkins:2017jig} for operators with mass dimension $d\leq 6$ (5963 hermitian operators), Ref.~\cite{Liao:2020zyx} for operators of dimension $7$ (5218 hermitian operators), and Ref.~\cite{Murphy:2020cly} for the ones at $d=8$ (35058 hermitian operators).

To tackle the overwhelming amount of more than $45\cdot 10^3$ operators up to dimension $8$ in LEFT, we restrict our investigation to operators that may possibly contribute ToPe forces in $\eta$ decays. Therefore we ignore operators including neutrinos,\footnote{The $\eta$ decay listed as $\Gamma_{20}$ in Ref.~\cite{Workman:2022ynf}, $\eta \to \pi^+ e^{-} \bar \nu_e$ + c.c., is in fact a $C$- and $T$-allowed decay.} drop all operators that violate lepton- and/or baryon-number conservation, and restrict the analysis to $C$- and $CP$-odd operators only that are at the same time flavor-conserving.\footnote{
We note that the requirement of $C$ and $CP$ violation places a very selective constraint, reducing the amount of $45\cdot 10^3$ operators tremendously.} Still, we allow for chirality-conserving and -violating operators.

%---------------------------------------
\subsection{Discrete space-time symmetries}
\label{sec:discrete_symm_LEFT}
%---------------------------------------
In this section we shortly summarize the discrete symmetries of several quantities that constitute the respective LEFT operators, to pick the correct $C$- and $CP$-violating operators from the LEFT bases.

For any combination of Dirac matrices $\Gamma$, the well-known transformations of fermion bilinears under $C$, $P$, and $T$ read 
\begin{aligneq}
    C:\quad &\bar\psi \Gamma \chi \quad\xlongrightarrow{\ C \ }\quad  -\, &&\bar\chi \gamma_0 \gamma_2\Gamma^T \gamma_2 \gamma_0\psi\,,\\
    P:\quad &\bar\psi \Gamma \chi \quad\xlongrightarrow{\ P \ }\quad  &&\bar\psi \gamma_0\Gamma \gamma_0\chi\,,\\
    T:\quad &\bar\psi \Gamma \chi \quad\xlongrightarrow{\ T \ }\quad  &&\bar\psi \gamma_1 \gamma_3\Gamma^\ast \gamma_3 \gamma_1\chi\,,
\end{aligneq}
 where $\Gamma^T$ denotes the transposed of $\Gamma$, $\Gamma^\ast$ is its complex conjugate, and  the factor $-1$ in the first line arises from the anticommutation of fermion creation and annihilation operators. As the fermions in each bilinear change places under $C$, it may not be evident at first glance that the original covariant derivative $\vec{D}_\mu$ contributes to both
 eigenstates of $C$, in contrast to $\dvec{D}_\mu$.\footnote{For the standard derivative $\vec{D}$ one needs the decomposition $\bar\psi \vec{D}_\mu \psi=\frac{1}{2}\bar \psi \left[(\cev{D}_\mu +\vec{D}_\mu) - (\cev{D}_\mu -\vec{D}_\mu)\right]\psi$ to obtain (two distinct) $C$ eigenstates.}

The discrete symmetries of the gauge fields can be deduced from the requirement that their interaction with the quark currents implied by Eq.~\eqref{eq:QED+QCD} preserves $C$, $P$, and $T$ separately. The hermitian generators $T_a$ of ${SU}(3)$ transform as $T_a\to T_a^T=T_a^\ast$ under $C$ and $T$. Hence the discrete symmetries of all color structures can be derived with~\cite{Branco:1999fs}

\begin{aligneq}\label{eq:discrete_sym_QCD}
    &T_a\quad&&\xlongrightarrow{\ C \ }\quad  && &&x_aT_a\,,\qquad
    &&T_a\quad&&\xlongrightarrow{\ P \ }\quad  && T_a\,,\qquad &&T_a\quad&&\xlongrightarrow{\ T \ }\quad  && && x_aT_a\,,\\
    &G_a^{\mu}\quad&&\xlongrightarrow{\ C \ }\quad  && -&&x_a G_a^{\mu}\,,\qquad
    &&G_a^{\mu}\quad&&\xlongrightarrow{\ P \ }\quad  && \varepsilon^\mu G_a^\mu\,,\qquad 
    &&G_a^{\mu}\quad&&\xlongrightarrow{\ T \ }\quad && &&x_a\varepsilon^\mu G_a^\mu\,,\\
    &G_a^{\munu}\quad&&\xlongrightarrow{\ C \ }\quad  && -&&x_aG_a^{\munu}\,,\qquad
    &&G_a^{\munu}\quad&&\xlongrightarrow{\ P \ }\quad  && \varepsilon^\mu\varepsilon^\nu G_a^\munu\,,\qquad 
    &&G_a^{\munu}\quad&&\xlongrightarrow{\ T \ }\quad && -&&x_a\varepsilon^\mu\varepsilon^\nu G_a^\munu\,,
\end{aligneq}
where we used $x_a=1$ for $a\in\{1,3,4,6,8\}$ and $x_a=-1$ for $a\in\{2,5,7\}$ to keep the notation short. The sign $\varepsilon^\mu$ equals $1$ for $\mu=0$ and $-1$ for $\mu\in\{1,2,3\}$. To compile the discrete symmetries of operators including ${SU}(3)$ structure constants note that the non-vanishing values of $\fabc$ ($\dabc$) contain an odd (even) number of indices picked from $\{2,5,7\}$. Let us illustrate this at a simple example: consider the $C$-transformation of the Weinberg-type term $\fabc G_\mu^{a\nu} G_\nu^{b\rho} G_\rho^{c\mu}$, which according to Eq.~\eqref{eq:discrete_sym_QCD} has the eigenvalue $(-1)^3x_ax_bx_c$. Due to the color contraction with $\fabc$, either one or all signs $x_{a,b,c}$ must be negative, such that the operator is $C$-even.  In complete analogy, one can compute the discrete symmetries of arbitrary color contractions. 

We summarize the $C$, $P$, and $T$ transformations of various quark bilinears, gauge fields, and color structures in Table~\ref{tab:discrete_symmetries}, which comes in handy when searching for $C$- and $CP$-violating operators in LEFT.  

\begin{table}[t]
	\centering
	\renewcommand{\arraystretch}{1.5}
	\setlength{\tabcolsep}{5pt}
	\setlength\extrarowheight{2pt}
	\resizebox{\columnwidth}{!}{
    \begin{tabular}{lcccccccccccc}
		\toprule
		 & $\bar\psi\psi$ & $\bar\psi i \gamma_5\psi$ & $\bar\psi\gamma_\mu\psi$ & $\bar\psi\gamma_\mu\gamma_5\psi$ & $\bar\psi\sigma_\munu\psi$& $\bar\psi i \sigma_\munu\gamma_5\psi$ & $X_\mu$ & $X_\munu$ & 
		 $\fabc G_\mu^{a\nu} G_\nu^{b\rho} G_\rho^{c\mu}$ 
		 & 
		 $f_{abc}T^aG_\munu^{b} G^{c\munu}$ 
		 & $f_{abc}T^aT^{b} G^c_{\munu}$ \\
 		\midrule
        $C$\quad 	
        & $+$ & $+$ & $-$ & $+$ & $-$ & $-$ 
        & $-$ & $-$ & $+$ & $-$ & $+$ \\
        $P$	& $+$ & $-$ & $\varepsilon^\mu$ & $-\varepsilon^\mu$ & $\phantom{-}\varepsilon^\mu\varepsilon^\nu$ & $-\varepsilon^\mu\varepsilon^\nu$
        & $\varepsilon^\mu$ & $\phantom{-}\varepsilon^\mu\varepsilon^\nu$ & $+$ & $+$& $\varepsilon^\mu\varepsilon^\nu$\\
        $T$	& $+$ & $-$ & $\varepsilon^\mu$ & $\phantom{-}\varepsilon^\mu$ & $-\varepsilon^\mu\varepsilon^\nu$ & $\phantom{-}\varepsilon^\mu\varepsilon^\nu$ 
        & $\varepsilon^\mu$ & $-\varepsilon^\mu\varepsilon^\nu$ & $+$ & $-$ & $\varepsilon^\mu\varepsilon^\nu$\\
	\bottomrule
	\end{tabular}
	}
\renewcommand{\arraystretch}{1.0}
	\caption{Discrete space-time symmetries of quark bilinears with a single flavor and several gauge terms. In this simplified notation each ${SU}(3)$ color generator $T_a$ is thought to be part of one quark bilinear. We furthermore introduce $X_\mu\in\{A_\mu, T_aG^a_\mu\}$ and $X_\munu\in\{F_\munu, T_aG^a_\munu\}$. Replacing any field-strength tensor by its representation in dual space $\tilde X_\munu$, i.e., any contraction with the Levi-Civita symbol $\epsilon_{\alpha\beta\mu\nu}$, flips the signs of $P$ and $T$. An exchange of the structure constants $\fabc$ and $\dabc$ changes signs of $C$ and $T$.
	When multiplying each term with the imaginary unit $i$ the sign of $T$ flips, while the inclusion of $\protect\dvec{D}_\mu$ in a bilinear flips the signs of $C$ and $T$. If necessary, factors of $i$ are multiplied to render quark bilinears hermitian. 
	}
	\label{tab:discrete_symmetries}
\end{table}

%----------------------------------------------------
\boldmath
\subsection{$C$- and $CP$-odd operators in LEFT}
\label{sec:LEFT_summary}
\unboldmath
%----------------------------------------------------

In this section we list our most convenient choice of linearly independent ToPe operators up to and including mass dimension 8 in LEFT, based on the operator bases in Refs.~\cite{Jenkins:2017jig, Liao:2020zyx, Murphy:2020cly}. Details on the derivation of the full set of $C$- and $CP$-violating operators can be found in App.~\ref{app:LEFT}, which also takes care of operators that are not hermitian in the first place. 

At this point we shortly summarize the straightforward but, given the vast number of operators, quite tedious procedure. First, multiply each LEFT operator by a complex Wilson coefficient and add the hermitian conjugate. All LEFT bases under consideration are formulated in terms of left- and right-handed fermions, such that the resulting multilinears do not necessarily have definite eigenvalues under $C$, $P$, and $T$. To remedy this issue we decompose the chiral fields into their (pseudo)scalar, (axial)vector, and (pseudo)tensor contributions and compile their eigenvalues with the aid of Table~\ref{tab:discrete_symmetries} and similar relations for different color structures. Finally we have to identify chirality-conserving and -violating operators. If both appear at the same mass dimension of LEFT, we can ignore the latter, which is in some more detail discussed in Sect.~\ref{sec:ToPe_dim_8}.

While the identification of chirality-violating fermion bilinears is straightforward, special care has to be taken for quadrilinear quark operators. 
Some four-quark operators that are naively found to be chirality-breaking
may be mediated 
by  gauge-invariant BSM couplings of left-handed $W^\pm$ bosons
to right-handed currents, induced via mixing of the
$W^\pm$ bosons with their right-handed BSM counterparts. 
For instance, the dimension-6  $P$- and $T$-violating quadrilinear quark operator of Ng and Tulin~\cite{Ng:2011ui}
\begin{equation}
    i\frac{C_\text{NT}}{\Lambda^2} 
     \left\{(\bar u_R \gamma^\mu d_R)(\bar d_L \gamma_\mu  u_L) 
            -(\bar d_R \gamma^\mu u_R)(\bar u_L \gamma_\mu  d_L) \right\} 
     \label{eq:NgTulin}
\end{equation}
(with Wilson coefficient $C_\text{NT}$)
can be traced back to the following gauge-invariant, manifestly chirally symmetric dimension-6 operator  (resulting,
e.g., in the reduction of minimal left-right-symmetric models~\cite{Zhang:2007da,Deshpande:1990ip}---a subclass of left-right models~\cite{Pati:1974yy,Mohapatra:1974hk})
\begin{equation}
 \frac{C_\text{LR}}{ \Lambda^2} (i\tilde H^\dagger D_\mu H) (\bar u_R \gamma^\mu d_R)
 +\hc\,.
\end{equation}
Here  $H$ with $\tilde H \equiv i\tau_2 H^\ast$ is the Higgs doublet field, 
$D_\mu =\partial_\mu + i g W^a \tau^a$, $a=1,2$,
and $C_\text{LR}$ the pertinent Wilson coefficient.
This term yields, after electroweak symmetry breaking, in unitary gauge
\begin{equation}
  -\frac{g v^2}{2\sqrt{2}} \left[\frac{C_\text{LR}}{\Lambda^2} \bar u_R \gamma^\mu d_R W_\mu^\dagger+\hc \right] \left(1 + \frac{h}{v} \right)^2 \,,
\end{equation}
with $h$ the lightest Higgs boson of the model, which corresponds to the physical
Higgs boson, and $v$ its vacuum expectation value. 
After integrating out the Higgs  and $W^\pm$ bosons, we obtain, just below the $W^\pm$ mass, 
the four-quark operator (\ref{eq:NgTulin}) to leading order.
Note that the  Higgs vev $v$ cancels against the mass of the $W^\pm$ bosons,
as the coupling to the right-handed current involved the Higgs kinetic term and  is therefore not
of Yukawa nature.
More details can be found in Refs.~\cite{Xu:2009nt,An:2009zh}, see also
Ref.~\cite{Dekens:2014jka}.

In accordance  with this argument, the chirality-conserving operators can also be easily identified by the fact that they must appear at the same order in both LEFT and SMEFT, if a consistent operator basis is used.

To shorten the notation in the following subsections, we omit the ratio of scales $v^4/\Lambda^4$, which according to Eq.~\eqref{eq:LEFT_suppression} is common to all our operators. Regarding Eq.~\eqref{eq:LEFT_expansion}, the suppression in terms of these heavy scales can explicitly be restored by multiplying each LEFT operator of dimension 7 by $v/\Lambda^4$ and the ones of dimension 8 by $1/\Lambda^4$.

%----------------------------------
\subsubsection{Dimension-7 operators}
\label{sec:ToPe_dim_7}
%----------------------------------
We show in App.~\ref{app:LEFT} that there are indeed no ToPe operators below dimension 7 in LEFT as already implicitly claimed decades ago~\cite{Khriplovich:1990ef}, but which was---to the best of our knowledge---not proven explicitly in the literature.\footnote{To consistently construct a $C$- and $CP$-odd operator from lower-dimensional ones, one could for instance include a $T$- and $P$-odd operator of dimension 6 of SMEFT in a $C$-violating electroweak loop. However, integrating out the weak gauge boson with mass dimension 1, i.e., replacing it by the quark current of dimension 3, effectively leads to a dimension-8 operator in LEFT. Due to the completeness of the LEFT operator bases used in this work, these contributions are automatically taken care of.} 
At dimension 7 we can confirm that exactly the operators already quoted in Eq.~\eqref{eq:ToPe_historic} contribute (except the one including the $Z$-boson, which obviously does not belong to LEFT), i.e.,
\begin{aligneq}\label{eq:ToPe_LEFT_dim7}
    & \O_{\psi\chi}^{(a)} &&\equiv&& \ c_{\psi\chi}^{(a)}\, \bar\psi \dvec D_\mu \gamma_5\psi\bar\chi\gamma^\mu\gamma_5\chi\,, \\[0.1cm]
    & \O_{\psi}^{(b)} &&\equiv&& \ c_{\psi}^{(b)}\, \bar\psi T^a \sigma^{\mu\nu}\psi F_{\mu\rho} G^{a\rho}_{\nu}\,,
\end{aligneq}
where $c_{\psi\chi}^{(a)}$ and $c_{\psi}^{(b)}$ denote \textit{real}-valued Wilson coefficients with flavor indices $\psi,\chi$ combined with superscripts $(a),(b)$ serving as labels to classify operators unambiguously. The quadrilinear in this equation can in principle appear with two different color contractions.\footnote{
We note that according to Refs.~\cite{Dekens:2013zca,Dekens:2014jka}, the four-quark operator in Eq.~\eqref{eq:ToPe_LEFT_dim7} should be valid just below the $W^\pm$ threshold. This means that one expects QCD corrections when running down to a scale $\mu$, $1\,\text{GeV} \ll \mu \ll M_{W\pm}$, such that  
the quadrilinear in Eq.~\eqref{eq:ToPe_LEFT_dim7} would mix with its corresponding different color contraction. These corrections are beyond the scope of our analysis and do not have any effect once the LEFT basis is matched onto $\chi$PT.
}
In addition, we find the (semi-leptonic) operators containing quarks and charged leptons 
\begin{aligneq}\label{eq:ToPe_LEFT_dim7_semi_leptonic}
    &\O_{\ell\psi}^{(c)} &&\equiv&& \ c_{\ell\psi}^{(c)}\, \bar\ell \dvec D_\mu \gamma_5\ell \bar\psi\gamma^\mu\gamma_5\psi\,,\\[0.1cm]
    &\O_{\ell\psi}^{(d)} &&\equiv&& \ c_{\ell\psi}^{(d)}\, \bar\ell\gamma^\mu\gamma_5\ell\bar\psi \dvec D_\mu \gamma_5\psi\,.
\end{aligneq}

%----------------------------------
\subsubsection{Dimension-8 operators}
\label{sec:ToPe_dim_8}
%---------------------------------- 
As all terms in Sect.~\ref{sec:ToPe_dim_7} are chirality-breaking, we can a priori not neglect chirality-conserving operators at dimension 8 in LEFT. Investigating the latter, we do not find any ToPe operators for pure gauge terms or terms including four fermions and two derivatives. For quark multilinears coupling to gluon field-strength tensors (henceforth called gluonic operators)  we identify
\begin{aligneq}\label{eq:ToPe_gluonic}
    &\O_{\psi}^{(e)} &&\equiv&& \ c_{\psi}^{(e)}\, \fabc \bar\psi\gamma^\mu i\dvec D^\nu T^a\psi G_{\mu\rho}^b G^{c\,\rho}_\nu \,,\\[0.1cm]
    & \O_{\psi\chi}^{(f)} &&\equiv&& \ c_{\psi\chi}^{(f)}\, \bar \psi \gamma^\mu \psi \bar \chi \gamma^\nu T^a \chi G^a_\munu \,,\\[0.1cm] 
    & \O_{\psi\chi}^{(g)} &&\equiv&& \ c_{\psi\chi}^{(g)}\, \bar \psi \gamma^\mu \gamma_5 \psi \bar \chi \gamma^\nu \gamma_5 T^a \chi G^a_\munu \,,\\[0.1cm]
    & \O_{\psi\chi}^{(h)} &&\equiv&& \ c_{\psi\chi}^{(h)}\, \fabc\bar \psi \gamma^\mu \gamma_5 T^a\psi \bar \chi \gamma^\nu T^b \chi \tilde G^c_\munu \,,\\[0.1cm]
    &\O_{\psi\chi}^{(i)} &&\equiv&& \ c_{\psi\chi}^{(i)}\, \dabc\bar \psi \gamma^\mu T^a\psi \bar \chi \gamma^\nu T^b \chi G^c_\munu\,,\\[0.1cm]
    &\O_{\psi\chi}^{(j)} &&\equiv&& \ c_{\psi\chi}^{(j)}\, \dabc\bar \psi \gamma^\mu \gamma_5 T^a\psi \bar \chi \gamma^\nu \gamma_5 T^b \chi G^c_\munu\,,\\[0.1cm]
    &\O_{\psi\chi}^{(k)} &&\equiv&& \ c_{\psi\chi}^{(k)}\, i\big[\bar\psi T^a \chi \bar\chi \sigma^\munu \psi + \bar\psi \gamma_5 T^a \chi \bar\chi \sigma^\munu \gamma_5 \psi -(\psi\leftrightarrow\chi)\big] G^a_{\mu\nu}\,,\\[0.1cm]
    &\O_{\psi\chi}^{(l)} &&\equiv&& \ c_{\psi\chi}^{(l)}\, i\big[\bar\psi \chi \bar\chi \sigma^\munu T^a \psi + \bar\psi \gamma_5 \chi \bar\chi \sigma^\munu \gamma_5 T^a \psi -(\psi\leftrightarrow\chi)\big] G^a_{\mu\nu}\,,\\[0.1cm]
    &\O_{\psi\chi}^{(m)} &&\equiv&& \ c_{\psi\chi}^{(m)}\, \big[\bar\psi  \sigma^{\lambda \mu} T^a \chi \bar\chi \sigma_\munu \psi + \bar\psi  \sigma^{\lambda \mu} \gamma_5 T^a \chi \bar\chi \sigma_\munu \gamma_5 \psi +(\psi\leftrightarrow\chi)\big] G^{a\, \nu}_{\lambda}\,,
\end{aligneq}
as $C$- and $CP$-odd. Among these operators, $\O_{\psi\chi}^{(i)},\O_{\psi\chi}^{(j)},\O_{\psi\chi}^{(k)},\O_{\psi\chi}^{(l)}$ are antisymmetric under flavor interchange, while $\O_{\psi\chi}^{(m)}$ is symmetric.

Similar to the operators in Eq.~\eqref{eq:ToPe_gluonic} we find quark quadrilinears including photon field-strength tensors (photonic operators), i.e.,
\begin{aligneq}\label{eq:ToPe_photonic}
    & \O_{\psi\chi}^{(n)} &&\equiv&& \ c_{\psi\chi}^{(n)}\, \bar \psi \gamma^\mu \psi \bar \chi \gamma^\nu \chi F_\munu\,,\\[0.1cm]
    &\O_{\psi\chi}^{(o)} &&\equiv&& \ c_{\psi\chi}^{(o)}\, \bar \psi \gamma^\mu \gamma_5 \psi \bar \chi \gamma^\nu \gamma_5 \chi F_\munu\,,\\[0.1cm]
    & \O_{\psi\chi}^{(p)} &&\equiv&& \ c_{\psi\chi}^{(p)}\, \bar \psi \gamma^\mu T^a \psi \bar \chi \gamma^\nu T^a \chi F_\munu\,,\\[0.1cm]
    & \O_{\psi\chi}^{(q)} &&\equiv&& \ c_{\psi\chi}^{(q)}\, \bar \psi \gamma^\mu \gamma_5 T^a \psi \bar \chi \gamma^\nu \gamma_5 T^a \chi F_\munu\,,\\[0.1cm]
    &\O_{\psi\chi}^{(r)} &&\equiv&& \ c_{\psi\chi}^{(r)}\, i\big[\bar\psi \chi \bar\chi \sigma^\munu \psi + \bar\psi \gamma_5\chi \bar\chi \sigma^\munu \gamma_5\psi -(\psi\leftrightarrow\chi)\big] F_{\mu\nu}\,,
\end{aligneq}
which, due to fewer possible color contractions, allow for fewer $C$- and $CP$-violating contributions than the gluonic operators. Each of these photonic operators is completely antisymmetric under interchange of quark flavors, up to the unknown Wilson coefficients. Therefore the operators $\O_{\chi\psi}^{(z)}$, with $z\in\{n,o,p,q,r\}$, can in principle always be absorbed by $\O_{\psi\chi}^{(z)}$ with an appropriate redefinition of the Wilson coefficients, leaving us with three independent flavor combinations to consider for the off-diagonal contributions, e.g., $\O_{ud}^{(z)}, \O_{us}^{(z)}$, and $\O_{ds}^{(z)}$. 
Note that the diagonal elements $\O_{uu}^{(z)}, \O_{dd}^{(z)}$, and $\O_{ss}^{(z)}$ vanish for all operators in Eq.~\eqref{eq:ToPe_photonic}.

There are only two ToPe operators in dimension 8 LEFT that contain quark bilinears, photons, and gluon field-strength tensors (photo-gluonic operators), which explicitly read
\begin{aligneq}\label{eq:ToPe_photo_gluonic}
    &\O_{\psi}^{(s)} &&\equiv&& \ c_{\psi}^{(s)}\, \bar\psi\gamma^\mu i\dvec D^\nu T^a \gamma_5\psi F_{\mu\rho}\tilde G^{a\,\rho}_\nu\,,\\[0.1cm]
    &\O_{\psi}^{(t)} &&\equiv&& \ c_{\psi}^{(t)}\, \bar\psi\gamma^\mu i\dvec D^\nu T^a \gamma_5\psi F_{\nu\rho}\tilde G^{a\,\rho}_\mu\,.
\end{aligneq}

Finally we quote our findings for semi-leptonic operators. The inclusion of leptonic bilinears reduces the amount of possible color contractions and hence the number of contributing ToPe operators enormously. Our results for gluonic operators at dimension 8 in LEFT read
\begin{aligneq}\label{eq:ToPe_LEFT_dim8_semi_leptonic_gluonic}
    &\O_{\ell\psi}^{(u)} &&\equiv&& \ c_{\ell\psi}^{(u)}\, \bar \ell \gamma^\mu \ell \bar \psi \gamma^\nu T^a \psi G^a_\munu\,,\\[0.1cm]
    &\O_{\ell\psi}^{(v)} &&\equiv&& \ c_{\ell\psi}^{(v)}\, \bar \ell \gamma^\mu \gamma_5 \ell \bar \psi \gamma^\nu \gamma_5 T^a \psi G^a_\munu\,,
\end{aligneq}
while the ones for photonic terms are
\begin{aligneq}\label{eq:ToPe_LEFT_dim8_semi_leptonic_photonic}
    &\O_{\ell\psi}^{(w)} &&\equiv&& \ c_{\ell\psi}^{(w)}\, \bar \ell \gamma^\mu \ell \bar \psi \gamma^\nu \psi F_\munu\,,\\[0.1cm]
    &\O_{\ell\psi}^{(x)} &&\equiv&& \ c_{\ell\psi}^{(x)}\, \bar \ell \gamma^\mu \gamma_5 \ell \bar \psi \gamma^\nu \gamma_5 \psi F_\munu\,.
\end{aligneq}

We list $C$- and $CP$-odd chirality-breaking quark quadrilinears in dimension 8 of LEFT, which do not gain any further consideration, in App.~\ref{app:ToPe_dim8_summary}. These can surely be neglected because, other than all the operators listed above, they do \textit{not} arise from dimension 8 in SMEFT and thus originate from higher-dimensional operators in the SMEFT power counting, which implies a corresponding suppression due to additional inverse powers of the BSM scale $\Lambda$.

%---------------------------------------------------------------
\boldmath
\section{Construction of effective $C$- and $CP$-violating chiral Lagrangians}
\label{sec:Matching}
\unboldmath
%---------------------------------------------------------------

In the following we summarize and extend the principles of matching between LEFT and $\chi$PT operators to obtain a model-independent effective $SU(3)$ theory for $C$ and $CP$ violation in flavor-conserving light-meson interactions, originating from the complete list of $C$- and $CP$-odd operators worked out in the previous chapter. We refer to this theory as $T$-odd, $P$-even chiral perturbation theory (ToPe$\chi$PT). In particular, we work out the non-trivial matching of quark multilinears with derivative character and couplings to gluons, which is in this sense not included in the current literature, while introducing the formalism in detail. 
Whenever possible, we restrict the matching to the leading order in the chiral power counting and $C$ and $CP$ violation.

%-----------------------------------------------------------------
\subsection{Chiral perturbation theory: notation and conventions}
\label{sec:Standard_ChPT}
%-----------------------------------------------------------------

According to Gasser and Leutwyler~\cite{Gasser:1983yg,Gasser:1984gg}, the massless QCD Lagrangian $\L^0_\text{QCD}$ can be extended by introducing external sources to obtain the most general non-kinetic quark operators, by means of
\beq\label{eq:Lagrange_Source_Fields}
	\L = \L^0_\text{QCD} + \bar q_L \gamma^\mu l_\mu q_L + \bar q_R \gamma^\mu r_\mu q_R - \bar q_R s q_L - \bar q_L s^\dagger q_R  
    + \bar q_L \sigma^{\mu\nu} t_{\mu\nu} q_R  + \bar q_R \sigma^{\mu\nu} t^\dagger_{\mu\nu} q_L   \,,
\eeq
with the light-quark triplet $q=(u,d,s)^T$, and external sources $r_\mu=r_\mu^\dagger$, $l_\mu=l_\mu^\dagger$, $s$, and $t_\munu$, which are three-dimensional quadratic matrices in flavor space. The tensor source $t_\munu$ was first introduced in Ref.~\cite{Cata:2007ns}. 
The spontaneous breakdown of the $SU(3)_L\times SU(3)_R\times U(1)_V$ global and continuous gauge group of this theory results in an $SU(3)_V\times U(1)_V$ symmetry, thus generating eight Goldstone bosons $\phi_a$ as the relevant degrees of freedom.

The chiral theory (for reviews, see, e.g., Refs.~\cite{Scherer:2002tk,Scherer:2012xha,Meissner:2022odx}), which exhibits a certain power counting in terms of soft momenta and light quark masses, can then be described by the unitary matrix $U$ defined as
\beq\label{eq:Goldstone}
U=\exp\left(\frac{i\Phi}{\Fpi}\right) \eqwith  \Phi\equiv \lambda_a\phi^a= 
\begin{pmatrix}
\pi^0 +\frac{1}{\sqrt{3}}\eta_8  & \sqrt{2}\pi^+  &\sqrt{2}K^+   \\
\sqrt{2}\pi^-& -\pi^0+\frac{1}{\sqrt{3}}\eta_8  & \sqrt{2}K^0 \\
\sqrt{2}K^- & \sqrt{2}\bar K^0& -\frac{2}{\sqrt{3}}\eta_8
\end{pmatrix}\,,
\eeq
where $F_0 \lesssim F_\pi=92.2\MeV$~\cite{Workman:2022ynf} is the pion decay constant in the chiral limit and $\eta_8$ the octet part of the $\eta$ mesons. The matrix $\chi=2B_0s$ includes the scalar source and a low-energy coefficient $B_0$, and the field-strength tensors are given as
\beq\label{eq:ChPT_field_strength}
 f_R^\munu= \partial^\mu r^\nu - \partial^\nu r^\mu- i[r^\mu,r^\nu]\,,
 \qquad 
 f_L^\munu= \partial^\mu l^\nu - \partial^\nu l^\mu- i[l^\mu,l^\nu]\,.
\eeq
The dynamics of the Goldstone bosons is driven by the gauge covariant derivative acting on $U$ and $U^\dagger$ defined as\footnote{Confusion with the LEFT covariant derivative, which includes gluons, should be avoided by the context and the fact we only use the LEFT derivative $\vec{D}_\mu$ in vector notation.}
\beq
    D_\mu U \equiv \partial_\mu U -ir_\mu U+iUl_\mu \,, \qquad D_\mu U^\dagger \equiv \partial_\mu U^\dagger +iU^\dagger r_\mu -il_\mu U^\dagger\,,
\eeq
which is necessary to ensure invariance under local extension of the global gauge transformations.\footnote{A \textit{local} chiral symmetry is required to ensure that proper chiral Ward identities hold, cf.\ Ref.~\cite{Leutwyler:1993iq}.}
In particular, note that the product rule applies to these derivatives~\cite{Fearing:1994ga}, thereby---together with the unitarity of $U$---inducing the important identity $D_\mu U U^\dagger =-U D_\mu U^\dagger$. To keep the notation as simple as possible, we use the convention that the covariant derivative only acts on the object immediately to its right, by means of $D_\mu U U^\dagger\equiv(D_\mu U) U^\dagger$ and $D_\mu D_\nu U U^\dagger\equiv(D_\mu D_\nu U) U^\dagger$. We remark that the covariant derivative may in principle also act on any combination of chiral building blocks that transforms in the same manner as $U$ or $U^\dagger$, respectively, for instance on $Uf_L^\munu$ or $U^\dagger f_R^\munu$.

Our fundamental building blocks, ordered according to their power counting in soft momenta, transform under $SU(3)_L\times SU(3)_R$ group actions as
\beq\label{eq:chiral_building_blocks}
\begin{alignedat}{10}
&\O(p^0):\qquad\qquad&&U              &&\to R U L^\dagger\,,   \qquad\qquad&& U^\dagger            &&\to L U^\dagger R^\dagger\,,\\[0.1cm]
&\O(p^1):\qquad&&D_\mu U   &&\to R D_\mu U L^\dagger\,,       && D_\mu U^\dagger &&\to L D_\mu U^\dagger R^\dagger\,,\\[0.1cm]
&\O(p^2):\qquad&& \chi          &&\to R\chi L^\dagger\,,                && \chi^\dagger         &&\to L\chi^\dagger R^\dagger \,, \\[0.1cm]
&\O(p^2):\qquad&& f_R^\munu          &&\to R f_R^\munu R^\dagger \,,    && f_L^\munu          &&\to L f_L^\munu L^\dagger  \,, 
\end{alignedat}
\eeq
where $L\in SU(3)_L$, $R\in SU(3)_R$.
Any mesonic operator in Standard Model $\chi$PT (SM$\chi$PT) can be built by coupling these building blocks in all Lorentz covariant ways that respect the conservation of discrete symmetries $C$, $P$, and $T$, and the invariance under $SU(3)_L\times SU(3)_R$ group actions. The latter condition demands the inclusion of traces in flavor space, which we indicate as $\vev{\ldots}$. The lowest-order SM$\chi$PT Lagrangian thus yields
\beq
\L^{(2)}_\chi= \frac{\Fpi^2}{4} \vev{D_\mu U D^\mu U^\dagger} + \frac{\Fpi^2}{4} \vev{\chi U^\dagger+\chi^\dagger U}\,.
\eeq
To access mesonic interactions encoded in this Lagrangian, the matrix $U$ can be expanded in a simple Taylor series according to
\beq\label{eq:U_expanded}
    U=\mathds{1}+\frac{i}{\Fpi}\Phi-\frac{1}{{2\Fpi^2}}\Phi^2-\frac{i}{{6\Fpi^3}}\Phi^3+\ldots\,.
\eeq
Once the chiral Lagrangian has been built, the external sources can be fixed to their physical values, i.e., 
\beq
	s \mapsto  M_q \, , \quad
	r_\mu \mapsto  -e Q A_\mu \, , \quad
	l_\mu \mapsto  -e Q A_\mu \, , \quad
	t_{\mu\nu} \mapsto 0 \,,
\eeq
with the matrices $M_q=\text{diag}(m_u,m_d,m_s)$  and $Q=\text{diag}(2/3,-1/3,-1/3)$.
Finally, we quote the equation of motion to leading order~\cite{Gasser:1984gg,Scherer:2002tk,Scherer:2012xha,Meissner:2022odx}
\beq\label{eq:ChPT_EoM}
    D^2 U U^\dagger - U D^2 U^\dagger - \chi U^\dagger + U \chi^\dagger +\frac{1}{3} \vev{\chi U^\dagger- U \chi^\dagger}=0\,,
\eeq
with $D^2=D_\mu D^\mu$, which proves useful to remove redundancies.

%---------------------------------------
\subsection{Discrete space-time symmetries}
\label{sec:discrete_symm_ChPT}
%---------------------------------------
Similar to Sect.~\ref{sec:discrete_symm_LEFT}, we now discuss the transformation properties of the fundamental chiral building blocks under discrete space-time symmetries, which can be derived from those of the underlying quark currents and densities. 

The discrete symmetries of the mesons matrix $\Phi$ and the external sources $r_\mu$ and $l_\mu$ are similar to the ones of the hermitian pseudoscalar quark density $i\bar q \gamma_5 q$ and the quark currents $\bar q_R \gamma_\mu q_R$ and $\bar q_L \gamma_\mu q_L$, respectively. This leads, upon suppressing the explicit dependencies on the space-time coordinates, to
\begin{aligneq}
&\Phi &&\xleftrightarrow{\ C\ } \phantom{-}\Phi^T\,, &&\qquad
\Phi &&\xleftrightarrow{\ P\ }  -\Phi\,, &&\qquad
\Phi &&\xleftrightarrow{\ T\ }  -\Phi\,, \\[0.1cm]
&r_\mu&&\xleftrightarrow{\  C \ } -l^T_\mu\,, &&\qquad
r_\mu&&\xleftrightarrow{\ P\ }  \phantom{-}l^\mu\,, &&\qquad
r_\mu&&\xleftrightarrow{\ T\ }  \phantom{-}r^\mu\,.
\end{aligneq}
To proceed, consider that $T$ is an anti-unitary operator, such that $T:i\xrightarrow{\phantom{T}}-i$, and that the derivative transforms as $T:\partial_\mu\xrightarrow{\phantom{T}}-\partial^\mu$. Hence we can conclude from the defining equations of $U$ and $f_{L/R}^\munu$, as well as the fact that $\chi$ has to have the same discrete symmetries as $U$, the following transformation properties of our building blocks:\footnote{Literature about the $T$-transformation of the chiral building blocks is scarce, cf.\ Refs.~\cite{Cronin:1967jq,Kambor:1989tz}. Unfortunately, Ref.~\cite{Kambor:1989tz} adapted the $T$ transformation from Ref.~\cite{Cronin:1967jq} erroneously, by choosing $\Phi\xleftrightarrow{\ T\ }\Phi$, which would imply that [according to Eq.~\eqref{eq:U_expanded}] $U$ is no eigenstate of $T$. Moreover, Ref.~\cite{Kambor:1989tz} quotes the wrong time reversal of $r_\mu$, $l_\mu$, and $f_{L/R}^\munu$, which should transform under $T$ like the physical photon and the photonic field-strength tensor, respectively.}
\begin{aligneq}\label{eq:ChPT_discrete_symmetries}
&U &&\xleftrightarrow{\ C\ } U^T\,, &&\qquad
U &&\xleftrightarrow{\ P\ }  U^\dagger\,, &&\qquad
U &&\xleftrightarrow{\ T\ }  U\,, \\[0.1cm]
&D_\mu U &&\xleftrightarrow{\ C\ }  D_\mu U^T\,, &&\qquad
D_\mu U &&\xleftrightarrow{\ P\ }  D^\mu U^\dagger\,, &&\qquad
D_\mu U &&\xleftrightarrow{\ T\ }  -D^\mu U\,, \\[0.1cm]
&\chi &&\xleftrightarrow{\ C\ } \chi^T\,, &&\qquad
\chi &&\xleftrightarrow{\ P\ }  \chi^\dagger\,, &&\qquad
\chi &&\xleftrightarrow{\ T\ }  \chi\,, \\[0.1cm]
&f^\munu_{L/R} &&\xleftrightarrow{\ C\ } -\big(f^\munu_{R/L}\big)^T\,, &&\qquad
f^\munu_{L/R} &&\xleftrightarrow{\ P\ }  f_\munu^{R/L}\,, &&\qquad
f^\munu_{L/R} &&\xleftrightarrow{\ T\ }  -f_\munu^{L/R}\,.
\end{aligneq}
Here and in the following we use the definitions $D_\mu U^T\equiv (D_\mu U)^T$ and $D_\mu U^\ast\equiv (D_\mu U)^\ast$, respectively.
Regarding products of chiral building blocks, these transformations apply to each matrix separately, while the algebraic properties of the trace play a vital role. For convenience we explicitly illustrate the transformation under discrete symmetries at the simple example of $\L_\chi^{(2)}$. Under charge conjugation the Lagrangian transforms as 
\begin{aligneq}
\L^{(2)}_\chi \xrightarrow{\ C \ } & \phantom{=}\ \, \frac{\Fpi^2}{4} \vev{D_\mu U^T D^\mu U^\ast} + \frac{\Fpi^2}{4} \vev{\chi^T U^\ast+\chi^\ast U^T}\\ 
& = \frac{\Fpi^2}{4} \vev{D^\mu U^\dagger D_\mu U} + \frac{\Fpi^2}{4} \vev{U^\dagger \chi+ U \chi^\dagger}\\
& = \frac{\Fpi^2}{4} \vev{D_\mu U D^\mu U^\dagger } + \frac{\Fpi^2}{4} \vev{\chi U^\dagger+\chi^\dagger U} \,,
\end{aligneq}
and is thus invariant. In the first equality we used that traces are invariant under matrix transposition, leading to the observation that for any operator consisting of the building blocks from Eq.~\eqref{eq:ChPT_discrete_symmetries} $C$ merely reverses the order of the matrices (and flips sign and handedness of $f^\munu_{L/R}$), whereas in the second equality we applied the cyclic property of the trace.
Analogously, cyclicity renders $\L_\chi^{(2)}$ parity-invariant, while its $T$ transformation is trivial. At this point, note that the terms $\chi U^\dagger$ and $\chi^\dagger U$ are summed without a relative factor to ensure that we have an eigenstate of $P$. We will make ample use of this observation in the following sections.

%--------------------------------------------------------------
\boldmath
\subsection{Matching LEFT and $\chi$PT: building the chiral basis}
\label{sec:principles_of_matching}
\unboldmath
%--------------------------------------------------------------

Having provided the  
fundamental building blocks as well as their transformation under $SU(3)_L\times SU(3)_R$ group actions and discrete space-time symmetries $C$, $P$, and $T$, we may now match the $C$- and $CP$-violating LEFT operators from Sect.~\ref{sec:LEFT_summary} onto $\chi$PT.

For this endeavour, we begin by regarding our LEFT operators as additional external sources, cf.\ Eq.~\eqref{eq:Lagrange_Source_Fields}.
These sources can be written as general chiral irreducible representations, which for an arbitrary quark multilinear consisting of $n$ bilinears takes the form
\beq \label{eq:chiral_irrep_general}
\O= T_{a_1b_1\ldots a_nb_n} (\bar q_{X_1} \Delta_1\Gamma_1 \hat \lambda^1_{a_1b_1} q_{Y_1} )\ldots(\bar q_{X_n} \Delta_n\Gamma_n \hat \lambda^n_{a_nb_n} q_{Y_n})\,,
\eeq
with $\hat \lambda^i_{a_ib_i}$ as $3\times3$ matrices (\textit{not} single matrix elements) projecting out the flavor $a_i, b_i\in\{u,d,s\}$ of each quark bilinear, 
chiralities $X_i,Y_i\in\{L,R\}$, any combination of Dirac matrices $\Gamma_i$, arbitrary operators $\Delta_i$ that leave the chiral structure invariant (these may include derivatives acting on quark fields, leptonic terms, as well as photonic or gluonic field-strength tensors), and a coefficient tensor $T\equiv T_{a_1b_1\ldots a_nb_n}\hat\lambda^1_{a_1b_1}\ldots\hat\lambda^n_{a_nb_n}$, which depends on the quark flavor and includes the Wilson coefficients of the respective LEFT operators.  
Upon treating the coefficients~$T$ of the external sources as spurions with well-defined transformation properties under $SU(3)_L\times SU(3)_R$ group actions, we can render the operators in Eq.~\eqref{eq:chiral_irrep_general}, in which the quarks triplets transform as
\beq
    q_L \to L\, q_L\,,\qquad \bar q_L \to  \bar q_L L^\dagger\,,\qquad 
    q_R \to R\, q_R\,,\qquad \bar q_R \to  \bar q_R R^\dagger\,,
\eeq
chirally invariant.\footnote{The matrices $L$ and $R$ should not be mixed up with the chiral projection operators $P_{L,R}=(\mathds{1}\pm\gamma_5)/2$.} This procedure is completely analogous to the inclusion of quark masses using the building block $\chi$.

The only terms in our LEFT operators, cf.\ Eq.~\eqref{eq:chiral_irrep_general}, that survive (at lowest order in the QED coupling) the transition from energies above the chiral scale $\Lambda_\chi$ to the ones below it, are photonic field-strength tensors and leptonic bilinears encoded in the $\Delta_i$ and the spurion $T$. All other quantites are either too heavy (already accounted for in LEFT, like $W$- and $Z$-bosons) or no observable degrees of freedom due to color confinement, as quarks---therefore also derivatives acting on them---and gluons. However, although the latter do not appear as observable quantities in the effective theory, we still have to account for the information on their discrete symmetries and Lorentz structure when constructing the effective theory.
For the explicit mapping of quark-level operators to the mesonic level we proceed as follows.
\begin{itemize}
    \item First, rewrite each LEFT operator in terms of chiral irreps, cf.\ Eq.~\eqref{eq:chiral_irrep_general}, and identify the spurions and their transformation properties under chiral group actions and $C$, $P$, and $T$. 
    \item Next, attach chiral building blocks to the spurions, respect the initial Lorentz structure of quark--gluon terms (which, at leading order, only includes the contraction with the metric tensor $g_\munu$, but at higher orders also with the Levi-Civita symbol $\epsilon_{\alpha\beta\mu\nu}$), and contract flavor indices to form chirally invariant traces in all possible ways at the lowest possible order in chiral power counting, cf.\ Eq.~\eqref{eq:chiral_building_blocks}. This also includes the product of invariant flavor traces. 
    \item Finally ensure hermiticity and the appropriate discrete symmetries by constraining respective coupling constants (multiplied, if needed, by a factor of $i$) to be equal up to a sign. These symmetries encode the remaining information from gluons, quark bilinears, and their derivatives that were all in some sense integrated out. 
    \item In order to establish operators at higher chiral orders, one may repeat the above procedure with further insertions of $D_\mu$, $\chi^{(\dagger)}$, and $f_{R,L}^\munu$ to build chiral invariants or multiply other chirally invariant traces to the operators obtained at lower orders. In either way, one has to ensure throughout that hermiticity and proper transformations under the discrete symmetries are respected. In principle, higher-order operators can also arise from products of the spurion $T$~\cite{Graesser:2016bpz} or loops of lower-order operators. We restrict the analysis to linear effects in the already strongly suppressed ToPe forces and only work at tree level. 
    \item Make sure to constantly get rid of redundancies by identifying independent and non-vanishing operators.
\end{itemize}
We furthermore remark that there is no one-to-one correspondence between quark operators with those at the mesonic level and that, as usual when building chiral theories to higher orders, there is no way to know the number of possible operators a priori. One still has to keep in mind that, after building the chiral bases as described above for each LEFT operator, there still remains the question how these operators can be distinguished in experiment, if this is possible at all.

In the following we will denote the mesonic counterpart of any LEFT operator $\O^{(z)}_{\psi\chi}$ by~$X^{(z)}_{\psi\chi}$.  

%--------------------------------------------------------------
\subsection{Matching dimension-7 LEFT operators}
\label{sec:Matching_LEFT_dim_7}
%--------------------------------------------------------------
This section is devoted to providing a detailed discussion of the $\chi$PT Lagrangian arising from the $C$- and $CP$-odd dimension-7 LEFT quark operators listed in Eq.~\eqref{eq:ToPe_LEFT_dim7} and the semi-leptonic ones from Eq.~\eqref{eq:ToPe_LEFT_dim7_semi_leptonic}.  

%--------------------------------------------
\subsubsection{The quark quadrilinear operator}
%--------------------------------------------
First we investigate the operator
\beq
    \O_{\psi\chi}^{(a)} = \ c_{\psi\chi}^{(a)}\, \bar\psi \dvec D_\mu \gamma_5\psi\bar\chi\gamma^\mu\gamma_5\chi\,.
\eeq
In terms of chiral irreps, this operator can analytically be rewritten as  
\begin{aligneq}\label{eq:chiral_irrep_ToPe_dim7_quadrilinear}
    \O_{\psi\chi}^{(a)} = c_{\psi\chi}^{(a)}\big[&
      (\bar q_L \dvec{D}_\mu \lambda^\dagger q_R )(\bar q_R \gamma^\mu \lambda_R q_R)
         -(\bar q_R \dvec{D}_\mu \lambda q_L ) (\bar q_R \gamma^\mu \lambda_R q_R) 
         \\[0.1cm]
     &+(\bar q_R \dvec{D}_\mu \lambda q_L ) (\bar q_L \gamma^\mu \lambda_L q_L)
         -(\bar q_L \dvec{D}_\mu \lambda^\dagger q_R ) (\bar q_L \gamma^\mu \lambda_L q_L)\big]\,,
\end{aligneq}
where, compared to Eq.~\eqref{eq:chiral_irrep_general}, we use the abbreviations  $\hat{\lambda}^{(\dagger)}_{\psi\psi}\equiv\lambda^{(\dagger)}$, $\hat\lambda^{\chi\chi}_{L/R}\equiv\lambda_{L/R}$ and hence keep the dependence on the quark flavor implicit.
For convenience we have chosen a notation such that the spurions---which do in our case not contribute to the chiral power counting---transform analogously to the building blocks from Eq.~\eqref{eq:chiral_building_blocks}, i.e., 
\begin{aligneq}\label{eq:chiral_trafo_spurions}
    &\O(p^0): \quad&&\lambda             &&\to R \lambda L^\dagger\,,   \qquad\qquad&& \lambda^\dagger            &&\to L \lambda^\dagger R^\dagger\,,\\[0.1cm]
    &\O(p^0): && \lambda_R          &&\to R \lambda_R R^\dagger \,,    && \lambda_L          &&\to L \lambda_L L^\dagger  \,,
\end{aligneq}
such that $\O_{\psi\chi}^{(a)}$ is a chiral invariant. In fact, $\lambda$ transforms analogously to $U$ or $\chi$, while $\lambda_R$ transforms, e.g., as $U\lambda^\dagger$ (or as $\lambda U^\dagger$) and
$\lambda_L$ as $\lambda^\dagger U$ (or as $U^\dagger \lambda$).

The discrete symmetries of the $\lambda$ yield
\begin{aligneq}\label{eq:discrete_symmetries_spurions}
&\lambda &&\xleftrightarrow{\ C\ } \lambda^T\,, &&\qquad
\lambda &&\xleftrightarrow{\ P\ }  \lambda^\dagger\,, &&\qquad
\lambda &&\xleftrightarrow{\ T\ }  \lambda\,, \\[0.1cm]
&\lambda_R &&\xleftrightarrow{\ C\ } \lambda_L^T\,, &&\qquad
\lambda_R &&\xleftrightarrow{\ P\ }  \lambda_L\,, &&\qquad
\lambda_R &&\xleftrightarrow{\ T\ }  \lambda_R\,, \\[0.1cm]
&\lambda_L &&\xleftrightarrow{\ C\ } \lambda_R^T\,, &&\qquad
\lambda_L &&\xleftrightarrow{\ P\ }  \lambda_R\,, &&\qquad
\lambda_L &&\xleftrightarrow{\ T\ }  \lambda_L\,.
\end{aligneq}

Once the chiral operator basis is established, each spurion can be set to its physical value, i.e., the respective $3\times 3$ matrix projecting out the correct flavor in each bilinear, or more explicitly $\lambda^{(\dagger)},\lambda_{L/R}\in\{\text{diag}(1,0,0),\,\text{diag}(0,1,0),\,\text{diag}(0,0,1)\}$. This gives rise to the conditions $\lambda=\lambda^\dagger,\, \lambda_L=\lambda_R$, and, for the case $\psi=\chi$, $\lambda=\lambda_L$. Furthermore note that $\lambda\lambda_L=\lambda^2=\lambda$ for $\psi=\chi$ and $\lambda\lambda_L=0$ for $\psi\neq\chi$.

Assigning the chiral irreps to each of the four summands in Eq.~\eqref{eq:chiral_irrep_ToPe_dim7_quadrilinear}, we can symbolically write 
\beq\label{eq:irreps_LEFT_dim7_quadrilinear}
    \O_{\psi\chi}^{(a)} 
    = \bar 3_L \otimes (15_R \oplus 3_R) 
    - 3_L \otimes (\bar{15}_R \oplus \bar 3_R) 
    + (15_L \oplus 3_L) \otimes\bar 3_R 
    - (\bar{15}_L \oplus \bar 3_L)\otimes 3_R\,.
\eeq 
At the mesonic level, these irreps come (a priori) with independent coupling constants, called low-energy constants (LECs), which encode all information about the involved non-perturbative QCD effects. The first summand in the equation above has, for instance, the LEC $g_{\bar 3_L \otimes (15_R \oplus 3_R)}\equiv g_{\bar 3_L \otimes 15_R}+g_{\bar 3_L \otimes 3_R}$. 
Each of these unknown LECs is common to all $\chi$PT operators arising from the same irrep operator of Eq.~\eqref{eq:irreps_LEFT_dim7_quadrilinear}. However, chiral symmetry does not fix the relative size of mesonic operators in each single irrep, thus generating additional LECs that have to be determined by external input.

We may now move on to the $\chi$PT operator basis and consider only the dominant linear effects in the spurions as higher orders in $\lambda$ would imply an additional suppression by the small expansion parameter of LEFT.\footnote{Translating each quark bilinear separately from LEFT to $\chi$PT has hidden complications. In Ref.~\cite{Shi:2017ffh} this was done to illustrate examples of operators in $\chi$PT. In general there is no guarantee for the completeness of the operator basis, neither for the correct assignment of independent LECs nor for finding the lowest contributing order. We thank Peter Stoffer for pointing this out to us. The last of these aspects is also mentioned in Ref.~\cite{He:2020jly}. Moreover, the translation of separate bilinears is doomed to fail if non-trivial color structures occur, as we can only translate color-neutral objects to $\chi$PT.}
Let us start with the lowest possible order $p^0$, which only allows for $\lambda^{(\dagger)}$, $\lambda_{L/R}$, and $U$ to occur in the traces. There are two possible ways of arranging the $\lambda$: either $\lambda^{(\dagger)}$ and $\lambda_{L/R}$ are part of the same trace, or of two different traces. We cannot multiply by any other traces, as there is no non-constant chirally invariant trace solely consisting of $U$ and $U^\dagger$.\footnote{In general, one can neither multiply any additional chirally invariant trace that contains only fields $U$ and less than two derivatives, as all such traces can be brought to the form $\vev{D_\mu U U^\dagger}=0$ using the unitarity of $U$ and the cyclic property of the trace.} 

Due to the unitarity of $U$, the only invariant traces that can be built in the first of the two cases are 
\begin{aligneq}\label{eq:LECs_example}
&g^{(a)}_{(15_L \oplus 3_L) \otimes\bar 3_R} g^{(a)}_0 \vev{\lambda\lambda_L U^\dagger}\,, \quad 
&&g^{(a)}_{\bar 3_L \otimes (15_R \oplus 3_R)}\hat g^{(a)}_0 \vev{\lambda^\dagger\lambda_R U}\,, \quad
\\[0.1cm]
&g^{(a)}_{(\bar{15}_L \oplus \bar 3_L)\otimes 3_R}\tilde g^{(a)}_0 \vev{\lambda_L\lambda^\dagger U}\,, \quad
&&g^{(a)}_{3_L \otimes (\bar{15}_R \oplus \bar 3_R)} \breve g^{(a)}_0 \vev{\lambda_R\lambda U^\dagger}\,,   
\end{aligneq}
where $g^{(a)}_0, \tilde g^{(a)}_0, \hat g^{(a)}_0, \breve g^{(a)}_0$ denote LECs, and the superscripts like $g^{(a)}_{(15_L \oplus 3_L) \otimes\bar 3_R}$ are introduced to distinguish LECs that correspond to the same irrep, but may in principle be different when they are derived from other operators of LEFT. Any linear combination of these traces gives a chiral term consistent with the $SU(3)_L\times SU(3)_R$ symmetry of the underlying LEFT operator, yet still does not account for the discrete space-time symmetries. The inclusion of the correct $C$, $P$, and $T$ transformations in the chiral operator implies an appropriate linear combination of these four traces, while the LECs are constrained to be the same up to a sign (multiplied by a factor of $i$ if necessary), as already done for the mass term $\vev{\chi U^\dagger+\chi^\dagger U}$ in the original work of Ref.~\cite{Gasser:1983yg}. Hence the corresponding $C$- and $CP$-odd operator takes the form
\beq\label{eq:ToPe_ChPT_vanishing_1}
    i\vev{(\lambda\lambda_L U^\dagger + \lambda^\dagger\lambda_R U) -(\lambda_L\lambda^\dagger U + \lambda_R\lambda U^\dagger)}\,,
\eeq
with only one overall LEC that can conventionally be chosen to be $g^{(a)}_{(15_L \oplus 3_L) \otimes\bar 3_R} g^{(a)}_0$ or any of the four combination of LECs listed in Eq.~\eqref{eq:LECs_example}.\footnote{
In the literature one often introduces a set of (anti-)hermitian chiral building blocks, cf.\ for instance Refs.~\cite{Scherer:2002tk,Cata:2007ns,Scherer:2012xha,Meissner:2022odx}, which make it easier to get rid of redundancies in higher-order operators. These building blocks are eigenstates of the discrete symmetries and thus already have constraints imposed on the LECs, at least to some extent, built in. However, as we are interested in leading contributions of $C$- and $CP$-violating effects, the more historic building blocks quoted in Eq.~\eqref{eq:chiral_building_blocks} do not have any major disadvantage.} The expression in each parenthesis is parity-invariant on its own, while the relative minus sign ensures $C$ violation. An imaginary unit in front of the trace is required by hermiticity and the $T$-odd nature of the initial LEFT operator.\footnote{The constraints of hermiticity and the correct $T$ transformation are often correlated.} Analogously, we find the operator
\beq\label{eq:ToPe_ChPT_vanishing_2}
    i\vev{\lambda U^\dagger-\lambda^\dagger U}\vev{\lambda_L-\lambda_R}
\eeq
for the case where the spurions appear in two different traces. Unfortunately, both of the operators in Eqs.~\eqref{eq:ToPe_ChPT_vanishing_1} and~\eqref{eq:ToPe_ChPT_vanishing_2} vanish once the spurions are set to their physical values described previously. Hence, we must move to the next higher order $\O(p^2)$ and proceed in the same manner. This order is obviously more intricate, as we have to consider more chiral building blocks from Eq.~\eqref{eq:chiral_building_blocks}. Since the strategy should be clear by now, we directly write down our results for all independent traces of combinations of spurions and chiral building blocks that form hermitian $C$-, $CP$-, and $T$-odd operators that are at the same time chiral and Lorentz invariants and do not vanish after the spurions acquire their physical realizations. Henceforth we will drop all LECs belonging to chiral irreps, like $g^{(a)}_{3_L \otimes (\bar{15}_R \oplus \bar 3_R)}$, as they can always be absorbed by the relative LECs between each operator. Up to $\O(p^2)$ we obtain:
\begin{aligneq}\label{eq:ChPT_LEFT_dim7_quadrilinear}
    X_{\psi\chi}^{(a)}= \frac{v}{\Lambda^4}c^{(a)}_{\psi\chi}
    \big[
    &
    \ \phantom{+}i g^{(a)}_1\vev{\lambda D_\mu U^\dagger + \lambda^\dagger D_\mu U}\vev{\lambda_L D^\mu U^\dagger U +\lambda_R D^\mu U U^\dagger}\\[0.1cm]
    &
    +ig^{(a)}_2\vev{(\lambda D^2 U^\dagger U \lambda_L U^\dagger +\lambda^\dagger D^2 U U^\dagger \lambda_R U )\\[0.1cm]
    & \hspace{3cm}
    -(\lambda^\dagger U \lambda_L D^2 U^\dagger U + \lambda U^\dagger \lambda_R D^2 U U^\dagger)}\\[0.1cm]
    &
    +ig^{(a)}_3\vev{(\lambda D_\mu U^\dagger D^\mu U \lambda_L U^\dagger +\lambda^\dagger D^\mu U D_\mu U^\dagger \lambda_R U )\\[0.1cm]
    & \hspace{3cm} -(\lambda^\dagger U \lambda_L D_\mu U^\dagger D^\mu U + \lambda U^\dagger \lambda_R D^\mu U D_\mu U^\dagger)}\\[0.1cm]
    &
    +ig^{(a)}_4\vev{(\lambda D_\mu U^\dagger U \lambda_L D^\mu U^\dagger +\lambda^\dagger D^\mu U U^\dagger \lambda_R D_\mu U )\\[0.1cm]
    & \hspace{3cm}-(\lambda^\dagger D_\mu U \lambda_L U^\dagger D^\mu U + \lambda D^\mu U^\dagger \lambda_R  U D_\mu U^\dagger)}\\[0.1cm]
    & +\O(p^4)\big]\,.
\end{aligneq}
Keeping the implicit dependence of $\lambda_{L/R}$ and $\lambda^{(\dagger)}$ on the flavor $\psi,\chi$ in mind, each of the for summands proportional to the $g_i^{(a)}$ gives in principle rise to nine operators, i.e., one for each combination of flavor indices.

There is also another color contraction for the underlying LEFT operator $\O_{\psi\chi}^{(a)}$; as stated in Ref.~\cite{Liao:2019gex} this leads to the same operator basis but merely with different LECs. However the latter can always be absorbed by redefining the LECs in Eq.~\eqref{eq:ChPT_LEFT_dim7_quadrilinear}. 

%-----------------------------------------
\subsubsection{The quark bilinear operator}
\label{sec:matching_dim7_LEFT_bilinear}
%-----------------------------------------
The matching of
\beq\label{eq:dim_7_LEFT_bilinear}
    \O_{\psi}^{(b)} = \ c_{\psi}^{(b)}\, \bar\psi T^a \sigma^{\mu\nu}\psi F_{\mu\rho} G^{a\rho}_{\nu}\,,
\eeq
proceeds in the same manner, but includes further subtleties due to the presence of the photonic and gluonic field-strength tensor.
Once more we start with the decomposition in chiral irreps by 
\beq
    \O_{\psi}^{(b)} = \ c_{\psi}^{(b)} \big[ \bar q_R \Delta_{\mu\nu}^{(b)} \sigma^\munu \lambda q_L + \bar q_L \Delta_{\mu\nu}^{(b)} \sigma^\munu \lambda^\dagger q_R \big] 
\eeq
with $\Delta_{\mu\nu}^{(b)}\equiv T^a G^{a\rho}_{\nu} F_{\mu\rho}$, $\hat \lambda^{(\dagger)}_{\psi\psi}\equiv \lambda^{(\dagger)}$
with the same transformations under the gauge group action and discrete symmetries as quoted in Eqs.~\eqref{eq:chiral_trafo_spurions} and~\eqref{eq:discrete_symmetries_spurions}. Again, when setting the spurions to their physical values we can apply $\lambda=\lambda^\dagger$.  

The quark--gluon structure of this operator is a Lorentz tensor coupling to the photonic field-strength tensor $F_\munu$.
If we allow for further interactions of the photon with quarks, $F_\munu$ has to be treated in the same manner as $G_\munu$ in the course of our matching procedure. 
However, these contributions are suppressed by the QED coupling $\alpha $. Working at lowest order in $\alpha $, $F_\munu$ on its own cannot contribute to hadronic states and can thus be considered a fixed external source.
In this case the corresponding $\chi$PT operator for $\O_{\psi}^{(b)}$ has to take the form 
\beq 
X_{\psi}^{(b)}=\frac{v}{\Lambda^4}c_{\psi}^{(b)} \big(X_{\psi}^{(b)}\big)^\munu F_\munu\,,
\eeq
with a mesonic Lorentz tensor $\big(X_{\psi}^{(b)}\big)^\munu$ that includes the traces over spurions and chiral building blocks.\footnote{As for any external source one can in general take derivatives of $F_\munu$. However, these derivatives can be shifted into the hadronic part of $X^{(b)}_\psi$ by partial integration.} As a consequence $\big(X_{\psi}^{(b)}\big)^\munu$ must not be symmetric under $\mu\leftrightarrow\nu$ and $X_{\psi}^{(b)}$ has to be at least of chiral order $p^4$.\footnote{We need at least two Lorentz indices, which according to Eq.~\eqref{eq:chiral_building_blocks} result in one power in the chiral counting each, to build  $\big(X_{\psi}^{(b)}\big)^\munu$. Moreover, we know that $F_\munu$ is of order $p^2$\,.} In accordance with Ref.~\cite{Unal:2021lhb}, the gluon in Eq.~\eqref{eq:dim_7_LEFT_bilinear}, which is a chiral singlet, merely enters $X_{\psi}^{(b)}$ as an overall constant that can---together with all unknown non-perturbative QCD effects---be absorbed in the LECs. Apart from this, the external source $F_\munu$ already reproduces the correct discrete symmetries of the operator $\O_\psi^{(b)}$, such that $\big(X_{\psi}^{(b)}\big)^\munu$ has to be matched to an operator that preserves $C$, $P$, and $T$ separately. 
As we search for a Lorentz tensor at lowest possible order, we either have to build the chiral operator with two derivatives acting on matrices $U^{(\dagger)}$ or one field-strength tensor~$f_{L/R}^\munu$. In the first of these cases both derivatives have to enter the same trace, as a trace without spurions and only one derivative, i.e., $\vev{D^\mu U U^\dagger}$, vanishes. However, as we have only one spurion, these single traces are always symmetric under the exchange $\mu\leftrightarrow\nu$, which can easily be seen using $D^\mu U D^\nu U^\dagger=-(U D^\mu U^\dagger)(U D^\nu U^\dagger)$. For the remaining option with $f_{L/R}^\munu$ the relative signs within each trace are fixed by the correct $C$ transformation. These relative signs lead to a cancellation of the respective traces, since $\lambda^{(\dagger)}$ and $f^\munu_{L/R}$ are diagonal and therefore commute, and furthermore for physical values $\lambda=\lambda^{\dagger}$ and $f^\munu_{L}=f^\munu_{R}$. Hence, there is no non-vanishing contribution to $X_{\psi}^{(b)}$ at order $p^4$. To find chiral analogs of the LEFT operator $\O^{(b)}_\psi$ we must extend our search to operators at order $p^6$. However, the derivation of the complete operator set at this order is beyond the scope of our analysis.
Although $X_{\psi}^{(b)}$ is suppressed in the chiral power counting compared to the one found in Eq.~\eqref{eq:ChPT_LEFT_dim7_quadrilinear}, it may still be of relevance for physical applications, as it can contribute with a different field content and different quantum numbers, e.g., partial waves.

Remember, throughout we will, for \textit{all} operators, work at the lowest order in $\alpha$ and hence treat $F_\munu$ and semi-leptonic bilinears as fixed external sources.\footnote{If we allow $F_\munu$ to hadronize, every of our remaining LEFT operators with a photonic field-strength tensor maps onto the same $\chi$PT expression as a corresponding gluonic one, but with different LECs and an additional suppression in $\alpha$.} 

%--------------------------------------------------------------
\subsubsection{Semi-leptonic operators}
%--------------------------------------------------------------
Starting with $\O^{(c)}_{\ell\psi}$, we can in this case match the quark--gluon structure, which is simply given by the axialvector current $\bar\psi\gamma^\mu\gamma_5\psi$, onto $\chi$PT by identifying traces built from the spurions $\lambda_{L/R}$ that are Lorentz vectors and have the signature $CPT=+-+$. The operator complying with these requirements is
\beq\label{eq:ChPT_LEFT_dim7_quadrilinear_semi-leptonic}
    X_{\ell\psi}^{(c)}= \frac{v}{\Lambda^4} c^{(c)}_{\ell\psi} g^{(c)}_1 \big(\bar\ell \dvec{D}_\mu \gamma_5 \ell\big) \, i\vev{\lambda_L D^\mu U^\dagger U -\lambda_R D^\mu U U^\dagger}+\O(p^4)\,.
\eeq
Regarding the mesonic analog of $\O^{(d)}_{\ell\psi}$, it can be easily checked that there is no contribution to $X_{\ell\psi}^{(d)}$ at $\O(p^2)$.

%--------------------------------------------------------------
\subsection{Matching dimension-8 LEFT operators}
\label{sec:Matching_LEFT_dim_8}
%--------------------------------------------------------------
In this section we quote our results for the $\chi$PT expressions derived from the dimension-8 quark level in LEFT. Again we restrict the chiral basis to linear contributions in $1/\Lambda^4$, lowest order in $\alpha $, as well as to $\O(p^2)$ for gluonic operators and $\O(p^4)$ for photonic and photo-gluonic ones, or in other words the lowest order for each contribution.

%--------------------------------------------------------------
\subsubsection{Gluonic operators}
\label{sec:ChPT_Gluonic_Dim8}
%--------------------------------------------------------------
There is only one leading-order contribution to $X_{\psi}^{(e)}$
\beq
    \frac{c^{(e)}_{\psi}}{\Lambda^4}\, i\vev{\big(\lambda_L D^2 U^\dagger U + \lambda_R D^2 U U^\dagger \big)-\big(\lambda_L  U^\dagger D^2U + \lambda_R U D^2 U^\dagger \big)} +\O(p^4), 
\eeq
which vanishes for physical values of the spurions as demanded by the equations of motion~\eqref{eq:ChPT_EoM}. Thus $X_{\psi}^{(e)}$ starts at the next higher order.
We refrain from deriving the numerous contributions of higher orders at this early stage in the analysis of ToPe operators and proceed similarly for all other operators.

The LEFT operators $\O^{(f,g,h,i,j)}$ differ only in their color contractions and their vector or axialvector Dirac structure and therefore map to the same $\chi$PT operator, but with different LECs. They cannot be distinguished at the mesonic level and give rise to the operator
\begin{aligneq}\label{eq:ChPT_LEFT_dim8_gluonic}
    X_{\psi\chi}^{(f,g,h,i,j)}= &\frac{1}{\Lambda^4}c^{(f,g,h,i,j)}_{\psi\chi}
    \\ 
    &\times \Big[ i g^{(f,g,h,i,j)}_1 \vev{\big( \lambda_L U^\dagger D^2 U U^\dagger \tilde \lambda_R U + \lambda_R U D^2 U^\dagger U \tilde \lambda_L U^\dagger \big)
    \\
    & \hspace{3cm}
    -\big( \tilde \lambda_R U D^2 U^\dagger U \lambda_L U^\dagger  + \tilde \lambda_L U^\dagger  D^2 U U^\dagger \lambda_R U \big)} 
    \\
    &
    \phantom{\times} +i g^{(f,g,h,i,j)}_2 \vev{\big(\lambda_L D_\mu U^\dagger U D^\mu U^\dagger \tilde\lambda_R U + \lambda_R D_\mu U U^\dagger D^\mu U \tilde\lambda_L U^\dagger\big)
    \\
    & \hspace{3cm}
    -\big(\lambda_L  U^\dagger \tilde\lambda_R D_\mu U U^\dagger D^\mu U  + \lambda_R  U \tilde\lambda_L D_\mu U^\dagger U D^\mu U^\dagger\big)
    }
    \\
    &
    \phantom{\times}+i g^{(f,g,h,i,j)}_3 \vev{\big(\lambda_L D^2 U^\dagger \tilde \lambda_R U + \lambda_R D^2 U \tilde \lambda_L U^\dagger \big)
    \\
    & \hspace{3cm}
    -\big( \tilde \lambda_R D^2 U\lambda_L U^\dagger  + \tilde \lambda_L   D^2 U^\dagger \lambda_R U \big)} 
    \\
    &\phantom{\times} +i g^{(f,g,h,i,j)}_4 \vev{\lambda_L U^\dagger \tilde\lambda_R U -\lambda_R U \tilde\lambda_L U^\dagger}
    \vev{\chi^\dagger U-\chi U^\dagger}
    \\
    &
    \phantom{\times}+\O(p^4)     \Big] \,,
\end{aligneq}
where we defined the products of Wilson coefficients and LECs by $c^{(f,g,h,i,j)}_{\psi\chi}g^{(f,g,h,i,j)}_i\equiv \sum_{z=f,g,h,i,j}c_{\psi\chi}^{(z)}g_i^{(z)}$.
We will use this notation throughout for any combination of indices. The diagonal matrices $\lambda_{L/R}(\tilde \lambda_{L/R})$ project out the flavor $\psi(\chi)$ in complete analogy to the definitions in the previous sections. Some terms can be discarded using $\vev{D^2UU^\dagger-D^2U^\dagger U}=D_\mu\big(\vev{D^\mu U U^\dagger-D^\mu U^\dagger U}\big)=0$, which can also easily be deduced from Eq.~\eqref{eq:ChPT_EoM}.

Analogously, $\O^{(k)}$ and $\O^{(l)}$ can be projected onto one single mesonic operator. The peculiarity of these quadrilinears is that the physical values of their spurions, let us call them $\hat\lambda_{\psi\chi}\equiv\lambda$ and $\hat\lambda_{\chi\psi}\equiv\tilde\lambda$, appearing in the two contributing bilinears are no eigenstates of hermitian conjugation. To simplify the evaluation we introduce the hermitian and antihermitian combinations $\lambda_\pm=\lambda\pm\tilde\lambda$.\footnote{Going back to the irreducible representation of a general quark multilinear from Eq.~\eqref{eq:chiral_irrep_general}, this redefinition of the spurions merely leads to another redefinition of the LECs.} 
When the spurions acquire their physical values, we can set $\lambda_+^\dagger=\lambda_+$, $\lambda_-^\dagger=-\lambda_-$, $\lambda_\pm\lambda_\pm^\dagger=\mathds{1}$, and $\lambda_+\lambda_-=-\lambda_-\lambda_+$, independently of their explicit flavor indices. Moreover, the $\lambda_\pm$ are real and thus $\lambda^T_\pm=\lambda^\dagger_\pm$. As $\lambda$ and $\tilde\lambda$ transform according to the first line in Eq.~\eqref{eq:discrete_symmetries_spurions}, the discrete symmetries of $\lambda_\pm$ become
\begin{aligneq}
&\lambda_\pm &&\xleftrightarrow{\ C\ } \lambda^T_\pm\,, &&\qquad
\lambda_\pm &&\xleftrightarrow{\ P\ }  \pm\lambda^\dagger_\pm\,, &&\qquad
\lambda_\pm &&\xleftrightarrow{\ T\ }  \lambda_\pm\,.
\end{aligneq}
The minus sign for the parity transform of $\lambda_-$ may be unintuitive, but compensates for the fact that $\lambda_-$ is anti-hermitian, i.e., after inserting the physical values $\lambda_-$ is invariant under parity as it should. With this new set of spurions---and noting that the antisymmetry of the LEFT operator under $\psi\leftrightarrow\chi$ demands that one bilinear includes $\lambda_+^{(\dagger)}$ while the other one has to contain $\lambda_-^{(\dagger)}$---we can evaluate the $\chi$PT analogs of $\O_{\psi\chi}^{(k,l)}$ in the familiar way.  However, we do not find any operator at chiral order $p^2$ and hence ignore the contribution of $X_{\psi\chi}^{(k,l)}$ for now. 

Analogously, the symmetry of the operator $\O_{\psi\chi}^{(m)}$ under $\psi\leftrightarrow\chi$ demands that either both quark bilinears include all possible combinations of the spurions $\lambda_-^{(\dagger)}\lambda_-^{(\dagger)}$ or of $\lambda_+^{(\dagger)}\lambda_+^{(\dagger)}$. Again there is no non-vanishing $C$- and $CP$-odd mesonic operator at leading order.

%--------------------------------------------------------------
\subsubsection{Photonic operators}
%--------------------------------------------------------------

As previously seen for the gluonic operators, we can again match several photonic operators from LEFT to the same $\chi$PT operator upon redefining the LECs. Hence the mesonic counterpart of the operator $\O_{\psi\chi}^{(n,o,p,q)}$ yields to lowest order
\begin{aligneq}\label{eq:ChPT_LEFT_dim8_photonic}
    X^{(n,o,p,q)}_{\psi\chi}=&\frac{1}{\Lambda^4}c_{\psi\chi}^{(n,o,p,q)}
    \\
    & \Big[ g_1^{(n,o,p,q)} \epsilon_{\alpha\beta\mu\nu}\vev{\big(
    \lambda_L U^\dagger \tilde\lambda_R f_R^{\alpha\beta} U 
    -\lambda_R U \tilde\lambda_L f_L^{\alpha\beta} U^\dagger\big)
    \\
    &\hspace{3cm}
    +\big(
    \tilde\lambda_R U \lambda_L U^\dagger f_R^{\alpha\beta} 
    -\tilde\lambda_L U^\dagger \lambda_R U f_L^{\alpha\beta} \big)}
    \\
    &
    + g_2^{(n,o,p,q)} \vev{\lambda_L D_\mu U^\dagger U + \lambda_R D_\mu U U^\dagger}\vev{\tilde\lambda_L D_\nu U^\dagger U + \tilde\lambda_R D_\nu U U^\dagger}
    \\
    &
    + g_3^{(n,o,p,q)} \vev{\lambda_L D_\mu U^\dagger U - \lambda_R D_\mu U U^\dagger}\vev{\tilde\lambda_L D_\nu U^\dagger U - \tilde\lambda_R D_\nu U U^\dagger} \Big]
    \\
    &
    \times 
    F^\munu +\O(p^6) \,.
\end{aligneq}
Note that the $\epsilon$-tensor flips sign under $P$ and $T$.
The photonic LEFT quadrilinear $\O^{(r)}_{\psi\chi}$, mixing quark flavors in each bilinear, can again be conveniently matched using the \mbox{(anti-)} hermi\-tian spurions $\lambda^{(\dagger)}_\pm$. But once more, we do not find any operator up to and including chiral order $p^4$.

%--------------------------------------------------------------
\subsubsection{Photo-gluonic operators}
%--------------------------------------------------------------
Both photo-gluonic operators $\O_{\psi}^{(s,t)}$ map onto the same $\chi$PT expression
\begin{aligneq}\label{eq:ChPT_LEFT_dim8_photo-gluonic}
    X_{\psi}^{(s,t)}= &\frac{1}{\Lambda^4} c^{(s,t)}_{\psi}g^{(s,t)}_1 \epsilon_{\alpha\beta\mu\nu}
    \vev{\lambda_L U^\dagger f_R^{\alpha\beta} U- \lambda_R U f_L^{\alpha\beta} U^\dagger} F^\munu
    +\O(p^6)\,.    
\end{aligneq}

%--------------------------------------------------------------
\subsubsection{Semi-leptonic operators}
%--------------------------------------------------------------
We find that the $C$- and $CP$-violating contributions from the gluonic semi-leptonic operators $\O^{(u)}_{\ell\psi}$, $\O^{(v)}_{\ell\psi}$ vanish at $\O(p^2)$, while the photonic ones $\O^{(w)}_{\ell\psi}$, $\O^{(x)}_{\ell\psi}$ give rise to 
\beq\label{eq:ChPT_LEFT_dim8_photonic_semi-leptonic_vector}
    X_{\ell\psi}^{(w)}= \frac{1}{\Lambda^4}c^{(w)}_{\ell\psi}g_1^{(w)} \, i\vev{\lambda_L D_\nu U^\dagger U +\lambda_R D_\nu U U^\dagger}\bar\ell \gamma_\mu \ell F^\munu+\O(p^6)
\eeq
and 
\beq\label{eq:ChPT_LEFT_dim8_photonic_semi-leptonic_pseudovector}
    X_{\ell\psi}^{(x)}= \frac{1}{\Lambda^4}g_1^{(x)}c^{(x)}_{\ell\psi} \, i\vev{\lambda_L D_\nu U^\dagger U -\lambda_R D_\nu U U^\dagger}\bar\ell \gamma_\mu\gamma_5 \ell F^\munu+\O(p^6)\,,
\eeq
respectively.

%-----------------------------------------------------------------
\boldmath
\subsection{Summary of the effective $C$- and $CP$-odd Lagrangian}
\label{sec:intermediate_summary}
\unboldmath
%-----------------------------------------------------------------
In the preceding sections we derived the lowest possible contributing order of mesonic operators 
for all flavor-conserving, neutrinoless $C$- and $CP$-violating sources (except purely leptonic ones) that preserve baryon and lepton number up to dimension 8 in LEFT. Working to lowest order in the QED coupling $\alpha$, i.e., treating photons and leptons as fixed external sources, these contributions start at $\O(p^2)$ for gluonic and semi-leptonic operators and at $\O(p^4)$ for photonic ones. We identified that the 24 LEFT operators $\O^{(a)}_{\psi\chi},\ldots, \O^{(x)}_{\ell\chi}$ (without counting different flavor combinations) listed in Eqs.~\eqref{eq:ToPe_LEFT_dim7}--\eqref{eq:ToPe_LEFT_dim8_semi_leptonic_gluonic} give in general rise to 15 different groups of operators $X^{(z)}$ on the mesonic level, which build the full chiral Lagrangian
\begin{aligneq}\label{eq:summary_ToPe_Lagrangian}
\L_{\slashed{C}P\slashed{T}}=\sum\limits_{\psi,\chi,\ell}\Big[ &X_{\psi\chi}^{(a)}+X_{\psi}^{(b)}+X_{\ell\psi}^{(c)}+X_{\ell\psi}^{(d)}+X_{\psi\chi}^{(e)}+X_{\psi\chi}^{(f,g,h,i,j)}+X_{\psi\chi}^{(k,l)}+X_{\psi\chi}^{(m)}
\\
&+X_{\psi\chi}^{(n,o,p,q)}+X_{\psi\chi}^{(r)}+X_{\psi}^{(s,t)}+X_{\ell\psi}^{(u)}+X_{\ell\psi}^{(v)}+X_{\ell\psi}^{(w)}+X_{\ell\psi}^{(x)}\Big]\,.
\end{aligneq}
In this Lagrange density, the terms $X_{\psi\chi}^{(a)}$, $X_{\psi}^{(b)}$, $X_{\ell\psi}^{(c)}$, $X_{\ell\psi}^{(d)}$ originate from dimension 7 of LEFT, the rest from dimension 8.
The contributions of order $p^2$ for the gluonic operators  $X_{\psi\chi}^{(a)}$, $X_{\psi\chi}^{(f,g,h,i,j)}$ can be found in Eqs.~\eqref{eq:ChPT_LEFT_dim7_quadrilinear} and~\eqref{eq:ChPT_LEFT_dim8_gluonic}, respectively, while $X_{\psi\chi}^{(e)}$, $X_{\psi\chi}^{(k,l)}$, $X_{\psi\chi}^{(m)}$ start at higher orders. The photonic operator $X_{\psi\chi}^{(n,o,p,q)}$ is mapped to its lowest possible order $p^4$ in Eq.~\eqref{eq:ChPT_LEFT_dim8_photonic}, whereas $X_{\psi\chi}^{(r)}$ first appears at $\O(p^6)$. For the photo-gluonic operators we find that $X_{\psi}^{(b)}$ only starts at $\O(p^6)$,
and $X_{\psi}^{(s,t)}$ at $\O(p^4)$ is given in Eq.~\eqref{eq:ChPT_LEFT_dim8_photo-gluonic}. Finally, the lowest possible contribution at $\O(p^2)$ to the semi-leptonic operators $X_{\ell\psi}^{(c)}$, $X_{\ell\psi}^{(w)}$, $X_{\ell\psi}^{(x)}$ are listed in Eqs.~\eqref{eq:ChPT_LEFT_dim7_quadrilinear_semi-leptonic}, \eqref{eq:ChPT_LEFT_dim8_photonic_semi-leptonic_vector}, and \eqref{eq:ChPT_LEFT_dim8_photonic_semi-leptonic_pseudovector}, while $X_{\ell\psi}^{(d)}$, $X_{\ell\psi}^{(u)}$, $X_{\ell\psi}^{(v)}$ start at higher orders.

%---------------------------------------------------------------
\boldmath
\section{The large-$N_c$ extension}
\label{sec:Large_Nc}
\unboldmath
%---------------------------------------------------------------
So far, the framework of ToPe$\chi$PT covers the sector of the meson octet.  It can, however, be generalized to include the singlet $\eta'$, whose mass $\metap$ remains non-vanishing in the chiral limit due to the $U(1)_A$ anomaly, in a straightforward manner. Taking the number of colors $N_c$ to be large, this anomaly is suppressed, so that the $\eta'$ is rendered massless and takes the role of the ninth Goldstone boson.

As a consequence for the perturbative treatment in the effective low-energy theory, not only the momentum $p$ but also $\metap$ needs to be considered as small. This can be achieved by simultaneously expanding the chiral Lagrangian in soft momenta, light quark masses, and powers of $1/N_c$. One hence introduces a small counting parameter $\delta$ and uses
\begin{aligneq}
        p=\O(\sqrt{\delta})\,,\qquad
        m=\O(\delta)\,,\qquad
    1/N_c=\O(\delta)\,.
\end{aligneq}
The large-$N_c$ extension of SM$\chi$PT has been subject to many previous analyses, see for instance Refs.~\cite{Rosenzweig:1979ay,DiVecchia:1980yfw, Witten:1980sp,Kawarabayashi:1980dp,Nath:1979ik,Leutwyler:1997yr,Herrera-Siklody:1996tqr,Kaiser:2000gs,Bickert:2016fgy} and the numerous references therein. As these considerations rely on general gluon dynamics, we can apply the large-$N_c$ extension in this section to ToPe$\chi$PT without much trouble, and refer to the abovementioned works for further details.

To include the singlet in our formalism at the level of the general Lagrangian from Eq.~\eqref{eq:Lagrange_Source_Fields}, we add the term $\L_{N_c}\equiv\theta \omega$ with a new external source $\theta$, whose physical value is the QCD vacuum angle $\theta_{\text{QCD}}$, and the winding number density $\omega=g^2/(32\pi^2)G^a_\munu \tilde G^{a\, \munu}$.\footnote{Note that none of the LEFT operators considered in this work can contribute to a singlet under $SU(3)_L\times SU(3)_R$, so that $\L_{N_c}$, which arises naturally from QCD, is indeed the only external source with the desired transformation property we can add to Eq.~\eqref{eq:Lagrange_Source_Fields}.} The first modifications we have to make in order to enhance the $SU(3)_L\times SU(3)_R$ symmetry to $U(3)_L\times U(3)_R$ is to replace
\beq
U\mapsto\bar U=e^{i\varphi}U\,,
\eeq
where $\varphi\equiv \sqrt{2/3}\,\eta_0/\Fpi$, and the new chiral transformation reads $\bar U\to R\bar U L^\dagger$ with $L\in U(3)_L$ and $R\in U(3)_R$. The remaining chiral building blocks stay unchanged, except that they transform with $L, R$ as elements of $U(3)_{L,R}$ instead of $SU(3)_{L,R}$. We shall not introduce a new notation for the large-$N_c$ case of these building blocks, which is unambiguously fixed by the use of either $U$ or $\bar U$ in each operator.
At leading order, the octet and singlet components $\eta_8$ and $\eta_0$ are related to the physical mass eigenstates by the single-angle mixing scheme
\beq\label{eq:eta-eta'_mixing}
    \begin{pmatrix}
        \eta_8 \\ \eta_0
    \end{pmatrix}
    =
    \begin{pmatrix}
        \cos\theta & \sin\theta\\
        -\sin\theta & \cos\theta
    \end{pmatrix}
    \begin{pmatrix}
        \eta \\ \eta'
    \end{pmatrix}\,.
\eeq
Henceforth we work with the ideal mixing angle $\theta=\arcsin(-1/3)$, so that  
\beq\label{eq:Goldstone_large_Nc}
\bar U=\exp\left(\frac{i\bar \Phi}{\Fpi}\right) \eqwith  \bar\Phi= 
\begin{pmatrix}
\frac{1}{\sqrt{3}}\eta'+\sqrt{\frac{2}{3}}\eta+\pi^0   & \sqrt{2}\pi^+  &\sqrt{2}K^+   \\
\sqrt{2}\pi^-& \frac{1}{\sqrt{3}}\eta'+\sqrt{\frac{2}{3}}\eta-\pi^0  & \sqrt{2}K^0 \\
\sqrt{2}K^- & \sqrt{2}\bar K^0& \frac{2}{\sqrt{3}}\eta'-\sqrt{\frac{2}{3}}\eta
\end{pmatrix}\,.
\eeq
For convenience and later use we quote $\varphi$ in terms of the physical $\eta$ and $\eta'$ fields, i.e., 
\beq
     \varphi = \frac{ \sqrt{2} } { 3 \sqrt{3} F_0 }\,  \eta  + \frac{ 4 }{ 3 \sqrt{3} F_0}\, \eta'\,.
\eeq
We have to introduce new building blocks from the pure singlet contribution, which are
\beq\label{eq:large_Nc_building_blocks}
(\varphi+\theta)\to (\varphi+\theta)\,, \qquad D_\mu \varphi\to D_\mu \varphi\,, \qquad D_\mu \theta \to D_\mu\theta\,,
\eeq
with 
\beq
    D_\mu \varphi\equiv\partial_\mu \varphi - 2 \vev{a_\mu} \eqand D_\mu \theta\equiv\partial_\mu \theta + 2 \vev{a_\mu}\,,
\eeq
where $2\vev{a_\mu}=\vev{r_\mu-l_\mu}$ is the singlet axial current discussed in more detail in Ref.~\cite{Kaiser:2000gs}.\footnote{More precisely, we have $\varphi\to \varphi-i\ln(\det R)+i\ln(\det L)$ and $\theta\to\theta+i\ln(\det R)-i\ln(\det L)$, so that $\varphi$ and $\theta$ are not invariant on their own. However, $D_\mu \varphi$ and $D_\mu \theta$ are still invariant as separate quantities.} With these new building blocks we cannot only construct completely new operators, e.g., by contracting a vector operator with $D_\mu \varphi$ or $D_\mu \theta$, but can also multiply any Lorentz invariant combination of them to any operator without affecting the transformation under $U(3)_L\times U(3)_R$. As $\varphi$ and $\theta$ transform as $CPT=+--$~\cite{Herrera-Siklody:1996tqr}, odd powers of them will change the discrete symmetries of the overall operator. 

However, we do not have to consider the infinite amount of all new operators that arise from insertions of the elements in Eq.~\eqref{eq:large_Nc_building_blocks}, as the latter affect the power counting in $\delta$, which can be summarized as follows.
Traceless operators are subject to purely gluonic interactions, which scale at leading order as $N_c^2$. 
Each trace in flavor space originates from one quark loop, leading to a suppression of $1/N_c$. Moreover each $\varphi$ and $\theta$ counts as another factor $1/N_c$. Hence the generalized power counting in large-$N_c$ $\chi$PT, i.e., the order of $\delta$, can be understood as
\beq\label{eq:large_Nc_power_counting}
    \O_\delta=-2+N_{\text{tr}}+\frac{1}{2}N_\chi+N_\varphi\,,
\eeq
where $N_{\text{tr}}$ denotes the number of traces, $N_\varphi$ indicates the power of $\varphi$ and $\theta$, and $N_\chi\equiv N_p + 2N_m$ is the power counting in standard  $\chi$PT as described in Sect.~\ref{sec:Standard_ChPT}, which keeps track of the power of soft momenta $N_p$ and the power of light quark masses $N_m$. This power counting allows for four different contributions at order $\delta^0$, namely 
$(N_{\text{tr}},N_\chi,N_\varphi) \in \{(2,0,0),(1,2,0),(1,0,1),(0,0,2)\}$, 
out of which only $(1,2,0)$ can contribute to a $C$-odd operator. Hence the leading contribution to all gluonic operators at large $N_c$ can directly be read off the respective contributions in standard ToPe$\chi$PT, which consist of one single trace. Similarly, all photonic and photo-gluonic operators start at $\O(\delta)$ in large-$N_c$, as they require $N_\chi\geq 4$ and can thus be obtained by the corresponding ToPe$\chi$PT operators with $N_{\text{tr}}=1$, $N_\varphi=0$. 
Considering the matching of LEFT operators that initially had no chiral counterpart at $\O(p^2)$ for $N_c=3$, the chiral singlets from Eq.~\eqref{eq:large_Nc_building_blocks} allow for new chirally invariant operators in the large-$N_c$ limit, so that these LEFT sources may indeed show up at $\O(p^2)$ but at higher order in $\delta$.

For convenience we quote the order $\delta^0$ analog to Eq.~\eqref{eq:summary_ToPe_Lagrangian} in the large-$N_c$ limit as
\begin{aligneq}\label{eq:Nc_Lagrangian}
    \bar{\L}_{\slashed{C}P\slashed{T}}= \frac{iv}{\Lambda^4}&\sum\limits_{\psi,\chi,\ell}\Big[
    \phantom{+\ } c_{\psi\chi}^{(a)}\bar g^{(a)}_2\vev{(\lambda D^2 \bar{U}^\dagger \bar{U} \lambda_L \bar{U}^\dagger +\lambda^\dagger  D^2 \bar{U} \bar{U}^\dagger \lambda_R \bar{U} )
    -\hc}\\
    &
    +c_{\psi\chi}^{(a)}\bar g^{(a)}_3\vev{(\lambda D_\mu \bar{U}^\dagger D^\mu \bar{U} \lambda_L \bar{U}^\dagger +\lambda^\dagger D^\mu \bar{U} D_\mu \bar{U}^\dagger \lambda_R \bar{U} )
    -\hc}\\[0.1cm]
    &
    +c_{\psi\chi}^{(a)}\bar g^{(a)}_4\vev{(\lambda D_\mu \bar{U}^\dagger \bar{U} \lambda_L D^\mu \bar{U}^\dagger +\lambda^\dagger D^\mu \bar{U} \bar{U}^\dagger \lambda_R D_\mu \bar{U} )
    -\hc}\\[0.1cm]
    &
    +c^{(c)}_{\ell\psi}\bar g^{(c)}_1 \, \bar\ell \dvec{D}_\mu \gamma_5 \ell \vev{\lambda_L D^\mu \bar{U}^\dagger \bar{U} -\lambda_R D^\mu \bar{U} \bar{U}^\dagger}\\[0.1cm]
    &
    +\frac{1}{v}c^{(f,g,h,i,j)}_{\psi\chi}\bar g^{(f,g,h,i,j)}_1 \vev{\big( \lambda_L \bar{U}^\dagger D^2 \bar{U} \bar{U}^\dagger \tilde \lambda_R \bar{U} + \lambda_R \bar{U} D^2 \bar{U}^\dagger \bar{U} \tilde \lambda_L \bar{U}^\dagger \big)
    -\hc}\\[0.1cm]
    &
    +\frac{1}{v}c^{(f,g,h,i,j)}_{\psi\chi} \bar{g}^{(f,g,h,i,j)}_2 \vev{\big(\lambda_L D_\mu \bar{U}^\dagger \bar{U} D^\mu \bar{U}^\dagger \tilde\lambda_R \bar{U} + \lambda_R D_\mu \bar{U} \bar{U}^\dagger D^\mu \bar{U} \tilde\lambda_L \bar{U}^\dagger\big)-\hc
    }\\[0.1cm]
    &
    +\frac{1}{v}c^{(f,g,h,i,j)}_{\psi\chi} \bar{g}^{(f,g,h,i,j)}_3 \vev{\big(\lambda_L D^2 \bar{U}^\dagger \tilde \lambda_R \bar{U} + \lambda_R D^2 \bar{U} \tilde \lambda_L \bar{U}^\dagger \big)-\hc}\\[0.1cm]
    &+\O(\delta)
    \Big]\,.
\end{aligneq}
Interactions at higher order in $\delta$ can be obtained by the procedure described above. In the following sections we will refer to the large-$N_c$ limit of a ToPe$\chi$PT operator $X_{\psi\chi}^{(z)}$ as $\bar X_{\psi\chi}^{(z)}$. As a final remark, all operators in ToPe$\chi$PT and its large-$N_c$ extension that only differ by the shift $U\to\bar U$ carry the same LECs at leading order, as is the case for the leading order in large-$N_c$ SM$\chi$PT~\cite{Kaiser:2000gs}. Nevertheless, we will still denote the LECs in the large-$N_c$ theory by $\bar g^{(z)}_i$ to be as general as possible.

%-------------------------------------------------------
\boldmath
\section{Application to $C$- and $CP$-violating decays}
\label{sec:applications}
\unboldmath
%-------------------------------------------------------
Up to now, various experiments have actively searched for $C$ violation in $\eta$ decays (in the following we will use the abbreviation $\etap$ to refer to both $\eta$ and $\eta'$), as in $\etap\to 3\gamma$~\cite{ Serpukhov-Brussels-LosAlamos-AnnecyLAPP:1987kiw,KLOE:2004ukf,Blik:2007ne}, $\eta\to \pi^0 \gamma$~\cite{Nefkens:2005ka}, $\etap\to\pi^0\ell^+\ell^-$~\cite{WASA-at-COSY:2018jdv, CLEO:1999nsy, Dzhelyadin:1980ti} and $\eta'\to\eta\ell^+\ell^-$~\cite{CLEO:1999nsy, Dzhelyadin:1980ti} driven by a single virtual photon, $\eta\to \pi^+\pi^-\gamma$~\cite{Thaler:1972ax, Gormley:1970qz, Jane:1974es}, $\eta \to 2\pi^0 \gamma$~\cite{CrystalBall:2005zrs,Blik:2007ne}, $\eta \to 3\pi^0 \gamma$~\cite{CrystalBall:2005zrs,Blik:2007ne}, and in $\pi^0\to 3\gamma$~\cite{McDonough:1988nf}, without strong empirical evidence for this kind of BSM physics.\footnote{Some of the listed decays may in principle also be driven by $C$- and $P$-odd operators, which are not covered by our framework. 
These contributions have less physical motivation, as they conserve $CP$, and are beyond the scope of this work. Therefore we assume all decay widths of $C$-violating amplitudes to originate solely from sources with additional $T$ violation and ignore possible $P$-violating effects at this stage.} However, in the foreseeable future the new experimental setups from the REDTOP~\cite{Gatto:2016rae,Gatto:2019dhj,REDTOP:2022slw} and JEF~\cite{Gan:2015nyc,JEF:2016,Gan:2017kfr} collaborations will search for rare $\etap$ decays with an increased accuracy that may allow us to set more stringent bounds on ToPe forces.

The model-independent effective theory derived in the previous sections provides the theoretical foundation to identify the most promising decays to observe and to figure out any, as yet unknown, correlation between different $C$- and $CP$-violating transitions. As the sources of the latter are rigorously worked out on the quark level, we can provide the explicit dependence of $C$- and $CP$-odd observables on the new-physics scale $\Lambda$. To this end, we not only restrict our analysis to pure BSM processes, but also investigate the interference of SM and $C$-violating contributions for suitable candidates. A list of all decays considered in this work is contained in Table~\ref{tab:Overview_C-odd_Decays}, which summarizes our results in a compact way.
Each $C$- and $CP$-odd contribution to these decays, except for $\eta\to 3\pi^0 \gamma$, exhibits a unique representation in terms of mesonic degrees of freedom. Identifying these mesonic operators first eases the search for a corresponding ToPe$\chi$PT operator that generates the desired transition. These chiral operators in turn can be related to the underlying quark operators. 
For a consistent treatment of $\eta$ and $\eta'$ decays, we will work with ToPe$\chi$PT in the large-$N_c$ limit with generalized power counting in $\delta$ throughout, as explained in Sect.~\ref{sec:Large_Nc}. However, we may still quote the power counting in soft momenta $p$, because it is directly visible in the operators at the mesonic level. As central numeric results of this work we give the theoretical estimates of observables in dependence on $\Lambda$ (explained in more detail below), while the limits that can be set on $\Lambda$ with the currently most precise measurements are quoted in the respective sections. In the rest of this manuscript we explain in detail the assumptions and simplifications entering Table~\ref{tab:Overview_C-odd_Decays}.
\begin{table}[t]
\centering
\renewcommand{\arraystretch}{1.5}
\resizebox{\columnwidth}{!}{
\begin{tabular}{l|ccccc}
\toprule
Decay & Mesonic operator & Lowest order & Current measurement & Theoretical estimate & Section\\[0.1cm] 
\midrule
$\etap\to \pi^0\pi^+\pi^-$  & $i\,\etap \partial^\mu\pi^0 (\pi^+\partial_\mu\pi^- - \pi^-\partial_\mu\pi^+)$  & $p^2\,(\delta^0)$ & $g_2= -9.3(4.5)\cdot 10^3/ \text{TeV}^2$~\cite{Akdag:2021efj} & $|g_2|\sim 3\cdot 10^{-4}\,\text{TeV}^{2}/\Lambda^4$ & \ref{sec:Eta3Pi}
\\[0.1cm]
$\eta'\to \eta\pi^+\pi^-$  & $i\,\eta' \partial^\mu\eta(\pi^+\partial_\mu\pi^- - \pi^-\partial_\mu\pi^+)$  & $p^2\,(\delta^1)$ &  $g_1=0.7(1.0)\cdot 10^6/ \text{TeV}^2$~\cite{Akdag:2021efj} & $|g_1|\sim 3\cdot 10^{-4}\,\text{TeV}^{2}/\Lambda^4$ & \ref{sec:EtapEtaPiPi}
\\[0.1cm]
$\etap\to \pi^0 \gamma^\ast$ & $\partial_\mu \etap\, \partial_\nu \pi^0 F^\munu$ & $p^4\,(\delta^2)$ & -- & -- & \ref{sec:EtaPiGamma}
\\[0.1cm]
$\eta'\to \eta \gamma^\ast$ & $\partial_\mu \eta'\, \partial_\nu \eta F^\munu$ & $p^4\,(\delta^2)$ & -- & -- & \ref{sec:EtaPiGamma}
\\[0.1cm] 
$\eta\to \pi^0 e^+e^-$ & $\eta \partial_\mu \pi^0\,\bar e \gamma^\mu e$ & $p^2\,(\delta^1)$ & $\text{BR}<7.5\cdot10^{-6}$~\cite{WASA-at-COSY:2018jdv} & $\text{BR}\sim7\cdot10^{-27}\,\rm{TeV}^8/\Lambda^8$ &\ref{sec:EtaPiLepLep}
\\[0.1cm] 
$\eta\to \pi^0 \mu^+\mu^-$ & $\eta \partial_\mu \pi^0\,\bar \mu \gamma^\mu \mu$ & $p^2\,(\delta^1)$ & $\text{BR}<5\cdot10^{-6}$~\cite{Dzhelyadin:1980ti} & $\text{BR}\sim2\cdot10^{-27}\,\rm{TeV}^8/\Lambda^8$ &\ref{sec:EtaPiLepLep}
\\[0.1cm] 
$\eta'\to \pi^0 e^+e^-$ & $\eta' \partial_\mu \pi^0\,\bar e \gamma^\mu e$ & $p^2\,(\delta^1)$ & $\text{BR}<1.4\cdot 10^{-3}$~\cite{CLEO:1999nsy} & $\text{BR}\sim9\cdot10^{-28}\,\rm{TeV}^8/\Lambda^8$ &\ref{sec:EtaPiLepLep}
\\[0.1cm] 
$\eta'\to \pi^0 \mu^+\mu^-$ & $\eta' \partial_\mu \pi^0\,\bar \mu \gamma^\mu \mu$ & $p^2\,(\delta^1)$ & $\text{BR}<6\cdot 10^{-5}$~\cite{Dzhelyadin:1980ti} & $\text{BR}\sim6\cdot10^{-28}\,\rm{TeV}^8/\Lambda^8$ &\ref{sec:EtaPiLepLep}
\\[0.1cm] 
$\eta'\to \eta e^+e^-$ & $\eta' \partial_\mu \eta\, \bar e \gamma^\mu e$ & $p^2\,(\delta^1)$ & $\text{BR}< 2.4 \cdot 10^{-3}$~\cite{CLEO:1999nsy} & $\text{BR}\sim9\cdot10^{-29}\,\rm{TeV}^8/\Lambda^8$ &\ref{sec:EtaPiLepLep}
\\[0.1cm] 
$\eta'\to \eta \mu^+\mu^-$ & $\eta' \partial_\mu \eta\, \bar \mu \gamma^\mu \mu$ & $p^2\,(\delta^1)$ & $\text{BR}<1.5\cdot 10^{-5}$~\cite{Dzhelyadin:1980ti} & $\text{BR}\sim3\cdot10^{-29}\,\rm{TeV}^8/\Lambda^8$ &\ref{sec:EtaPiLepLep}
\\[0.1cm] 
$\eta\to \pi^+\pi^-\gamma$  & $\epsilon_{\alpha\beta\mu\nu}\,\eta \big(\partial^\nu \pi^+ \partial^\rho \partial^\mu \pi^- + \partial^\nu \pi^- \partial^\rho \partial^\mu \pi^+\big) \partial_\rho F^{\alpha\beta}$ & $p^6\,(\delta^2)$ & $A_{LR}=0.009(4)$~\cite{Workman:2022ynf} & $|A_{LR}|\sim5\cdot 10^{-16}\,\text{TeV}^4/\Lambda^4$ & \ref{sec:EtaPiPiGamma_charged}
\\[0.1cm] 
$\eta'\to \pi^+\pi^-\gamma$  & $\epsilon_{\alpha\beta\mu\nu}\,\eta' \big(\partial^\nu \pi^+ \partial^\rho \partial^\mu \pi^- + \partial^\nu \pi^- \partial^\rho \partial^\mu \pi^+\big) \partial_\rho F^{\alpha\beta}$ & $p^6\,(\delta^2)$ & $A_{LR}=0.03(4)$~\cite{Workman:2022ynf} & $|A_{LR}|\sim1\cdot 10^{-14}\,\text{TeV}^4/\Lambda^4$ & \ref{sec:EtaPiPiGamma_charged}
\\[0.1cm] 
$\eta \to \pi^0\pi^0 \gamma$ & $\epsilon_{\alpha\beta\mu\nu}\,\eta \big(\partial^\nu \pi^0 \partial^\rho \partial^\mu \pi^0 + \partial^\nu \pi^0 \partial^\rho \partial^\mu \pi^0\big) \partial_\rho F^{\alpha\beta}$ & $p^6\,(\delta^3)$ & $\text{BR}< 5\cdot 10^{-4}$~\cite{CrystalBall:2005zrs} & $\text{BR}\sim1\cdot10^{-29}\,\rm{TeV}^8/\Lambda^8$  &\ref{sec:EtaPiPiGamma_neutral}
\\[0.1cm] 
$\eta' \to \pi^0\pi^0 \gamma$ & $\epsilon_{\alpha\beta\mu\nu}\,\eta' \big(\partial^\nu \pi^0 \partial^\rho \partial^\mu \pi^0 + \partial^\nu \pi^0 \partial^\rho \partial^\mu \pi^0\big) \partial_\rho F^{\alpha\beta}$ & $p^6\,(\delta^3)$ & -- & $\text{BR}\sim2\cdot10^{-28}\,\rm{TeV}^8/\Lambda^8$
&\ref{sec:EtaPiPiGamma_neutral}
\\[0.1cm] 
$\eta' \to \eta\pi^0 \gamma$ & $\epsilon_{\alpha\beta\mu\nu}\,\eta' \partial^\mu \eta \partial^\nu \pi^0 F^{\alpha\beta}$ & $p^4\,(\delta^3)$ & -- & $\text{BR}\sim2\cdot10^{-28}\,\rm{TeV}^8/\Lambda^8$ &\ref{sec:EtapEtaPiGamma}
\\[0.1cm] 
$\eta'\to \eta\pi^0\pi^0 \gamma$ & $\eta'\partial_\mu\eta\pi^0\partial_\nu\pi^0 F^\munu$ & $p^4\,(\delta^2)$ & -- & $\text{BR}\sim2\cdot10^{-32}\,\rm{TeV}^8/\Lambda^8$ & \ref{sec:EtapEtaPiPiGamma}
\\[0.1cm] 
$\eta\to 3\pi^0 \gamma$ & $\partial_\mu\eta\partial_\nu\pi^0\partial_\alpha\pi^0\pi^0 \partial^\alpha F^\munu$ & $p^6\,(\delta^3)$ & $\text{BR}<6\cdot 10^{-5}$~\cite{CrystalBall:2005zrs} & $\text{BR}\sim1\cdot10^{-35}\,\rm{TeV}^8/\Lambda^8$ & \ref{sec:Eta3PiGamma}
\\[0.1cm] 
$\eta'\to 3\gamma$ & $\epsilon^{\mu\nu\rho\sigma}\partial_\alpha \eta' (\partial^\gamma F^{\alpha\beta})(\partial_\gamma \partial_\beta F_{\rho\sigma})F_\munu$ & $p^{10}\,(\delta^4)$ & $\text{BR}<1 \cdot10^{-4}$~\cite{Serpukhov-Brussels-LosAlamos-AnnecyLAPP:1987kiw} & $\text{BR}\sim 3\cdot 10^{-35}\,\text{TeV}^8/\Lambda^8$ &\ref{sec:Eta3Gamma}
\\[0.1cm] 
$\eta\to 3\gamma$ & $\epsilon^{\mu\nu\rho\sigma}\partial_\alpha \eta (\partial^\gamma F^{\alpha\beta})(\partial_\gamma \partial_\beta F_{\rho\sigma})F_\munu$ & $p^{10}\,(\delta^4)$ & $\text{BR}<4 \cdot10^{-5}$~\cite{Nefkens:2005ka} & $\text{BR}\sim 1\cdot 10^{-36}\,\text{TeV}^8/\Lambda^8$ &\ref{sec:Eta3Gamma}
\\[0.1cm] 
$\pi^0\to 3\gamma$ & $\epsilon^{\mu\nu\rho\sigma}\partial_\alpha \pi^0 (\partial^\gamma F^{\alpha\beta})(\partial_\gamma \partial_\beta F_{\rho\sigma})F_\munu$ & $p^{10}\,(\delta^4)$ & $\text{BR}<3.1 \cdot10^{-8}$~\cite{McDonough:1988nf} & $\text{BR}\sim 2\cdot 10^{-43}\,\text{TeV}^8/\Lambda^8$ & \ref{sec:Eta3Gamma}
\\[0.1cm] 
\bottomrule
\end{tabular}
}
\renewcommand{\arraystretch}{1.0}
\caption{
Overview of $C$- and $CP$-odd decays analyzed in this work. At the lowest possible order in soft momenta $p$, each process exhibits a unique representation in terms of mesonic degrees of freedom (up to overall normalizations and partial integration) as quoted in the second column, except for 
the decay $\eta\to 3\pi^0 \gamma$, for which we list only one possible momentum assignment.
Each operator can be seen as part of a Lagrangian once multiplied with a real-valued coupling constant. 
The decays are ordered according to increasing number of photons (the dilepton decays are assumed to proceed via single virtual photons), and furthermore according to increasing number of mesons involved.
As numerical results we quote the explicit dependence on the BSM scale $\Lambda$ derived from the LEFT operator $\bar\psi \dvec D_\mu \gamma_5\psi\bar\chi\gamma^\mu\gamma_5\chi$ in the fifth column. The  assumptions and simplifications these results (i.e., coupling constants $g_{1,2}$, left-right asymmetries $A_{LR}$, and branching ratios BR) rely on, can be found in the main text and referenced sections.}
\label{tab:Overview_C-odd_Decays}
\end{table}

First of all, we need to emphasize that the computation of most of the considered decays with ToPe$\chi$PT is rather meant to be a proof of principle. As we will see in the following, a rigorous evaluation would require the complete construction of the chiral basis for all $C$- and $CP$-odd LEFT operators also including higher orders in ToPe$\chi$PT, which leads to a large number of free LECs that cannot be fixed at the present stage. Therefore we do not investigate each single ToPe$\chi$PT operator. Instead, we focus on one set of operators that stands out, namely the ones derived from the LEFT source 
\begin{aligneq}\label{eq:ToPe_LEFT_Master}
    & \O_{\psi\chi}^{(a)} &&=&& \frac{v}{\Lambda^4}\ c_{\psi\chi}^{(a)}\, \bar\psi \dvec D_\mu \gamma_5\psi\bar\chi\gamma^\mu\gamma_5\chi\,, 
\end{aligneq}
for which, in comparison to Eq.~\eqref{eq:ToPe_LEFT_dim7}, we restored the explicit dependence on the EFT scale. This is the \textit{only} LEFT operator able to generate the $C$- and $CP$-violating contributions to \textit{all} mesonic decays listed in Table~\ref{tab:Overview_C-odd_Decays}
at the corresponding leading orders in $p$. The special feature that makes this operator unique in our analysis and allows us to in particular generate $\etap\to \pi^0\pi^+\pi^-$ 
at lowest order is its compositeness of both spurions $\lambda^{(\dagger)}$ and $\lambda_{L,R}$. These are the only decays, according to the third row in Table~\ref{tab:Overview_C-odd_Decays}, that occur at lowest order in $\delta$ and $p$.\footnote{The remaining non-vanishing terms in the leading-order Lagrangian from Eq.~\eqref{eq:Nc_Lagrangian} contribute either to interactions of the type $\eta'\eta (\pi^+\partial^\mu\pi^-+\pi^-\partial^\mu\pi^+)A_\mu$ or to operators with a larger number of mesons.}

Comparing the fourth and fifth columns of Table~\ref{tab:Overview_C-odd_Decays} we see that in order to set a realistic lower limit on $\Lambda$ in the TeV range, the biggest chance to find evidence for ToPe forces in future experiments is given by processes including an interference of SM and BSM contributions. This is no surprising result, as the respective observables scale linearly with BSM physics~\cite{Gardner:2019nid}, i.e., with $1/\Lambda^4$, as opposed to purely $C$-odd decays that can only be observed by quadratic effects scaling with $1/\Lambda^8$. Still, one can judge from our numerical results which pure BSM processes are more suitable candidates for experimental setups than others, e.g., $\pi^0\to3\gamma$ is the least suitable one since a realistic limit on $\Lambda\sim1\,\text{TeV}$ would require the experiment to measure a branching ratio that is roughly $10^{35}$ times smaller than the currently most stringent limit.

Before investigating each decay appearing in Table~\ref{tab:Overview_C-odd_Decays} in detail, we would like to comment on the method we use to estimate the included coupling constants. As a rough order-of-magnitude estimate we rely on naive dimensional analysis (NDA)~\cite{Manohar:1983md, Weinberg:1989dx, Georgi:1992dw, Jenkins:2013sda, Gavela:2016bzc}. The latter describes a method to estimate the scale of coupling constants of an effective field theory by simply counting powers of the mass dimension and keeping track of factors of $4\pi$ in each operator. This simple kind of power counting already proved to be very successful when estimating the order of magnitude of the LECs at $\O(p^4)$ in SM$\chi$PT, as illustrated in Ref.~\cite{Gavela:2016bzc}. 
To properly account for the matching of LEFT and ToPe$\chi$PT at the renormalization scale $\Lambda_\chi=4\pi \Fpi$ we pursue the following strategy:
for a generic coupling $g$ in any of our EFTs we introduce, in accordance to Weinberg's power counting scheme~\cite{Weinberg:1989dx}, a reduced coupling constant
\beq\label{eq:NDA_Weinberg}
    \tilde g\equiv (4\pi)^{2-n}\Lambda_\chi^{d-4} g\,,
\eeq
where $n$ indicates the number of involved fields and $d$ is the canonical dimension of the operator, i.e., the overall mass dimension of fields and derivatives but without counting couplings. This procedure renders the reduced coupling $\tilde g$ dimensionless and approximately of order unity. We first consider the case without dynamical photons and apply the rescaling to the coupling constant $C^{(a)}_{\psi\chi}\equiv\frac{v}{\Lambda^4}c^{(a)}_{\psi\chi}$ in Eq.~\eqref{eq:ToPe_LEFT_Master}, yielding
\beq\label{eq:NDA_example}
    C^{(a)}_{\psi\chi}\bar\psi (\vec \partial_\mu-\cev \partial_\mu) \gamma_5\psi\bar\chi\gamma^\mu\gamma_5\chi = \frac{(4\pi)^2}{\Lambda_\chi^3} \tilde C^{(a)}_{\psi\chi} \bar\psi (\vec \partial_\mu-\cev \partial_\mu) \gamma_5\psi\bar\chi\gamma^\mu\gamma_5\chi\,,
\eeq
where we obtained $\tilde C^{(a)}_{\psi\chi}=  C^{(a)}_{\psi\chi}\Lambda_\chi^3/(4\pi)^2\sim\O(1)$ from Eq.~\eqref{eq:NDA_Weinberg} with $n=4$ and $d=7$. When matching to a ToPe$\chi$PT operator with $m$ photons, we have to include additional factors of the reduced QED coupling, i.e., multiply the LEFT operator by $e^m/(4\pi)^m$.\footnote{This statement is consistent with Weinberg's power counting. If we take for instance the minimal coupling of a photon via the covariant derivative, i.e., the part left out in Eq.~\eqref{eq:NDA_example}, the NDA rule demands the reduced coupling $e\tilde C^{(a)}_{\psi\chi}=e C^{(a)}_{\psi\chi}\Lambda_\chi^3/(4\pi)^3\sim\O(1)$, as $n=5$ and $d=7$.} 
In this way we continuously keep track of what causes the $\chi$PT operator to occur, as necessary for a consistent description by NDA. 
At the level of ToPe$\chi$PT we proceed analogously for any given operator that is derived from this LEFT source and relate the LECs ($\bar{g}^{(a)}_i$) to reduced ones ($\tilde g^{(a)}_i$). To connect the corresponding reduced LECs $\tilde g^{(a)}_i\sim\O(1)$ to LEFT as the underlying theory we can set $\tilde g^{(a)}_i\sim\tilde C^{(a)}_{\psi\chi}$ for mesonic operators without dynamical photons (and similarly for ToPe$\chi$PT operators with additional photons). This is justified as NDA merely provides an order of magnitude estimate for coupling constants, which may well differ by a factor of a few. With this matching between reduced couplings in LEFT and ToPe$\chi$PT we can read off the approximate order of magnitude for the LECs $\bar{g}^{(a)}_i$, which can be used as a numerical input for the chiral theory.

%--------------------------------------------------------------
\boldmath
\subsection{$\etap\to \pi^0\pi^+\pi^-$}
\label{sec:Eta3Pi}
\unboldmath
%--------------------------------------------------------------
In this section we investigate possible $C$- and $CP$-violating contributions to the three-body decay $\etap\to \pi^0\pi^+\pi^-$ with ToPe$\chi$PT. As already pointed out in Ref.~\cite{Gardner:2019nid} these BSM contributions are driven by transitions of total isospin $I=0$ and $I=2$. Hence the amplitude can be decomposed as
\beq\label{eq:eta3Pi_decomposition}
    \M^{\not C}(s,t,u)= \M^{\not C}_0(s,t,u) + \M^{\not C}_2(s,t,u)\,.
\eeq
These contributions were constructed in Ref.~\cite{Akdag:2021efj} using dispersion-theoretical methods. Regression to the respective Dalitz-plot distribution~\cite{Anastasi:2016cdz} resulted in limits on the BSM coupling constants $g_0,g_2\in\mathbb{R}$ defined via
\begin{aligneq}\label{eq:Eta3Pi_Couplings}
    &\M^{\not C}_0(s,t,u) &&\approx i g_0(s-t)(t-u)(u-s)\,,\\
    &\M^{\not C}_2(s,t,u) &&\approx i g_2(t-u)\,.
\end{aligneq}
While $T$ violation arises naturally by the imaginary unit $i$, $C$ violation is encoded in the antisymmetry in the Mandelstam variables. In the following we investigate how to reconstruct these amplitudes with ToPe$\chi$PT, so that $g_0$ and $g_2$ serve as input for this effective theory allowing us to set limits on the BSM scale $\Lambda$.

%--------------------------------------------------------------
\subsubsection{Kinematics and isospin projections}
\label{sec:isospin_decomp}
%--------------------------------------------------------------
We define the $C$- and $CP$-odd contribution to the $T$-matrix element of $\etap \to \pi^+ \pi^- \pi^0$ by
\beq
    \big\langle \pi^+(p_+)\,\pi^-(p_-)\,\pi^0(p_0)\big|iT\big| \etap(P_{\etap})\big\rangle
    =(2\pi)^4\,\delta^{(4)}( P_{\etap}-p_+-p_--p_0)\,i\M^{\not C}(s,t,u)
\eeq
and work in the isospin limit, i.e., $\mpi\equiv M_{\pi^\pm}=M_{\pi^0}$. The Mandelstam variables are chosen to be 
\beq
    s = ( P_{\etap}-p_0)^2\,, \qquad
    t = ( P_{\etap}-p_+)^2\,, \qquad
    u = ( P_{\etap}-p_-)^2\,,  
\eeq
which are related to each other by
\beq
    s+t+u=M_{\etap}^2+3M_\pi^2\equiv 3r\,.
\eeq
The isospin decomposition of the isoscalar and isotensor three-pion final states, i.e., $|I=0\rangle$ and $|I=2\rangle$ respectively, are given by~\cite{Gaspero:2008rs}
\beq\label{eq:3pi_isospin_decomposition}
\begin{alignedat}{6}
&|2(2)\rangle &=&\  \frac{1}{2} &&\Big[\big(|\pi^+\pi^0\pi^-\rangle - |\pi^-\pi^0\pi^+\rangle\big) + \big(|\pi^0\pi^+\pi^-\rangle 
                                          - |\pi^0\pi^-\pi^+\rangle\big)\Big],\\
&|2(1)\rangle&=&\  \frac{1}{2\sqrt{3}} &&\Big[\big(|\pi^0\pi^-\pi^+\rangle - |\pi^0\pi^+\pi^-\rangle\big)
                                             - 2 \big( |\pi^-\pi^+\pi^0\rangle- |\pi^+\pi^-\pi^0\rangle \big)\\
                                            &&&&&\qquad\qquad\qquad\qquad\qquad
                                             +  \big(|\pi^+\pi^0\pi^-\rangle - |\pi^-\pi^0\pi^+\rangle\big)\Big]\,,\\
&|0(1)\rangle &=&\  \frac{1}{\sqrt{6}}&&\Big[  \big(|\pi^0\pi^-\pi^+\rangle -  |\pi^0\pi^+\pi^-\rangle\big)
                                         +\big(|\pi^-\pi^+\pi^0\rangle - |\pi^+\pi^-\pi^0\rangle\big)\\
                                            &&&&&\qquad\qquad\qquad\qquad\qquad
                                          + \big(|\pi^+\pi^0\pi^-\rangle -|\pi^-\pi^0\pi^+\rangle\big) \Big]\,,
\end{alignedat}
\eeq
where the integer in parenthesis denote the isospin of the first two pions. From this Clebsch--Gordan series one can already judge that the isoscalar and isotensor contributions are antisymmetric under exchange of the charged pions and thus $C$-violating (the $|0\rangle$ state has also an enhanced symmetry under exchange of any two pions). Hence each ToPe$\chi$PT operator that contributes to $\etap\to \pi^0\pi^+\pi^-$ can for instance be associated with an isospin state of the form $|\pi^-\pi^+\pi^0\rangle-|\pi^+\pi^-\pi^0\rangle$. With Eq.~\eqref{eq:3pi_isospin_decomposition} we can project out the single isospin contributions by means of\footnote{One could as well use $|\pi^+\pi^0\pi^-\rangle-|\pi^-\pi^0\pi^+\rangle=\frac{1}{\sqrt{3}}(\sqrt{2}|0(1)\rangle +|2(1)\rangle+\sqrt{3}|2(2)\rangle )$ or the order $|\pi^0\pi^-\pi^+\rangle-|\pi^0\pi^+\pi^-\rangle=\frac{1}{\sqrt{3}}(\sqrt{2}|0(1)\rangle +|2(1)\rangle-\sqrt{3}|2(2)\rangle)$. As we cannot distinguish between the states $|2(1)\rangle$ and $|2(2)\rangle$ we can only make a statement for the overall isospin $2$ contribution. The latter, as well as the overall isoscalar contribution, is the same for all of the three sequences $|\pi^-\pi^+\pi^0\rangle-|\pi^+\pi^-\pi^0\rangle\,, |\pi^+\pi^0\pi^-\rangle-|\pi^-\pi^0\pi^+\rangle$, and $|\pi^0\pi^-\pi^+\rangle-|\pi^0\pi^+\pi^-\rangle$, as long as we stay consistent in notation. Hence, it does not matter which order we choose for the pions in the isospin state for our ToPe$\chi$PT operators. }
\beq\label{eq:ChPT_eta3pi_isosin_decomp}
    |\pi^-\pi^+\pi^0\rangle-|\pi^+\pi^-\pi^0\rangle=\frac{1}{\sqrt{3}}\big(\sqrt{2}\,|0(1)\rangle -2\,|2(1)\rangle\big)\,.
\eeq

%--------------------------------------------------------------
\subsubsection{Limits on the BSM physics scale}
\label{sec:Eta3Pi_scale}
%--------------------------------------------------------------
Starting from the large-$N_c$ Lagrangian at leading order $\delta^0$, cf.\  Eq.~\eqref{eq:Nc_Lagrangian}, we can evaluate the matrix element $\M^{\not C}$ upon expanding $\bar U$ up to second order in $\bar \Phi$ and neglecting photons. We will first investigate the decay of the $\eta$ meson and discuss the $\eta'$ at the end of this section.

Whenever possible, we conventionally eliminate derivatives acting on the decay particle by partial integration, helping us to find a more compact notation of our operators.
The operator generating the $C$- and $CP$-odd contributions to the desired decay is  
\begin{aligneq}\label{eq:ChPT_evaluation_eta3pi}
 \bar{\L}_{\slashed{C}P\slashed{T}}=i\frac{v}{\Lambda^4\Fpi^4}\, 2\N_{\eta\to3\pi}\, \eta\partial^\mu\pi^0(\pi^+\partial_\mu\pi^- - \pi^-\partial_\mu\pi^+)+\ldots\,,
\end{aligneq}
where the ellipsis includes operators that cannot generate the desired transition at $\O(\delta^0)$ and the normalization, given as a linear combinations of Wilson coefficients and LECs, reads
\beq
     \N_{\eta\to3\pi}=4\sqrt{\frac{2}{3}} \big(c^{(a)}_{uu}-c^{(a)}_{ud}-c^{(a)}_{du}+c^{(a)}_{dd}\big)\big(\bar g^{(a)}_3-\bar g^{(a)}_2\big)\,.
\eeq
We see that the leading-order contributions to $\eta\to \pi^0\pi^+\pi^-$ arises solely from $\bar{X}_{\psi\chi}^{(a)}$ and furthermore note that all contributions proportional to $c^{(a)}_{\psi\chi}$ with $\psi=s$ and/or $\chi=s$ vanish. The evaluation of the corresponding amplitude $\M^{\not C}$ yields
\begin{aligneq}
     \M^{\not C} &=i\frac{v}{\Lambda^4\Fpi^4}2\N_{\eta\to3\pi}\,
    p_0(p_--p_+)=i\frac{v}{\Lambda^4\Fpi^4}\N_{\eta\to3\pi}\,(t-u)\,.
\end{aligneq}

In order to match the included coupling constants to known observables, we first have to separate the different isospin contributions to this matrix element.
According to Eq.~\eqref{eq:ChPT_eta3pi_isosin_decomp}, $\M^{\not C}$ decomposes into\footnote{The decomposition in Eq.~\eqref{eq:eta3Pi_decomposition}, used for the dispersive approach in Ref.~\cite{Akdag:2021efj}, absorbs the relative factors $\sqrt{2/3}$ and $2/\sqrt{3}$ directly in the coupling constants $g_0$ and $g_2$, respectively.} 
\beq
\M^{\not C}=\frac{1}{\sqrt{3}} \big( \sqrt{2}\M^{\not C}_0 - 2\M^{\not C}_2 \big)\,.
\eeq
The isoscalar and isotensor contributions in this equation can be evaluated in compliance with Eq.~\eqref{eq:3pi_isospin_decomposition} by taking the appropriate linear combinations with interchange of pions. Noting that under $\pi^+\leftrightarrow \pi^-$ ($\pi^+\leftrightarrow \pi^0$, $\pi^-\leftrightarrow \pi^0$)  the Mandelstam variables exchange as $t\leftrightarrow u$, ($t\leftrightarrow s$, $s\leftrightarrow u$), the amplitudes $\M^{\not C}_{0,2}$ become 
\begin{aligneq}
    &\M^{\not C}_0 &&= i\N_{\eta\to3\pi}\frac{v}{\Lambda^4\Fpi^4}\frac{1}{\sqrt{6}}\big[(s-t)+(t-u)+(u-s)\big]=0\,,\\[0.1cm]
    &\M^{\not C}_2 &&= i\N_{\eta\to3\pi}\frac{v}{\Lambda^4\Fpi^4}\frac{1}{2\sqrt{3}}\big[(s-t)-2(t-u)-(s-u)\big]=-i\N_{\eta\to3\pi}\frac{v}{\Lambda^4\Fpi^4}\frac{\sqrt{3}}{2}(t-u)\,,
\end{aligneq}
and hence $\M^{\not C}=-2/\sqrt{3}\, \M^{\not C}_2$ at chiral order $\delta^0$. 

Similarly, all contributions of $\M^{\not C}_0$ vanish in the isospin limit up to $\O(\delta)$
and we thus need at least six derivatives to have enough freedom to reproduce the totally antisymmetric behaviour of the isospin $0$ state. This fact was already known decades ago, cf.\ Ref.~\cite{Prentki:1965tt}. Hence, to evaluate a possible contribution of $\M_0^{\not C}$ we have to construct the contributions to $X^{(a)}_{\chi\psi}$ at order $\delta^2$. Unfortunately this involves the construction of an overwhelming amount of operators with
independent free parameters to fix the already strongly suppressed amplitude $\M_0^{\not C}$. One can bring all of these numerous operators to the form of Eq.~\eqref{eq:Eta3Pi_Couplings} and effectively fix one overall normalization that would correspond to $g_0$. However, this would not provide new physical insights, because these operators at order $\delta^2$ are less likely to contribute to any other process in a meaningful way and the theory does thus not gain any predictive power by fixing this normalization. 
Nevertheless, we give one arbitrarily chosen example how the $I=0$ contribution arises from $\bar{X}_{\psi\chi}^{(a)}$, i.e., 
\begin{aligneq}
    \bar{X}_{\psi\chi}^{(a)}\supset \frac{v}{\Lambda^4}c_{\psi\chi}^{(a)} \bar g^{(a)}_0i\vev{\big(
    \lambda\lambda_L\partial_\mu\partial_\nu\partial_\alpha \bar U^\dagger\partial^\mu\partial^\nu \bar U \partial^\alpha \bar U^\dagger
    +
    \lambda^\dagger\lambda_R\partial_\mu\partial_\nu\partial_\alpha \bar U\partial^\mu\partial^\nu \bar U^\dagger \partial^\alpha \bar U\big)-\hc}\,,
\end{aligneq}
as a proof of concept.

However, as we control the dominant isotensor contribution, we can use the result of Ref.~\cite{Akdag:2021efj}, i.e., $g_2=-0.0093(46)\, \text{GeV}^{-2}$, to place bounds on the BSM scale $\Lambda$. For simplicity, we first consider the NDA estimate of a generic meson operator included in Eq.~\eqref{eq:ChPT_evaluation_eta3pi} like
\beq
i\frac{v}{\Lambda^4\Fpi^4}c^{(a)}_{\psi\chi}\bar g^{(a)}_i\eta \partial_\mu\pi^0 (\pi^+\partial_\mu\pi^- - \pi^-\partial_\mu\pi^+)\,.
\eeq
According to Weinberg's power counting from Eq.~\eqref{eq:NDA_Weinberg} the reduced coupling for this generic operator with $n=4$ and $d=6$ reads
\beq
    \tilde G_i^{(a)}\equiv \frac{v}{\Lambda^4\Fpi^4}c^{(a)}_{\psi\chi}\bar g^{(a)}_i \frac{\Lambda_\chi^2}{(4\pi)^2} \sim\O(1)\,.
\eeq
On the other side, we have already seen in Eq.~\eqref{eq:NDA_example} that the reduced coupling for the underlying LEFT operator $\bar X_{\psi\chi}^{(a)}$, with $n=4$ and $d=7$, is
\beq
    \tilde C^{(a)}_{\psi\chi}\equiv \frac{v}{\Lambda^4} c^{(a)}_{\psi\chi}\frac{\Lambda_\chi^3}{(4\pi)^2}\sim\O(1)\,.
\eeq
As both couplings are by construction of the same order of magnitude, we can set $\tilde C^{(a)}_{\psi\chi}\sim\tilde G_i^{(a)}$ to obtain $\bar g_i^{(a)}\sim \Lambda_\chi  \Fpi^4$. As long as there is no unexpected fine tuning of the Wilson coefficients or LECs, we can apply the same NDA estimate to their linear combination encoded in the normalization $\N_{\eta\to3\pi}$ and therefore obtain $\N_{\eta\to3\pi}\sim \Lambda_\chi  \Fpi^4$. 
Combining this estimate with the external input for the isotensor coupling (for simplicity we will only consider the respective central value) by means of
\beq
    \M^{\not C} = i\frac{v}{\Lambda^4\Fpi^4} \N_{\eta\to3\pi}(t-u)\overset{!}{=} i g_2(t-u)\,,
\eeq 
the currently best experimental precision for the $\eta\to\pi^0\pi^+\pi^-$ Dalitz plot~\cite{Anastasi:2016cdz} can merely set
\beq\label{eq:scale_et3pi}
    \Lambda\sim\left(\frac{v}{|g_2|} \Lambda_\chi\right)^{1/4}> 13\,\text{GeV}
\eeq
as the lower limit on the BSM scale $\Lambda$.\footnote{Note that NDA does not fix the sign of the normalization $\N_{\eta\to3\pi}$. In order to pick the correct sign of the latter and thereby ensure that $\Lambda\in\mathds{R}$ we take the absolute value of $g_2$.} 
This result depends of course strongly on the validity of NDA, but should give a reasonable approximation for the order of magnitude. If we were to estimate a more realistic limit on $\Lambda$, i.e., a value in the TeV range, one should expect an increase by a factor of $10^2$ (staying in the framework of naive dimensional analysis). Hence, to set a reasonable limit on $\Lambda$, let us take for instance $\Lambda\sim 1\,\text{TeV}$, the charge asymmetry in the $\eta\to\pi^0\pi^+\pi^-$ Dalitz-plot distribution, which is proportional to $g_2$, has to be roughly $10^{8}$ times smaller than the current value of Ref.~\cite{Anastasi:2016cdz}, which can be readily obtained from Eq.~\eqref{eq:scale_et3pi}:\footnote{Similarly, the more suppressed isoscalar coupling would have to take a value $g_0\sim \Lambda_\chi^4 g_2$ as demanded by NDA.}
\beq
    |g_2|\sim\frac{v}{\Lambda^4} \Lambda_\chi \approx 3\cdot 10^{-4}\,\text{TeV}^{2}/\Lambda^4\,.
\eeq

We now turn the focus on the decay amplitude $\eta'\to\pi^0\pi^+\pi^-$ that can be computed with the same Lagrangian of Eq.~\eqref{eq:Nc_Lagrangian} and reads
\begin{aligneq}
     \M^{\not C} &= i\frac{v}{\Lambda^4\Fpi^4}\N_{\eta'\to3\pi}(t-u)\,,
\end{aligneq}
where $\N_{\eta'\to3\pi}=\N_{\eta\to3\pi}/\sqrt{2}$. Hence, the $C$- and $CP$-violating contributions to the decays $\eta\to\pi^0\pi^+\pi^-$ and $\eta'\to\pi^0\pi^+\pi^-$ are maximally correlated at leading order in large-$N_c$. They merely differ by their available phase space and an overall factor $\sqrt{2}$. 
Unfortunately, the current data situation~\cite{BESIII:2016tdb} does not allow for a rigorous regression to the respective Dalitz-plot distribution, cf.\ Ref.~\cite{Akdag:2021efj}. Therefore we can at this point not cross-check the limit on $g_2$ set above.

%--------------------------------------------------------------------------
\boldmath
\subsection{$\eta'\to\eta\pi^+\pi^-$}
\label{sec:EtapEtaPiPi}
\unboldmath
%--------------------------------------------------------------------------
In this section we focus on another interference of SM contributions and ToPe forces. The decay $\eta'\to\eta\pi^+\pi^-$ is driven by a transition of total isospin $I=1$ and is at leading order of the form 
\begin{aligneq}\label{eq:EtapEtaPiPi_Coupling}
    &\M^{\not C}_1(s,t,u) &&= i g_1(t-u)\,,
\end{aligneq}
with the same reasoning as for $\etap\to\pi^0\pi^+\pi^-$. A result for the isovector coupling $g_1$ (within the scope of current experimental precision) can again be found in Ref.~\cite{Akdag:2021efj}. In the following we use the same kinematics as in the previous chapter but replace $p_0 \to P_\eta$ and $P_{\etap}\to P_{\eta'}$.

There is no leading-order contribution from Eq.~\eqref{eq:Nc_Lagrangian} that does not vanish after partial integration.
Non-vanishing contributions could be generated at subleading order in $\delta$, i.e., in the $N_c$ counting; at higher orders in the chiral expansion, which, given that we look for the exact energy dependence of Eq.~\eqref{eq:EtapEtaPiPi_Coupling}, would amount to quark-mass suppression; or via isospin-breaking mixing of $\pi^0$ and $\etap$, surely the smallest and most negligible effect.
We thus consider operators at $\O(\delta^1)$ but with $\O(p^2)$ and use the freedom of the large-$N_c$ expansion to include the $\eta'$ via the chiral singlet $(\varphi+\theta)$. Henceforth we will directly drop the contribution of $\theta$ entering this chiral building block. Note that $\varphi$ includes a linear combination of $\eta$ and $\eta'$. There are only a couple of operators that generate the desired transition at the given order, as for instance 
\begin{aligneq}\label{eq:chiral_etap_eta_pi_pi}
\bar{X}_{\psi\chi}^{(a)} \supset \frac{v}{\Lambda^4}c_{\psi\chi}^{(a)} \bar g^{(a)}_5\cdot \varphi\, \vev{
\big(\lambda \bar U^\dagger\lambda_R \partial_\mu \bar U\partial^\mu\bar U^\dagger
-\lambda^\dagger\bar U\lambda_L \partial_\mu\bar U^\dagger\partial^\mu\bar U\big)+\hc
}\,,
\end{aligneq}
which gives rise to
\beq
    \sum_{\psi,\chi} \bar{X}_{\psi\chi}^{(a)}\supset i\frac{v}{\Lambda^4\Fpi^4}2\N_{\eta'\to\eta\pi\pi}\,
\eta'\partial_\mu\eta(\pi^+\partial_\mu\pi^- - \pi^-\partial_\mu\pi^+)\,.
\eeq 
The normalization of this operator is 
\beq
    \N_{\eta'\to\eta\pi\pi}=4\frac{\sqrt{2}}{3} \big(c_{uu}^{(a)}-c_{ud}^{(a)}+c_{du}^{(a)}-c_{dd}^{(a)}\big)\bar g^{(a)}_5\,.
\eeq
We note that every other operator able to generate $\eta'\to\eta\pi^+\pi^-$ at $\O(p^2)$, which can be obtained by letting one of the two derivatives in Eq.~\eqref{eq:chiral_etap_eta_pi_pi} act on another chiral buildung block, can analytically be written in this form and absorbed by a shift in the normalization. The respective matrix element reads
\beq
    \M^{\not C}_1=i\frac{v}{\Lambda^4\Fpi^4}2\N_{\eta'\to\eta\pi\pi}\,P_\eta(p_--p_+)=i\frac{v}{\Lambda^4\Fpi^4}\N_{\eta'\to\eta\pi\pi}(t-u)\,.
\eeq
With the NDA prediction $\N_{\eta'\to\eta\pi\pi}\sim \Lambda_\chi  \Fpi^4$, this result can be compared to the isovector coupling $g_1=0.7(1.0)\, \text{GeV}^{-2}$ of Ref.~\cite{Akdag:2021efj}. This reveals that the current experimental limit on the Dalitz-plot asymmetries~\cite{BESIII:2017djm} constrains the new-physics scale roughly as
\beq\label{eq:scale_EtapEtaPiPi}
    \Lambda\sim\left(\frac{v}{g_1} \Lambda_\chi\right)^{1/4}> 4\,\text{GeV}\,,
\eeq
where we applied the central value of $g_1$.
A scale $\Lambda\sim 1\,\text{TeV}$ could be tested if the experiment restricted the isovector coupling $g_1$ and thus the corresponding mirror asymmetry to a value that is approximately $10^{-8}$ times the current value.

%--------------------------------------------------------------
\boldmath
\subsection{$\etap\to \pi^0\gamma^\ast$ and $\eta'\to\eta\gamma^\ast$}
\label{sec:EtaPiGamma}
\unboldmath
%--------------------------------------------------------------
In this section we consider the simplest $C$-violating decays of the $\etap$ into an odd number of photons. To shorten the notation we will refer to $\etap\to \pi^0\gamma^{\ast}$ and $\eta'\to\eta\gamma^{\ast}$ by $X\to Y\gamma^{\ast}$ and ignore the decay into a real photon, as it has to either violate gauge invariance or does not preserve angular momentum~\cite{Sakurai:1964, Gan:2020aco}. 
The latter enforces a relative $P$-wave between the pion and photon, which moreover demands that parity is conserved and hence $CP$ is violated. Already in the 1960s it was proposed that the Lagrangian driving the $\eta\to \pi^0\gamma^\ast$ transition starts at chiral order $p^4$~\cite{Barrett:1965ia}, by means of
\beq\label{eq:eta_pi_gamma_1960}
    \L_{\eta\to \pi^0\gamma^\ast}\propto \partial_\mu \eta\, \partial_\nu \pi^0 F^\munu+\O(p^6)\,.
\eeq
This manifestation of gauge invariance was also applied in the SM contributions to kaon decays~\cite{Ecker:1987, DAmbrosio:1998gur} and holds similarly for all processes $X\to Y\gamma^{\ast}$ with pseudoscalars $X$, $Y$.

Without deriving the full set of mesonic operators for $\bar{X}_{\psi\chi}^{(a)}$ at next-to-leading order, we just give one example of how this operator contributes to $X\to Y\gamma^\ast$. As similarly argued in Sect.~\ref{sec:matching_dim7_LEFT_bilinear}, a single-trace operator with the correct discrete symmetries that includes both derivatives vanishes due to the antisymmetry of $F^\munu$. Therefore we have to increase the order of $\delta$ by either using $\partial_\mu \varphi$, i.e., the derivative of the chiral singlet, or simply writing down a double-trace operator. To recover the form of Eq.~\eqref{eq:eta_pi_gamma_1960} we stick to the latter strategy and obtain
\beq\label{eq:ChPT_eta_pi_gamma}
    \bar{X}_{\psi\chi}^{(a)}\supset \frac{v}{\Lambda^4}c_{\psi\chi}^{(a)} \bar{g}^{(a)}_6\vev{\big(\lambda f_L^\munu \partial_\mu \bar U^\dagger-\lambda^\dagger f_R^\munu \partial_\mu \bar U\big)-\hc}\vev{\lambda_L\partial_\nu \bar U^\dagger \bar U-\lambda_R\partial_\nu \bar U\bar U^\dagger}
\eeq
at $\O(\delta^2)$, yielding 
\begin{aligneq}\label{eq:ChPT_evaluation_eta_pi_gamma}
    \sum_{\psi,\chi}  \bar{X}_{\psi\chi}^{(a)} \supset  e\frac{v}{\Lambda^4\Fpi^2}\N_{X\to Y\gamma^\ast}\partial_\mu X\, \partial_\nu Y F^\munu\equiv\L_{X\to Y\gamma^\ast}\,,
\end{aligneq}
with
\begin{aligneq}\label{eq:normalizations_XYGamma}
&\N_{\eta\to\pi^0\gamma^\ast}&&=\bar{g}^{(a)}_6 \frac{8}{3}\sqrt{\frac{2}{3}}\Big(-4c^{(a)}_{ud}+2c^{(a)}_{us}-2c^{(a)}_{du}+c^{(a)}_{ds}+c^{(a)}_{su}-c^{(a)}_{sd}\Big)\,,\\[0.1cm]
&\N_{\eta'\to\pi^0\gamma^\ast}&&=-\bar{g}^{(a)}_6 \frac{16}{3\sqrt{3}}\Big(2c^{(a)}_{ud}+2c^{(a)}_{us}+c^{(a)}_{du}+c^{(a)}_{ds}+c^{(a)}_{su}-c^{(a)}_{sd}\Big)\,,\\[0.1cm]
&\N_{\eta'\to\eta\gamma^\ast}&&=\bar{g}^{(a)}_6 \frac{8\sqrt{2}}{3}\Big(-2c^{(a)}_{us}+c^{(a)}_{ds}-c^{(a)}_{su}-c^{(a)}_{sd}\Big)\,.
\end{aligneq}
One has to keep in mind that each of our LEFT operators, once taken to $\O(\delta^2)$, may in principle also contribute at the same order of magnitude as $\bar{X}_{\psi\chi}^{(a)}$. But again, the full set of NLO expression derived from all $C$- and $CP$-violating LEFT operators is beyond the scope of this work. However, we already proved at this point that the decays at hand provide orthogonal probes of ToPe forces as their normalizations in Eq.~\eqref{eq:normalizations_XYGamma} are linearly independent.

Still, we would like to comment on the contribution of the second original dimension-7 LEFT operator from Eq.~\eqref{eq:ToPe_LEFT_dim7}, i.e., the bilinear $\O^{(b)}_\psi$.
We note that the leading-order contribution of the latter in the $SU(3)$ case does not contribute to the desired decays of order $p^4$. However, we can still consider the $U(3)$ version of this operator by including $\varphi$ using
\beq
    \bar X^{(b)}_{\psi}\supset i\frac{v}{\Lambda^4}c_{\psi}^{(b)}\bar{g}^{(b)}_2\,(\partial_\mu\varphi)\vev{\lambda \partial_\nu \bar U^\dagger-\lambda^\dagger \partial_\nu \bar U} F^{\mu\nu}\,,
\eeq
which involves only one trace and is therefore of the
order $\O(\delta^2)$ and $\O(p^4)$. 

We continue with $\L_{X\to Y\gamma^\ast}$ from Eq.~\eqref{eq:ChPT_evaluation_eta_pi_gamma} and consider the decay of the virtual photon in Sect.~\ref{sec:EtaPiLepLep} to extract physical observables. To this end, we quote the normalization according to NDA as similarly derived in Sect.~\ref{sec:Eta3Pi_scale}, i.e.,
$\N_{X\to Y\gamma^\ast}\sim \Fpi^4/\Lambda_\chi$, and remark that any other leading-order contribution derived from $\O^{(a)}_{\psi\chi}$ just leads to additional linear combinations of LECs and Wilson coefficients, which can be absorbed by a redefinition of $\N_{X\to Y\gamma^\ast}$ but do not affect the naive power counting.
For further calculations it is convenient to describe the decay $X\to Y\gamma^\ast$ in terms of a singularity-free electromagnetic transition form factor $F_{XY}(s)$.
Using Poincar\'e invariance and current conservation, the amplitude can be decomposed as~\cite{Bernstein:1965, DAmbrosio:1998gur}
\beq\label{eq:formfac}
    \langle Y(p)|J^{em}_\mu(0)|X(P)\rangle=
    -i\left[ s(P+p)_\mu -(P^2-p^2)q_\mu\right]\,F_{XY}(s),
\eeq
with electromagnetic current $J^{em}_\mu$, $q_\mu=(P-p)_\mu$, and $s=q^2$.

%--------------------------------------------------------------
\boldmath
\subsection{$\etap\to \pi^0\ell^+\ell^-$ and $\eta'\to \eta\ell^+\ell^-$}
\label{sec:EtaPiLepLep}
\unboldmath
%--------------------------------------------------------------
The framework presented in this paper allows us to consider the decays $\etap\to \pi^0\ell^+\ell^-$ and $\eta'\to \eta\ell^+\ell^-$ (abbreviated with $X\to Y\ell^+\ell^-$) in two ways: we can either compute the decay chain $X\to Y \gamma^\ast\to Y\ell^+\ell^-$ or even directly access it as a point interaction originating from semi-leptonic operators. 
Note that in the decay chain the photon 
pole $1/q^2$ cancels against a necessary $q^2$ term in the numerator if the coupling to $\ell^+ \ell^-$ respects gauge invariance. As a consequence, the
single-photon and the direct amplitude cannot be separated by searching for a photon pole~\cite{Bernstein:1965,Barrett:1965ia,Bazin:1968zz}.
Note that within the SM, these decays can be generated via two-photon intermediate states~\cite{Cheng:1967zza,Ng:1992yg,Escribano:2020rfs}.

The operator from Eq.~\eqref{eq:ChPT_evaluation_eta_pi_gamma} coupling to a conserved lepton current gives a dominant contribution to $X\to Y\ell^+\ell^-$ if the underlying $C$- and $CP$-violating mechanism is driven by a one-photon exchange. 
The only semi-leptonic operator at order $\delta^0$ is the one from Eq.~\eqref{eq:Nc_Lagrangian}, which does not generate the desired transition. 
Instead of deriving the full set of operators at $\O(\delta)$ for all six semi-leptonic LEFT sources, we can easily discard most of them with the following considerations.
First of all, the photonic semi-leptonic operators are obviously not involved at lowest order in $\alpha$ as we have no photon in the initial or final state. Moreover, operators including a pseudoscalar or axialvector lepton bilinear must couple to an hadronic operator that is $P$-odd to preserve parity. On the hadronic level, a $P$-odd operator that involves an even number of pseudoscalars (in our case $\etap,\pi^0$) requires a contraction with the Levi-Civita symbol, as explained in more detail Sect.~\ref{sec:EtaPiPiGamma_charged}. The only Lorentz structure left that can contract with the $\epsilon$-tensor includes three derivatives, which have to act on different $U^{(\dagger)}$ to generate a non-vanishing operator. However, this goes along with an interaction containing at least three pseudoscalars. Hence, the only LEFT operator that can contribute to $X\to Y\ell^+\ell^-$ at lowest order is the one involving the $P$-even lepton bilinear, i.e., $\O_{\ell\psi}^{(u)}$. 
Using partial integration and the Dirac equation for the leptons, one can pin down the requested leading-order semi-leptonic four-point interaction to only one operator at $\O(\delta)$ and $\O(p^2)$: 
\begin{aligneq}\label{eq:ChPT_LEFT_dim8_semi-leptonic_NLO}
    \bar X_{\ell\psi}^{(u)}\supset \frac{c^{(u)}_{\ell\psi}}{\Lambda^4} \,\bar g_1^{(u)}\,i\varphi
    &\vev{\lambda_L\partial_\mu \bar{U}^\dagger \bar{U}-\lambda_R \partial_\mu \bar{U} \bar{U}^\dagger}\bar\ell \gamma^\mu \ell\,.
\end{aligneq}
This chiral operator gives rise to 
an expression of the form
\beq\label{eq:XYll_dim8}
   (X \partial_\mu Y)\bar\ell \gamma^\mu \ell\,.
\eeq
For now we continue with the one-photon exchange driven by the LEFT operator $\O_{\psi\chi}^{(a)}$ in focus of this chapter and define the corresponding $T$-matrix element as 
\beq
    \langle Y(p) \ell^+(p_{\ell^+})\ell^-(p_{\ell^-})| iT | X(P) \rangle \equiv (2\pi)^4 \delta^{(4)}(P-p-p_{\ell^+}-p_{\ell^-}) i\M(s,t,u)\,,
\eeq
where the amplitude $\M$ depends on the three Mandelstam variables
\beq
    s=(P-p)^2\,, \hspace{.5cm} t_\ell=(P-p_{\ell^+})^2\,, \hspace{.5cm}  u_\ell=(P-p_{\ell^-})^2\,,
\eeq
which obey $s+t+u=M_X^2+M_Y^2+2m_\ell^2$.
Starting from Eq.~\eqref{eq:ChPT_evaluation_eta_pi_gamma} we allow the photon with momentum $q\equiv p_{\ell^+}+p_{\ell^-}$ to decay into a lepton pair, so that the amplitude becomes
\begin{aligneq}\label{eq:matrix_element_eta_pi_2ell}
    i\M(s,t,u)&=\frac{v}{\Lambda^4 \Fpi^2}e^2 \N_{X \to Y\gamma^\ast}\frac{1}{s} \,(P_\mu p_\nu - P_\nu p_\mu) q^\mu \,\Bar{u}_r( p_{\ell^-})\gamma^\nu v_{r'}( p_{\ell^+})\\
    &=\frac{v}{\Lambda^4\Fpi^2}e^2\N_{X \to Y\gamma^\ast}\,P_\nu \,\Bar{u}_r( p_{\ell^-})\gamma^\nu v_{r'}( p_{\ell^+})
    \\
    &
   = e^2(P+p)_\nu F_{XY}(s)\,\Bar{u}_r( p_{\ell^-})\gamma^\nu v_{r'}( p_{\ell^+})
    \,.
\end{aligneq}
In the second line we simplified the expression using $q^2=s$ and the fact that $q_\nu$ contracted with the lepton current vanishes as demanded by the Dirac equation. As a consistency check, we expressed the amplitude in terms of the transition form factor $F_{XY}(s)=v\N_{X \to Y\gamma^\ast}/(2e\Lambda^4\Fpi^2)$ from Eq.~\eqref{eq:formfac}. We observe that the form factor is a constant at leading chiral order, which meets our expectations. 
Note that the second line in Eq.~\eqref{eq:matrix_element_eta_pi_2ell},
which comes from a LEFT operator of dimension 7, gives the same structure as the chiral operator~\eqref{eq:ChPT_LEFT_dim8_semi-leptonic_NLO}, which comes from an LEFT operator of dimension 8.

In analogy to Ref.~\cite{Kubis:2010mp}, the doubly differential decay width reads
\beq
     \frac{\diff\Gamma_{X\to Y \ell^+ \ell^-}}{\diff s\, \diff \tau}
     =\left(\frac{v}{\Lambda^4\Fpi^2}\right)^2\frac{\alpha^2}{64\pi M_X^3} \, \N^2_{X \to Y\gamma^\ast}\, \left(\lambda(s,M_X^2,M_{Y}^2)-\tau^2\right),
\eeq
with the electromagnetic fine structure constant $\alpha=e^2/4\pi$, the K\"all\'en function $\lambda(x,y,z)=x^2+y^2+z^2-2(xy+xz+yz)$, and the Lorentz invariant $\tau=t_\ell-u_\ell$. An analytic integration over $\tau$ yields
\beq
    \frac{\diff\Gamma_{X\to Y \ell^+ \ell^-}}{\diff s}
    =\left(\frac{v}{\Lambda^4\Fpi^2}\right)^2\frac{\alpha^2}{32\pi M_X^3}\,\N^2_{X \to Y\gamma^\ast}\, \lambda^{3/2}(s,M_X^2,M_{Y}^2)\, \sigma_\ell(s)\left(1-\frac{\sigma^2_\ell(s)}{3}\right),
\eeq
where $\sigma_\ell(s)=\sqrt{1-4m_\ell^2/s}$ and the physical range is restricted to $4m_\ell^2\leq s \leq (M_X-M_{Y})^2$. After an additional numeric integration over $s$ we can investigate how rigorously the bounds on the new-physics scale $\Lambda$ can be placed with measurements of the electronic and muonic decay channels. With the shorthand  notation
\beq
    \tilde \Lambda_{X\to Y \ell^+ \ell^-}\equiv \frac{v^2}{\Fpi^4\, \Gamma_{X\to Y \ell^+ \ell^-}}\frac{\alpha^2}{32\pi M_X^3}\N^2_{X \to Y\gamma^\ast}\,\cdot 10^{-2}\text{GeV}^8
\eeq
we obtain the limits 
\begin{aligneq}
    &\Lambda\sim (0.087\,\tilde \Lambda_{\eta\to \pi^0 e^+ e^-}&&)^{1/8}&&>2.3\,\text{GeV}\,, \qquad
    &&\Lambda\sim (0.027\,\tilde \Lambda_{\eta\to \pi^0 \mu^+ \mu^-}&&)^{1/8}&&>2.1\,\text{GeV}\,,
    \\[0.1cm]
    &\Lambda\sim (10.1\,\tilde \Lambda_{\eta'\to \pi^0 e^+ e^-}&&)^{1/8}&&>3.5\,\text{GeV}\,, \qquad
    &&\Lambda\sim (7.4\,\tilde \Lambda_{\eta'\to \pi^0 \mu^+ \mu^-}&&)^{1/8}&&>3.5\,\text{GeV}\,,
    \\[0.1cm]
    &\Lambda\sim (1.0\,\tilde \Lambda_{\eta'\to \eta e^+ e^-}&&)^{1/8}&&>0.7\,\text{GeV}\,, \qquad
    &&\Lambda\sim (0.3\,\tilde \Lambda_{\eta'\to \eta \mu^+ \mu^-}&&)^{1/8}&&>1.1\,\text{GeV}\,,\\[0.1cm]
\end{aligneq}
where we applied the NDA estimate $\N_{\eta\to\pi^0\gamma^\ast}\sim \Fpi^4/\Lambda_\chi$, used $M_\eta=547.86\,\text{MeV}$, $M_{\pi^0}=134.98\,\text{MeV}$, $m_e=0.51\,\text{MeV}$, $m_\mu=105.67\,\text{MeV}$~\cite{Workman:2022ynf}, neglected their errors with respect to the dominating uncertainty from NDA, and inserted the branching ratios from Table~\ref{tab:Overview_C-odd_Decays}.

We can again reverse this argument, i.e., the semi-leptonic branching ratios in 
explicit dependence of $\Lambda$ read
\begin{aligneq}
       & \text{BR}_{\eta\to \pi^0 e^+ e^-}&&\sim7\cdot10^{-27}\,\text{TeV}^8/\Lambda^8\,, \qquad &&\text{BR}_{\eta\to \pi^0 \mu^+ \mu^-}&&\sim2\cdot10^{-27}\,\text{TeV}^8/\Lambda^8\,,
    \\[0.1cm]
        &\text{BR}_{\eta'\to \pi^0 e^+ e^-}&&\sim9\cdot10^{-28}\,\text{TeV}^8/\Lambda^8\,, \qquad &&\text{BR}_{\eta'\to \pi^0 \mu^+ \mu^-}&&\sim6\cdot10^{-28}\,\text{TeV}^8/\Lambda^8\,,
    \\[0.1cm]
        &\text{BR}_{\eta'\to \eta e^+ e^-}&&\sim9\cdot10^{-29}\,\text{TeV}^8/\Lambda^8\,, \qquad &&\text{BR}_{\eta'\to \eta \mu^+ \mu^-}&&\sim3\cdot10^{-29}\,\text{TeV}^8/\Lambda^8\,,
\end{aligneq}
respectively.\footnote{Note that, here and henceforth, we use the total decay width $\Gamma_{\eta'}=0.23\,\text{MeV}$ indicated as \textit{PDG average} in Ref.~\cite{Workman:2022ynf}.} At this point we once more underline that these estimates are only valid for the mechanism $X\to Y\gamma^\ast\to Y \ell^+ \ell^-$ driven by $\O^{(a)}_{\psi\chi}$. A more thorough investigation of $X\to Y \ell^+ \ell^-$ including the remaining LEFT sources, semi-leptonic four-point interactions, as well as hadronic contributions to the $X\to Y \gamma^\ast$ form factor is left for future work.

%--------------------------------------------------------------
\boldmath
\subsection{$\etap\to \pi^+\pi^-\gamma$}
\label{sec:EtaPiPiGamma_charged}
\unboldmath
%--------------------------------------------------------------
While the SM contribution to the anomalous decay $\etap\to \pi^+\pi^-\gamma$ is well known and has been studied extensively in particular using dispersion-theoretical approaches~\cite{Stollenwerk:2011zz,Hanhart:2013vba,Kubis:2015sga,Hanhart:2016pcd,Holz:2015tcg,Holz:2022hwz}, the considerations of $C$ violation in $\eta\to \pi^+\pi^-\gamma$ date back to the 1960s~\cite{Bernstein:1965,Barrett:1966} and 1970s~\cite{Thaler:1972ax, Gormley:1970qz, Jane:1974es}. Let us define the respective matrix element by
\begin{aligneq}
 \langle \pi^+(p_+) \pi^-(p_-) \gamma(q) | iT | \etap(P) \rangle&=(2\pi)^4\delta^{(4)}(P-p_+-p_--q)i\M_c(s,t_c,u_c)\,,
\end{aligneq}
with Mandelstam variables
\beq
    s=(P-q)^2\,, \hspace{.3cm} t_c=(P-p_+)^2\,, \hspace{.3cm}  u_c=(P-p_-)^2
\eeq
obeying $s+t_c+u_c=M_{\etap}^2+2M_{\pi}^2$. Unless otherwise stated, we work in the isospin limit.
We begin our discussion by relaxing the constraint of $C$-invariance and split the amplitude according to
\beq
    i\M_c(s,t_c,u_c)\equiv \M_c^C(s,t_c,u_c) + \M_c^{\not C}(s,t_c,u_c)\,.
\eeq
The SM contribution $\M_c^C(s,t_c,u_c)$ is, at leading order, given by the Wess--Zumino--Witten (WZW) term~\cite{Wess:1971yu,Witten:1983tw} and can be described by
\beq\label{eq:eta_2pi_gamma_dispersive}
    \M_c^{C}(s,t_c,u_c)= i\epsilon_{\alpha\beta\mu\nu} \varepsilon^{\ast\,\alpha} q^\beta p_+^\mu p_-^\nu F_c^C(s,t_c,u_c)\,.
\eeq
The invariant function $F_c^C$ can be expanded in terms of pion--pion partial waves according to~\cite{Jacob:1959at}
\beq
F_c^C(s,t_c,u_c) = \sum_\ell P'_\ell(z_s) f_\ell(s) \,, \qquad
z_s=\frac{s (t-u)+(M_1^2-M_2^2) M_{\etap}^2}{\lambda^{1/2}(s,M_1^2,M_2^2)(M_{\etap}^2-s)} \,,
\label{eq:PWE}
\eeq
where $z_s$ is the cosine of the scattering angle, $P'_\ell(z_s)$ refers to the first derivatives of the Legendre polynomials,
and for convenience and later use we keep the dependence on the masses of the two mesons in the final state explicit. 
For the application at hand we can simply set $M_1=M_2=M_\pi$.
For the $C$-even SM amplitude, only partial waves of odd $\ell$ contribute.
Accounting for $s$-channel final-state rescattering and restricting the calculation to the dominant $P$-wave, the scalar function $F_c^C$ becomes
\beq
    F_c^C(s,t_c,u_c)= P(s)\,\Omega(s) \,, \qquad\Omega(s)=\exp\left(\frac{s}{\pi}\int_{4M_\pi^2}^{\infty}\diff x\, \frac{\delta(x)}{x(x-s)}\right)\,,
\eeq
where $\Omega(s)$ is the Omn\`es function~\cite{Omnes:1958hv}, $\delta(s)$ is the $\pi\pi$ P-wave phase shift, for which we employ the parameterization of Ref.~\cite{Garcia-Martin:2011iqs}, and $P(s)$ is a real-valued subtraction polynomial, for which we employ $P(s)=5.09/\text{GeV}^{3}(1+2.40s/\text{GeV}^{2}-2.42s^2/\text{GeV}^{4})$ for the decay of the $\eta$ and $P(s)=5.05/\text{GeV}^{3}(1+0.99s/\text{GeV}^{2}-0.55s^2/\text{GeV}^{4})$ for the $\eta'$~\cite{Akdag:2018}. For our purposes we can neglect all parameter uncertainties, left-hand cuts, and higher partial waves.

\begin{sloppypar}
In contrast, we only work at leading order for the $\chi$PT analog of the $C$-violating contribution $\M_c^{\not C}(s,t_c,u_c)$, which was found in Ref.~\cite{Barrett:1966} to be $\O(p^6)$. It is commonly known that an interaction with an odd number of pseudoscalars requires an $\epsilon$-tensor to render the Lagrangian invariant under parity.\footnote{This statement is also manifest in the construction of chirally invariant traces: a parity-violating trace, i.e., a trace with a relative minus sign between its parity transformed as for instance $\vev{\lambda \bar{U}^\dagger-\lambda^\dagger \bar{U}}$, always includes an odd number of pseudoscalars according to Eq.~\eqref{eq:U_expanded}. The only freedom we have in the construction of $\chi$PT operators to restore parity invariance without flipping this relative sign or multiplying other parity-violating traces (which both lead to an overall even number of pseudoscalars) is the inclusion of an $\epsilon$-tensor. We remark that this argument does not hold for semi-leptonic interactions, as the multiplication with a parity-flipping lepton current or density does not change the number of mesons.} Thus we naively start at $\O(p^4)$ like the WZW term. To furthermore violate $C$ the dipion system must have an even orbital angular momentum $l$. 
Hence, when interchanging the pions, we find $|\pi^+\pi^-\rangle=(-1)^l|\pi^-\pi^+\rangle=|\pi^-\pi^+\rangle$. Finally demanding Bose statistics, i.e., symmetrizing under interchange of the pions, the amplitude at $\O(p^4)$ vanishes due to contraction with the  $\epsilon$-tensor. Thus we need to equip the matrix element by another momentum configuration that is antisymmetric under $\pi^+\leftrightarrow\pi^-$, which leads to 
\beq\label{eq:eta_2pi_gamma_1960s}
    \M_c^{\not C}(s,t_c,u_c)\sim \epsilon_{\alpha\beta\mu\nu} \varepsilon^{\ast\,\alpha} q^\beta p_+^\mu p_-^\nu q_\rho (p_-^\rho-p_+^\rho)
\eeq
in consistency with Ref.~\cite{Barrett:1966}. Note that this matrix element also differs from the WZW term by a relative factor $i$, ensuring $T$ violation and hence $CPT$-invariance. For better comparability of Eq.~\eqref{eq:eta_2pi_gamma_1960s} with the SM amplitude in Eq.~\eqref{eq:eta_2pi_gamma_dispersive}, we can also define a scalar function in the $C$-violating case, i.e., 
\beq
F_c^{\not{C}}(s,t_c,u_c)\equiv q_\rho (p_-^\rho-p_+^\rho)+\ldots=(t_c-u_c)/2+\ldots\,, \label{eq:CV-pipigamma}
\eeq 
where the ellipsis denotes higher-order terms in the chiral expansion. Comparing to Eq.~\eqref{eq:PWE}, we see that the amplitude~\eqref{eq:CV-pipigamma} indeed corresponds to the leading $C$-odd partial wave, a $D$-wave.
\end{sloppypar}

We now wish to reconstruct Eq.~\eqref{eq:eta_2pi_gamma_1960s} with ToPe$\chi$PT and again pick one arbitrary operator that may generate this matrix element at lowest order. One contribution at $\O(\delta^2)$ originates from
\beq\label{eq:eta_2pi_gamma_ChPT_example}
    \bar{X}_{\psi\chi}^{(a)}\supset i\frac{v}{\Lambda^4}c_{\psi\chi}^{(a)} \bar{g}^{(a)}_{7} \epsilon_{\alpha\beta\mu\nu} \vev{\big(\lambda\lambda_L \partial^\nu\bar{U}^\dagger \partial_\rho f_R^{\alpha\beta}\partial^\mu \bar{U}\partial^\rho \bar{U}^\dagger 
    - \lambda^\dagger\lambda_R \partial^\nu\bar{U} \partial_\rho f_L^{\alpha\beta}\partial^\mu \bar{U}^\dagger\partial^\rho \bar{U}\big)-\hc}\,.
\eeq
If we only consider contributions to $\eta\to\pi^+\pi^-\gamma$, use partial integration, and make use of the amplitude's symmetry, this operator evaluates to the compact expression
\begin{aligneq}
    \sum_{\psi,\chi} \bar{X}_{\psi\chi}^{(a)}&\supset e\frac{v}{\Lambda^4 \Fpi^3}\N_{\etap\to\pi^+\pi^-\gamma} \,\epsilon_{\alpha\beta\mu\nu}\,\etap(\partial^\nu\pi^+\partial^\rho\partial^\mu\pi^-
    +\partial^\nu\pi^-\partial^\rho\partial^\mu\pi^+) \partial_\rho F^{\alpha\beta}\,.
\end{aligneq}
The constants
\beq
    \N_{\eta\to\pi^+\pi^-\gamma}=\sqrt{2}\N_{\eta'\to\pi^+\pi^-\gamma}\,,\qquad\N_{\eta'\to\pi^+\pi^-\gamma}\equiv -\frac{4}{\sqrt{3}}\bar{g}^{(a)}_{7} \big(c^{(a)}_{uu}-c^{(a)}_{dd}\big)
\eeq 
serve as the normalizations. We cannot claim at hand of this single example that $\eta\to\pi^+\pi^-\gamma$ and $\eta'\to\pi^+\pi^-\gamma$ are maximally correlated. From this operator we can compute the matrix element
\beq\label{eq:eta_2pi_gamma_BSM_charged}
    \M_c^{\not C}(s,t_c,u_c)= e\frac{v}{\Lambda^4 \Fpi^3}\N_{\etap\to\pi^+\pi^-\gamma} \,\epsilon_{\alpha\beta\mu\nu}\, \varepsilon^{\ast\,\alpha} q^\beta p_+^\mu p_-^\nu (t_c-u_c)
\eeq
in consistency with the previous considerations. The respective NDA estimate yields $\N_{\etap\to\pi^+\pi^-\gamma}\sim \Fpi^4/\Lambda_\chi^3$.

The interference of the SM and BSM amplitudes $M_c^C$ and $M_c^{\not C}$ gives rise to an asymmetry in the distribution of charged pion momenta. To quantify this so-called left--right asymmetry, we introduce the ratio
\beq
    A_{LR}\equiv \frac{\Gamma(t_c>u_c)-\Gamma(u_c>t_c)}{\Gamma_{\etap\to \pi^+ \pi^- \gamma}}\,,
\eeq
where the $\Gamma$ denote the phase space integrals over $|\M_c(s,t_c,u_c)|^2$ for $t_c>u_c$, $u_c>t_c$, and the full range, respectively. 
These integrals are explicitly defined by
\begin{aligneq}
    \Gamma&= \int_{4M_{\pi}^2}^{M_{\etap}^2} \diff s \, \Gamma_0(s)  \int_{z_s^{\text{min}}}^{z_s^{\text{max}}} \diff z_s\, (1-z_s^2)\,|i F_c^C(s,t_c,u_c)+F_c^{\not C}(s,t_c,u_c)|^2\,, 
\end{aligneq}
with 
\begin{aligneq}\label{eq:phasespace_factors_eta_pi_pi_gamma}
\Gamma_0(s)\equiv \frac{(M_{\etap}^2-s)^3\lambda^{3/2}(s,M_1^2,M_2^2)}{16(8\pi M_{\etap})^3s^2} \,,
\end{aligneq}
again keeping the final-state masses $M_1$ and $M_2$ general for generalization in the coming sections.
The limits of the angular integration $z_s^{\text{min}}$, $z_s^{\text{max}}$ are fixed by $0\leq z_s\leq 1$ for $\Gamma(t_c>u_c)$, $-1\leq z_s\leq 0$ for $\Gamma(u_c>t_c)$, and $-1\leq z_s\leq 1$ for $\Gamma_{\etap\to \pi^+ \pi^- \gamma}$. Note that only the contribution of the interference term, i.e., 
\beq\label{eq:eta_2pi_gamma_interference}
    2\text{Re}\left[iF_c^C(s,t_c,u_c)\left(F_c^{\not C}(s,t_c,u_c)\right)^\ast\right]\subset \left|i F_c^C(s,t_c,u_c)+F_c^{\not C}(s,t_c,u_c) \right|^2\,,
\eeq
survives in the numerator of $A_{LR}$, while the denominator is dominated by the SM part.
We can now express $F_c^{\not C}(s,t_c,u_c)$ in terms of $z_s$ and carry out the $\diff z_s$ integral analytically, yielding
\begin{aligneq}
    \Gamma(t_c>u_c)-\Gamma(u_c>t_c)&= -\frac{e\,v\,\N_{\eta\to\pi^+\pi^-\gamma}}{2\Lambda^4 \Fpi^3}\int_{4M_{\pi}^2}^{M_\eta^2} \diff s \, \Gamma_0(s)\sigma(s)\big(M_{\etap}^2-s\big)P(s)\text{Im}\left(\Omega(s)\right)\,.
\end{aligneq}
The last factor demonstrates a crucial aspect about the $C$-odd asymmetry: due to the relative factor of $i$ between $C$-conserving and $C$-violating amplitude, their interference would actually vanish, were it not for strong rescattering phases.
For the two different decays of the $\eta$ and $\eta'$ we obtain 
\begin{aligneq}
    &\Gamma(t_c>u_c)-\Gamma(u_c>t_c)\big|_{\eta\to\pi^+\pi^-\gamma}&&=-6.6\cdot 10^{-12}\,\text{GeV}^6\times e\frac{v}{\Lambda^4 \Fpi^3}\N_{\eta\to\pi^+\pi^-\gamma}\,,\\[0.1cm]
    &\Gamma(t_c>u_c)-\Gamma(u_c>t_c)\big|_{\eta'\to\pi^+\pi^-\gamma}&&=-1.5\cdot 10^{-7}\,\text{GeV}^6\times e\frac{v}{\Lambda^4 \Fpi^3}\N_{\eta'\to\pi^+\pi^-\gamma}\,,
\end{aligneq}
respectively.
The polynomial $P(s)$ is already normalized such that the integral over the full decay range reproduces the experimental decay width, i.e., $\Gamma_{\eta\to \pi^+ \pi^- \gamma}\approx 55 \,\text{eV}$ and $\Gamma_{\eta'\to \pi^+ \pi^- \gamma}\approx 56 \,\text{keV}$, respectively.
Finally, the lower bound on the new-physics scale as a function of $A_{LR}=0.009(4)$~\cite{Workman:2022ynf} for the decay of the $\eta$ and $A_{LR}=0.03(4)$~\cite{Workman:2022ynf} for the $\eta'$ under the abovementioned NDA approximation becomes
\begin{aligneq}
    &\Lambda|_{\eta\to\pi^+\pi^-\gamma}&&\sim \left(1.2\cdot 10^{-4}\,\text{GeV}^5\, e\frac{v}{A_{LR} \Lambda_\chi^3}\Fpi\right)^{1/4}> 0.5 \,\text{GeV}\,,\\[0.1cm]
    &\Lambda|_{\eta'\to\pi^+\pi^-\gamma}&&\sim \left(1\cdot 10^{-7}\,\text{GeV}^5\, e\frac{v}{A_{LR} \Lambda_\chi^3}\Fpi\right)^{1/4}> 0.8\,\text{GeV}\,.
\end{aligneq}
Both results were computed with the central values of the empirical asymmetries.
In terms of the BSM scale, the left-right asymmetries become
\begin{aligneq}
&|A_{LR}|_{\eta\to\pi^+\pi^-\gamma}&&\sim5\cdot 10^{-16}\,\text{TeV}^4/\Lambda^4\,,\\[0.1cm]
&|A_{LR}|_{\eta'\to\pi^+\pi^-\gamma}&&\sim1\cdot 10^{-14}\,\text{TeV}^4/\Lambda^4\,,
\end{aligneq}
respectively. The significantly larger asymmetry in the $\eta'$ decay is mainly due to the fact that the phase space covers the whole region of the $\rho(770)$ resonance in the $\pi^+\pi^-$ invariant mass, with its associated phase motion and peaking imaginary part. --- In principle, the $D$-wave phase motion of the $C$-violating amplitude would induce another contribution to the asymmetry, which we have neglected in the above.  However, this is strongly suppressed relative to the $P$-wave in the near-threshold region covered in the $\eta$ decay, staying well below $1^\circ$, while it rises only up to about $10^\circ$ at the $\eta'$ mass~\cite{Garcia-Martin:2011iqs}, where it competes with the resonating $P$-wave.  Neither effect is relevant at the present level of accuracy.

%--------------------------------------------------------------
\boldmath
\subsection{$\etap\to \pi^0\pi^0\gamma$}
\label{sec:EtaPiPiGamma_neutral}
\unboldmath
%--------------------------------------------------------------
 In full analogy to the charged $\pi^+\pi^-\gamma$ final state from the previous section, we will investigate $C$ violation via the neutral one $\eta\to \pi^0\pi^0\gamma$, as was suggested by Refs.~\cite{Nefkens:2002sa, Jarlskog:2002zz}, and furthermore extend the analysis straightforwardly to $\eta'\to \pi^0\pi^0\gamma$. The $T$-matrix element
\begin{aligneq}
 \langle \pi^0(p_1) \pi^0(p_2) \gamma(q) | iT | \etap(P) \rangle&=(2\pi)^4\delta^{(4)}(P-p_1-p_2-q)i\M_n(s,t_n,u_n)
\end{aligneq}
is described by the Mandelstam variables
\beq
    s=(P-q)^2\,,\hspace{.3cm} t_n=(P-p_1)^2\,, \hspace{.3cm}  u_n=(P-p_2)^2\,,
\eeq
fulfilling the relation $s+t_n+u_n=M_\eta^2+2M_{\pi^0}^2$. The matrix element $\M_n$ has the same structure as given in Eq.~\eqref{eq:eta_2pi_gamma_1960s}. 
The ToPe$\chi$PT operator from Eq.~\eqref{eq:eta_2pi_gamma_ChPT_example} we found in the charged channel is not able to generate non-vanishing contributions to the uncharged one. This is rooted in the fact that all interactions in which no charged mesons participate are located in the diagonal entries of matrices $\bar{U}^{(\dagger)}$. Hence, any product of the latter commutes with the spurions and $f^\munu_{L,R}$ upon setting all charged mesons to zero. This fact rules out single-trace operators at $\O(\delta^2)$ derived from $\bar X_{\psi\chi}^{(a)}$. Thus we once more consider a double-trace operator (although the chiral singlet $\partial_\mu\varphi$ multiplied with a single trace might work as well), so that the lowest-order operator we find occurs at $\O(\delta^3)$ and reads\footnote{Note we have not explicitly checked whether a contribution at lower order in $\delta$ can be derived from one of the other numerous $C$- and $CP$-odd LEFT operators. However, the lowest possible order in soft momenta must still be $p^6$.}
\beq
    \bar X_{\psi\chi}^{(a)}=i\frac{v}{\Lambda^4}c_{\psi\chi}^{(a)}\bar{g}_{8}^{(a)}\epsilon_{\alpha\beta\mu\nu}\,
    \vev{\bar{U}\partial^\gamma\partial^\mu\bar{U}^\dagger+\bar{U}^\dagger\partial^\gamma\partial^\mu\bar{U}}
    \vev{\big(
    \lambda\lambda_L\partial_\gamma f_L^{\alpha\beta}\partial^\nu\bar{U}^\dagger
    -\lambda^\dagger\lambda_R\partial_\gamma f_R^{\alpha\beta}\partial^\nu\bar{U}
    \big)-\hc}
\eeq
which yields  
\begin{aligneq}
    \sum_{\psi,\chi} \bar{X}_{\psi\chi}^{(a)}&\supset e\frac{v}{\Lambda^4 \Fpi^3}\N_{\etap\to\pi^0\pi^0\gamma} \epsilon_{\alpha\beta\mu\nu}\,\etap\partial^\nu\pi^0\partial^\rho\partial^\mu\pi^0\partial_\rho F^{\alpha\beta}\,.
\end{aligneq}
The normalizations 
\begin{aligneq}
    \N_{\eta\to\pi^0\pi^0\gamma}=\frac{16}{3}\sqrt{\frac{2}{3}}\bar{g}^{(a)}_{8} \big(2c^{(a)}_{uu}-c^{(a)}_{dd}+c^{(a)}_{ss}\big)\,,\quad 
    &\N_{\eta'\to\pi^0\pi^0\gamma}= \frac{16}{3\sqrt{3}}\bar{g}^{(a)}_{8} \big(2c^{(a)}_{uu}-c^{(a)}_{dd}-2c^{(a)}_{ss}\big)
\end{aligneq}
show that both decays are uncorrelated. 
In particular, as the LECs involved differ from the ones relevant for the $\pi^+\pi^-\gamma$ final state studied in the previous section, we note that the $C$-violating operators do not relate to pion pairs of definite isospin.
Finally, the decay amplitude of the neutral channel becomes
\beq
    \M_n(s,t_n,u_n)= e\frac{v}{\Lambda^4 \Fpi^3}\N_{X\to Y\pi^0\gamma} \epsilon_{\alpha\beta\mu\nu} \varepsilon^{\ast\,\alpha} q^\beta p_1^\mu p_2^\nu (t_n-u_n)\,,
\eeq
where NDA presumes that $\N_{\etap\to \pi^0\pi^0\gamma}\sim \Fpi^4/\Lambda_\chi^3$.  As in Eq.~\eqref{eq:CV-pipigamma}, this corresponds to a $D$-wave amplitude: for two identical neutral pions, only even partial waves are allowed, the odd ones are forbidden by Bose symmetry.

As the decay at hand has no contribution by SM physics, the relevant observable is the full decay width 
\begin{aligneq}
    \Gamma_{\etap\to \pi^0 \pi^0 \gamma}&= \frac{1}{2}\left(e\frac{v}{\Lambda^4 \Fpi^3}\N_{\etap\to \pi^0\pi^0\gamma}\right)^2\int_{4\mpii^2}^{M_X^2} \diff s \,\Gamma_0(s) \int_{-1}^{1} \diff z_s\, (1-z_s^2)(t_n-u_n)^2 
    \\[0.1cm]
    &
    =\frac{2}{15}\left(e\frac{v}{\Lambda^4 \Fpi^3}\N_{\etap\to \pi^0\pi^0\gamma}\right)^2\int_{4\mpii^2}^{M_X^2} \diff s \,\Gamma_0(s) \frac{(M_X^2-s)^2(s-4\mpii^2)}{s}\,,
\end{aligneq}
where the kinematical functions can be adapted from Eq.~\eqref{eq:phasespace_factors_eta_pi_pi_gamma} and the additional factor $1/2$ accounts for Bose symmetry as we have two identical particles in the final state. 
The numeric values of the phase space integrals yield
\begin{aligneq}
&\Gamma_{\eta\to\pi^0\pi^0\gamma}&&= \left(e\frac{v}{\Lambda^4 \Fpi^3}\N_{\eta\to\pi^0\pi^0\gamma}\right)^2 &&\times 6.4\cdot 10^{-13} \,\text{GeV}^{11} \,,\\[0.1cm]
&\Gamma_{\eta'\to\pi^0\pi^0\gamma}&&= \left(e\frac{v}{\Lambda^4 \Fpi^3}\N_{\eta'\to\pi^0\pi^0\gamma}\right)^2 &&\times2.5\cdot 10^{-9} \,\text{GeV}^{11} \,.
\end{aligneq}
With the NDA estimate quoted above, the current experimental measurements of the decay widths listed in Table~\ref{tab:Overview_C-odd_Decays} set the limits
\begin{aligneq}
    \Lambda&\sim \left(\frac{6.4\cdot 10^{-13}\,\text{GeV}^{11}}{\Gamma_{\eta\to \pi^0 \pi^0 \gamma}}\alpha v^2 \frac{\Fpi}{\Lambda_\chi^5}\right)^{1/8}&&>0.6\, \text{GeV}\,,
\end{aligneq}
while no search has been performed for $\eta'\to\pi^0\pi^0\gamma$ to date.
For arbitrary $\Lambda$ the respective branching ratios behave as
\begin{aligneq}
    &\text{BR}_{\eta \to \pi^0 \pi^0 \gamma}&&\sim1\cdot 10^{-29}\,\text{TeV}^8/\Lambda^8 \,,\\[0.1cm]
    &\text{BR}_{\eta'\to \pi^0 \pi^0 \gamma}&&\sim2\cdot 10^{-28}\,\text{TeV}^8/\Lambda^8 \,.
\end{aligneq}
This tremendous suppression is due to the $\Lambda^{-8}$ dependence of the decay width and underlines that decays allowing for an interference of SM and BSM amplitudes---as the charged channel $\eta\to\pi^+\pi^-\gamma$---are much more suitable to search for this kind of new physics, as they scale with $\Lambda^{-4}$.

%--------------------------------------------------------------
\boldmath
\subsection{$\eta'\to \eta\pi^0\gamma$}
\label{sec:EtapEtaPiGamma}
\unboldmath
%--------------------------------------------------------------
In this section we focus on the decay $\eta'\to \eta\pi^0\gamma$, for which no measurement has been recorded so far. We define the corresponding matrix element via
\begin{aligneq}
 \langle \eta(p_1) \pi^0(p_2) \gamma(q) | iT | \etap(P) \rangle&=(2\pi)^4\delta^{(4)}(P-p_1-p_2-q)i\M(s,t,u)\,,
\end{aligneq}
with Mandelstam variables
\beq
    s=(P-q)^2\,,\hspace{.3cm} t=(P-p_1)^2\,, \hspace{.3cm}  u=(P-p_2)^2\,,
\eeq
obeying $s+t+u=M_{\eta'}^2+M_\eta^2+M_{\pi^0}^2$. At the mesonic level, the driving operator must have the form
\beq
    \epsilon_{\alpha\beta\mu\nu}\,\eta'\partial^\mu\eta\partial^\nu\pi^0 F^{\alpha\beta}
\eeq
in compliance with Sect.~\ref{sec:EtaPiPiGamma_charged}. 
Similar to the arguments given in Sect.~\ref{sec:EtaPiPiGamma_neutral} we cannot build an operator at $\O(\delta^1)$. The lowest order contribution we find is
\beq
    \bar X_{\psi\chi}^{(a)}=\frac{v}{\Lambda^4}c_{\psi\chi}^{(a)}\bar{g}_{9}^{(a)}
    \,\epsilon_{\alpha\beta\mu\nu}\,\varphi\,\vev{\big(\lambda_L f_L^{\alpha\beta}\partial^\mu\bar{U}^\dagger\bar{U}
    -\lambda_R f_R^{\alpha\beta}\partial^\mu\bar{U}\bar{U}^\dagger
    \big)-\hc}\vev{\lambda\partial^\nu\bar{U}^\dagger-\lambda^\dagger\partial^\nu\bar{U}}
\eeq
at $\O(\delta^3)$.\footnote{We have not explicitly checked whether any of the remaining LEFT operators can generate $\eta'\to \eta\pi^0\gamma$ at $\O(\delta^1)$ or $\O(\delta^2)$.}
The corresponding Lagrangian
\begin{aligneq}
    \sum_{\psi,\chi} \bar{X}_{\psi\chi}^{(a)}&\supset e\frac{v}{\Lambda^4 \Fpi^3}\frac{1}{2}\N_{\eta'\to\eta\pi^0\gamma}\, \epsilon_{\alpha\beta\mu\nu}\,\eta'\partial^\mu\eta\partial^\nu\pi^0 F^{\alpha\beta}\,,
\end{aligneq}
with normalization
\begin{aligneq}
    \N_{\eta'\to\eta\pi^0\gamma}=\frac{32\sqrt{2}}{9}\big(-c^{(a)}_{ud}+c^{(a)}_{us}-2c^{(a)}_{du}-c^{(a)}_{ds}+2c^{(a)}_{su}+c^{(a)}_{sd}\big)\,\bar{g}^{(a)}_{9}\,,
\end{aligneq}
results in the matrix element
\beq
    i\M=e\frac{v}{\Lambda^4 \Fpi^3}\N_{\eta'\to\eta\pi^0\gamma}\, \epsilon_{\alpha\beta\mu\nu}\,p_1^\mu p_2^\nu q^\alpha \epsilon^\beta\,. \label{eq:M-eta'-etapi0gamma}
\eeq
The lower number of derivatives/momenta involved in this amplitude as compared to the decays $\etap\to\pi^0\pi^0\gamma$ discussed in the previous section can again be understood in terms of the contributing leading partial waves: while all of these decays violate $C$ and do not allow for a SM decay amplitude as the similar ones with a $\pi^+\pi^-$ pair in the final state, there are no restrictions from Bose symmetry on the $\eta\pi^0$ final state, and hence the leading contribution~\eqref{eq:M-eta'-etapi0gamma} is a $P$-, not a $D$-wave; 
note how the $\eta\pi$ $P$-wave combines to $J^{PC}$ quantum numbers $1^{-+}$.
The respective decay width can be evaluated in the same manner as in the previous sections and becomes
\begin{aligneq}
    \Gamma_{\eta'\to \eta \pi^0 \gamma}&= \left(e\frac{v}{\Lambda^4 \Fpi^3}\N_{\eta'\to \eta\pi^0\gamma}\right)^2\frac{4}{3}\int_{M^2_\text{min}}^{M_{\eta'}^2}\diff s \, \Gamma_0(s)\,,
\end{aligneq}
with $M_\text{min}\equiv M_\eta+\mpii$.
A numeric integration yields
\beq
    \Gamma_{\eta'\to\eta\pi^0\gamma}= \left(e\frac{v}{\Lambda^4 \Fpi^3}\N_{\eta'\to\eta\pi^0\gamma}\right)^2 \times 1.6\cdot 10^{-9} \,\text{GeV}^{7}\,,
\eeq
so that the NDA estimate $\N_{\eta'\to\eta\pi^0\gamma}\sim \Fpi^4/\Lambda_\chi$ finally results in
\beq
    \text{BR}_{\eta'\to \eta \pi^0 \gamma}\sim 2\cdot 10^{-28}\,\text{TeV}^8/\Lambda^8\,.
\eeq

%----------------------------------------------------------------------------------
\boldmath
\subsection{$\eta'\to \eta \pi^0\pi^0\gamma$}
\label{sec:EtapEtaPiPiGamma}
\unboldmath
%---------------------------------------------------------------------------------
Another $C$-violating decay that has not yet been searched for is $\eta'\to \eta \pi^0\pi^0\gamma$. Let us define the corresponding $T$-matrix element as 
\beq
    \langle \pi^0(p_1) \pi^0(p_2)\eta(p_3) \gamma(p_4)| iT | \eta'(P) \rangle \equiv (2\pi)^4 \delta^{(4)}(P-p_1-p_2-p_3-p_4) i\M(p_1,p_2,p_3,p_4)\,.
\eeq
On the mesonic level this decay requires an operator coupling uncharged pseudoscalars to a photon. As the covariant derivative only couples the photon to charged mesons, the desired operator has to include one $F_\munu$, similar to the Lagrangian in Eq.~\eqref{eq:eta_pi_gamma_1960}. This leaves us with
\begin{aligneq}\label{eq:EtapEtaPiPiGamma_basic}
    \eta' \partial_\mu \eta \pi^0\partial_\nu\pi^0 F^\munu
\end{aligneq}
as the only possible assignment of derivatives that does not vanish for an on-shell photon respecting gauge invariance, i.e., upon setting $q^2=0$ and $q^\mu\varepsilon_\mu=0$. Any operator with a derivative acting on $\eta'$ can be brought to the same form as the one above using partial integration. 

We thus arbitrarily choose the chiral operator
\beq
    \bar{X}_{\psi\chi}^{(a)} \supset \frac{v}{\Lambda^4}c_{\psi\chi}^{(a)}
   \bar g_{10}^{(a)} \,i(\partial_\mu \varphi) \vev{(\lambda\lambda_L f_L^\munu \partial_\nu \bar U^\dagger -\lambda^\dagger\lambda_R f_R^\munu \partial_\nu \bar U)-\hc}
\eeq
as a contribution at lowest possible order.
Only keeping non-vanishing terms, the corresponding Lagrangian at $\O(\delta^2)$ in mesonic degrees of freedom reads
\begin{aligneq}
    \sum_{\psi,\chi} \bar{X}_{\psi\chi}^{(a)} \supset e\frac{v}{\Lambda^4\Fpi^4}
      \N_{\eta'\to \eta\pi^0\pi^0\gamma}  \,\eta'\partial_\mu\eta\pi^0\partial_\nu\pi^0 F^\munu\,,
\end{aligneq}
with
\begin{aligneq}
    \N_{\eta'\to \eta\pi^0\pi^0\gamma}\equiv-\frac{4\sqrt{2}}{9}\bar g_{10}^{(a)}\big(2c^{(a)}_{uu}-c^{(a)}_{dd}\big)\,.
\end{aligneq}
The resulting matrix element evaluates to
\beq
    i\M(p_1,p_2,p_3,p_4)=e\frac{v}{\Lambda^4\Fpi^4}
      \N_{\eta'\to\eta\pi^0\pi^0\gamma}(p_3^\mu (p_1^\nu+p_2^\nu) - p_3^\nu (p_1^\mu+p_2^\mu))p_{4\mu}\epsilon^\ast_\nu\,,
\eeq
and is related to the decay width by
\begin{aligneq}
    \Gamma_{\eta'\to \eta\pi^0\pi^0\gamma}&=(2\pi)^4\frac{S}{2M}\int\diff\Phi_4\sum_{\text{pol.}}|\M(p_1,p_2,p_3,p_4)|^2\,.
\end{aligneq}
Here $\diff \Phi_4$ is the four-body phase space, $M$ is the mass of the decaying particle, and we explicitly accounted for a symmetry factor $S$.
We now turn the focus on the computation of the four-body phase space and divide the final state into the two-body subsystems, with momenta $q=p_1+p_2$ and $k=p_3+p_4$. At this point we will keep the mass assignments of the particles as general as possible in order to be able to re-use the calculation at a later stage. Introducing $s_{12}=q^2$ and $s_{34}=k^2$, the absolute values of the occurring three-momenta read
\beq
|\bm{q}|=\frac{\lambda^{1/2}(M^2,s_{12},s_{34})}{2M}\,,\quad
|\bm{p}_1^{12}|=\frac{\lambda^{1/2}(s_{12},m_1^2,m_2^2)}{2\sqrt{s_{12}}}\,,\quad
|\bm{p}_3^{34}|=\frac{\lambda^{1/2}(s_{34},m_3^2,m_4^2)}{2\sqrt{s_{34}}}\,,\quad
\eeq
where the additional indices $12$ and $34$ indicate the respective center-of-mass systems chosen for the evaluation and $\bm{q}$ is taken in the rest frame of the decaying particle. The explicit expressions for the four-momenta are
\begin{aligneq}
    p_1&=\Big(
    \gamma_{12}(E_1^{12}+\beta_{12}|\bm{p}_1^{12}|\cos\theta_{12})\,,\ 
    \phantom{-}|\bm{p}_1^{12}|\sin\theta_{12}\,,\ 
    0\,,\ 
    \gamma_{12}(\beta_{12}E_1^{12}+|\bm{p}_1^{12}|\cos\theta_{12})
    \Big)^T \,,\\[0.1cm]
    p_2&=\Big(
    \gamma_{12}(E_2^{12}-\beta_{12}|\bm{p}_1^{12}|\cos\theta_{12})\,,\ 
    -|\bm{p}_1^{12}|\sin\theta_{12}\,,\ 
    0\,,\ 
    \gamma_{12}(\beta_{12}E_2^{12}-|\bm{p}_1^{12}|\cos\theta_{12})
    \Big)^T \,,\\[0.1cm]
    p_3&=\Big(
    \gamma_{34}(E_3^{34}+\beta_{34}|\bm{p}_3^{34}|\cos\theta_{34})\,,\ 
    -|\bm{p}_3^{34}|\sin\theta_{34}\cos\phi_{34}\,,\ 
    -|\bm{p}_3^{34}|\sin\theta_{34}\sin\phi_{34}\,,\ \\ &\hphantom{\ =\Big(} 
    \gamma_{34}(-\beta_{34}E_3^{34}-|\bm{p}_3^{34}|\cos\theta_{34})
    \Big)^T \,,\\[0.1cm]
    p_4&=\Big(
    \gamma_{34}(E_4^{34}-\beta_{34}|\bm{p}_3^{34}|\cos\theta_{34})\,,\ 
    \phantom{-}|\bm{p}_3^{34}|\sin\theta_{34}\cos\phi_{34}\,,\ 
    \phantom{-}|\bm{p}_3^{34}|\sin\theta_{34}\sin\phi_{34}\,,\ \\ &\hphantom{\ =\Big(}  
    \gamma_{34}(-\beta_{34}E_4^{34}+|\bm{p}_3^{34}|\cos\theta_{34})
    \Big)^T \,,
\end{aligneq}
with $E_n^{ij}=\sqrt{m_n^2+|\bm{p}_n^{ij}|^2}$, $\beta_{12}=|\bm{q}|/E_q$, $\beta_{34}=|\bm{k}|/E_k$, $E_q^2=|\bm{q}|^2+s_{12}$, $E_k^2=|\bm{k}|^2+s_{34}$, and $\gamma_{ij}=1/\sqrt{1-\beta_{ij}^2}$. The four-body phase space in terms of the five independent variables reads
\beq
    \diff\Phi_4=\frac{1}{32(2\pi)^{10}}\diff s_{12}\,\diff s_{34} \,\diff \theta_{12}\,\diff \theta_{34}\,\diff \phi_{34}\frac{|\bm{q}|}{M}\frac{|\bm{p}_1^{12}|}{\sqrt{s_{12}}}\frac{|\bm{p}_3^{34}|}{\sqrt{s_{34}}}\sin\theta_{12}\sin\theta_{34}\,, 
\eeq
where the non-trivial integration limits are
\beq
    (m_1+m_2)^2\leq s_{12}\leq (M-m_3-m_4)^2\,,\qquad (m_3+m_4)^2\leq s_{34}\leq (M-\sqrt{s_{12}})^2\,.
\eeq
For the remaining details regarding the kinematics we refer to Ref.~\cite{Guo:2011ir}, while equivalent formulations can be found in Refs.~\cite{Cabibbo:1965zzb, Barker:2002ib, Kampf:2018wau}. 

Inserting the explicit masses of the contributing particles and applying $S=1/2$,
we finally find with the NDA estimate $\N_{\eta'\to \eta\pi^0\pi^0\gamma}\sim \Fpi^4/\Lambda_\chi$ that the branching ratio yields
\beq
    \text{BR}_{\eta'\to \eta\pi^0\pi^0\gamma}\sim 4\pi\alpha\frac{v^2}{\Gamma_{\eta'}\Lambda_\chi^2\Lambda^8}
      \cdot 1\cdot10^{-15}\approx2\cdot10^{-32}\,\text{TeV}^8/\Lambda^8\,.
\eeq

%--------------------------------------------------------------
 \boldmath
 \subsection{$\eta\to 3\pi^0\gamma$}
 \label{sec:Eta3PiGamma}
 \unboldmath
%--------------------------------------------------------------

In complete analogy to Sect.~\ref{sec:EtapEtaPiPiGamma}, we can derive the transition $\eta\to 3\pi^0\gamma$ by appropriately replacing the four-momenta and masses of $\eta'$ and $\eta$ by the ones for $\eta$ and $\pi^0$.\footnote{We do not consider the decay $\eta'\to 3\pi^0\gamma$ here, as the increased phase space allows for an $\omega$ in the intermediate state. As a consequence, we would expect this to rather test the $C$-violating vector-meson decay $\omega\to3\pi^0$, analogously to how $\eta'\to\pi^+\pi^-\pi^0\gamma$ is dominated by $\eta'\to\omega\gamma$ in the Standard Model.} However, the enhanced symmetry of this process prohibits any operator whose derivatives on pion fields contract with the field-strength tensor. Thus, we require a term with at least four derivatives that does not lead to mass terms, as for instance
\beq
    \partial_\mu\eta\partial_\nu\pi^0\partial_\alpha\pi^0
    \pi^0 \partial^\alpha F^\munu\,,
\eeq
which we consider as an example for this decay.
The corresponding chiral Lagrangian from $\bar{X}_{\psi\chi}^{(a)}$ at lowest order (i.e., $\delta^3$) obtains a contribution from\footnote{Again, we have not considered other LEFT operators that may contribute at order $\delta^2$.}
\beq
     \bar{X}_{\psi\chi}^{(a)}\supset \frac{v}{\Lambda^4} c_{\psi\chi}^{(a)}\,\bar{g}_{11} \,i(\partial_\mu \varphi) \vev{(\lambda\lambda_L \partial^\alpha f_L^\munu \partial_\alpha \bar U^\dagger \bar U \partial_\nu \bar U^\dagger-\lambda^\dagger\lambda_R \partial^\alpha f_R^\munu \partial_\alpha \bar U \bar U^\dagger \partial_\nu \bar U)-\hc} \,,
\eeq
leading to the operator
\begin{aligneq}
    \sum_{\psi\chi} \bar{X}_{\psi\chi}^{(a)} \supset e\frac{v}{\Lambda^4\Fpi^4} \N_{\eta\to 3\pi^0\gamma}\, 
    \partial_\mu\eta\partial_\nu\pi^0\partial_\alpha\pi^0
    \pi^0 \partial^\alpha F^\munu\,,
\end{aligneq}
with the normalization
\beq
 \N_{\eta\to 3\pi^0\gamma}\equiv-\frac{4}{3}\sqrt{\frac{2}{3}}\, \bar{g}^{(a)}_{11}(2c_{uu}^{(a)}+c_{dd}^{(a)})\,.
\eeq
Accordingly, the overall matrix element becomes
\begin{aligneq}
    i\M(p_1,p_2,p_3,p_4)&=e\frac{v}{\Lambda^4\Fpi^4} \N_{\eta\to 3\pi^0\gamma}\, P^\mu \big(p_1^\nu p_2^\alpha + p_2^\nu p_1^\alpha+p_1^\nu p_3^\alpha+p_3^\nu p_1^\alpha+p_2^\nu p_3^\alpha+p_3^\nu p_2^\alpha\big) \\ 
    &\quad\times p_{4\alpha}\big(p_{4\mu} \epsilon^\ast_\nu - p_{4\nu} \epsilon^\ast_\mu \big)\,.
\end{aligneq}
With the same four-body phase space as in Sect.~\ref{sec:EtapEtaPiPiGamma}, but with appropriately re-assigned masses, a symmetry factor $S=1/6$, the NDA prediction $\N_{\eta\to 3\pi^0\gamma}\sim \Fpi^4/\Lambda_\chi^3$, and the experimental width $\Gamma_{\eta\to3\pi^0\gamma}$ listed in Table~\ref{tab:Overview_C-odd_Decays} we find the lower limit
\beq
    \Lambda\sim\left( 4\pi\alpha\frac{v^2}{\Gamma_{\eta\to3\pi^0\gamma}\Lambda_\chi^6} \cdot 6\cdot 10^{-21}\,\text{GeV}^{13}\right)^{1/8}>140\,\text{MeV} 
\eeq
or
\beq
    \text{BR}_{\eta\to3\pi^0\gamma}\sim  
    1\cdot10^{-35}\,\text{TeV}^8/\Lambda^8
\eeq
for the theoretically estimated branching ratio, respectively.

%--------------------------------------------------------------
\boldmath
\subsection{$\etap\to 3\gamma$ and $\pi^0\to 3\gamma$}
\label{sec:Eta3Gamma}
\unboldmath
%--------------------------------------------------------------
In this section we investigate the $CP$-odd contributions of the $C$-violating decays 
$\etap\to 3\gamma$ and $\pi^0\to 3\gamma$, which have been considered in Refs.~\cite{Berends:1965ftl,Tarasov:1967,Jarlskog:2002zz} while possible $C$- and $P$-violating contributions through weak interactions (in the case of $\pi^0$) have been discussed in Ref.~\cite{Dicus:1975cz}. For this purpose we introduce the $T$-matrix element
\beq
    \langle \gamma(q_1)\gamma(q_2)\gamma(q_3)|iT| X(P) \rangle= (2\pi)^4\delta^{(4)}(P-q_1-q_2-q_3)i\M(s,t,u)
\eeq
with $X=\eta', \eta,\pi^0$ and define the Mandelstam variables
\beq
    s=(P-q_1)^2\,, \hspace{.3cm} t=(P-q_2)^2\,, \hspace{.3cm}  u=(P-q_3)^2\,,
\eeq
with $s+t+u=M_X^2$. The three-photon final state sets demanding constraints on the amplitude. 
First of all, the covariant derivative only couples charged mesons to the photon, thus we need exactly three field-strength tensors picked from $f_L^\munu,f_R^\munu,F_\munu$ to generate the $3\gamma$ final state and can use $\partial_\mu$ instead of $D_\mu$. Since we have an odd number of pseudoscalars, a Levi-Civita symbol has to be involved when contracting the Lorentz indices to respect parity invariance. Moreover Bose statistics demands a symmetrized $3\gamma$ final state, so that non-vanishing operators require at least four additional derivatives~\cite{Berends:1965ftl}. Finally, to obtain a coupling with a single meson (multiple) derivatives can only act on one single $\bar{U}$ or $\bar{U}^\dagger$ at a time. We can now pick one ToPe$\chi$PT operator that meets all these requirements (which demand an operator starting at $\O(\delta^4)$) and arbitrarily choose
\begin{aligneq}
    &\bar{X}_{\psi\chi}^{(a)}&&\supset \frac{v \bar{g}^{(a)}_{12} c_{\psi\chi}^{(a)}}{\Lambda^4 \Fpi}&& i\epsilon^{\mu\nu\rho\sigma} \vev{(\lambda\lambda_L f^L_\munu \partial^\gamma f_L^{\alpha\beta} \partial_\gamma \partial_\beta f^L_{\rho\sigma} \partial_\alpha \bar U^\dagger
    -\lambda^\dagger\lambda_R f^R_\munu \partial^\gamma f_R^{\alpha\beta} \partial_\gamma \partial_\beta f^R_{\rho\sigma} \partial_\alpha \bar U) -\hc}\,,
\end{aligneq}
giving rise to 
\begin{aligneq}\label{eq:Lagrange_3gamma}
    \sum_{\psi,\chi} \bar{X}_{\psi\chi}^{(a)}\supset e^3\frac{v}{\Lambda^4\Fpi}2\N_{X\to 3\gamma}
    \, \epsilon^{\mu\nu\rho\sigma}\partial_\alpha X (\partial^\gamma F^{\alpha\beta})(\partial_\gamma \partial_\beta F_{\rho\sigma})F_\munu\,,
\end{aligneq}
with 
\begin{aligneq}
\N_{\pi^0\to3\gamma}=\bar{g}^{(a)}_{12} \frac{2}{27}\big(-8c_{uu}^{(a)}-c_{dd}^{(a)}\big)\,, \qquad 
&\N_{\eta \to3\gamma}=\bar{g}^{(a)}_{12} \frac{2}{27}\sqrt{\frac{2}{3}}\big(-8c_{uu}^{(a)}+c_{dd}^{(a)}-c_{ss}^{(a)}\big)\,, \\[0.1cm] 
&\hspace{-3cm}\N_{\eta'\to3\gamma}=\bar{g}^{(a)}_{12} \frac{2}{27\sqrt{3}}\big(-8c_{uu}^{(a)}+c_{dd}^{(a)}+2c_{ss}^{(a)}\big)\,.
\end{aligneq}
This result is consistent with Refs.~\cite{Berends:1965ftl,Jarlskog:2002zz} who claimed that the only contribution to $X\to3\gamma$ arises at order $p^{10}$ in soft momenta.

Although the derivation of the full set of ToPe$\chi$PT operators up to chiral order $p^{10}$ is far beyond the scope of this work, every operator that contributes to $X\to3\gamma$ at this order has to have the same functional form as in Eq.~\eqref{eq:Lagrange_3gamma}, modulo partial integrations, so that all contributions from the genuine LEFT operator $\O^{(a)}_{\psi\chi}$ lead to the same NDA estimate. We continue the computation of the matrix element following Ref.~\cite{Berends:1965ftl} and write
\begin{aligneq}
    \sum_{\text{pol}} |\M(s,t,u)|^2&=\bigg(e^3\frac{v}{\Lambda^4\Fpi}2\N_{X\to 3\gamma}\bigg)^2 32\,(q_1q_2)(q_2q_3)(q_3q_1)\\
    &\,\, \times\left[ (q_1q_2)^2 \left(q_1q_3-q_2q_3\right)^2+(q_1q_3)^2  \left(q_1q_2-q_3q_2\right)^2 +(q_2q_3)^2\left(q_2q_1-q_3q_1\right)^2 \right]\\
    &=\bigg(e^3\frac{v}{\Lambda^4\Fpi}\N_{X\to 3\gamma}\bigg)^2s\,t\,u\left[ u^2 \left(t-s\right)^2+t^2  \left(u-s\right)^2 +s^2\left(u-t\right)^2 \right]\,.
\end{aligneq}
Inserting this result in the decay width 
\beq
    \Gamma_{X\to 3\gamma}=\frac{S}{256\pi^3 M_X^3}\int_0^{M_X^2}\diff s\int_{0}^{M_X^2-s}\diff t \sum_{\text{pol}} |\M(s,t,u)|^2 
\eeq
with symmetry factor $S=1/6$ and carrying out the integrals over $s$ and $t$ analytically, we obtain
\begin{aligneq}
    \Gamma_{X\to 3\gamma}
    &=\frac{\alpha^3 M_X^{15}}{24}\frac{1}{5040}\bigg(\frac{v}{\Lambda^4\Fpi}\N_{X\to 3\gamma}\bigg)^2\,.
\end{aligneq}
With the NDA estimate $\N_{X\to 3\gamma}\sim \Lambda_\chi^{-3}(4\pi)^{-4}$, the experimental decay widths, cf.\ Table~\ref{tab:Overview_C-odd_Decays}, 
and the abbreviation 
\beq
    \tilde\Lambda_{3\gamma}\equiv \frac{1}{5040}\frac{v^2\alpha^3}{24\Fpi^2\Lambda_\chi^{6}(4\pi)^{8}} \,,
\eeq
we set the following lower limits on $\Lambda$:
\begin{aligneq}
    \Lambda&\sim\bigg[\frac{M_{\eta'}^{15}}{\Gamma_{\eta'\to 3\gamma}}&&\tilde\Lambda_{3\gamma}\bigg]^{1/8}&&>160     \,\text{MeV} \,,\\[0.1cm]
    \Lambda&\sim\bigg[\frac{M_{\eta}^{15}}{\Gamma_{\eta \to 3\gamma}}&&\tilde\Lambda_{3\gamma}\bigg]^{1/8}&&>120 \,\text{MeV} \,,\\[0.1cm]
    \Lambda&\sim\bigg[\frac{M_{\pi^0}^{15}}{\Gamma_{\pi^0\to 3\gamma}}&&\tilde\Lambda_{3\gamma}\bigg]^{1/8}&&>40 \,\text{MeV}\,.
\end{aligneq}
Reversing the argument, the branching ratios as functions of $\Lambda$ are 
\begin{aligneq}
    &\text{BR}_{\eta'\to 3\gamma}&&\sim 3\cdot 10^{-35}\,\text{TeV}^8/\Lambda^8\,,\\[0.1cm]
    &\text{BR}_{\eta \to 3\gamma}&&\sim 1\cdot 10^{-36}\,\text{TeV}^8/\Lambda^8\,,\\[0.1cm]
    &\text{BR}_{\pi^0\to 3\gamma}&&\sim 2\cdot 10^{-43}\,\text{TeV}^8/\Lambda^8\,.
\end{aligneq}
%--------------------------------------------------------------------------
\section{Summary and outlook}
\label{sec:summary}
%--------------------------------------------------------------------------

In this article, we provided a complete set of fundamental neutrinoless, flavor-preserving, lepton- and baryon-number-conserving  $C$- and $CP$-odd quark-level operators in LEFT up to and including mass dimension 8. 
We have verified the operators from dimension-7 LEFT that were known before, but have also tackled the issue that these operators are chirality-violating, hence carefully taking chirality-conserving operators of mass dimension 8 into account. These may in principle be of the same numerical size as those of dimension 7, because both can arise from operators of dimension 8 of SMEFT; similar observations were made previously for dimension-5 and -6 operators in nucleon EDM analyses. As a consequence, as long as SMEFT is accepted as the universal starting point of our investigation, every $C$- and $CP$-odd operator we identified is suppressed by $1/\Lambda^4$, with $\Lambda$ indicating the new-physics scale.

By matching these LEFT operators thoroughly onto $\chi$PT we established a new rigorous and model-independent framework to access possible $C$- and $CP$-violating effects in flavor-conserving decays of $\eta$, $\eta'$, and $\pi^0$, which solely relies on the conjecture that these BSM effects arise from phenomena at scales of yet unknown high energies, for which QCD plus QED provide an appropriate low-energy approximation. In this context,  novel $3 \times 3$ spurion matrices were applied to ensure the transfer of the $u, d, s$ flavor degrees of 
freedom and  chirality structure of the quark bilinears from the SMEFT and/or LEFT levels to chiral operators at the $\chi$PT level. Knowing the underlying mechanisms at the level of LEFT and $\chi$PT, we derived mesonic operators, amplitudes, and observables for more than 20 decays in total.

Furthermore, we estimated that the currently most precise experiments searching for $C$ and $CP$ violation in the light-meson sector can merely restrict the SMEFT scale $\Lambda$ to the few GeV range.
Due to the lack of sufficient input to fix the numerous low-energy constants and Wilson coefficients entering the effective chiral theory, these estimates are based on naive dimensional analysis, which does not only require knowledge about the mesonic operators (some of them were already known in the 1960s) but also about the particular sources of these operators on the quark level. 

As the central numerical results of our analysis we reversed this argument and expressed the observable branching ratios for pure BSM processes as well as asymmetry parameters for interferences of SM and BSM contributions in terms of the new-physics scale. While the former scale with $1/\Lambda^8$, the interference effects are proportional to $1/\Lambda^4$ and are thus more suitable candidates for experimental searches. Hence, the most promising of our investigated decays to find evidence for ToPe forces are $\etap\to\pi^0\pi^+\pi^-$, $\eta'\to\eta\pi^+\pi^-$, and $\etap\to\pi^+\pi^-\gamma$. In addition, our estimates for the pure $C$-violating decays allow us to weed out those that require significantly higher experimental precision than others. 

We find that the currently most rigorous experimental limits on ToPe forces in the light-meson sector must become more stringent by roughly a factor of $10^7$ in order to test this scenario for a BSM scale of $\Lambda\sim 1\,\mathrm{TeV}$. Although these theoretical bounds cannot be reached by experiments in the near future, the search for the decays proposed in this work---prospectively conducted, for instance, by the REDTOP~\cite{Gatto:2016rae,Gatto:2019dhj,REDTOP:2022slw} and JEF~\cite{Gan:2015nyc,JEF:2016,Gan:2017kfr} collaborations---can still provide important insights to understand the sources of possible $C$ and $CP$ violation. 
Any experimental evidence for these decays could imply, for instance, that a simultaneous violation of $C$ and $CP$ violation  originates from (weakly coupled) light degrees of freedom, or that the SMEFT and/or LEFT power counting is bypassed by another, yet unknown mechanism.

Obviously, this casts doubt at the possibility to
interpret any observable $C$- and $CP$-odd signals in terms of SMEFT. However, we can
relax the constraints obtained here to some extent by concentrating on LEFT without any
reference to SMEFT at all; this would effectively replace the TeV scale $\Lambda$ by an electroweak
scale of the order of $100\,\text{GeV}$, and therefore reduce the discrepancy between our theoretical
expectation and current experimental sensitivity 
by a factor of about $\sfrac{1}{4} \times 10^3$ -- $10^4$ for interferences with SM amplitudes and 
 $10^5$ -- $10^8$ for pure BSM transitions.

Our analysis opens a new window to model-independent theoretical analyses of $C$ and $CP$ violation with a vast number of possible future extensions. These are not only restricted to applications to meson scattering and decays not covered in this work, especially to ones that include interferences of SM and BSM physics, but also to extensions to flavor-changing transitions~\cite{Shi:2017ffh}, to heavy-quark physics, to processes in the baryon or nuclear sector~\cite{Conzett:1992dn,Beyer:1993zw,Uzikov:2015aua,Uzikov:2016lsc,Eversheim:2017zxl,Aksentyev:2017dnk}, and cross-relations to EDMs~\cite{Haxton:1994bq}.

%---------------------------------------------------------------------------------------------------
\acknowledgments
We thank Christopher Murphy, Yi Liao, Xiao-Dong Ma, Hao-Lin Wang, Susan Gardner, and Jun Shi for useful discussions.  
We are particularly grateful for Peter Stoffer's advices about the matching between LEFT and $\chi$PT. Moreover, we thank Jordy de Vries for most helpful explanations about the correct application of naive dimensional analysis, and Daniel Severt for related discussions.
Financial support by the Avicenna-Studienwerk e.V.\ with funds from the BMBF,
as well as by the DFG (CRC 110, ``Symmetries and the Emergence of Structure in QCD''),
is gratefully acknowledged.

%---------------------------------------------------------------------------------------------------

%---------------------------------------------------------------------------------------------------
\appendix
%---------------------------------------------------------------------------------------------------

%--------------------------------------------------------------------------
\section{Characterization of discrete symmetries in LEFT operators up to dimension 8}
\label{app:LEFT}
%--------------------------------------------------------------------------
In this appendix we comment on our classification of $C$- and $CP$-violating LEFT operators presented in Sect.~\ref{sec:LEFT_classification}, based on the complete sets from Ref.~\cite{Jenkins:2017jig} up to and including mass dimension 6, Ref.~\cite{Liao:2020zyx} for dimension 7, and Ref.~\cite{Murphy:2020cly} for dimension-8 operators. For simplicity we do not quote all of the numerous contributing operators in these bases, but go directly to their characterization in terms of $C$-, $P$-, and $T$-eigenstates. 
The genuine LEFT operators are written in terms of chiral projectors, i.e.,
\beq
    \psi_{L/R}\equiv P_{L/R}\psi \eqwith P_L=\frac{1-\gamma_5}{2}\,, \quad P_R=\frac{1+\gamma_5}{2} \,,
\eeq
and thus include in general superpositions of states with different discrete symmetries. Our $C$- and $CP$-odd operators will be identified as linear combinations of these LEFT operators, such that
the separated (pseudo)scalar, (axial)vector, and (pseudo)tensor contributions have definite eigenvalues under $C$, $P$, and $T$. Technically, we merely write each projector explicitly in terms of Dirac matrices and separate the summands with different discrete symmetries. We will rephrase the quark portion of each LEFT operator in this way and drop possible field-strength tensors and $SU(3)$ generators in the first place, which can be restored in most cases straightforwardly. 

To keep the notation as short and simple as possible we use a rather sloppy notation and refer to generic Wilson coefficients by $c\in \mathbb{C}$, whose numerical value may be different in each operator. This abuse of notation shall not bother us, as we are solely interested in LEFT and not in any matching between operators in SMEFT and LEFT. The notes on the following pages are all restricted to the flavor-conserving case. However, flavor-violating LEFT operators that may contribute to ToPe forces, which are less relevant for our analysis of $\eta$ decays, can be derived in a similar manner. Other than that, we will follow the strategy already sketched in Sect.~\ref{sec:LEFT_classification}.

The outline of this appendix is as follows.
 In Sect.~\ref{sec:LEFT_dim_leq_6} we explicitly confirm at hand of well-known operator bases of LEFT that there are no ToPe interactions of dimension~$\leq 6$. Subsequently we investigate the operators at dimension 7 and 8 LEFT in  Sects.~\ref{sec:LEFT_dim_7} and~\ref{sec:LEFT_dim_8}, respectively, and carefully distinguish between the ones that are chirality-violating and chirality-conserving.

%-----------------------------------
\boldmath
\subsection{Dimension $\leq 6$ LEFT}
\label{sec:LEFT_dim_leq_6}
\unboldmath
%-----------------------------------

In this section we consider LEFT operators carefully worked out by Ref.~\cite{Jenkins:2017jig} and explicitly show that there are indeed no ToPe operators below dimension 7 in LEFT. We directly discard the dimension-3 operators, which are solely given by neutrino bilinears~\cite{Jenkins:2017jig}.

%-------------------------
\subsubsection{Dimension 5 LEFT}
\label{sec:LEFT_dim_5}
%-------------------------
The modest number of dimension-5 operators only allows for quarks with the sole Dirac structure $\bar\psi_L \sigma_\munu \psi_R$. We can multiply this structure with the Wilson coefficient and respect hermiticity, to obtain
\beq
    c\,\bar\psi \sigma_\munu P_R \psi + \hc=\Re c\,\bar\psi \sigma_\munu \psi+i \,\Im c\,\bar\psi \sigma_\munu \gamma_5 \psi\,.
    \label{eq:sigma_PR}
\eeq
After contracting with the gluon or photon field-strength tensor, we can read off from Table~\ref{tab:discrete_symmetries} that the resulting terms preserve $C$.   

%-------------------------
\subsubsection{Dimension 6 LEFT}
\label{sec:LEFT_dim_6}
%-------------------------
In dimension 6 LEFT we encounter operators including the quadrilinear
\beq
\begin{alignedat}{4}
    c\,\bar\psi\gamma_\mu\PRL\psi\bar\chi\gamma^\mu\PRL\chi+\hc=\frac{1}{2}\Re c\,
    \big[
    \bar\psi\gamma_\mu\psi\bar\chi\gamma^\mu\chi &+\bar\psi\gamma_\mu\gamma_5\psi\bar\chi\gamma^\mu\gamma_5\chi
    \\
    &\pm \bar\psi\gamma_\mu\gamma_5\psi\bar\chi\gamma^\mu\chi \pm\bar\psi\gamma_\mu\psi\bar\chi\gamma^\mu\gamma_5\chi
    \big]\,.
\end{alignedat}
\eeq
While the first two summands have the signature $CPT=+++$, the last two have the eigenvalues $CPT=--+$. Therefore these cannot contribute to ToPe interactions. The same holds for $\bar\psi\gamma_\mu\PL\psi\bar\chi\gamma^\mu\PR\chi$, which just distinguishes by relative signs from the case discussed above.
Next, consider
\beq
\begin{alignedat}{4}
    c\,\bar\psi\PR\chi\bar\chi\PR\psi +\hc=\frac{1}{2}\Re c\,
    \big[
    \bar\psi\chi\bar\chi\psi +  \bar\psi\gamma_5\chi\bar\chi\gamma_5\psi\big]+
    \frac{i}{2}\Im c
    \big[
    \bar\psi\gamma_5\chi\bar\chi\psi+ \bar\psi\chi\bar\chi\gamma_5\psi
    \big]\,.
\end{alignedat}
\eeq
While the summand scaling with $\Re c$ has $CPT=+++$ the one proportional to $\Im c$ is $CPT=+--$. Analogously, this is also true for $\bar\psi\PL\psi\bar\chi\PL\chi$.  

Using $\sigma_\munu^\dagger=\gamma_0\sigma_\munu\gamma_0$ we can easily derive
\beq
\begin{alignedat}{4}
    c\,\bar\psi\sigma_\munu\PR\chi\bar\chi\sigma^\munu\PR\psi +\hc=\frac{1}{2}\Re c\,
    \big[
    \bar\psi&\sigma_\munu\chi\bar\chi\sigma^\munu\psi +  \bar\psi\sigma_\munu\gamma_5\chi\bar\chi\sigma^\munu\gamma_5\psi\big]
    \\
    &
    +
    \frac{i}{2}\Im c
    \big[
    \bar\psi\sigma_\munu\gamma_5\chi\bar\chi\sigma^\munu\psi+ \bar\psi\sigma_\munu\chi\bar\chi\sigma^\munu\gamma_5\psi
    \big]\,,
\end{alignedat}
\eeq
where the two summands have $CPT=+++$ and $CPT=+--$, respectively. Again, the term $\bar\psi\sigma_\munu\PL\psi\bar\chi\sigma^\munu\PL\chi$ proceeds in the same manner.  The remaining dimension-6 operators under consideration are the triple gauge terms
\beq
\fabc G_\mu^{\nu a} G_\nu^{\rho b} G_\rho^{\mu c} \eqand \fabc \tilde G_\mu^{\nu a} G_\nu^{\rho b} G_\rho^{\mu c}\,, 
\eeq
which, according to Table~\ref{tab:discrete_symmetries}, have the symmetries $CPT=+++$ and $CPT=+--$, respectively.

Thus no operator in dimension-6 LEFT can create ToPe effects. Furthermore note that one can neither build a loop consisting of two dimension-6 LEFT operators that results in a $C$- and $CP$-odd transition.

%-------------------------
\subsection{Dimension-7 LEFT}
\label{sec:LEFT_dim_7}
%-------------------------

Considering the---for our purposes relevant---lepton- and baryon-number-conserving operators, there occur two different types of fermion bilinears: $\bar\psi_R\psi_L$ and $\bar\psi_R\sigma_\munu\psi_L$.
Let us again neglect the (hermitian) product of field-strength tensors accompanying these bilinears for now.
Accounting for the respective Wilson coefficients, the hermitian bilinears can be rewritten as 
\beq
\begin{split}
    c\,\bar\psi_R\psi_L + \text{h.c.} 
    = \Re c\, \,\bar\psi\psi -  \Im c\,\,\bar\psi i\gamma_5\psi
\end{split}
\eeq
and 
\beq
\begin{split}
    c \,\bar\psi_R\sigma_\munu\psi_L + \text{h.c.} 
    = \Re c\, \,\bar\psi\sigma_\munu\psi - \Im c\,\,\bar\psi i\sigma_\munu\gamma_5\psi\,.
\end{split}
\eeq
In complete analogy, the two types of fermion quadrilinears  $(\bar\psi_L \gamma^\mu \psi_L)(\bar\chi_L i \dvec D_\mu \chi_R)$ and $(\bar\psi_R \gamma^\mu \psi_R)(\bar\chi_L i \dvec D_\mu \chi_R)$ become 
\beq\label{eq:Wilson_c1ffff}
\begin{split}
    c \,(\bar\psi_L \gamma^\mu \psi_L)(\bar\chi_L i \dvec D_\mu \chi_R) + \text{h.c.} 
    &= \frac{1}{2}\Re c\, \,\left[ (\bar\chi i \dvec D_\mu \chi) (\bar\psi\gamma^\mu\psi)-(\bar\chi i \dvec D_\mu \chi) (\bar\psi\gamma^\mu\gamma_5\psi) \right]\\
    &\phantom{=} -\frac{1}{2}\Im c\,\,\left[ (\bar\chi \dvec D_\mu \gamma_5 \chi) (\bar\psi\gamma^\mu\psi)-(\bar\chi \dvec D_\mu\gamma_5 \chi) (\bar\psi\gamma^\mu\gamma_5\psi) \right]\,,
\end{split}
\eeq
where in the second step $P_L+P_R=1$ and $P_R-P_L=\gamma_5$ were applied. Similarly, we obtain
\beq
\begin{split}
    c \,(\bar\psi_R \gamma^\mu \psi_R)(\bar\chi_L i \dvec D_\mu \chi_R) + \text{h.c.} 
    &= \frac{1}{2}\Re c\, \,\left[ (\bar\chi i \dvec D_\mu \chi) (\bar\psi\gamma^\mu\psi)+(\bar\chi i \dvec D_\mu \chi) (\bar\psi\gamma^\mu\gamma_5\psi) \right]\\
    &\phantom{=} -\frac{1}{2}\Im c\,\,\left[ (\bar\chi \dvec D_\mu \gamma_5 \chi) (\bar\psi\gamma^\mu\psi)+(\bar\chi \dvec D_\mu\gamma_5 \chi) (\bar\psi\gamma^\mu\gamma_5\psi) \right]\,,
\end{split}
\eeq
with the same operators as Eq.~\eqref{eq:Wilson_c1ffff}, but with different real-valued prefactors.
Note that the factor $i$ from the Wilson coefficients flips the sign of time reversal, while $\gamma_5$ changes the one of parity.
Attaching the products of field-strength tensors to the fermion bilinears, like explicitly done in Table~\ref{tab:Liao_Dim7_LEFT_ToPe}, we can formulate the operators of Ref.~\cite{Liao:2020zyx} in a way that allows us to directly read off the transformation properties under the discrete symmetries $C$, $P$, and $T$. According to Table~\ref{tab:Liao_Dim7_LEFT_ToPe}, there are only two operators at dimension 7 in LEFT which violate $C$ and $CP$, namely
\begin{aligneq}\label{eq:dim7_4quark}
&\bar\psi T^A \sigma^{\mu\nu}\psi F_{\mu\rho} G^{A\rho}_{\nu}\,,\\
&\bar\chi \dvec D_\mu \gamma_5\chi\bar\psi\gamma^\mu\gamma_5\psi\,, 
\end{aligneq}
with the same form as already proposed decades ago, cf.\ Eq.~\eqref{eq:ToPe_historic}.
There is in principle also another color contraction for the quark quadrilinear allowed, but we refrain from quoting it explicitly because it leads to the same effective operator on the mesonic level and in this way we are as consistent as possible with the original operators from Eq.~\eqref{eq:ToPe_historic}. Furthermore, as already stated in Sect.~\ref{sec:ToPe_dim_7}, in this work we do not consider corrections due to QCD running, which arise from possible mixing of different color contractions.

\begin{table}[htbp!]
\centering
\renewcommand{\arraystretch}{1.5}
\begin{tabular}{ll|ccc}
\toprule
&  & $C$ & $P$ & $T$ \\
\midrule
1a) & $\bar\psi\psi F_\munu F^{\mu\nu}$ & ~~$+$~~ & ~~$+$~~ & ~~$+$~~ \\
1b) & $\bar\psi i\gamma_5\psi F_\munu F^{\mu\nu}$ & ~~$+$~~ & ~~$-$~~ & ~~$-$~~ \\
2a) & $\bar\psi T^A\psi F^{\mu\nu} G^A_\munu $ & ~~$+$~~ & ~~$+$~~ & ~~$+$~~ \\
2b) & $\bar\psi T^A i\gamma_5\psi F^{\mu\nu} G^A_\munu$ & ~~$+$~~ & ~~$-$~~ & ~~$-$~~ \\
3a) & $\bar\psi T^A \sigma^{\mu\nu}\psi F_{\mu\rho} G^{A\rho}_{\nu}$ & ~$-$~ & ~$+$~ & ~$-$~ \\
3b) & $\bar\psi T^A \sigma^{\mu\nu}i\gamma_5\psi F_{\mu\rho} G^{A\rho}_{\nu} $ & ~~$-$~~ & ~~$-$~~ & ~~$+$~~ \\
4a) & $\dabc\bar\psi T^A\psi G^B_\munu G^{C\mu\nu}$ & ~~$+$~~ & ~~$+$~~ & ~~$+$~~ \\
4b) & $\dabc\bar\psi T^A i\gamma_5\psi G^B_\munu G^{C\mu\nu}$ & ~~$+$~~ & ~~$-$~~ & ~~$-$~~ \\
5a) & $\fabc\bar\psi T^A \sigma^{\mu\nu}\psi G^B_{\mu\rho} G_{\nu}^{C\rho}$ & ~~$+$~~ & ~~$+$~~ & ~~$+$~~ \\
5b) & $\fabc\bar\psi T^A \sigma^{\mu\nu}i\gamma_5\psi G^B_{\mu\rho} G_{\nu}^{C\rho}$ & ~~$+$~~ & ~~$-$~~ & ~~$-$~~ \\
6a) & $\bar\psi\psi F_\munu \tilde F^{\mu\nu}$ & ~~$+$~~ & ~~$-$~~ & ~~$-$~~ \\
6b) & $\bar\psi i\gamma_5\psi F_\munu \tilde F^{\mu\nu}$ & ~~$+$~~ & ~~$+$~~ & ~~$+$~~ \\
7a) & $\bar\psi T^A\psi  F^{\mu\nu} \tilde G^A_\munu$ & ~~$+$~~ & ~~$-$~~ & ~~$-$~~ \\
7b) & $\bar\psi T^A i\gamma_5\psi  F^{\mu\nu} \tilde G^A_\munu $ & ~~$+$~~ & ~~$+$~~ & ~~$+$~~ \\
8a) & $\dabc\bar\psi T^A\psi G^B_\munu \tilde G^{C\mu\nu}$ & ~~$+$~~ & ~~$-$~~ & ~~$-$~~ \\
8b) & $\dabc\bar\psi T^A i\gamma_5\psi G^B_\munu \tilde G^{C\mu\nu}$ & ~~$+$~~ & ~~$+$~~ & ~~$+$~~ \\
9) & $(\bar\chi i \dvec D_\mu \chi) (\bar\psi\gamma^\mu\psi)$ & ~~$+$~~ & ~~$+$~~ & ~~$+$~~ \\
10) & $(\bar\chi \dvec D_\mu \gamma_5\chi) (\bar\psi\gamma^\mu\psi)$ & ~~$+$~~ & ~~$-$~~ & ~~$-$~~ \\
11) & $(\bar\chi i \dvec D_\mu \chi) (\bar\psi\gamma^\mu\gamma_5\psi)$ & ~~$-$~~ & ~~$-$~~ & ~~$+$~~ \\
12) & $(\bar\chi \dvec D_\mu \gamma_5\chi) (\bar\psi\gamma^\mu\gamma_5\psi)$ & ~~$-$~~ & ~~$+$~~ & ~~$-$~~ \\
\bottomrule
\end{tabular}
\renewcommand{\arraystretch}{1.0}
\caption{Operators in dimension 7 LEFT with well defined discrete space-time symmetries. All operators listed are hermitian and have to be multiplied by a real-valued coefficient corresponding to real or imaginary parts of the respective Wilson coefficients. This table covers the operators that are not discarded beforehand, as described in Sect.~\ref{sec:LEFT_classification}, and can be generalized in future analyses. The operators 3a) and 12) are $C$- and $CP$-odd. }
\label{tab:Liao_Dim7_LEFT_ToPe}
\end{table}

%-------------------------
\subsection{Dimension-8 LEFT}
\label{sec:LEFT_dim_8}
%-------------------------
We proceed with the classification of dimension-8 operators from Ref.~\cite{Murphy:2020cly}. For the sake of simplicity, we categorize the operators according to the number $n$ of contributing quarks ($\psi^n$),  derivatives ($D^n$), and gauge field-strength tensors ($X^n$). Once more, we first focus on the Dirac structure of each operator and characterize different combinations (given in the original basis of LEFT operators) with $X^n$, structure constants $\dabc$, $\fabc$, and $SU(3)$ generators $T^a$ in Tables~\ref{tab:LEFT_dim_8_2psi_2X_D}--\ref{tab:LEFT_dim_8_4psi_2D_4}. Operators that are not listed in these tables do either not appear in the original LEFT basis of Ref.~\cite{Murphy:2020cly} or are beforehand identified as irrelevant for our analysis, cf.\ Sect.~\ref{sec:LEFT_classification}. The results, i.e., chirality-conserving and -violating dimension-8 LEFT operators, are summarized in Sect.~\ref{app:ToPe_dim8_summary}.

%----------------------------------------
\boldmath
\subsubsection{Operator class $X^4$}
\unboldmath
%----------------------------------------
Each field-strength tensor $X_\munu$ has the signature $CPT=-+-$, while the ones in dual space, i.e., $\tilde X_\munu$, obey $CPT=--+$. Hence any combination of four field-strength tensors with no or trivial color contractions, i.e., contractions without structure constants $\fabc$ or $\dabc$, conserves $C$.
Having a look at these non-trivial color structures, there appear either terms including three gluons and one photon or four gluons and no photon. In the LEFT basis under consideration the former always involve the symmetric structure constant $\dabc$, so that terms of the form 
\beq
    \dabc G_\munu^{a} G^{b\mu\nu}  G_{\alpha\beta}^{c}F^{\alpha\beta} 
\eeq
are always $C$-even, as can be read off from Table~\ref{tab:discrete_symmetries}. The possible four-gluon operators include structures like
\beq
    d_{abe} d_{cde} G_\munu^{a} G^{b\mu\nu}  G_{\alpha\beta}^{c}G^{d\alpha\beta}\,, 
\eeq
which conserve each of the three fundamental discrete symmetries separately. This can easily be checked analogously to the example given in Sect.~\ref{sec:discrete_symm_LEFT}. The exchange of any of these field-strength tensors with its dual representation preserves the $C$-even nature of the operators. Thus there is no $C$-violating operator in this class.

%----------------------------------------
\boldmath
\subsubsection{Operator class $\psi^2X^2D$}
\unboldmath
%----------------------------------------

In this operator class we encounter
\beq
\begin{alignedat}{4}
    &c\,\bar\psi \gamma_\mu i\dvec D_\nu \PRL \psi&&+\hc&&=\Re c\,\big(\bar\psi\gamma_\mu i\dvec D_\nu \psi \pm\bar\psi\gamma_\mu \gamma_5 i\dvec D_\nu \psi\big)\,.
\end{alignedat}
\eeq
Taking appropriate linear combinations, these Dirac structures can be reduced to
\beq\label{eq:class_4psi_2X_2D_intermediate_1}
    \bar\psi\gamma_\mu i\dvec D_\nu \psi \eqand \bar\psi\gamma_\mu \gamma_5 i\dvec D_\nu \psi\,,
\eeq
which have to be multiplied with a real-valued linear combination of Wilson coefficients that can be absorbed in a single overall normalization for each of these two operators. All combinations with attached field strengths are listed in Table~\ref{tab:LEFT_dim_8_2psi_2X_D}.

%----------------------------------------
\boldmath
\subsubsection{Operator class $\psi^4X$}
\unboldmath
%----------------------------------------

The simplest quadrilinears occurring in the class $\psi^4X$ read
\beq
\begin{alignedat}{4}
    &c\,\bar\psi \gamma^\mu \PL \psi\bar\chi \gamma^\nu \PR \chi+\hc
    =\frac{1}{2}\Re c\,\big(&&
    \bar\psi \gamma^\mu \psi\bar\chi \gamma^\nu\chi
    -\bar\psi \gamma^\mu  \gamma_5\psi\bar\chi \gamma^\nu \gamma_5\chi
    \\[0.1cm]
    &&&
    -\bar\psi \gamma^\mu  \gamma_5\psi\bar\chi \gamma^\nu \chi
    +\bar\psi \gamma^\mu \psi\bar\chi \gamma^\nu \gamma_5 \chi\big)\,
\end{alignedat}
\eeq
and
\beq
\begin{alignedat}{4}
    &c\,\bar\psi \gamma^\mu \PRL \psi\bar\chi \gamma^\nu \PRL \chi+\hc
    =\frac{1}{2}\Re c\,\big(&&
     \bar\psi \gamma^\mu \psi\bar\chi \gamma^\nu\chi
    +\bar\psi \gamma^\mu \gamma_5\psi\bar\chi \gamma^\nu \gamma_5\chi
    \\[0.1cm]
    &&&
    \pm\bar\psi \gamma^\mu \gamma_5\psi\bar\chi \gamma^\nu \chi
    \pm\bar\psi \gamma^\mu \psi\bar\chi \gamma^\nu \gamma_5 \chi\big)\,.
\end{alignedat}
\eeq
Note that the expression $\bar\psi \gamma^\mu \PL \psi\bar\chi \gamma^\nu \PR \chi$ is (up to a sign) the same as $\bar\psi \gamma^\mu \PR \psi\bar\chi \gamma^\nu \PL \chi$ after a contraction with the field-strength tensor and a re-labelling $\psi\leftrightarrow\chi$. Therefore it suffices to consider only one of them.
In analogy to Eq.~\eqref{eq:class_4psi_2X_2D_intermediate_1} we can write these operators as linearly independent combinations with the same eigenvalue of $C$ by means of   
\beq
\begin{alignedat}{4}
    \bar\psi \gamma^\mu \psi\bar\chi \gamma^\nu\chi 
    \pm \bar\psi \gamma^\mu \gamma_5\psi\bar\chi \gamma^\nu \gamma_5\chi
    \eqand
    \bar\psi \gamma^\mu \gamma_5\psi\bar\chi \gamma^\nu \chi
    \pm\bar\psi \gamma^\mu \psi\bar\chi \gamma^\nu \gamma_5 \chi\,.
\end{alignedat}
\eeq
After contracting with $F_\munu$ or $G^a_\munu$ and attaching the respective color structures, the discrete symmetries of these operators can be read off straightforwardly.

Next, we have a look at 
\beq
\begin{alignedat}{4}
    &c\,\bar\psi \gamma^\mu \PL \chi\bar\chi \gamma^\nu \PR \psi+\hc&&=\Re c\,\big(\bar\psi \gamma^\mu \PL \chi\bar\chi \gamma^\nu \PR \psi+\bar\psi \gamma^\nu \PR \chi\bar\chi \gamma^\mu \PL \psi\big)\\[0.1cm]
    & && \phantom{=}+ i\,\Im c\,\big(\bar\psi \gamma^\mu \PL \chi\bar\chi \gamma^\nu \PR \psi-\bar\psi \gamma^\nu \PR \chi\bar\chi \gamma^\mu \PL \psi\big)\,.
\end{alignedat}
\eeq
To simplify the expression after expanding the projectors, we need to contract the operator with the field-strength tensors. The antisymmetry of $F_\munu$ under interchange of the Lorentz indices leads to 
\beq\label{eq:class_4psi_X_intermediate_1}
\begin{alignedat}{4}
    &\big(c\,\bar\psi \gamma^\mu \PL \chi\bar\chi \gamma^\nu \PR \psi+\hc\big)F_\munu&&=\frac{1}{2}\Re c\,\big(\bar\psi \gamma^\mu \chi\bar\chi \gamma^\nu \gamma_5 \psi-\bar\psi \gamma^\mu  \gamma_5\chi\bar\chi \gamma^\nu \psi\big)F_\munu\\[0.1cm]
    & && \phantom{=}+\frac{i}{2}\Im c\,\big(\bar\psi \gamma^\mu \chi\bar\chi \gamma^\nu\psi-\bar\psi \gamma^\mu  \gamma_5\chi\bar\chi \gamma^\nu \gamma_5 \psi\big)F_\munu\,.
\end{alignedat}
\eeq
Special care has to be taking when working out the $C$-transformation of operators mixing different flavors in a single bilinear, as is the case for the operator above. The charge conjugate of the first summand in Eq.~\eqref{eq:class_4psi_X_intermediate_1} reads
\beq
\begin{alignedat}{4}
    &C\left[\big(\bar\psi \gamma^\mu \chi\bar\chi \gamma^\nu \gamma_5 \psi-\bar\psi \gamma^\mu  \gamma_5\chi\bar\chi \gamma^\nu \psi\big)F_\munu\right]&&
    =\big(\bar\psi \gamma^\nu \gamma_5 \chi\bar\chi \gamma^\mu \psi-\bar\psi \gamma^\nu \chi\bar\chi \gamma^\mu  \gamma_5\psi\big)F_\munu\\[0.1cm]
    & &&     
    =\big(-\bar\psi \gamma^\mu \gamma_5 \chi\bar\chi \gamma^\nu \psi+\bar\psi \gamma^\mu \chi\bar\chi \gamma^\nu  \gamma_5\psi\big)F_\munu\,,
\end{alignedat}
\eeq
where, in the last step, we renamed $\mu\leftrightarrow\nu$ and again used the antisymmetry of $F_\munu$. For the second summand one can proceed analogously. Hence the operator in Eq.~\eqref{eq:class_4psi_X_intermediate_1} is $C$-even. 
The case when contracting the quadrilinear with $G^a_\munu$ instead of $F_\munu$ is slightly more intricate, as we need to account for the $SU(3)$ color generator $T^a$:
\beq
\begin{alignedat}{4}
    &c\,\bar\psi \gamma^\mu \PL &&T^a \chi\bar\chi \gamma^\nu \PR \psi G^a_\munu+\hc=
    \\[0.1cm]
    & &&
    \frac{1}{4}\Re c\,\big[
     \bar\psi \gamma^\mu T^a \chi\bar\chi \gamma^\nu \psi
    +\bar\psi \gamma^\mu T^a \chi\bar\chi \gamma^\nu \gamma_5 \psi
    -\bar\psi \gamma^\mu\gamma_5 T^a\chi\bar\chi \gamma^\nu \psi  
    \\[0.1cm]
    & && 
    \phantom{\frac{1}{4}\Re c\,\big[
     \bar\psi \gamma^\mu T^a \chi\bar\chi \gamma^\nu \psi
    \chi\bar\chi \gamma^\nu \gamma_5 \psi
    }
    -\bar\psi \gamma^\mu\gamma_5 T^a\chi\bar\chi \gamma^\nu \gamma_5\psi 
    +(\psi\leftrightarrow\chi)\big]G^a_\munu
    \\[0.1cm]
    & &&
    +\frac{i}{4}\Im c\,\big[
     \bar\psi \gamma^\mu T^a \chi\bar\chi \gamma^\nu \psi
    +\bar\psi \gamma^\mu T^a \chi\bar\chi \gamma^\nu \gamma_5 \psi
    -\bar\psi \gamma^\mu\gamma_5 T^a\chi\bar\chi \gamma^\nu \psi  
    \\[0.1cm]
    & && 
    \phantom{\frac{i}{4}\Im c\,\big[
     \bar\psi \gamma^\mu T^a \chi\bar\chi \gamma^\nu \psi
    \chi\bar\chi \gamma^\nu \gamma_5 \psi
    }
    -\bar\psi \gamma^\mu\gamma_5 T^a\chi\bar\chi \gamma^\nu \gamma_5\psi 
    -(\psi\leftrightarrow\chi)\big]G^a_\munu\,.
\end{alignedat}
\eeq
As already done several times, one can conveniently split this operator into its $C$-even and $C$-odd eigenstates.
In complete analogy, we now evaluate 
\beq
\begin{alignedat}{4}
    &c\,\fabc\,\bar\psi \gamma^\mu \PL T^a \chi\bar\chi \gamma^\nu &&\PR T^b\psi G^c_\munu+\hc=
    \\[0.1cm]
    & &&
    \frac{1}{2}\Re c\,\fabc\,\big[
     \bar\psi \gamma^\mu T^a \chi\bar\chi \gamma^\nu T^b\psi 
    -\bar\psi \gamma^\mu\gamma_5 T^a\chi\bar\chi \gamma^\nu \gamma_5 T^b\psi\big]G^c_\munu
    \\[0.1cm]
    & &&
    +\frac{i}{2}\Im c\,\fabc\,\big[
    \bar\psi \gamma^\mu T^a \chi\bar\chi \gamma^\nu T^b\gamma_5\psi
    -\bar\psi \gamma^\mu\gamma_5 T^a\chi\bar\chi \gamma^\nu T^b\psi  
    \big]G^c_\munu\,,
\end{alignedat}
\eeq
where we simplified the expression using the antisymmetry of the structure constant $\fabc$ and $G^c_\munu$. Similarly, we find
\beq
\begin{alignedat}{4}
    &c\,\dabc\,\bar\psi \gamma^\mu \PL T^a \chi\bar\chi \gamma^\nu &&\PR T^b\psi G^c_\munu+\hc
    \\[0.1cm]
    & &&
    =\frac{1}{2}\Re c\,\dabc\,\big[
    \bar\psi \gamma^\mu T^a \chi\bar\chi \gamma^\nu \gamma_5 T^b\psi
    -\bar\psi \gamma^\mu\gamma_5 T^a\chi\bar\chi \gamma^\nu T^b\psi  
    \big]G^c_\munu\,,
\end{alignedat}
\eeq
where terms symmetric under $\psi\leftrightarrow\chi$ drop out, because the operator is symmetric under $a\leftrightarrow b$ and antisymmetric under $\mu\leftrightarrow\nu$.
We continue with
\beq
\begin{alignedat}{4}
    &c\,\bar\psi \PR \psi\bar\chi \sigma^\munu \PR \chi+\hc
    &&=\frac{1}{2}\Re c\,\big(
    \bar\psi \psi\bar\chi \sigma^\munu\chi
    +\bar\psi \gamma_5\psi\bar\chi \sigma^\munu \gamma_5\chi\big)
    \\[0.1cm]
    &&&\phantom{=}
    +\frac{i}{2}\Im c\,\big(
    \bar\psi \gamma_5\psi\bar\chi \sigma^\munu\chi
    +\bar\psi\psi\bar\chi \sigma^\munu \gamma_5\chi
    \big)\,.
\end{alignedat}
\eeq
The next quadrilinear under consideration has the form 
\beq
\begin{alignedat}{4}
    c\,\bar\psi \PR &\chi\bar\chi \sigma^\munu \PR \psi+\hc
    \\[0.1cm]
    & 
    =\Re c\,\big(
     \bar\psi \PR \chi\bar\chi \sigma^\munu \PR \psi
    +\bar\chi \PL \psi \bar\psi \sigma^\munu \PL \chi\big)
    \\[0.1cm]
    & 
    \phantom{=}+i\Im c\,\big(
    \bar\psi \PR \chi\bar\chi \sigma^\munu \PR \psi
    -\bar\chi \PL \psi\bar\psi \sigma^\munu \PL \chi\big)
    \\[0.1cm]
    & 
    = \frac{1}{4}\Re c\,\left[
    \big(
    \bar\psi \chi\bar\chi \sigma^\munu \psi
    +\bar\chi \psi\bar\psi \sigma^\munu \chi
    +\bar\psi \gamma_5\chi\bar\chi \sigma^\munu\gamma_5 \psi
    +\bar\chi \gamma_5\psi\bar\chi \sigma^\munu\gamma_5 \chi
    \big)
    \right.
    \\[0.1cm]
    & 
    \phantom{==\frac{1}{4}\Re c\,}
    \left.
    +\big(
    \bar\psi \gamma_5\chi\bar\chi \sigma^\munu \psi
    +\bar\psi \chi\bar\chi \sigma^\munu\gamma_5 \psi
    -\bar\chi \gamma_5\psi\bar\chi \sigma^\munu \chi
    -\bar\chi \psi\bar\psi \sigma^\munu\gamma_5 \chi
    \big)
    \right]
    \\[0.1cm]
    & 
    \phantom{=}+\frac{i}{4}\Im c\,\left[
    \big(
    \bar\psi \chi\bar\chi \sigma^\munu \psi
    -\bar\chi \psi\bar\psi \sigma^\munu \chi
    +\bar\psi \gamma_5\chi\bar\chi \sigma^\munu\gamma_5 \psi
    -\bar\chi \gamma_5\psi\bar\chi \sigma^\munu\gamma_5 \chi
    \big)
    \right.
    \\[0.1cm]
    & 
    \phantom{==\frac{1}{4}\Re c\,}
    \left.
    +\big(
     \bar\psi \gamma_5\chi\bar\chi \sigma^\munu \psi
    +\bar\psi \chi\bar\chi \sigma^\munu\gamma_5 \psi
    +\bar\chi \gamma_5\psi\bar\chi \sigma^\munu \chi
    +\bar\chi \psi\bar\psi \sigma^\munu\gamma_5 \chi
    \big)
    \right]\,.
\end{alignedat}
\eeq
In this equation, the terms are ordered such that each expression in parenthesis has the same eigenvalue under charge conjugation. 
In the same manner, the last operator occurring in this class is
\beq
\begin{alignedat}{4}
    &c\,\bar\psi \sigma^{\mu}_{\ \lambda} \PR \chi\bar\chi \sigma^{\lambda\nu} \PR \psi+\hc
    \\[0.1cm]
    & 
    =\Re c \big(
     \bar\psi \sigma^{\mu}_{\ \lambda}\PR \chi\bar\chi \sigma^{\lambda\nu} \PR \psi
    +\bar\psi \sigma^{\mu}_{\ \lambda}\PL \chi\bar\chi \sigma^{\lambda\nu} \PL \psi\big)
    \\[0.1cm]
    & 
    \phantom{=}+i\Im c\,\big(
    \bar\psi \sigma^{\mu}_{\ \lambda}\PR \chi\bar\chi \sigma^{\lambda\nu} \PR \psi
    -\bar\psi \sigma^{\mu}_{\ \lambda}\PL \chi\bar\chi \sigma^{\lambda\nu} \PL \psi\big)
    \\[0.1cm]
    & 
    = \frac{1}{4}\Re c\left[
    \big(
    \bar\psi \sigma^{\mu}_{\ \lambda}\chi\bar\chi \sigma^{\lambda\nu} \psi
    +\bar\chi \sigma^{\mu}_{\ \lambda}\psi\bar\psi \sigma^{\lambda\nu} \chi
    +\bar\psi \sigma^{\mu}_{\ \lambda}\gamma_5\chi\bar\chi \sigma^{\lambda\nu}\gamma_5 \psi
    +\bar\chi \sigma^{\mu}_{\ \lambda}\gamma_5\psi\bar\chi \sigma^{\lambda\nu}\gamma_5 \chi
    \big)
    \right.
    \\[0.1cm]
    & 
    \phantom{==\frac{1}{4}\Re c\,}
    \left.
    +\big(
    \bar\psi \sigma^{\mu}_{\ \lambda}\gamma_5\chi\bar\chi \sigma^{\lambda\nu} \psi
    +\bar\psi \sigma^{\mu}_{\ \lambda}\chi\bar\chi \sigma^{\lambda\nu}\gamma_5 \psi
    -\bar\chi \sigma^{\mu}_{\ \lambda}\gamma_5\psi\bar\chi \sigma^{\lambda\nu} \chi
    -\bar\chi \sigma^{\mu}_{\ \lambda}\psi\bar\psi \sigma^{\lambda\nu}\gamma_5 \chi
    \big)
    \right]
    \\[0.1cm]
    & 
    \phantom{=}+\frac{i}{4}\Im c\left[
    \big(
    \bar\psi \sigma^{\mu}_{\ \lambda}\chi\bar\chi \sigma^{\lambda\nu} \psi
    -\bar\chi \sigma^{\mu}_{\ \lambda}\psi\bar\psi \sigma^{\lambda\nu} \chi
    +\bar\psi \sigma^{\mu}_{\ \lambda}\gamma_5\chi\bar\chi \sigma^{\lambda\nu}\gamma_5 \psi
    -\bar\chi \sigma^{\mu}_{\ \lambda}\gamma_5\psi\bar\chi \sigma^{\lambda\nu}\gamma_5 \chi
    \big)
    \right.
    \\[0.1cm]
    & 
    \phantom{==\frac{1}{4}\Re c\,}
    \left.
    +\big(
     \bar\psi \sigma^{\mu}_{\ \lambda}\gamma_5\chi\bar\chi \sigma^{\lambda\nu} \psi
    +\bar\psi \sigma^{\mu}_{\ \lambda}\chi\bar\chi \sigma^{\lambda\nu}\gamma_5 \psi
    +\bar\chi \sigma^{\mu}_{\ \lambda}\gamma_5\psi\bar\chi \sigma^{\lambda\nu} \chi
    +\bar\chi \sigma^{\mu}_{\ \lambda}\psi\bar\psi \sigma^{\lambda\nu}\gamma_5 \chi
    \big)
    \right]\,.
\end{alignedat}
\eeq
Once multiplied with the field-strength tensors, many operators of this class simplify depending on their color contractions.
We list the operators presented in this section with all allowed (non-vanishing) contractions with field-strength tensors in Tables~\ref{tab:LEFT_dim_8_4psi_X_1}--\ref{tab:LEFT_dim_8_4psi_X_4}.

%----------------------------------------
\boldmath
\subsubsection{Operator class $\psi^4D^2$}
\unboldmath
%----------------------------------------

The fermion multilinears in the class $\psi^4D^2$ have the simplest structure and (ignoring the derivatives for now) either appear as quadrilinears consisting of two quark currents,~i.e.,
\beq
\begin{alignedat}{4}
    &c\,\bar\psi \gamma^\mu \PL \psi\bar\chi \gamma_\mu \PR \chi+\hc
    =\frac{1}{2}\Re c\,\big(&&
    \bar\psi \gamma^\mu \psi\bar\chi \gamma_\mu\chi
    -\bar\psi \gamma^\mu  \gamma_5\psi\bar\chi \gamma_\mu \gamma_5\chi
    \\[0.1cm]
    & &&
    -\bar\psi \gamma^\mu  \gamma_5\psi\bar\chi \gamma_\mu \chi
    +\bar\psi \gamma^\mu \psi\bar\chi \gamma_\mu \gamma_5 \chi\big)\,,\\[0.1cm]
\end{alignedat}
\eeq
\beq
\begin{alignedat}{4}
    &c\,\bar\psi \gamma^\mu \PRL \psi\bar\chi \gamma_\mu \PRL \chi+\hc
    =\frac{1}{2}\Re c\,\big(&&
     \bar\psi \gamma^\mu \psi\bar\chi \gamma_\mu\chi
    +\bar\psi \gamma^\mu \gamma_5\psi\bar\chi \gamma_\mu \gamma_5\chi
    \\[0.1cm]
    & &&
    \pm\bar\psi \gamma^\mu \gamma_5\psi\bar\chi \gamma_\mu \chi
    \pm\bar\psi \gamma^\mu \psi\bar\chi \gamma_\mu \gamma_5 \chi\big)\,,
\end{alignedat}
\eeq
and
\beq
\begin{alignedat}{4}
    &c\,\bar\psi \gamma_\mu\PL \chi\bar\chi \gamma^\mu \PR \psi+\hc
    &&=\frac{1}{2}\Re c\,\big(
     \bar\psi\gamma_\mu\chi\bar\chi\gamma^\mu\psi
    -\bar\psi\gamma_\mu\gamma_5\chi\bar\chi\gamma^\mu\gamma_5\psi\big)
    \\[0.1cm]
    &&&\phantom{=}
    +\frac{i}{2}\Im c\,\big(
     \bar\psi\gamma_\mu\chi\bar\chi\gamma^\mu\gamma_5\psi
    -\bar\psi\gamma_\mu\gamma_5\chi\bar\chi\gamma^\mu\psi\big)\,,
\end{alignedat}
\eeq
or as a product of two densities like
\beq
\begin{alignedat}{4}
    &c\,\bar\psi\PR \chi\bar\chi\PR \psi+\hc
    &&=\frac{1}{2}\Re c\,\big(
     \bar\psi\chi\bar\chi\psi
    +\bar\psi\gamma_5\chi\bar\chi\gamma_5\psi\big)
    +\frac{i}{2}\Im c\,\big(
     \bar\psi\chi\bar\chi\gamma_5\psi
    +\bar\psi\gamma_5\chi\bar\chi\psi\big) 
\end{alignedat}
\eeq
and
\beq
\begin{alignedat}{4}
    &c\,\bar\psi\PR \psi\bar\chi\PR \chi+\hc
    &&=\frac{1}{2}\Re c\,\big(
     \bar\psi\psi\bar\chi\chi
    +\bar\psi\gamma_5\psi\bar\chi\gamma_5\chi\big)
    +\frac{i}{2}\Im c\,\big(
     \bar\psi\psi\bar\chi\gamma_5\chi
    +\bar\psi\gamma_5\psi\bar\chi\chi\big)\,.
\end{alignedat}
\eeq
These Dirac structures are categorized with all allowed combinations of derivatives in Tables~\ref{tab:LEFT_dim_8_4psi_2D_1}--\ref{tab:LEFT_dim_8_4psi_2D_4}.
%-----------------------------------------
\boldmath
\subsubsection{Summary of $C$- and $CP$-odd operators}
\label{app:ToPe_dim8_summary}
\unboldmath
%-----------------------------------------
For LEFT dimension-8 ToPe operators that are chirality-breaking, i.e., they  do \textit{not} arise at dimension 8 in SMEFT and are hence suppressed with respect to the chirality-breaking dimension-7 LEFT and chirality-conserving dimension-8 LEFT operators by at least one additional inverse power of the new-physics scale $\Lambda$, we find (note that all these operators vanish for $\psi=\chi$) 
\begin{aligneq}\label{eq:ToPe_dim8_suppressed}
&\big[\bar\psi \gamma^\mu \chi \bar\chi \gamma^\nu T^a \psi - \bar\psi \gamma^\mu \gamma_5 \chi \bar\chi \gamma^\nu T^a \gamma_5 \psi + (\psi\leftrightarrow\chi)\big] G^a_{\mu\nu}\,,\\[0.1cm]
&
i\big[\bar\psi \gamma^\mu \gamma_5 \chi \bar\chi \gamma^\nu T^a \psi - \bar\psi \gamma^\mu \chi \bar\chi \gamma^\nu T^a \gamma_5 \psi - (\psi\leftrightarrow\chi)\big] \tilde  G^a_{\mu\nu}\,,\\[0.1cm]
&i\big[\bar\psi T^a \chi \bar\chi \sigma^\munu T^a \psi + \bar\psi \gamma_5 T^a \chi \bar\chi \sigma^\munu \gamma_5 T^a \psi -(\psi\leftrightarrow\chi)\big] F_{\mu\nu}\,,\\[0.1cm]
&i\dabc\big[\bar\psi T^a \chi \bar\chi \sigma^\munu T^b \psi + \bar\psi \gamma_5 T^a \chi \bar\chi \sigma^\munu \gamma_5 T^b \psi -(\psi\leftrightarrow\chi)\big] G^c_{\mu\nu}\,,\\[0.1cm]
&\fabc\big[\bar\psi T^a \chi \bar\chi \sigma^\munu T^b \psi + \bar\psi \gamma_5 T^a \chi \bar\chi \sigma^\munu \gamma_5 T^b \psi +(\psi\leftrightarrow\chi)\big] G^c_{\mu\nu}\,.
\end{aligneq}
The fact that these operators are indeed chirality-violating can also be understood as follows. Quark quadrilinears in which both bilinears contain an $SU(3)_C$ generator cannot originate from a coupling to a $W$-boson as described in detail in Sect.~\ref{sec:LEFT_summary}. Although less obvious, the same holds for the first two operators in Eq.~\eqref{eq:ToPe_dim8_suppressed} as they both arise from quadrilinears with the handedness 
$\bar\psi_L \gamma^\mu \chi_L\bar\chi_R \gamma^\nu\psi_R$. Thus, all of the operators listed above are point interactions that convert left-handed $\psi$ and $\chi$ to respective right-handed ones.

Our results for LEFT dimension-8 ToPe operators that are chirality-conserving and a priori not necessarily suppressed by the chirality-violating dimension-7 ToPe operators read
\begin{aligneq}\label{eq:ToPe_dim8_quark_level}
&\fabc\bar\psi \gamma^\mu i \dvec D^\nu T^a\psi \, G^b_{\mu\rho}G^{c\,\rho}_\nu\,,\\[0.1cm]
&\bar\psi \gamma^\mu i \dvec D^\nu T^a\gamma_5\psi \, \big(F_{\mu\rho}\tilde G^{a\,\rho}_\nu \pm F_{\nu\rho}\tilde G^{a\,\rho}_\mu\big)\,,\\[0.1cm]
&\big(\bar\psi \gamma^\mu \psi \bar\chi \gamma^\nu\chi \pm \bar\psi \gamma^\mu \gamma_5 \psi \bar\chi \gamma^\nu \gamma_5 \chi \big)F_{\mu\nu}\,,\\[0.1cm]
&\big(\bar\psi \gamma^\mu T^a\psi \bar\chi \gamma^\nu T^a\chi \pm \bar\psi \gamma^\mu \gamma_5 T^a \psi \bar\chi \gamma^\nu \gamma_5 T^a \chi \big)F_{\mu\nu}\,,\\[0.1cm]
&\big(\bar\psi \gamma^\mu \psi \bar\chi \gamma^\nu T^a\chi \pm \bar\psi \gamma^\mu \gamma_5 \psi \bar\chi \gamma^\nu \gamma_5 T^a \chi \big)G^a_{\mu\nu}\,,\\[0.1cm]
&\fabc\big(\bar\psi \gamma^\mu \gamma_5 T^a \psi \bar\chi \gamma^\nu T^b \chi \pm \bar\psi \gamma^\mu T^a\psi  \bar\chi \gamma^\nu \gamma_5 T^b \chi\big)\tilde G^c_{\mu\nu}\,,\\[0.1cm]
&\dabc\big(\bar\psi \gamma^\mu T^a \psi \bar\chi \gamma^\nu T^b\chi \pm \bar\psi \gamma^\mu \gamma_5 T^a \psi \bar\chi \gamma^\nu \gamma_5 T^b \chi \big) G^c_{\mu\nu}\,,\\[0.1cm]
&i\big[\bar\psi \chi \bar\chi \sigma^\munu \psi + \bar\psi \gamma_5\chi \bar\chi \sigma^\munu \gamma_5\psi -(\psi\leftrightarrow\chi)\big] F_{\mu\nu}\,,\\[0.1cm]
&i\big[\bar\psi T^a \chi \bar\chi \sigma^\munu \psi + \bar\psi \gamma_5 T^a \chi \bar\chi \sigma^\munu \gamma_5 \psi -(\psi\leftrightarrow\chi)\big] G^a_{\mu\nu}\,,\\[0.1cm]
&i\big[\bar\psi \chi \bar\chi \sigma^\munu T^a \psi + \bar\psi \gamma_5 \chi \bar\chi \sigma^\munu \gamma_5 T^a \psi -(\psi\leftrightarrow\chi)\big] G^a_{\mu\nu}\,,\\[0.1cm]
&\big[\bar\psi  \sigma^{\lambda \mu} T^a \chi \bar\chi \sigma_\munu \psi + \bar\psi  \sigma^{\lambda \mu} \gamma_5 T^a \chi \bar\chi \sigma_\munu \gamma_5 \psi +(\psi\leftrightarrow\chi)\big] G^{a\, \nu}_{\lambda} 
\,,\\[0.1cm]
&\big[\bar\psi  \sigma^{\lambda \mu} \chi \bar\chi \sigma_\munu T^a \psi + \bar\psi \sigma^{\lambda \mu}\gamma_5 \chi \bar\chi \sigma_\munu  \gamma_5 T^a \psi +(\psi\leftrightarrow\chi)\big] G^{a\, \nu}_{\lambda}\,.
\end{aligneq}
Note that these operators are not unique, as they depend on the LEFT operator basis and the linear combinations chosen to group respective operators together with an appropriate redefinition of the Wilson coefficients. For example, the last two operators in Eq.~\eqref{eq:ToPe_dim8_quark_level} are linearly dependent, which can be shown using the antisymmetry of $\sigma_\munu$ and $G_{\lambda}^{a\, \nu}$, so that they can be understood as only one operator with an appropriately redefined Wilson coefficient. In a similar manner one can decompose all LEFT operators with `$\pm$' into two linearly independent operators each.
Moreover, we remark
that operators that only differ (up to an overall sign) in the interchange of $\psi$ and $\chi$ can be summarized as one operator by adding flavor indices.

\begin{table}[t!]
\centering
\renewcommand{\arraystretch}{1.5}
\begin{tabular}{ll|ccc}
\toprule
\multicolumn{2}{c|}{  $\bar\psi_{L/R} \gamma^\mu i \dvec D^\nu\psi_{L/R}$} & $C$ & $P$ & $T$\\
\midrule
1a) & $\bar\psi \gamma^\mu i \dvec D^\nu \psi \, F_{\mu\rho}F^{\rho}_\nu$ & ~~$+$~~ & ~~$+$~~ & ~~$+$~~ \\
1b) & $\bar\psi \gamma^\mu i \dvec D^\nu \gamma_5 \psi \, F_{\mu\rho}F^{\rho}_\nu$ 
& ~~$-$~~ & ~~$-$~~ & ~~$+$~~ \\
2a) & $\bar\psi \gamma^\mu i \dvec D^\nu \psi \, G^a_{\mu\rho}G^{a\, \rho}_\nu$
& ~~$+$~~ & ~~$+$~~ & ~~$+$~~ \\
2b) & $\bar\psi \gamma^\mu i \dvec D^\nu \gamma_5 \psi \, G^a_{\mu\rho}G^{a\,\rho}_\nu$ 
& ~~$-$~~ & ~~$-$~~ & ~~$+$~~ \\
3a) & $\fabc\bar\psi \gamma^\mu i \dvec D^\nu T^a\psi \, G^b_{\mu\rho}G^{c\,\rho}_\nu$ 
& ~~$-$~~ & ~~$+$~~ & ~~$-$~~ \\
3b) & $\fabc\bar\psi \gamma^\mu i \dvec D^\nu T^a\gamma_5\psi \, G^b_{\mu\rho}G^{c\,\rho}_\nu$ 
& ~~$+$~~ & ~~$-$~~ & ~~$-$~~ \\
4a) & $\dabc\bar\psi \gamma^\mu i \dvec D^\nu T^a\psi \, G^b_{\mu\rho}G^{c\,\rho}_\nu$ 
& ~~$+$~~ & ~~$+$~~ & ~~$+$~~ \\
4b) & $\dabc\bar\psi \gamma^\mu i \dvec D^\nu T^a\gamma_5\psi \, G^b_{\mu\rho}G^{c\,\rho}_\nu$ 
& ~~$-$~~ & ~~$-$~~ & ~~$+$~~ \\
5a) & $\bar\psi \gamma^\mu i \dvec D^\nu T^a\psi \, \big(F_{\mu\rho}G^{a\,\rho}_\nu \pm F_{\nu\rho}G^{a\,\rho}_\mu\big)$ 
& ~~$+$~~ & ~~$+$~~ & ~~$+$~~ \\
5b) & $\bar\psi \gamma^\mu i \dvec D^\nu T^a\gamma_5\psi \, \big(F_{\mu\rho}G^{a\,\rho}_\nu \pm F_{\nu\rho}G^{a\,\rho}_\mu\big)$ 
& ~~$-$~~ & ~~$-$~~ & ~~$+$~~ \\
6a) & $\bar\psi \gamma^\mu i \dvec D^\nu T^a\psi \, \big(F_{\mu\rho}\tilde G^{a\,\rho}_\nu \pm F_{\nu\rho}\tilde G^{a\,\rho}_\mu\big)$ 
& ~~$+$~~ & ~~$-$~~ & ~~$-$~~  \\
6b) & $\bar\psi \gamma^\mu i \dvec D^\nu T^a\gamma_5\psi \, \big(F_{\mu\rho}\tilde G^{a\,\rho}_\nu \pm F_{\nu\rho}\tilde G^{a\,\rho}_\mu\big)$ 
& ~~$-$~~ & ~~$+$~~ & ~~$-$~~ \\
7a) & $\fabc\bar\psi \gamma^\mu i \dvec D^\nu T^a\psi \, \big(G^b_{\mu\rho}\tilde G^{c\,\rho}_\nu \pm \tilde G^b_{\nu\rho}G^{c\,\rho}_\mu\big)$ 
& ~~$-$~~ & ~~$-$~~ & ~~$+$~~  \\
7b) & $\fabc\bar\psi \gamma^\mu i \dvec D^\nu T^a\gamma_5\psi \, \big(G^b_{\mu\rho}\tilde G^{c\,\rho}_\nu \pm \tilde G^b_{\nu\rho}G^{c\,\rho}_\mu\big)$ 
& ~~$+$~~ & ~~$+$~~ & ~~$+$~~ \\
\bottomrule
\end{tabular}
\renewcommand{\arraystretch}{1.0}
\caption{Operators of the class $\psi^2X^2D$ with well defined discrete space-time symmetries. All operators follow the description given in Table~\ref{tab:Liao_Dim7_LEFT_ToPe}. The operators 3a) and 6b) are $C$- and $CP$-odd. Note that for the cases 1a) to 4b) there are no operators with the
dual field-strength tensor $\tilde X_\nu^\rho$ in the considered LEFT basis of Ref.~\cite{Murphy:2020cly}.} 
\label{tab:LEFT_dim_8_2psi_2X_D}
\end{table}

\begin{table}[t!]
\centering
\renewcommand{\arraystretch}{1.5}
\begin{tabular}{ll|ccc}
\toprule
\multicolumn{2}{c|}{  $\bar\psi_{L/R} \gamma^\mu \psi_{L/R}\bar\chi_{L/R} \gamma^\nu\chi_{L/R} \ \& \ \bar\psi_{L} \gamma^\mu \psi_{L}\bar\chi_{R} \gamma^\nu\chi_{R}$} & $C$ & $P$ & $T$ \\
\midrule
1a) & $\big(\bar\psi \gamma^\mu \psi \bar\chi \gamma^\nu\chi \pm \bar\psi \gamma^\mu \gamma_5 \psi \bar\chi \gamma^\nu \gamma_5 \chi \big)F_{\mu\nu}$ 
& ~~$-$~~ & ~~$+$~~ & ~~$-$~~ \\
1b) & $\big(\bar\psi \gamma^\mu \gamma_5 \psi \bar\chi \gamma^\nu \chi \pm \bar\psi \gamma^\mu \psi \bar\chi \gamma^\nu \gamma_5 \chi\big)F_{\mu\nu}$
&~~$+$~~ & ~~$-$~~ & ~~$-$~~\\
2a) & $\big(\bar\psi \gamma^\mu \psi \bar\chi \gamma^\nu\chi \pm \bar\psi \gamma^\mu \gamma_5 \psi \bar\chi \gamma^\nu \gamma_5 \chi \big)\tilde F_{\mu\nu}$ 
& ~~$-$~~ & ~~$-$~~ & ~~$+$~~ \\
2b) & $\big(\bar\psi \gamma^\mu \gamma_5 \psi \bar\chi \gamma^\nu \chi \pm \bar\psi \gamma^\mu \psi \bar\chi \gamma^\nu \gamma_5 \chi\big)\tilde F_{\mu\nu}$
&~~$+$~~ & ~~$+$~~ & ~~$+$~~\\
3a) & $\big(\bar\psi \gamma^\mu T^a\psi \bar\chi \gamma^\nu T^a\chi \pm \bar\psi \gamma^\mu \gamma_5 T^a \psi \bar\chi \gamma^\nu \gamma_5 T^a \chi \big)F_{\mu\nu}$ 
& ~~$-$~~ & ~~$+$~~ & ~~$-$~~ \\
3b) & $\big(\bar\psi \gamma^\mu \gamma_5 T^a \psi \bar\chi \gamma^\nu T^a \chi \pm \bar\psi \gamma^\mu T^a \psi \bar\chi \gamma^\nu \gamma_5 T^a \chi\big)F_{\mu\nu}$
&~~$+$~~ & ~~$-$~~ & ~~$-$~~\\
4a) & $\big(\bar\psi \gamma^\mu T^a \psi \bar\chi \gamma^\nu T^a \chi \pm \bar\psi \gamma^\mu \gamma_5 T^a \psi \bar\chi \gamma^\nu \gamma_5 T^a \chi \big)\tilde F_{\mu\nu}$ 
& ~~$-$~~ & ~~$-$~~ & ~~$+$~~ \\
4b) & $\big(\bar\psi \gamma^\mu \gamma_5 T^a  \psi \bar\chi \gamma^\nu T^a \chi \pm \bar\psi \gamma^\mu T^a \psi \bar\chi \gamma^\nu \gamma_5 T^a \chi\big)\tilde F_{\mu\nu}$
&~~$+$~~ & ~~$+$~~ & ~~$+$~~\\
5a) & $\big(\bar\psi \gamma^\mu \psi \bar\chi \gamma^\nu T^a\chi \pm \bar\psi \gamma^\mu \gamma_5 \psi \bar\chi \gamma^\nu \gamma_5 T^a \chi \big)G^a_{\mu\nu}$ 
& ~~$-$~~ & ~~$+$~~ & ~~$-$~~ \\
5b) & $\big(\bar\psi \gamma^\mu \gamma_5 \psi \bar\chi \gamma^\nu T^a \chi \pm \bar\psi \gamma^\mu \psi \bar\chi \gamma^\nu \gamma_5 T^a \chi\big)G^a_{\mu\nu}$
&~~$+$~~ & ~~$-$~~ & ~~$-$~~\\
6a) & $\big(\bar\psi \gamma^\mu \psi \bar\chi \gamma^\nu T^a\chi \pm \bar\psi \gamma^\mu \gamma_5 \psi \bar\chi \gamma^\nu \gamma_5 T^a \chi \big)\tilde G^a_{\mu\nu}$ 
& ~~$-$~~ & ~~$-$~~ & ~~$+$~~ \\
6b) & $\big(\bar\psi \gamma^\mu \gamma_5 \psi \bar\chi \gamma^\nu T^a \chi \pm \bar\psi \gamma^\mu \psi \bar\chi \gamma^\nu \gamma_5 T^a \chi\big)\tilde G^a_{\mu\nu}$
&~~$+$~~ & ~~$+$~~ & ~~$+$~~\\
7a) & $\fabc\big(\bar\psi \gamma^\mu T^a \psi \bar\chi \gamma^\nu T^b\chi \pm \bar\psi \gamma^\mu \gamma_5 T^a \psi \bar\chi \gamma^\nu \gamma_5 T^b \chi \big) G^c_{\mu\nu}$ 
& ~~$+$~~ & ~~$+$~~ & ~~$+$~~ \\
7b) & $\fabc\big(\bar\psi \gamma^\mu \gamma_5 T^a \psi \bar\chi \gamma^\nu T^b \chi \pm \bar\psi \gamma^\mu  T^a\psi \bar\chi \gamma^\nu \gamma_5 T^b \chi\big) G^c_{\mu\nu}$
&~~$-$~~ & ~~$-$~~ & ~~$+$~~ \\
8a) & $\fabc\big(\bar\psi \gamma^\mu T^a \psi \bar\chi \gamma^\nu T^b\chi \pm \bar\psi \gamma^\mu \gamma_5 T^a \psi \bar\chi \gamma^\nu \gamma_5 T^b \chi \big)\tilde G^c_{\mu\nu}$ 
& ~~$+$~~ & ~~$-$~~ & ~~$-$~~ \\
8b) & $\fabc\big(\bar\psi \gamma^\mu \gamma_5 T^a \psi \bar\chi \gamma^\nu T^b \chi \pm \bar\psi \gamma^\mu T^a \psi  \bar\chi \gamma^\nu \gamma_5 T^b \chi\big)\tilde G^c_{\mu\nu}$
&~~$-$~~ & ~~$+$~~ & ~~$-$~~\\
9a) & $\dabc\big(\bar\psi \gamma^\mu T^a \psi \bar\chi \gamma^\nu T^b\chi \pm \bar\psi \gamma^\mu \gamma_5 T^a \psi \bar\chi \gamma^\nu \gamma_5 T^b \chi \big) G^c_{\mu\nu}$ 
& ~~$-$~~ & ~~$+$~~ & ~~$-$~~ \\
9b) & $\dabc\big(\bar\psi \gamma^\mu \gamma_5 T^a \psi \bar\chi \gamma^\nu T^b \chi \pm \bar\psi \gamma^\mu  T^a \psi \bar\chi \gamma^\nu \gamma_5 T^b \chi\big) G^c_{\mu\nu}$
&~~$+$~~ & ~~$-$~~ & ~~$-$~~\\
10a) & $\dabc\big(\bar\psi \gamma^\mu T^a \psi \bar\chi \gamma^\nu T^b\chi \pm \bar\psi \gamma^\mu \gamma_5 T^a \psi \bar\chi \gamma^\nu \gamma_5 T^b \chi \big)\tilde G^c_{\mu\nu}$ 
& ~~$-$~~ & ~~$-$~~ & ~~$+$~~ \\
10b) & $\dabc\big(\bar\psi \gamma^\mu \gamma_5 T^a \psi \bar\chi \gamma^\nu T^b \chi \pm \bar\psi \gamma^\mu T^a\psi  \bar\chi \gamma^\nu \gamma_5 T^b \chi\big)\tilde G^c_{\mu\nu}$
&~~$+$~~ & ~~$+$~~ & ~~$+$~~\\
\bottomrule
\end{tabular}
\renewcommand{\arraystretch}{1.0}
\caption{Operators of the class $\psi^4X$ with well defined discrete space-time symmetries. All operators follow the description given in Table~\ref{tab:Liao_Dim7_LEFT_ToPe}. The operators 1a), 3a), 5a), 8b), and 9a) are $C$- and $CP$-odd.}
\label{tab:LEFT_dim_8_4psi_X_1}
\end{table}

\begin{table}[htbp!]
\centering
\renewcommand{\arraystretch}{1.5}
\begin{tabular}{ll|ccc}
\toprule
\multicolumn{2}{c|}{$\bar\psi_L \gamma^\mu \chi_L\bar\chi_R \gamma^\nu\psi_R$} & $C$ & $P$ & $T$ \\
\midrule
1a) & $(\bar\psi \gamma^\mu \gamma_5 \chi \bar\chi \gamma^\nu \psi- \bar\psi \gamma^\mu \chi \bar\chi \gamma^\nu \gamma_5 \psi) F_{\mu\nu}$ 
& ~~$+$~~ & ~~$-$~~ & ~~$-$~~ \\
1b) & $i(\bar\psi \gamma^\mu \chi \bar\chi \gamma^\nu \psi - \bar\psi \gamma^\mu \gamma_5 \chi \bar\chi \gamma^\nu \gamma_5 \psi) F_{\mu\nu}$ 
& ~~$+$~~ & ~~$+$~~ & ~~$+$~~ \\
2a) & $(\bar\psi \gamma^\mu \gamma_5 \chi \bar\chi \gamma^\nu \psi- \bar\psi \gamma^\mu \chi \bar\chi \gamma^\nu \gamma_5 \psi) \tilde F_{\mu\nu}$ 
& ~~$+$~~ & ~~$+$~~ & ~~$+$~~ \\
2b) & $i(\bar\psi \gamma^\mu \chi \bar\chi \gamma^\nu \psi - \bar\psi \gamma^\mu \gamma_5 \chi \bar\chi \gamma^\nu \gamma_5 \psi) \tilde F_{\mu\nu}$ 
& ~~$+$~~ & ~~$-$~~ & ~~$-$~~ \\
3a) & $(\bar\psi \gamma^\mu \gamma_5 T^a \chi \bar\chi \gamma^\nu T^a \psi - \bar\psi \gamma^\mu T^a \chi \bar\chi \gamma^\nu \gamma_5 T^a \psi) F_{\mu\nu}$ 
& ~~$+$~~ & ~~$-$~~ & ~~$-$~~ \\
3b) & $i(\bar\psi \gamma^\mu T^a \chi \bar\chi \gamma^\nu T^a \psi - \bar\psi \gamma^\mu \gamma_5 T^a \chi \bar\chi \gamma^\nu \gamma_5 T^a \psi) F_{\mu\nu}$ 
& ~~$+$~~ & ~~$+$~~ & ~~$+$~~ \\
4a) & $(\bar\psi \gamma^\mu \gamma_5 T^a \chi \bar\chi \gamma^\nu T^a \psi- \bar\psi \gamma^\mu T^a \chi \bar\chi \gamma^\nu \gamma_5 T^a \psi) \tilde F_{\mu\nu}$ 
& ~~$+$~~ & ~~$+$~~ & ~~$+$~~ \\
4b) & $i(\bar\psi \gamma^\mu T^a \chi \bar\chi \gamma^\nu T^a \psi - \bar\psi \gamma^\mu \gamma_5 T^a \chi \bar\chi \gamma^\nu \gamma_5 T^a \psi) \tilde F_{\mu\nu}$ 
& ~~$+$~~ & ~~$-$~~ & ~~$-$~~ \\
5a) & $\big[\bar\psi \gamma^\mu \chi \bar\chi \gamma^\nu T^a \psi - \bar\psi \gamma^\mu \gamma_5 \chi \bar\chi \gamma^\nu T^a \gamma_5 \psi + (\psi\leftrightarrow\chi)\big] G^a_{\mu\nu}$ 
& ~~$-$~~ & ~~$+$~~ & ~~$-$~~ \\
5b) & $\big[\bar\psi \gamma^\mu \gamma_5 \chi \bar\chi \gamma^\nu T^a \psi - \bar\psi \gamma^\mu \chi \bar\chi \gamma^\nu T^a \gamma_5 \psi + (\psi\leftrightarrow\chi)\big] G^a_{\mu\nu}$ 
& ~~$+$~~ & ~~$-$~~ & ~~$-$~~ \\
5c) & $i\big[\bar\psi \gamma^\mu \chi \bar\chi \gamma^\nu T^a \psi - \bar\psi \gamma^\mu \gamma_5 \chi \bar\chi \gamma^\nu T^a \gamma_5 \psi - (\psi\leftrightarrow\chi)\big] G^a_{\mu\nu}$ 
& ~~$+$~~ & ~~$+$~~ & ~~$+$~~  \\
5d) & $i\big[\bar\psi \gamma^\mu \gamma_5 \chi \bar\chi \gamma^\nu T^a \psi - \bar\psi \gamma^\mu \chi \bar\chi \gamma^\nu T^a \gamma_5 \psi - (\psi\leftrightarrow\chi)\big] G^a_{\mu\nu}$ 
& ~~$-$~~ & ~~$-$~~ & ~~$+$~~  \\
6a) & $\big[\bar\psi \gamma^\mu \chi \bar\chi \gamma^\nu T^a \psi - \bar\psi \gamma^\mu \gamma_5 \chi \bar\chi \gamma^\nu T^a \gamma_5 \psi + (\psi\leftrightarrow\chi)\big] \tilde G^a_{\mu\nu}$ 
& ~~$-$~~ & ~~$-$~~ & ~~$+$~~ \\
6b) & $\big[\bar\psi \gamma^\mu \gamma_5 \chi \bar\chi \gamma^\nu T^a \psi - \bar\psi \gamma^\mu \chi \bar\chi \gamma^\nu T^a \gamma_5 \psi + (\psi\leftrightarrow\chi)\big] \tilde  G^a_{\mu\nu}$ 
& ~~$+$~~ & ~~$+$~~ & ~~$+$~~ \\
6c) & $i\big[\bar\psi \gamma^\mu \chi \bar\chi \gamma^\nu T^a \psi - \bar\psi \gamma^\mu \gamma_5 \chi \bar\chi \gamma^\nu T^a \gamma_5 \psi - (\psi\leftrightarrow\chi)\big] \tilde  G^a_{\mu\nu}$ 
& ~~$+$~~ & ~~$-$~~ & ~~$-$~~  \\
6d) & $i\big[\bar\psi \gamma^\mu \gamma_5 \chi \bar\chi \gamma^\nu T^a \psi - \bar\psi \gamma^\mu \chi \bar\chi \gamma^\nu T^a \gamma_5 \psi - (\psi\leftrightarrow\chi)\big] \tilde  G^a_{\mu\nu}$ 
& ~~$-$~~ & ~~$+$~~ & ~~$-$~~   \\
7a) & $i \fabc(\bar\psi \gamma^\mu T^a\chi \bar\chi \gamma^\nu T^b \gamma_5 \psi - \bar\psi \gamma^\mu T^a \gamma_5 \chi \bar\chi \gamma^\nu T^b \psi ) G^c_{\mu\nu}$ 
 & ~~$+$~~ & ~~$-$~~ & ~~$-$~~  \\
7b) & $\fabc(\bar\psi \gamma^\mu T^a\chi \bar\chi \gamma^\nu T^b \psi - \bar\psi \gamma^\mu T^a \gamma_5 \chi \bar\chi \gamma^\nu T^b \gamma_5 \psi ) G^c_{\mu\nu}$ 
& ~~$+$~~ & ~~$+$~~ & ~~$+$~~ \\
 8a) & $i\fabc(\bar\psi \gamma^\mu T^a\chi \bar\chi \gamma^\nu T^b \gamma_5 \psi - \bar\psi \gamma^\mu T^a \gamma_5 \chi \bar\chi \gamma^\nu T^b \psi ) \tilde G^c_{\mu\nu}$ 
 & ~~$+$~~ & ~~$+$~~ & ~~$+$~~ \\
8b) & $\fabc(\bar\psi \gamma^\mu T^a\chi \bar\chi \gamma^\nu T^b \psi - \bar\psi \gamma^\mu T^a \gamma_5 \chi \bar\chi \gamma^\nu T^b \gamma_5 \psi )  \tilde G^c_{\mu\nu}$ 
& ~~$+$~~ & ~~$-$~~ & ~~$-$~~ \\
9) & $\dabc(\bar\psi \gamma^\mu T^a\chi \bar\chi \gamma^\nu T^b \gamma_5 \psi - \bar\psi \gamma^\mu T^a \gamma_5 \chi \bar\chi \gamma^\nu T^b \psi ) G^c_{\mu\nu}$ 
& ~~$+$~~ & ~~$-$~~ & ~~$-$~~\\
10) & $\dabc(\bar\psi \gamma^\mu T^a\chi \bar\chi \gamma^\nu T^b \gamma_5 \psi - \bar\psi \gamma^\mu T^a \gamma_5 \chi \bar\chi \gamma^\nu T^b \psi ) \tilde G^c_{\mu\nu}$ 
& ~~$+$~~ & ~~$+$~~ & ~~$+$~~\\
\bottomrule
\end{tabular}
\renewcommand{\arraystretch}{1.0}
\caption{Operators of the class $\psi^4X$ with well defined discrete space-time symmetries. All operators follow the description given in Table~\ref{tab:Liao_Dim7_LEFT_ToPe}. The operators 5a) and 6d) are $C$- and $CP$-odd.}
\label{tab:LEFT_dim_8_4psi_X_2}
\end{table}

\begin{table}
\centering
\renewcommand{\arraystretch}{1.45}
\begin{tabular}{ll|ccc}
\toprule
\multicolumn{2}{c|}{  $\bar\psi_L \chi_R\bar\chi_L \sigma^\munu\psi_R$} & $C$ & $P$ & $T$ \\
\midrule
1a) & $i\big[\bar\psi \chi \bar\chi \sigma^\munu \psi + \bar\psi \gamma_5\chi \bar\chi \sigma^\munu \gamma_5\psi -(\psi\leftrightarrow\chi)\big] F_{\mu\nu}$ 
& ~~$-$~~ & ~~$+$~~ & ~~$-$~~ \\
1b) & $i\big[\bar\psi \gamma_5\chi \bar\chi \sigma^\munu \psi + \bar\psi \chi \bar\chi \sigma^\munu \gamma_5\psi +(\psi\leftrightarrow\chi)\big] F_{\mu\nu}$ 
& ~~$+$~~ & ~~$-$~~ & ~~$-$~~ \\
1c) & $\big[\bar\psi \chi \bar\chi \sigma^\munu \psi + \bar\psi \gamma_5 \chi \bar\chi \sigma^\munu \gamma_5\psi +(\psi\leftrightarrow\chi)\big] F_{\mu\nu}$ 
& ~~$+$~~ & ~~$+$~~ & ~~$+$~~ \\
1d) & $\big[\bar\psi \gamma_5 \chi \bar\chi \sigma^\munu \psi + \bar\psi \chi \bar\chi \sigma^\munu \gamma_5\psi -(\psi\leftrightarrow\chi)\big] F_{\mu\nu}$ 
& ~~$-$~~ & ~~$-$~~ & ~~$+$~~ \\
2a) & $i\big[\bar\psi T^a \chi \bar\chi \sigma^\munu T^a \psi + \bar\psi \gamma_5 T^a \chi \bar\chi \sigma^\munu \gamma_5 T^a \psi -(\psi\leftrightarrow\chi)\big] F_{\mu\nu}$ 
& ~~$-$~~ & ~~$+$~~ & ~~$-$~~ \\
2b) & $i\big[\bar\psi \gamma_5 T^a \chi \bar\chi \sigma^\munu T^a \psi + \bar\psi T^a \chi \bar\chi \sigma^\munu \gamma_5 T^a \psi +(\psi\leftrightarrow\chi)\big] F_{\mu\nu}$ 
& ~~$+$~~ & ~~$-$~~ & ~~$-$~~ \\
2c) & $\big[\bar\psi T^a \chi \bar\chi \sigma^\munu T^a \psi + \bar\psi \gamma_5 T^a \chi \bar\chi \sigma^\munu \gamma_5 T^a \psi +(\psi\leftrightarrow\chi)\big] F_{\mu\nu}$ 
& ~~$+$~~ & ~~$+$~~ & ~~$+$~~ \\
2d) & $\big[\bar\psi \gamma_5 T^a \chi \bar\chi \sigma^\munu T^a \psi + \bar\psi T^a \chi \bar\chi \sigma^\munu \gamma_5 T^a \psi -(\psi\leftrightarrow\chi)\big] F_{\mu\nu}$ 
& ~~$-$~~ & ~~$-$~~ & ~~$+$~~ \\
3a) & $i\big[\bar\psi T^a \chi \bar\chi \sigma^\munu \psi + \bar\psi \gamma_5 T^a \chi \bar\chi \sigma^\munu \gamma_5 \psi -(\psi\leftrightarrow\chi)\big] G^a_{\mu\nu}$ 
& ~~$-$~~ & ~~$+$~~ & ~~$-$~~ \\
3b) & $i\big[\bar\psi \gamma_5 T^a \chi \bar\chi \sigma^\munu \psi + \bar\psi T^a \chi \bar\chi \sigma^\munu \gamma_5 \psi +(\psi\leftrightarrow\chi)\big] G^a_{\mu\nu}$ 
& ~~$+$~~ & ~~$-$~~ & ~~$-$~~ \\
3c) & $\big[\bar\psi T^a \chi \bar\chi \sigma^\munu \psi + \bar\psi \gamma_5 T^a \chi \bar\chi \sigma^\munu \gamma_5 \psi +(\psi\leftrightarrow\chi)\big] G^a_{\mu\nu}$ 
& ~~$+$~~ & ~~$+$~~ & ~~$+$~~ \\
3d) & $\big[\bar\psi \gamma_5 T^a \chi \bar\chi \sigma^\munu \psi + \bar\psi T^a \chi \bar\chi \sigma^\munu \gamma_5 \psi -(\psi\leftrightarrow\chi)\big] G^a_{\mu\nu}$ 
& ~~$-$~~ & ~~$-$~~ & ~~$+$~~ \\
4a) & $i\big[\bar\psi \chi \bar\chi \sigma^\munu T^a \psi + \bar\psi \gamma_5 \chi \bar\chi \sigma^\munu \gamma_5 T^a \psi -(\psi\leftrightarrow\chi)\big] G^a_{\mu\nu}$ 
& ~~$-$~~ & ~~$+$~~ & ~~$-$~~ \\
4b) & $i\big[\bar\psi \gamma_5 \chi \bar\chi \sigma^\munu T^a \psi + \bar\psi \chi \bar\chi \sigma^\munu \gamma_5 T^a \psi +(\psi\leftrightarrow\chi)\big] G^a_{\mu\nu}$ 
& ~~$+$~~ & ~~$-$~~ & ~~$-$~~ \\
4c) & $\big[\bar\psi \chi \bar\chi \sigma^\munu T^a \psi + \bar\psi \gamma_5 \chi \bar\chi \sigma^\munu \gamma_5 T^a \psi +(\psi\leftrightarrow\chi)\big] G^a_{\mu\nu}$ 
& ~~$+$~~ & ~~$+$~~ & ~~$+$~~ \\
4d) & $\big[\bar\psi \gamma_5 \chi \bar\chi \sigma^\munu T^a \psi + \bar\psi \chi \bar\chi \sigma^\munu \gamma_5 T^a \psi -(\psi\leftrightarrow\chi)\big] G^a_{\mu\nu}$ 
& ~~$-$~~ & ~~$-$~~ & ~~$+$~~ \\
5a) & $i\dabc\big[\bar\psi T^a \chi \bar\chi \sigma^\munu T^b \psi + \bar\psi \gamma_5 T^a \chi \bar\chi \sigma^\munu \gamma_5 T^b \psi -(\psi\leftrightarrow\chi)\big] G^c_{\mu\nu}$ 
& ~~$-$~~ & ~~$+$~~ & ~~$-$~~ \\
5b) & $i\dabc\big[\bar\psi \gamma_5 T^a \chi \bar\chi \sigma^\munu T^b \psi + \bar\psi T^a \chi \bar\chi \sigma^\munu \gamma_5 T^b \psi +(\psi\leftrightarrow\chi)\big] G^c_{\mu\nu}$ 
& ~~$+$~~ & ~~$-$~~ & ~~$-$~~ \\
5c) & $\dabc\big[\bar\psi T^a \chi \bar\chi \sigma^\munu T^b \psi + \bar\psi \gamma_5 T^a \chi \bar\chi \sigma^\munu \gamma_5 T^b \psi +(\psi\leftrightarrow\chi)\big] G^c_{\mu\nu}$ 
& ~~$+$~~ & ~~$+$~~ & ~~$+$~~ \\
5d) & $\dabc\big[\bar\psi \gamma_5 T^a \chi \bar\chi \sigma^\munu T^b \psi + \bar\psi T^a \chi \bar\chi \sigma^\munu \gamma_5 T^b \psi -(\psi\leftrightarrow\chi)\big] G^c_{\mu\nu}$ 
& ~~$-$~~ & ~~$-$~~ & ~~$+$~~ \\
6a) & $i\fabc\big[\bar\psi T^a \chi \bar\chi \sigma^\munu T^b \psi + \bar\psi \gamma_5 T^a \chi \bar\chi \sigma^\munu \gamma_5 T^b \psi -(\psi\leftrightarrow\chi)\big] G^c_{\mu\nu}$ 
& ~~$+$~~ & ~~$+$~~ & ~~$+$~~ \\
6b) & $i\fabc\big[\bar\psi \gamma_5 T^a \chi \bar\chi \sigma^\munu T^b \psi + \bar\psi T^a \chi \bar\chi \sigma^\munu \gamma_5 T^b \psi +(\psi\leftrightarrow\chi)\big] G^c_{\mu\nu}$ 
& ~~$-$~~ & ~~$-$~~ & ~~$+$~~ \\
6c) & $\fabc\big[\bar\psi T^a \chi \bar\chi \sigma^\munu T^b \psi + \bar\psi \gamma_5 T^a \chi \bar\chi \sigma^\munu \gamma_5 T^b \psi +(\psi\leftrightarrow\chi)\big] G^c_{\mu\nu}$ 
& ~~$-$~~ & ~~$+$~~ & ~~$-$~~ \\
6d) & $\fabc\big[\bar\psi \gamma_5 T^a \chi \bar\chi \sigma^\munu T^b \psi + \bar\psi T^a \chi \bar\chi \sigma^\munu \gamma_5 T^b \psi -(\psi\leftrightarrow\chi)\big] G^c_{\mu\nu}$ 
& ~~$+$~~ & ~~$-$~~ & ~~$-$~~ \\
\bottomrule
\end{tabular}
\renewcommand{\arraystretch}{1.0}
\caption{Operators of the class $\psi^4X$ with well-defined discrete space-time symmetries. All operators follow the description given in Table~\ref{tab:Liao_Dim7_LEFT_ToPe}. The operators 1a), 2a), 3a), 4a), 5a), and 6c) are $C$- and $CP$-odd. Note that there are no operators with the
dual field-strength tensor $\tilde X_\nu^\rho$ in the considered LEFT basis of Ref.~\cite{Murphy:2020cly}.}
\label{tab:LEFT_dim_8_4psi_X_3}
\end{table}

\begin{table}
\centering
\renewcommand{\arraystretch}{1.5}
\resizebox{\textwidth}{!}{
\begin{tabular}{ll|ccc}
\toprule
\multicolumn{2}{c|}{  $\bar\psi_L \sigma^{\lambda \mu} \chi_R\bar\chi_L \sigma_\munu\psi_R$} & $C$ & $P$ & $T$ \\
\midrule
1a) & $i\big[\bar\psi \sigma^{\lambda \mu} \chi \bar\chi \sigma_\munu \psi + \bar\psi \sigma^{\lambda \mu} \gamma_5\chi \bar\chi \sigma_\munu \gamma_5\psi )\big] F^\nu_{\lambda}$ 
& ~~$+$~~ & ~~$+$~~ & ~~$+$~~ \\
1b) & $\big[\bar\psi  \sigma^{\lambda \mu} \gamma_5 \chi \bar\chi \sigma_\munu \psi + \bar\psi  \sigma^{\lambda \mu} \chi \bar\chi \sigma_\munu \gamma_5\psi \big] F^\nu_{\lambda}$ 
& ~~$+$~~ & ~~$-$~~ & ~~$-$~~ \\
2a) & $i\big[\bar\psi  \sigma^{\lambda \mu} T^a \chi \bar\chi \sigma_\munu T^a \psi + \bar\psi  \sigma^{\lambda \mu} \gamma_5 T^a \chi \bar\chi \sigma_\munu \gamma_5 T^a \psi \big] F^\nu_{\lambda}$ 
& ~~$+$~~ & ~~$+$~~ & ~~$+$~~ \\
2b) & $\big[\bar\psi  \sigma^{\lambda \mu} \gamma_5 T^a \chi \bar\chi \sigma_\munu T^a \psi + \bar\psi  \sigma^{\lambda \mu} T^a \chi \bar\chi \sigma_\munu \gamma_5 T^a \psi \big] F^\nu_{\lambda}$ 
& ~~$+$~~ & ~~$-$~~ & ~~$-$~~ \\
3a) & $i\big[\bar\psi  \sigma^{\lambda \mu} T^a \chi \bar\chi \sigma_\munu \psi + \bar\psi  \sigma^{\lambda \mu} \gamma_5 T^a \chi \bar\chi \sigma_\munu \gamma_5 \psi -(\psi\leftrightarrow\chi)\big] G^{a\, \nu}_{\lambda}$ 
& ~~$+$~~ & ~~$+$~~ & ~~$+$~~ \\
3b) & $i\big[\bar\psi  \sigma^{\lambda \mu} \gamma_5 T^a \chi \bar\chi \sigma_\munu \psi + \bar\psi  \sigma^{\lambda \mu} T^a \chi \bar\chi \sigma_\munu \gamma_5 \psi +(\psi\leftrightarrow\chi)\big] G^{a\, \nu}_{\lambda}$ 
& ~~$-$~~ & ~~$-$~~ & ~~$+$~~ \\
3c) & $\big[\bar\psi  \sigma^{\lambda \mu} T^a \chi \bar\chi \sigma_\munu \psi + \bar\psi  \sigma^{\lambda \mu} \gamma_5 T^a \chi \bar\chi \sigma_\munu \gamma_5 \psi +(\psi\leftrightarrow\chi)\big] G^{a\, \nu}_{\lambda}$ 
& ~~$-$~~ & ~~$+$~~ & ~~$-$~~ \\
3d) & $\big[\bar\psi  \sigma^{\lambda \mu} \gamma_5 T^a \chi \bar\chi \sigma_\munu \psi + \bar\psi  \sigma^{\lambda \mu} T^a \chi \bar\chi \sigma_\munu \gamma_5 \psi -(\psi\leftrightarrow\chi)\big] G^{a\, \nu}_{\lambda}$ 
& ~~$+$~~ & ~~$-$~~ & ~~$-$~~ \\
4a) & $i\big[\bar\psi  \sigma^{\lambda \mu} \chi \bar\chi \sigma_\munu T^a \psi + \bar\psi  \sigma^{\lambda \mu} \gamma_5 \chi \bar\chi \sigma_\munu \gamma_5 T^a \psi -(\psi\leftrightarrow\chi)\big] G^{a\, \nu}_{\lambda}$ 
& ~~$+$~~ & ~~$+$~~ & ~~$+$~~ \\
4b) & $i\big[\bar\psi  \sigma^{\lambda \mu} \gamma_5 \chi \bar\chi \sigma_\munu T^a \psi + \bar\psi  \sigma^{\lambda \mu} \chi \bar\chi \sigma_\munu \gamma_5 T^a \psi +(\psi\leftrightarrow\chi)\big] G^{a\, \nu}_{\lambda}$ 
& ~~$-$~~ & ~~$-$~~ & ~~$+$~~ \\
4c) & $\big[\bar\psi  \sigma^{\lambda \mu} \chi \bar\chi \sigma_\munu T^a \psi + \bar\psi \gamma_5 \chi \bar\chi \sigma_\munu  \gamma_5 T^a \psi +(\psi\leftrightarrow\chi)\big] G^{a\, \nu}_{\lambda}$ 
& ~~$-$~~ & ~~$+$~~ & ~~$-$~~ \\
4d) & $\big[\bar\psi  \sigma^{\lambda \mu} \gamma_5 \chi \bar\chi \sigma_\munu T^a \psi + \bar\psi  \sigma^{\lambda \mu} \chi \bar\chi \sigma_\munu \gamma_5 T^a \psi -(\psi\leftrightarrow\chi)\big] G^{a\, \nu}_{\lambda}$ 
& ~~$+$~~ & ~~$-$~~ & ~~$-$~~ \\
5a) & $i\dabc\big[\bar\psi  \sigma^{\lambda \mu} T^a \chi \bar\chi \sigma_\munu T^b \psi + \bar\psi  \sigma^{\lambda \mu} \gamma_5 T^a \chi \bar\chi \sigma_\munu \gamma_5 T^b \psi \big] G^{c\, \nu}_{\lambda}$ 
& ~~$+$~~ & ~~$+$~~ & ~~$+$~~ \\
5b) & $\dabc\big[\bar\psi  \sigma^{\lambda \mu} \gamma_5 T^a \chi \bar\chi \sigma_\munu T^b \psi + \bar\psi  \sigma^{\lambda \mu} T^a \chi \bar\chi \sigma_\munu \gamma_5 T^b \psi \big] G^{c\, \nu}_{\lambda}$ 
& ~~$+$~~ & ~~$-$~~ & ~~$-$~~ \\
6a) & $i\fabc\big[\bar\psi  \sigma^{\lambda \mu} \gamma_5 T^a \chi \bar\chi \sigma_\munu T^b \psi + \bar\psi  \sigma^{\lambda \mu} T^a \chi \bar\chi \sigma_\munu \gamma_5 T^b \psi )\big] G^{c\, \nu}_{\lambda}$ 
& ~~$+$~~ & ~~$-$~~ & ~~$-$~~ \\
6b) & $\fabc\big[\bar\psi  \sigma^{\lambda \mu} T^a \chi \bar\chi \sigma_\munu T^b \psi + \bar\psi  \sigma^{\lambda \mu} \gamma_5 T^a \chi \bar\chi \sigma_\munu \gamma_5 T^b \psi \big] G^{c\, \nu}_{\lambda}$ 
& ~~$+$~~ & ~~$+$~~ & ~~$+$~~ \\
\bottomrule
\end{tabular}
}
\renewcommand{\arraystretch}{1.0}
\caption{Operators of the class $\psi^4X$ with well defined discrete space-time symmetries. All operators follow the description given in Table~\ref{tab:Liao_Dim7_LEFT_ToPe}. The operators 3c) and 4c) are $C$- and $CP$-odd. Note that there are no operators with the
dual field-strength tensor $\tilde X_\nu^\rho$ in the considered LEFT basis of Ref.~\cite{Murphy:2020cly}.}
\label{tab:LEFT_dim_8_4psi_X_4}
\end{table}

\begin{table}
\centering
\renewcommand{\arraystretch}{1.5}
\begin{tabular}{ll|ccc}
\toprule
\multicolumn{2}{c|}{  $\bar\psi_{L/R}\gamma_\mu\psi_{L/R} \bar\chi_{L/R}\gamma^\mu\chi_{L/R} \ \& \ \bar\psi_{L/R}\gamma_\mu\psi_{L/R} \bar\chi_{R/L}\gamma^\mu\chi_{R/L}$} & $C$ & $P$ & $T$ \\
\midrule
1a) & $ D_\nu(\bar\psi\gamma_\mu\psi) D^\nu(\bar\chi\gamma^\mu\chi) \pm D_\nu(\bar\psi\gamma_\mu\gamma_5\psi) D^\nu(\bar\chi\gamma^\mu\gamma_5\chi)$ 
& ~~$+$~~ & ~~$+$~~ & ~~$+$~~ \\
1b) & $ D_\nu(\bar\psi\gamma_\mu\psi) D^\nu(\bar\chi\gamma^\mu\gamma_5\chi) \pm D_\nu(\bar\psi\gamma_\mu\gamma_5\psi) D^\nu(\bar\chi\gamma^\mu\chi)$ 
& ~~$-$~~ & ~~$-$~~ & ~~$+$~~ \\
2a) & $ \bar\psi\gamma_\mu\dvec D_\nu\psi \bar\chi\gamma^\mu\dvec D^\nu\chi \pm \bar\psi\gamma_\mu\gamma_5\dvec D_\nu\psi \bar\chi\gamma^\mu\gamma_5\dvec D^\nu\chi$ 
& ~~$+$~~ & ~~$+$~~ & ~~$+$~~ \\
2b) & $ \bar\psi\gamma_\mu\dvec D_\nu\psi \bar\chi\gamma^\mu\dvec D^\nu\gamma_5\chi \pm \bar\psi\gamma_\mu\gamma_5\dvec D_\nu\psi \bar\chi\gamma^\mu\dvec D^\nu\chi$ 
& ~~$-$~~ & ~~$-$~~ & ~~$+$~~ \\
3a) & $ D_\nu(\bar\psi\gamma_\mu T^a \psi) D^\nu(\bar\chi\gamma^\mu T^a \chi) \pm D_\nu(\bar\psi\gamma_\mu\gamma_5 T^a \psi) D^\nu(\bar\chi\gamma^\mu\gamma_5 T^a \chi)$ 
& ~~$+$~~ & ~~$+$~~ & ~~$+$~~ \\
3b) & $ D_\nu(\bar\psi\gamma_\mu T^a \psi) D^\nu(\bar\chi\gamma^\mu\gamma_5 T^a \chi) \pm D_\nu(\bar\psi\gamma_\mu\gamma_5 T^a \psi) D^\nu(\bar\chi\gamma^\mu T^a \chi)$ 
& ~~$-$~~ & ~~$-$~~ & ~~$+$~~ \\
4a) & $ \bar\psi\gamma_\mu\dvec D_\nu T^a \psi \bar\chi\gamma^\mu\dvec D^\nu T^a \chi \pm \bar\psi\gamma_\mu\gamma_5\dvec D_\nu T^a \psi \bar\chi\gamma^\mu\gamma_5\dvec D^\nu T^a \chi$ 
& ~~$+$~~ & ~~$+$~~ & ~~$+$~~ \\
4b) & $ \bar\psi\gamma_\mu\dvec D_\nu T^a \psi \bar\chi\gamma^\mu\dvec D^\nu\gamma_5 T^a \chi \pm \bar\psi\gamma_\mu\gamma_5\dvec D_\nu T^a \psi \bar\chi\gamma^\mu\dvec D^\nu T^a \chi$ 
& ~~$-$~~ & ~~$-$~~ & ~~$+$~~ \\
\bottomrule
\end{tabular}
\renewcommand{\arraystretch}{1.0}
\caption{Operators of the class $\psi^4D^2$ with well defined discrete space-time symmetries. All operators follow the description given in Table~\ref{tab:Liao_Dim7_LEFT_ToPe}. None of these operators is $C$- and $CP$-odd.
}
\label{tab:LEFT_dim_8_4psi_2D_1}
\end{table}

\begin{table}
\centering
\renewcommand{\arraystretch}{1.5}
\resizebox{\textwidth}{!}{
\begin{tabular}{ll|ccc}
\toprule
\multicolumn{2}{c|}{  $\bar\psi_{L}\gamma_\mu\chi_{L} \bar\chi_{R}\gamma^\mu\psi_{R}$} & $C$ & $P$ & $T$ \\
\midrule
1a) & $ D_\nu(\bar\psi\gamma_\mu\chi) D^\nu(\bar\chi\gamma^\mu\psi) - D_\nu(\bar\psi\gamma_\mu\gamma_5\chi) D^\nu(\bar\chi\gamma^\mu\gamma_5\psi)$ 
& ~~$+$~~ & ~~$+$~~ & ~~$+$~~ \\
1b) & $ i\big[D_\nu(\bar\psi\gamma_\mu\chi) D^\nu(\bar\chi\gamma^\mu\gamma_5\psi) - D_\nu(\bar\psi\gamma_\mu\gamma_5\chi) D^\nu(\bar\chi\gamma^\mu\psi)\big]$ 
& ~~$+$~~ & ~~$-$~~ & ~~$-$~~ \\
2a) & $ \bar\psi\gamma_\mu\dvec D_\nu\chi \bar\chi\gamma^\mu\dvec D^\nu\psi - \bar\psi\gamma_\mu\gamma_5\dvec D_\nu\chi \bar\chi\gamma^\mu\gamma_5\dvec D^\nu\psi$ 
& ~~$+$~~ & ~~$+$~~ & ~~$+$~~ \\
2b) & $ i\big[\bar\psi\gamma_\mu\dvec D_\nu\chi \bar\chi\gamma^\mu\dvec D^\nu\gamma_5\psi - \bar\psi\gamma_\mu\gamma_5\dvec D_\nu\chi \bar\chi\gamma^\mu\dvec D^\nu\psi\big]$ 
& ~~$+$~~ & ~~$-$~~ & ~~$-$~~ \\
3a) & $ D_\nu(\bar\psi\gamma_\mu T^a \chi) D^\nu(\bar\chi\gamma^\mu T^a \psi) - D_\nu(\bar\psi\gamma_\mu\gamma_5 T^a \chi) D^\nu(\bar\chi\gamma^\mu\gamma_5 T^a \psi)$ 
& ~~$+$~~ & ~~$+$~~ & ~~$+$~~ \\
3b) & $ i\big[D_\nu(\bar\psi\gamma_\mu T^a \chi) D^\nu(\bar\chi\gamma^\mu\gamma_5 T^a \psi) - D_\nu(\bar\psi\gamma_\mu\gamma_5 T^a \chi) D^\nu(\bar\chi\gamma^\mu T^a \psi)\big]$ 
& ~~$+$~~ & ~~$-$~~ & ~~$-$~~ \\
4a) & $ \bar\psi\gamma_\mu\dvec D_\nu T^a \chi \bar\chi\gamma^\mu\dvec D^\nu T^a \psi - \bar\psi\gamma_\mu\gamma_5\dvec D_\nu T^a \chi \bar\chi\gamma^\mu\gamma_5\dvec D^\nu T^a \psi$ 
& ~~$+$~~ & ~~$+$~~ & ~~$+$~~ \\
4b) & $ i\big[\bar\psi\gamma_\mu\dvec D_\nu T^a \chi \bar\chi\gamma^\mu\dvec D^\nu\gamma_5 T^a \psi - \bar\psi\gamma_\mu\gamma_5\dvec D_\nu T^a \chi \bar\chi\gamma^\mu\dvec D^\nu T^a \psi\big]$ 
& ~~$+$~~ & ~~$-$~~ & ~~$-$~~ \\
\bottomrule
\end{tabular}
}
\renewcommand{\arraystretch}{1.0}
\caption{Operators of the class $\psi^4D^2$ with well defined discrete space-time symmetries. All operators follow the description given in Table~\ref{tab:Liao_Dim7_LEFT_ToPe}. None of these operators is $C$- and $CP$-odd.}
\label{tab:LEFT_dim_8_4psi_2D_2}
\end{table}

\begin{table}[htbp!]
\centering
\renewcommand{\arraystretch}{1.5}
\begin{tabular}{ll|ccc}
\toprule
\multicolumn{2}{c|}{  $\bar\psi_{L} \chi_{R} \bar\chi_{L}  \psi_{R}$ } & $C$ & $P$ & $T$ \\
\midrule
1a) & $ D_ \mu(\bar\psi\chi) D^ \mu(\bar\chi\psi) + D_ \mu(\bar\psi\gamma_5\chi) D^ \mu(\bar\chi\gamma_5\psi)$ 
& ~~$+$~~ & ~~$+$~~ & ~~$+$~~ \\
1b) & $ i\big[D_ \mu(\bar\psi \chi) D^ \mu(\bar\chi\gamma_5\psi) + D_ \mu(\bar\psi\gamma_5\chi) D^ \mu(\bar\chi  \psi)\big]$ 
& ~~$+$~~ & ~~$-$~~ & ~~$-$~~ \\
2a) & $ \bar\psi \dvec D_ \mu\chi \bar\chi\dvec D^ \mu\psi + \bar\psi \gamma_5\dvec D_ \mu\chi \bar\chi  \gamma_5\dvec D^ \mu\psi$
& ~~$+$~~ & ~~$+$~~ & ~~$+$~~ \\
2b) & $ i\big[\bar\psi \dvec D_ \mu\chi \bar\chi  \dvec D^ \mu\gamma_5\psi + \bar\psi \gamma_5\dvec D_ \mu\chi \bar\chi  \dvec D^ \mu\psi\big]$ 
& ~~$+$~~ & ~~$-$~~ & ~~$-$~~ \\
3a) & $ D_ \mu(\bar\psi  T^a \chi) D^ \mu(\bar\chi T^a \psi) + D_ \mu(\bar\psi \gamma_5 T^a \chi) D^ \mu(\bar\chi  \gamma_5 T^a \psi)$ 
& ~~$+$~~ & ~~$+$~~ & ~~$+$~~ \\
3b) & $ i\big[D_ \mu(\bar\psi  T^a \chi) D^ \mu(\bar\chi  \gamma_5 T^a \psi) + D_ \mu(\bar\psi \gamma_5 T^a \chi) D^ \mu(\bar\chi   T^a \psi)\big]$ 
& ~~$+$~~ & ~~$-$~~ & ~~$-$~~ \\
4a) & $ \bar\psi \dvec D_ \mu T^a \chi \bar\chi  \dvec D^ \mu T^a \psi + \bar\psi \gamma_5\dvec D_ \mu T^a \chi \bar\chi  \gamma_5\dvec D^ \mu T^a \psi$ 
& ~~$+$~~ & ~~$+$~~ & ~~$+$~~ \\
4b) & $ i\big[\bar\psi \dvec D_ \mu T^a \chi \bar\chi  \dvec D^ \mu\gamma_5 T^a \psi + \bar\psi \gamma_5\dvec D_ \mu T^a \chi \bar\chi \dvec D^\mu T^a \psi\big]$ 
& ~~$+$~~ & ~~$-$~~ & ~~$-$~~ \\
\bottomrule
\end{tabular}
\renewcommand{\arraystretch}{1.0}
\caption{Operators of the class $\psi^4D^2$ with well defined discrete space-time symmetries. All operators follow the description given in Table~\ref{tab:Liao_Dim7_LEFT_ToPe}. None of these operators is $C$- and $CP$-odd.
}
\label{tab:LEFT_dim_8_4psi_2D_3}
\end{table}

\begin{table}
\centering
\renewcommand{\arraystretch}{1.5}
\begin{tabular}{ll|ccc}
\toprule
\multicolumn{2}{c|}{  $\bar\psi_{L} \psi_{R} \bar\chi_{L}  \chi_{R}$} & $C$ & $P$ & $T$ \\
\midrule
1a) & $ D_ \mu(\bar\psi\psi) D^ \mu(\bar\chi\chi) + D_ \mu(\bar\psi\gamma_5\psi) D^ \mu(\bar\chi\gamma_5\chi)$ 
& ~~$+$~~ & ~~$+$~~ & ~~$+$~~ \\
1b) & $ i\big[D_ \mu(\bar\psi \psi) D^ \mu(\bar\chi\gamma_5\chi) + D_ \mu(\bar\psi\gamma_5\psi) D^ \mu(\bar\chi  \chi)\big]$ 
& ~~$+$~~ & ~~$-$~~ & ~~$-$~~ \\
2a) & $ \bar\psi \dvec D_ \mu\psi \bar\chi\dvec D^ \mu\chi + \bar\psi \gamma_5\dvec D_ \mu\psi \bar\chi  \gamma_5\dvec D^ \mu\chi$
& ~~$+$~~ & ~~$+$~~ & ~~$+$~~ \\
2b) & $ i\big[\bar\psi \dvec D_ \mu\psi \bar\chi  \dvec D^ \mu\gamma_5\chi + \bar\psi \gamma_5\dvec D_ \mu\psi \bar\chi  \dvec D^ \mu\chi\big]$ 
& ~~$+$~~ & ~~$-$~~ & ~~$-$~~ \\
3a) & $ D_ \mu(\bar\psi  T^a \psi) D^ \mu(\bar\chi   T^a \chi) + D_ \mu(\bar\psi \gamma_5 T^a \psi) D^ \mu(\bar\chi  \gamma_5 T^a \chi)$ 
& ~~$+$~~ & ~~$+$~~ & ~~$+$~~ \\
3b) & $ i\big[D_ \mu(\bar\psi  T^a \psi) D^ \mu(\bar\chi  \gamma_5 T^a \chi) + D_ \mu(\bar\psi \gamma_5 T^a \psi) D^ \mu(\bar\chi   T^a \chi)\big]$ 
& ~~$+$~~ & ~~$-$~~ & ~~$-$~~ \\
4a) & $ \bar\psi \dvec D_ \mu T^a \psi \bar\chi  \dvec D^ \mu T^a \chi + \bar\psi \gamma_5\dvec D_ \mu T^a \psi \bar\chi  \gamma_5\dvec D^ \mu T^a \chi$ 
& ~~$+$~~ & ~~$+$~~ & ~~$+$~~ \\
4b) & $ i\big[\bar\psi \dvec D_ \mu T^a \psi \bar\chi  \dvec D^ \mu\gamma_5 T^a \chi + \bar\psi \gamma_5\dvec D_ \mu T^a \psi \bar\chi \dvec D^\mu T^a \chi\big]$ 
& ~~$+$~~ & ~~$-$~~ & ~~$-$~~ \\
\bottomrule
\end{tabular}
\renewcommand{\arraystretch}{1.0}
\caption{Operators of the class $\psi^4D^2$ with well defined discrete space-time symmetries. All operators follow the description given in Table~\ref{tab:Liao_Dim7_LEFT_ToPe}. None of these operators is $C$- and $CP$-odd.
}
\label{tab:LEFT_dim_8_4psi_2D_4}
\end{table}

\clearpage
%---------------------------------------------------------------------------------------------------
\bibliographystyle{JHEP_mod}
\bibliography{base}

\providecommand{\href}[2]{#2}\begingroup\raggedright\begin{thebibliography}{100}

\bibitem{ATLAS:2012yve}
{\scshape ATLAS} collaboration, \emph{{Observation of a new particle in the
  search for the Standard Model Higgs boson with the ATLAS detector at the
  LHC}}, \href{https://doi.org/10.1016/j.physletb.2012.08.020}{\emph{Phys.
  Lett. B} {\bfseries 716} (2012) 1}
  [\href{https://arxiv.org/abs/1207.7214}{{\ttfamily 1207.7214}}].

\bibitem{CMS:2012zhx}
{\scshape CMS} collaboration, \emph{{Combined results of searches for the
  standard model Higgs boson in $pp$ collisions at $\sqrt{s}=7$ TeV}},
  \href{https://doi.org/10.1016/j.physletb.2012.02.064}{\emph{Phys. Lett. B}
  {\bfseries 710} (2012) 26} [\href{https://arxiv.org/abs/1202.1488}{{\ttfamily
  1202.1488}}].

\bibitem{Appelquist:1974tg}
T.~Appelquist and J.~Carazzone, \emph{{Infrared Singularities and Massive
  Fields}}, \href{https://doi.org/10.1103/PhysRevD.11.2856}{\emph{Phys. Rev. D}
  {\bfseries 11} (1975) 2856}.

\bibitem{Murphy:2020rsh}
C.W.~Murphy, \emph{{Dimension-8 operators in the Standard Model Eective Field
  Theory}}, \href{https://doi.org/10.1007/JHEP10(2020)174}{\emph{JHEP}
  {\bfseries 10} (2020) 174}
  [\href{https://arxiv.org/abs/2005.00059}{{\ttfamily 2005.00059}}].

\bibitem{Weinberg:1979sa}
S.~Weinberg, \emph{{Baryon and Lepton Nonconserving Processes}},
  \href{https://doi.org/10.1103/PhysRevLett.43.1566}{\emph{Phys. Rev. Lett.}
  {\bfseries 43} (1979) 1566}.

\bibitem{Buchmuller:1985jz}
W.~Buchm{\"u}ller and D.~Wyler, \emph{{Effective Lagrangian Analysis of New
  Interactions and Flavor Conservation}},
  \href{https://doi.org/10.1016/0550-3213(86)90262-2}{\emph{Nucl. Phys. B}
  {\bfseries 268} (1986) 621}.

\bibitem{Grzadkowski:2010es}
B.~Grzadkowski, M.~Iskrzynski, M.~Misiak and J.~Rosiek, \emph{{Dimension-Six
  Terms in the Standard Model Lagrangian}},
  \href{https://doi.org/10.1007/JHEP10(2010)085}{\emph{JHEP} {\bfseries 10}
  (2010) 085} [\href{https://arxiv.org/abs/1008.4884}{{\ttfamily 1008.4884}}].

\bibitem{Lehman:2014jma}
L.~Lehman, \emph{{Extending the Standard Model Effective Field Theory with the
  Complete Set of Dimension-7 Operators}},
  \href{https://doi.org/10.1103/PhysRevD.90.125023}{\emph{Phys. Rev. D}
  {\bfseries 90} (2014) 125023}
  [\href{https://arxiv.org/abs/1410.4193}{{\ttfamily 1410.4193}}].

\bibitem{Henning:2015alf}
B.~Henning, X.~Lu, T.~Melia and H.~Murayama, \emph{{2, 84, 30, 993, 560, 15456,
  11962, 261485, ...: Higher dimension operators in the SM EFT}},
  \href{https://doi.org/10.1007/JHEP08(2017)016}{\emph{JHEP} {\bfseries 08}
  (2017) 016} [\emph{Erratum}
  \href{https://doi.org/https://doi.org/10.1007/JHEP09(2019)019}{\emph{JHEP}
  {\bfseries 09} (2019) 019}]
  [\href{https://arxiv.org/abs/1512.03433}{{\ttfamily 1512.03433}}].

\bibitem{Liao:2016hru}
Y.~Liao and X.-D.~Ma, \emph{{Renormalization Group Evolution of Dimension-seven
  Baryon- and Lepton-number-violating Operators}},
  \href{https://doi.org/10.1007/JHEP11(2016)043}{\emph{JHEP} {\bfseries 11}
  (2016) 043} [\href{https://arxiv.org/abs/1607.07309}{{\ttfamily
  1607.07309}}].

\bibitem{Li:2020gnx}
H.-L.~Li, Z.~Ren, J.~Shu, M.-L.~Xiao, J.-H.~Yu and Y.-H.~Zheng, \emph{{Complete
  set of dimension-eight operators in the standard model effective field
  theory}}, \href{https://doi.org/10.1103/PhysRevD.104.015026}{\emph{Phys. Rev.
  D} {\bfseries 104} (2021) 015026}
  [\href{https://arxiv.org/abs/2005.00008}{{\ttfamily 2005.00008}}].

\bibitem{Liao:2020jmn}
Y.~Liao and X.-D.~Ma, \emph{{An explicit construction of the dimension-9
  operator basis in the standard model effective field theory}},
  \href{https://doi.org/10.1007/JHEP11(2020)152}{\emph{JHEP} {\bfseries 11}
  (2020) 152} [\href{https://arxiv.org/abs/2007.08125}{{\ttfamily
  2007.08125}}].

\bibitem{Li:2020xlh}
H.-L.~Li, Z.~Ren, M.-L.~Xiao, J.-H.~Yu and Y.-H.~Zheng, \emph{{Complete set of
  dimension-nine operators in the standard model effective field theory}},
  \href{https://doi.org/10.1103/PhysRevD.104.015025}{\emph{Phys. Rev. D}
  {\bfseries 104} (2021) 015025}
  [\href{https://arxiv.org/abs/2007.07899}{{\ttfamily 2007.07899}}].

\bibitem{delAguila:2008ir}
F.~del Aguila, S.~Bar-Shalom, A.~Soni and J.~Wudka, \emph{{Heavy Majorana
  Neutrinos in the Effective Lagrangian Description: Application to Hadron
  Colliders}},
  \href{https://doi.org/10.1016/j.physletb.2008.11.031}{\emph{Phys. Lett. B}
  {\bfseries 670} (2009) 399}
  [\href{https://arxiv.org/abs/0806.0876}{{\ttfamily 0806.0876}}].

\bibitem{Aparici:2009fh}
A.~Aparici, K.~Kim, A.~Santamaria and J.~Wudka, \emph{{Right-handed neutrino
  magnetic moments}},
  \href{https://doi.org/10.1103/PhysRevD.80.013010}{\emph{Phys. Rev. D}
  {\bfseries 80} (2009) 013010}
  [\href{https://arxiv.org/abs/0904.3244}{{\ttfamily 0904.3244}}].

\bibitem{Bhattacharya:2015vja}
S.~Bhattacharya and J.~Wudka, \emph{{Dimension-seven operators in the standard
  model with right handed neutrinos}},
  \href{https://doi.org/10.1103/PhysRevD.94.055022}{\emph{Phys. Rev. D}
  {\bfseries 94} (2016) 055022} [\emph{Erratum}
  \href{https://doi.org/https://doi.org/10.1103/PhysRevD.95.039904}{\emph{Phys.
  Rev. D} {\bfseries 95} (2017) 039904}]
  [\href{https://arxiv.org/abs/1505.05264}{{\ttfamily 1505.05264}}].

\bibitem{Liao:2016qyd}
Y.~Liao and X.-D.~Ma, \emph{{Operators up to Dimension Seven in Standard Model
  Effective Field Theory Extended with Sterile Neutrinos}},
  \href{https://doi.org/10.1103/PhysRevD.96.015012}{\emph{Phys. Rev. D}
  {\bfseries 96} (2017) 015012}
  [\href{https://arxiv.org/abs/1612.04527}{{\ttfamily 1612.04527}}].

\bibitem{Dekens:2020ttz}
W.~Dekens, J.~de~Vries, K.~Fuyuto, E.~Mereghetti and G.~Zhou, \emph{{Sterile
  neutrinos and neutrinoless double beta decay in effective field theory}},
  \href{https://doi.org/10.1007/JHEP06(2020)097}{\emph{JHEP} {\bfseries 06}
  (2020) 097} [\href{https://arxiv.org/abs/2002.07182}{{\ttfamily
  2002.07182}}].

\bibitem{Li:2021tsq}
H.-L.~Li, Z.~Ren, M.-L.~Xiao, J.-H.~Yu and Y.-H.~Zheng, \emph{{Operator bases
  in effective field theories with sterile neutrinos: d \ensuremath{\leq} 9}},
  \href{https://doi.org/10.1007/JHEP11(2021)003}{\emph{JHEP} {\bfseries 11}
  (2021) 003} [\href{https://arxiv.org/abs/2105.09329}{{\ttfamily
  2105.09329}}].

\bibitem{Galda:2021hbr}
A.M.~Galda, M.~Neubert and S.~Renner, \emph{{ALP \textemdash{} SMEFT
  interference}}, \href{https://doi.org/10.1007/JHEP06(2021)135}{\emph{JHEP}
  {\bfseries 06} (2021) 135}
  [\href{https://arxiv.org/abs/2105.01078}{{\ttfamily 2105.01078}}].

\bibitem{Murphy:2020cly}
C.W.~Murphy, \emph{{Low-Energy Effective Field Theory below the Electroweak
  Scale: Dimension-8 Operators}},
  \href{https://doi.org/10.1007/JHEP04(2021)101}{\emph{JHEP} {\bfseries 04}
  (2021) 101} [\href{https://arxiv.org/abs/2012.13291}{{\ttfamily
  2012.13291}}].

\bibitem{Grinstein:2007iv}
B.~Grinstein and M.~Trott, \emph{{A Higgs-Higgs bound state due to new physics
  at a TeV}}, \href{https://doi.org/10.1103/PhysRevD.76.073002}{\emph{Phys.
  Rev. D} {\bfseries 76} (2007) 073002}
  [\href{https://arxiv.org/abs/0704.1505}{{\ttfamily 0704.1505}}].

\bibitem{Alonso:2012px}
R.~Alonso, M.B.~Gavela, L.~Merlo, S.~Rigolin and J.~Yepes, \emph{{The Effective
  Chiral Lagrangian for a Light Dynamical ``Higgs Particle''}},
  \href{https://doi.org/10.1016/j.physletb.2013.04.037}{\emph{Phys. Lett. B}
  {\bfseries 722} (2013) 330} [\emph{Erratum}
  \href{https://doi.org/https://doi.org/10.1016/j.physletb.2013.09.028}{\emph{Phys.
  Lett. B} {\bfseries 726} (2013) 926}]
  [\href{https://arxiv.org/abs/1212.3305}{{\ttfamily 1212.3305}}].

\bibitem{Buchalla:2013rka}
G.~Buchalla, O.~Cat\`a and C.~Krause, \emph{{Complete Electroweak Chiral
  Lagrangian with a Light Higgs at NLO}},
  \href{https://doi.org/10.1016/j.nuclphysb.2014.01.018}{\emph{Nucl. Phys. B}
  {\bfseries 880} (2014) 552} [\emph{Erratum}
  \href{https://doi.org/https://doi.org/10.1016/j.nuclphysb.2016.09.010}{\emph{Nucl.
  Phys. B} {\bfseries 913} (2016) 475}]
  [\href{https://arxiv.org/abs/1307.5017}{{\ttfamily 1307.5017}}].

\bibitem{Gavela:2014uta}
M.B.~Gavela, K.~Kanshin, P.A.N.~Machado and S.~Saa, \emph{{On the
  renormalization of the electroweak chiral Lagrangian with a Higgs}},
  \href{https://doi.org/10.1007/JHEP03(2015)043}{\emph{JHEP} {\bfseries 03}
  (2015) 043} [\href{https://arxiv.org/abs/1409.1571}{{\ttfamily 1409.1571}}].

\bibitem{Brivio:2016fzo}
I.~Brivio, J.~Gonzalez-Fraile, M.C.~Gonzalez-Garcia and L.~Merlo, \emph{{The
  complete HEFT Lagrangian after the LHC Run I}},
  \href{https://doi.org/10.1140/epjc/s10052-016-4211-9}{\emph{Eur. Phys. J. C}
  {\bfseries 76} (2016) 416}
  [\href{https://arxiv.org/abs/1604.06801}{{\ttfamily 1604.06801}}].

\bibitem{Sun:2022ssa}
H.~Sun, M.-L.~Xiao and J.-H.~Yu, \emph{{Complete NLO Operators in the Higgs
  Effective Field Theory}},  \href{https://arxiv.org/abs/2206.07722}{{\ttfamily
  2206.07722}}.

\bibitem{Sun:2022snw}
H.~Sun, M.-L.~Xiao and J.-H.~Yu, \emph{{Complete NNLO Operator Bases in Higgs
  Effective Field Theory}},  \href{https://arxiv.org/abs/2210.14939}{{\ttfamily
  2210.14939}}.

\bibitem{Sun:2022aag}
H.~Sun, Y.-N.~Wang and J.-H.~Yu, \emph{{Hilbert Series and Operator Counting on
  the Higgs Effective Field Theory}},
  \href{https://arxiv.org/abs/2211.11598}{{\ttfamily 2211.11598}}.

\bibitem{Brivio:2017vri}
I.~Brivio and M.~Trott, \emph{{The Standard Model as an Effective Field
  Theory}}, \href{https://doi.org/10.1016/j.physrep.2018.11.002}{\emph{Phys.
  Rept.} {\bfseries 793} (2019) 1}
  [\href{https://arxiv.org/abs/1706.08945}{{\ttfamily 1706.08945}}].

\bibitem{Jenkins:2017jig}
E.E.~Jenkins, A.V.~Manohar and P.~Stoffer, \emph{{Low-Energy Effective Field
  Theory below the Electroweak Scale: Operators and Matching}},
  \href{https://doi.org/10.1007/JHEP03(2018)016}{\emph{JHEP} {\bfseries 03}
  (2018) 016} [\href{https://arxiv.org/abs/1709.04486}{{\ttfamily
  1709.04486}}].

\bibitem{Jenkins:2017dyc}
E.E.~Jenkins, A.V.~Manohar and P.~Stoffer, \emph{{Low-Energy Effective Field
  Theory below the Electroweak Scale: Anomalous Dimensions}},
  \href{https://doi.org/10.1007/JHEP01(2018)084}{\emph{JHEP} {\bfseries 01}
  (2018) 084} [\href{https://arxiv.org/abs/1711.05270}{{\ttfamily
  1711.05270}}].

\bibitem{Dekens:2019ept}
W.~Dekens and P.~Stoffer, \emph{{Low-energy effective field theory below the
  electroweak scale: matching at one loop}},
  \href{https://doi.org/10.1007/JHEP10(2019)197}{\emph{JHEP} {\bfseries 10}
  (2019) 197} [\href{https://arxiv.org/abs/1908.05295}{{\ttfamily
  1908.05295}}].

\bibitem{Li:2020tsi}
H.-L.~Li, Z.~Ren, M.-L.~Xiao, J.-H.~Yu and Y.-H.~Zheng, \emph{{Low energy
  effective field theory operator basis at d \ensuremath{\leq} 9}},
  \href{https://doi.org/10.1007/JHEP06(2021)138}{\emph{JHEP} {\bfseries 06}
  (2021) 138} [\href{https://arxiv.org/abs/2012.09188}{{\ttfamily
  2012.09188}}].

\bibitem{Liao:2020zyx}
Y.~Liao, X.-D.~Ma and Q.-Y.~Wang, \emph{{Extending low energy effective field
  theory with a complete set of dimension-7 operators}},
  \href{https://doi.org/10.1007/JHEP08(2020)162}{\emph{JHEP} {\bfseries 08}
  (2020) 162} [\href{https://arxiv.org/abs/2005.08013}{{\ttfamily
  2005.08013}}].

\bibitem{Li:2020lba}
T.~Li, X.-D.~Ma and M.A.~Schmidt, \emph{{General neutrino interactions with
  sterile neutrinos in light of coherent neutrino-nucleus scattering and meson
  invisible decays}},
  \href{https://doi.org/10.1007/JHEP07(2020)152}{\emph{JHEP} {\bfseries 07}
  (2020) 152} [\href{https://arxiv.org/abs/2005.01543}{{\ttfamily
  2005.01543}}].

\bibitem{Chala:2020vqp}
M.~Chala and A.~Titov, \emph{{One-loop matching in the SMEFT extended with a
  sterile neutrino}},
  \href{https://doi.org/10.1007/JHEP05(2020)139}{\emph{JHEP} {\bfseries 05}
  (2020) 139} [\href{https://arxiv.org/abs/2001.07732}{{\ttfamily
  2001.07732}}].

\bibitem{Dekens:2022gha}
W.~Dekens, J.~de~Vries and S.~Shain, \emph{{CP-violating axion interactions in
  effective field theory}},
  \href{https://doi.org/10.1007/JHEP07(2022)014}{\emph{JHEP} {\bfseries 07}
  (2022) 014} [\href{https://arxiv.org/abs/2203.11230}{{\ttfamily
  2203.11230}}].

\bibitem{Fermi:1934sk}
E.~Fermi, \emph{{Tentativo di una Teoria Dei Raggi $\beta$}},
  \href{https://doi.org/10.1007/BF02959820}{\emph{Nuovo Cim.} {\bfseries 11}
  (1934) 1}.

\bibitem{Fermi:1934hr}
E.~Fermi, \emph{{Versuch einer Theorie der $\beta$-Strahlen. I}},
  \href{https://doi.org/10.1007/BF01351864}{\emph{Z. Phys.} {\bfseries 88}
  (1934) 161}.

\bibitem{Shi:2017ffh}
J.~Shi, \emph{{Theoretical Studies of C and CP Violation in $\eta \to \pi^+
  \pi^- \pi^0$ Decay}}, Ph.D. thesis, Kentucky University, 2017.
\newblock \href{https://doi.org/10.13023/etd.2020.388}{10.13023/etd.2020.388}.

\bibitem{Gardner:prep}
S.~Gardner and J.~Shi, \emph{{Leading-dimension, effective operators with $CP$
  and $C$ or $P$ violation in Standard Model effective field theory}},  to be
  published, 2023.

\bibitem{Sakharov:1967dj}
A.D.~Sakharov, \emph{{Violation of $CP$ invariance, $C$ asymmetry, and baryon
  asymmetry of the universe}},
  \href{https://doi.org/10.1070/PU1991v034n05ABEH002497}{\emph{Pisma Zh. Eksp.
  Teor. Fiz.} {\bfseries 5} (1967) 32}.

\bibitem{Khriplovich:1990ef}
I.B.~Khriplovich, \emph{{What do we know about T odd but P even interaction?}},
  \href{https://doi.org/10.1016/0550-3213(91)90448-7}{\emph{Nucl. Phys. B}
  {\bfseries 352} (1991) 385}.

\bibitem{Conti:1992xn}
R.S.~Conti and I.B.~Khriplovich, \emph{{New limits on T odd, P even
  interactions}},
  \href{https://doi.org/10.1103/PhysRevLett.68.3262}{\emph{Phys. Rev. Lett.}
  {\bfseries 68} (1992) 3262}.

\bibitem{Engel:1995vv}
J.~Engel, P.H.~Frampton and R.P.~Springer, \emph{{Effective Lagrangians and
  parity conserving time reversal violation at low-energies}},
  \href{https://doi.org/10.1103/PhysRevD.53.5112}{\emph{Phys. Rev. D}
  {\bfseries 53} (1996) 5112}
  [\href{https://arxiv.org/abs/nucl-th/9505026}{{\ttfamily nucl-th/9505026}}].

\bibitem{Ramsey-Musolf:1999cub}
M.J.~Ramsey-Musolf, \emph{{Electric dipole moments and the mass scale of new T
  violating, P conserving interactions}},
  \href{https://doi.org/10.1103/PhysRevLett.83.3997}{\emph{Phys. Rev. Lett.}
  {\bfseries 83} (1999) 3997} [\emph{Erratum}
  \href{https://doi.org/https://doi.org/10.1103/PhysRevLett.84.5681}{\emph{Phys.
  Rev. Lett.} {\bfseries 84} (2000) 5681}]
  [\href{https://arxiv.org/abs/hep-ph/9905429}{{\ttfamily hep-ph/9905429}}].

\bibitem{Kurylov:2000ub}
A.~Kurylov, G.C.~McLaughlin and M.J.~Ramsey-Musolf, \emph{{Constraints on T
  odd, P even interactions from electric dipole moments, revisited}},
  \href{https://doi.org/10.1103/PhysRevD.63.076007}{\emph{Phys. Rev. D}
  {\bfseries 63} (2001) 076007}
  [\href{https://arxiv.org/abs/hep-ph/0011185}{{\ttfamily hep-ph/0011185}}].

\bibitem{Maekawa:2011vs}
C.M.~Maekawa, E.~Mereghetti, J.~de~Vries and U.~van Kolck, \emph{{The
  Time-Reversal- and Parity-Violating Nuclear Potential in Chiral Effective
  Theory}}, \href{https://doi.org/10.1016/j.nuclphysa.2011.09.020}{\emph{Nucl.
  Phys. A} {\bfseries 872} (2011) 117}
  [\href{https://arxiv.org/abs/1106.6119}{{\ttfamily 1106.6119}}].

\bibitem{deVries:2012ab}
J.~de~Vries, E.~Mereghetti, R.G.E.~Timmermans and U.~van Kolck, \emph{{The
  Effective Chiral Lagrangian From Dimension-Six Parity and Time-Reversal
  Violation}}, \href{https://doi.org/10.1016/j.aop.2013.05.022}{\emph{Annals
  Phys.} {\bfseries 338} (2013) 50}
  [\href{https://arxiv.org/abs/1212.0990}{{\ttfamily 1212.0990}}].

\bibitem{Dekens:2013zca}
W.~Dekens and J.~de~Vries, \emph{{Renormalization Group Running of
  Dimension-Six Sources of Parity and Time-Reversal Violation}},
  \href{https://doi.org/10.1007/JHEP05(2013)149}{\emph{JHEP} {\bfseries 05}
  (2013) 149} [\href{https://arxiv.org/abs/1303.3156}{{\ttfamily 1303.3156}}].

\bibitem{Gasser:1983yg}
J.~Gasser and H.~Leutwyler, \emph{{Chiral Perturbation Theory to One Loop}},
  \href{https://doi.org/10.1016/0003-4916(84)90242-2}{\emph{Annals Phys.}
  {\bfseries 158} (1984) 142}.

\bibitem{Gasser:1984gg}
J.~Gasser and H.~Leutwyler, \emph{{Chiral Perturbation Theory: Expansions in
  the Mass of the Strange Quark}},
  \href{https://doi.org/10.1016/0550-3213(85)90492-4}{\emph{Nucl. Phys. B}
  {\bfseries 250} (1985) 465}.

\bibitem{Prezeau:2003xn}
G.~Prezeau, M.~Ramsey-Musolf and P.~Vogel, \emph{{Neutrinoless double beta
  decay and effective field theory}},
  \href{https://doi.org/10.1103/PhysRevD.68.034016}{\emph{Phys. Rev. D}
  {\bfseries 68} (2003) 034016}
  [\href{https://arxiv.org/abs/hep-ph/0303205}{{\ttfamily hep-ph/0303205}}].

\bibitem{Graesser:2016bpz}
M.L.~Graesser, \emph{{An electroweak basis for neutrinoless double $\beta$
  decay}}, \href{https://doi.org/10.1007/JHEP08(2017)099}{\emph{JHEP}
  {\bfseries 08} (2017) 099}
  [\href{https://arxiv.org/abs/1606.04549}{{\ttfamily 1606.04549}}].

\bibitem{Cirigliano:2017ymo}
V.~Cirigliano, W.~Dekens, M.~Graesser and E.~Mereghetti, \emph{{Neutrinoless
  double beta decay and chiral $SU(3)$}},
  \href{https://doi.org/10.1016/j.physletb.2017.04.020}{\emph{Phys. Lett. B}
  {\bfseries 769} (2017) 460}
  [\href{https://arxiv.org/abs/1701.01443}{{\ttfamily 1701.01443}}].

\bibitem{Cirigliano:2018yza}
V.~Cirigliano, W.~Dekens, J.~de~Vries, M.L.~Graesser and E.~Mereghetti,
  \emph{{A neutrinoless double beta decay master formula from effective field
  theory}}, \href{https://doi.org/10.1007/JHEP12(2018)097}{\emph{JHEP}
  {\bfseries 12} (2018) 097}
  [\href{https://arxiv.org/abs/1806.02780}{{\ttfamily 1806.02780}}].

\bibitem{Dekens:2018pbu}
W.~Dekens, E.E.~Jenkins, A.V.~Manohar and P.~Stoffer, \emph{{Non-perturbative
  effects in $\mu \to e \gamma$}},
  \href{https://doi.org/10.1007/JHEP01(2019)088}{\emph{JHEP} {\bfseries 01}
  (2019) 088} [\href{https://arxiv.org/abs/1810.05675}{{\ttfamily
  1810.05675}}].

\bibitem{Liao:2019gex}
Y.~Liao, X.-D.~Ma and H.-L.~Wang, \emph{{Effective field theory approach to
  lepton number violating decays $K^\pm\rightarrow \pi^\mp l^{\pm}l^{\pm}$:
  short-distance contribution}},
  \href{https://doi.org/10.1007/JHEP01(2020)127}{\emph{JHEP} {\bfseries 01}
  (2020) 127} [\href{https://arxiv.org/abs/1909.06272}{{\ttfamily
  1909.06272}}].

\bibitem{Liao:2020roy}
Y.~Liao, X.-D.~Ma and H.-L.~Wang, \emph{{Effective field theory approach to
  lepton number violating decays $K^\pm\rightarrow \pi^\mp l^{\pm}_\alpha
  l^{\pm}_\beta$: long-distance contribution}},
  \href{https://doi.org/10.1007/JHEP03(2020)120}{\emph{JHEP} {\bfseries 03}
  (2020) 120} [\href{https://arxiv.org/abs/2001.07378}{{\ttfamily
  2001.07378}}].

\bibitem{He:2020jly}
X.-G.~He, X.-D.~Ma, J.~Tandean and G.~Valencia, \emph{{Evading the Grossman-Nir
  bound with $\Delta I=3/2$ new physics}},
  \href{https://doi.org/10.1007/JHEP08(2020)034}{\emph{JHEP} {\bfseries 08}
  (2020) 034} [\href{https://arxiv.org/abs/2005.02942}{{\ttfamily
  2005.02942}}].

\bibitem{Liao:2021qfj}
Y.~Liao, X.-D.~Ma and H.-L.~Wang, \emph{{Effective field theory approach to
  lepton number violating \ensuremath{\tau} decays}},
  \href{https://doi.org/10.1088/1674-1137/abf72e}{\emph{Chin. Phys. C}
  {\bfseries 45} (2021) 073102}
  [\href{https://arxiv.org/abs/2102.03491}{{\ttfamily 2102.03491}}].

\bibitem{He:2021mrt}
X.-G.~He and X.-D.~Ma, \emph{{An EFT toolbox for baryon and lepton number
  violating dinucleon to dilepton decays}},
  \href{https://doi.org/10.1007/JHEP06(2021)047}{\emph{JHEP} {\bfseries 06}
  (2021) 047} [\href{https://arxiv.org/abs/2102.02562}{{\ttfamily
  2102.02562}}].

\bibitem{Bijnens:2017xrz}
J.~Bijnens and E.~Kofoed, \emph{{Chiral perturbation theory for
  neutron\textendash{}antineutron oscillations}},
  \href{https://doi.org/10.1140/epjc/s10052-017-5411-7}{\emph{Eur. Phys. J. C}
  {\bfseries 77} (2017) 867}
  [\href{https://arxiv.org/abs/1710.04383}{{\ttfamily 1710.04383}}].

\bibitem{deVries:2010ah}
J.~de~Vries, R.G.E.~Timmermans, E.~Mereghetti and U.~van Kolck, \emph{{The
  Nucleon Electric Dipole Form Factor From Dimension-Six Time-Reversal
  Violation}},
  \href{https://doi.org/10.1016/j.physletb.2010.11.042}{\emph{Phys. Lett. B}
  {\bfseries 695} (2011) 268}
  [\href{https://arxiv.org/abs/1006.2304}{{\ttfamily 1006.2304}}].

\bibitem{Bsaisou:2014oka}
J.~Bsaisou, U.-G.~Mei\ss{}ner, A.~Nogga and A.~Wirzba, \emph{{P- and
  T-Violating Lagrangians in Chiral Effective Field Theory and Nuclear Electric
  Dipole Moments}},
  \href{https://doi.org/10.1016/j.aop.2015.04.031}{\emph{Annals Phys.}
  {\bfseries 359} (2015) 317}
  [\href{https://arxiv.org/abs/1412.5471}{{\ttfamily 1412.5471}}].

\bibitem{Kamand:2016xhv}
R.~Kamand, B.~Altschul and M.R.~Schindler, \emph{{Hadronic Lorentz Violation in
  Chiral Perturbation Theory}},
  \href{https://doi.org/10.1103/PhysRevD.95.056005}{\emph{Phys. Rev. D}
  {\bfseries 95} (2017) 056005}
  [\href{https://arxiv.org/abs/1608.06503}{{\ttfamily 1608.06503}}].

\bibitem{Kamand:2017bzl}
R.~Kamand, B.~Altschul and M.R.~Schindler, \emph{{Hadronic Lorentz Violation in
  Chiral Perturbation Theory Including the Coupling to External Fields}},
  \href{https://doi.org/10.1103/PhysRevD.97.095027}{\emph{Phys. Rev. D}
  {\bfseries 97} (2018) 095027}
  [\href{https://arxiv.org/abs/1712.00838}{{\ttfamily 1712.00838}}].

\bibitem{Altschul:2019beo}
B.~Altschul and M.R.~Schindler, \emph{{Lorentz- and CPT -violating standard
  model extension in chiral perturbation theory}},
  \href{https://doi.org/10.1103/PhysRevD.100.075031}{\emph{Phys. Rev. D}
  {\bfseries 100} (2019) 075031}
  [\href{https://arxiv.org/abs/1907.02490}{{\ttfamily 1907.02490}}].

\bibitem{Gan:2020aco}
L.~Gan, B.~Kubis, E.~Passemar and S.~Tulin, \emph{{Precision tests of
  fundamental physics with \ensuremath{\eta} and \ensuremath{\eta}' mesons}},
  \href{https://doi.org/10.1016/j.physrep.2021.11.001}{\emph{Phys. Rept.}
  {\bfseries 945} (2022) 1} [\href{https://arxiv.org/abs/2007.00664}{{\ttfamily
  2007.00664}}].

\bibitem{Simonius:1975ve}
M.~Simonius, \emph{{On Time Reversal Violation in the Nucleon--Nucleon
  System}}, \href{https://doi.org/10.1016/0370-2693(75)90624-3}{\emph{Phys.
  Lett. B} {\bfseries 58} (1975) 147}.

\bibitem{Gatto:2016rae}
{\scshape REDTOP} collaboration, \emph{{The REDTOP project: Rare Eta Decays
  with a TPC for Optical Photons}},
  \href{https://doi.org/10.22323/1.282.0812}{\emph{PoS} {\bfseries ICHEP2016}
  (2016) 812}.

\bibitem{Gatto:2019dhj}
{\scshape REDTOP} collaboration, \emph{{The REDTOP experiment}},
  \href{https://arxiv.org/abs/1910.08505}{{\ttfamily 1910.08505}}.

\bibitem{REDTOP:2022slw}
{\scshape REDTOP} collaboration, \emph{{The REDTOP experiment: Rare
  $\eta/\eta^{\prime}$ Decays To Probe New Physics}},
  \href{https://arxiv.org/abs/2203.07651}{{\ttfamily 2203.07651}}.

\bibitem{Gan:2015nyc}
L.~Gan, \emph{{Probes for Fundamental QCD Symmetries and a Dark Gauge Boson via
  Light Meson Decays}}, \href{https://doi.org/10.22323/1.253.0017}{\emph{PoS}
  {\bfseries CD15} (2015) 017}.

\bibitem{JEF:2016}
{\scshape GlueX} collaboration, L.~Gan et~al., ``\textit{Eta Decays with
  Emphasis on Rare Neutral Modes: The JLab Eta Factory (JEF) Experiment}.''
  \url{https://www.jlab.org/exp_prog/proposals/14/PR12-14-004.pdf}, 2014.

\bibitem{Gan:2017kfr}
L.~Gan, \emph{{Test Fundamental Symmetries via Precision Measurements of
  $\pi^{0}$, $\eta$, and $\eta^{\prime}$ Decays}},
  \href{https://doi.org/10.7566/JPSCP.13.020063}{\emph{JPS Conf. Proc.}
  {\bfseries 13} (2017) 020063}.

\bibitem{Sanchez-Puertas:2018tnp}
P.~S\'anchez-Puertas, \emph{{$CP$ violation in $\eta$ muonic decays}},
  \href{https://doi.org/10.1007/JHEP01(2019)031}{\emph{JHEP} {\bfseries 01}
  (2019) 031} [\href{https://arxiv.org/abs/1810.13228}{{\ttfamily
  1810.13228}}].

\bibitem{Escribano:2022wug}
R.~Escribano, E.~Royo and P.~S\'anchez-Puertas, \emph{{New-physics signatures
  via $C\!P$ violation in $\eta^{(\prime)}\to\pi^0\mu^+\mu^-$ and
  $\eta^\prime\to\eta\mu^+\mu^-$ decays}},
  \href{https://doi.org/10.1007/JHEP05(2022)147}{\emph{JHEP} {\bfseries 05}
  (2022) 147} [\href{https://arxiv.org/abs/2202.04886}{{\ttfamily
  2202.04886}}].

\bibitem{Zillinger:2022eva}
M.~Zillinger, B.~Kubis and P.~S\'anchez-Puertas, \emph{{$CP$ violation in
  $\eta^{(\prime)}\to\pi^+\pi^-\mu^+\mu^-$ decays}},
  \href{https://doi.org/10.1007/JHEP12(2022)001}{\emph{JHEP} {\bfseries 12}
  (2022) 001} [\href{https://arxiv.org/abs/2210.14925}{{\ttfamily
  2210.14925}}].

\bibitem{Workman:2022ynf}
{\scshape Particle Data Group} collaboration, \emph{{Review of Particle
  Physics}}, \href{https://doi.org/10.1093/ptep/ptac097}{\emph{PTEP} {\bfseries
  2022} (2022) 083C01}.

\bibitem{Branco:1999fs}
G.C.~Branco, L.~Lavoura and J.P.~Silva, \emph{{CP Violation}}, vol.~103 of
  \emph{Int. Ser. Monogr. Phys.}, Oxford University Press (1999).

\bibitem{Ng:2011ui}
J.~Ng and S.~Tulin, \emph{{D versus d: CP Violation in Beta Decay and Electric
  Dipole Moments}},
  \href{https://doi.org/10.1103/PhysRevD.85.033001}{\emph{Phys. Rev. D}
  {\bfseries 85} (2012) 033001}
  [\href{https://arxiv.org/abs/1111.0649}{{\ttfamily 1111.0649}}].

\bibitem{Zhang:2007da}
Y.~Zhang, H.~An, X.~Ji and R.N.~Mohapatra, \emph{{General CP Violation in
  Minimal Left-Right Symmetric Model and Constraints on the Right-Handed
  Scale}}, \href{https://doi.org/10.1016/j.nuclphysb.2008.05.019}{\emph{Nucl.
  Phys. B} {\bfseries 802} (2008) 247}
  [\href{https://arxiv.org/abs/0712.4218}{{\ttfamily 0712.4218}}].

\bibitem{Deshpande:1990ip}
N.G.~Deshpande, J.F.~Gunion, B.~Kayser and F.I.~Olness, \emph{{Left-right
  symmetric electroweak models with triplet Higgs}},
  \href{https://doi.org/10.1103/PhysRevD.44.837}{\emph{Phys. Rev. D} {\bfseries
  44} (1991) 837}.

\bibitem{Pati:1974yy}
J.C.~Pati and A.~Salam, \emph{{Lepton Number as the Fourth Color}},
  \href{https://doi.org/10.1103/PhysRevD.10.275}{\emph{Phys. Rev. D} {\bfseries
  10} (1974) 275} [\emph{Erratum}
  \href{https://doi.org/https://doi.org/10.1103/PhysRevD.11.703.2}{\emph{Phys.
  Rev. D} {\bfseries 11} (1975) 703}].

\bibitem{Mohapatra:1974hk}
R.N.~Mohapatra and J.C.~Pati, \emph{{Left-Right Gauge Symmetry and an
  Isoconjugate Model of CP Violation}},
  \href{https://doi.org/10.1103/PhysRevD.11.566}{\emph{Phys. Rev. D} {\bfseries
  11} (1975) 566}.

\bibitem{Xu:2009nt}
F.~Xu, H.~An and X.~Ji, \emph{{Neutron Electric Dipole Moment Constraint on
  Scale of Minimal Left-Right Symmetric Model}},
  \href{https://doi.org/10.1007/JHEP03(2010)088}{\emph{JHEP} {\bfseries 03}
  (2010) 088} [\href{https://arxiv.org/abs/0910.2265}{{\ttfamily 0910.2265}}].

\bibitem{An:2009zh}
H.~An, X.~Ji and F.~Xu, \emph{{P-odd and CP-odd Four-Quark Contributions to
  Neutron EDM}}, \href{https://doi.org/10.1007/JHEP02(2010)043}{\emph{JHEP}
  {\bfseries 02} (2010) 043} [\href{https://arxiv.org/abs/0908.2420}{{\ttfamily
  0908.2420}}].

\bibitem{Dekens:2014jka}
W.~Dekens, J.~de~Vries, J.~Bsaisou, W.~Bernreuther, C.~Hanhart,
  U.-G.~Mei\ss{}ner et~al., \emph{{Unraveling models of CP violation through
  electric dipole moments of light nuclei}},
  \href{https://doi.org/10.1007/JHEP07(2014)069}{\emph{JHEP} {\bfseries 07}
  (2014) 069} [\href{https://arxiv.org/abs/1404.6082}{{\ttfamily 1404.6082}}].

\bibitem{Cata:2007ns}
O.~Cat\`a and V.~Mateu, \emph{{Chiral perturbation theory with tensor
  sources}}, \href{https://doi.org/10.1088/1126-6708/2007/09/078}{\emph{JHEP}
  {\bfseries 09} (2007) 078} [\href{https://arxiv.org/abs/0705.2948}{{\ttfamily
  0705.2948}}].

\bibitem{Scherer:2002tk}
S.~Scherer, \emph{{Introduction to chiral perturbation theory}},
  \href{https://doi.org/10.1007/0-306-47916-8_2}{\emph{Adv. Nucl. Phys.}
  {\bfseries 27} (2003) 277}
  [\href{https://arxiv.org/abs/hep-ph/0210398}{{\ttfamily hep-ph/0210398}}].

\bibitem{Scherer:2012xha}
S.~Scherer and M.R.~Schindler, \emph{{A Primer for Chiral Perturbation
  Theory}}, vol.~830 of \emph{Lect. Notes Phys.}, Springer (2012),
  \href{https://doi.org/10.1007/978-3-642-19254-8}{10.1007/978-3-642-19254-8}.

\bibitem{Meissner:2022odx}
U.-G.~Meißner and A.~Rusetsky, \emph{Effective Field Theories}, Cambridge
  University Press (2022),
  \href{https://doi.org/10.1017/9781108689038}{10.1017/9781108689038}.

\bibitem{Leutwyler:1993iq}
H.~Leutwyler, \emph{{On the foundations of chiral perturbation theory}},
  \href{https://doi.org/10.1006/aphy.1994.1094}{\emph{Annals Phys.} {\bfseries
  235} (1994) 165} [\href{https://arxiv.org/abs/hep-ph/9311274}{{\ttfamily
  hep-ph/9311274}}].

\bibitem{Fearing:1994ga}
H.W.~Fearing and S.~Scherer, \emph{{Extension of the chiral perturbation theory
  meson Lagrangian to order $p^6$}},
  \href{https://doi.org/10.1103/PhysRevD.53.315}{\emph{Phys. Rev. D} {\bfseries
  53} (1996) 315} [\href{https://arxiv.org/abs/hep-ph/9408346}{{\ttfamily
  hep-ph/9408346}}].

\bibitem{Cronin:1967jq}
J.A.~Cronin, \emph{{Phenomenological model of strong and weak interactions in
  chiral $U(3) \times U(3)$}},
  \href{https://doi.org/10.1103/PhysRev.161.1483}{\emph{Phys. Rev.} {\bfseries
  161} (1967) 1483}.

\bibitem{Kambor:1989tz}
J.~Kambor, J.H.~Missimer and D.~Wyler, \emph{{The Chiral Loop Expansion of the
  Nonleptonic Weak Interactions of Mesons}},
  \href{https://doi.org/10.1016/0550-3213(90)90236-7}{\emph{Nucl. Phys. B}
  {\bfseries 346} (1990) 17}.

\bibitem{Unal:2021lhb}
Y.~\"Unal, D.~Severt, J.~de~Vries, C.~Hanhart and U.-G.~Mei\ss{}ner,
  \emph{{Electric dipole moments of baryons with bottom quarks}},
  \href{https://doi.org/10.1103/PhysRevD.105.055026}{\emph{Phys. Rev. D}
  {\bfseries 105} (2022) 055026}
  [\href{https://arxiv.org/abs/2111.13000}{{\ttfamily 2111.13000}}].

\bibitem{Rosenzweig:1979ay}
C.~Rosenzweig, J.~Schechter and C.G.~Trahern, \emph{{Is the Effective
  Lagrangian for QCD a Sigma Model?}},
  \href{https://doi.org/10.1103/PhysRevD.21.3388}{\emph{Phys. Rev. D}
  {\bfseries 21} (1980) 3388}.

\bibitem{DiVecchia:1980yfw}
P.~Di~Vecchia and G.~Veneziano, \emph{{Chiral Dynamics in the Large n Limit}},
  \href{https://doi.org/10.1016/0550-3213(80)90370-3}{\emph{Nucl. Phys. B}
  {\bfseries 171} (1980) 253}.

\bibitem{Witten:1980sp}
E.~Witten, \emph{{Large N Chiral Dynamics}},
  \href{https://doi.org/10.1016/0003-4916(80)90325-5}{\emph{Annals Phys.}
  {\bfseries 128} (1980) 363}.

\bibitem{Kawarabayashi:1980dp}
K.~Kawarabayashi and N.~Ohta, \emph{{The Problem of $\eta$ in the Large $N$
  Limit: Effective Lagrangian Approach}},
  \href{https://doi.org/10.1016/0550-3213(80)90024-3}{\emph{Nucl. Phys. B}
  {\bfseries 175} (1980) 477}.

\bibitem{Nath:1979ik}
P.~Nath and R.L.~Arnowitt, \emph{{The U(1) Problem: Current Algebra and the
  Theta Vacuum}}, \href{https://doi.org/10.1103/PhysRevD.23.473}{\emph{Phys.
  Rev. D} {\bfseries 23} (1981) 473}.

\bibitem{Leutwyler:1997yr}
H.~Leutwyler, \emph{{On the 1/N expansion in chiral perturbation theory}},
  \href{https://doi.org/10.1016/S0920-5632(97)01065-7}{\emph{Nucl. Phys. B
  Proc. Suppl.} {\bfseries 64} (1998) 223}
  [\href{https://arxiv.org/abs/hep-ph/9709408}{{\ttfamily hep-ph/9709408}}].

\bibitem{Herrera-Siklody:1996tqr}
P.~Herrera-Sikl\'ody, J.I.~Latorre, P.~Pascual and J.~Taron, \emph{{Chiral
  effective Lagrangian in the large $N_c$ limit: The Nonet case}},
  \href{https://doi.org/10.1016/S0550-3213(97)00260-5}{\emph{Nucl. Phys. B}
  {\bfseries 497} (1997) 345}
  [\href{https://arxiv.org/abs/hep-ph/9610549}{{\ttfamily hep-ph/9610549}}].

\bibitem{Kaiser:2000gs}
R.~Kaiser and H.~Leutwyler, \emph{{Large $N_c$ in chiral perturbation theory}},
  \href{https://doi.org/10.1007/s100520000499}{\emph{Eur. Phys. J. C}
  {\bfseries 17} (2000) 623}
  [\href{https://arxiv.org/abs/hep-ph/0007101}{{\ttfamily hep-ph/0007101}}].

\bibitem{Bickert:2016fgy}
P.~Bickert, P.~Masjuan and S.~Scherer, \emph{{$\eta$-$\eta'$ Mixing in
  Large-$N_c$ Chiral Perturbation Theory}},
  \href{https://doi.org/10.1103/PhysRevD.95.054023}{\emph{Phys. Rev. D}
  {\bfseries 95} (2017) 054023}
  [\href{https://arxiv.org/abs/1612.05473}{{\ttfamily 1612.05473}}].

\bibitem{Serpukhov-Brussels-LosAlamos-AnnecyLAPP:1987kiw}
{\scshape Serpukhov-Brussels-Los Alamos-Annecy (LAPP)} collaboration,
  \emph{{Neutral Decays of $\eta^\prime$ (958)}},
  \href{https://doi.org/10.1007/BF01630597}{\emph{Z. Phys. C} {\bfseries 36}
  (1987) 603}.

\bibitem{KLOE:2004ukf}
{\scshape KLOE} collaboration, \emph{{Upper limit on the $\eta \to \gamma
  \gamma \gamma$ branching ratio with the KLOE detector}},
  \href{https://doi.org/10.1016/j.physletb.2004.04.012}{\emph{Phys. Lett. B}
  {\bfseries 591} (2004) 49}
  [\href{https://arxiv.org/abs/hep-ex/0402011}{{\ttfamily hep-ex/0402011}}].

\bibitem{Blik:2007ne}
A.M.~Blik et~al., \emph{{Searches for rare and forbidden neutral decays of eta
  mesons at the GAMS-4pi facility}},
  \href{https://doi.org/10.1134/S1063778807040102}{\emph{Phys. Atom. Nucl.}
  {\bfseries 70} (2007) 693}.

\bibitem{Nefkens:2005ka}
B.M.K.~Nefkens et~al., \emph{{Search for the forbidden decays $\eta\to 3
  \gamma$ and $\eta \to \pi^0 \gamma$ and the rare decay $\eta \to \pi^0 \pi^0
  \gamma \gamma$}},
  \href{https://doi.org/10.1103/PhysRevC.72.035212}{\emph{Phys. Rev. C}
  {\bfseries 72} (2005) 035212}.

\bibitem{WASA-at-COSY:2018jdv}
{\scshape WASA-at-COSY} collaboration, \emph{{Search for $C$ violation in the
  decay $\eta\rightarrow\pi^0 e^+ e^-$ with WASA-at-COSY}},
  \href{https://doi.org/10.1016/j.physletb.2018.07.017}{\emph{Phys. Lett. B}
  {\bfseries 784} (2018) 378}
  [\href{https://arxiv.org/abs/1802.08642}{{\ttfamily 1802.08642}}].

\bibitem{CLEO:1999nsy}
{\scshape CLEO} collaboration, \emph{{Search for rare and forbidden $\eta'$
  decays}}, \href{https://doi.org/10.1103/PhysRevLett.84.26}{\emph{Phys. Rev.
  Lett.} {\bfseries 84} (2000) 26}
  [\href{https://arxiv.org/abs/hep-ex/9907046}{{\ttfamily hep-ex/9907046}}].

\bibitem{Dzhelyadin:1980ti}
R.I.~Dzhelyadin et~al., \emph{{Search for Rare Decays of $\eta$ and
  $\eta^\prime$ Mesons and for Light Higgs Particles}},
  \href{https://doi.org/10.1016/0370-2693(81)91031-5}{\emph{Phys. Lett. B}
  {\bfseries 105} (1981) 239}.

\bibitem{Thaler:1972ax}
J.J.~Thaler, J.A.~Appel, A.~Kotlewski, J.G.~Layter, W.-Y.~Lee and S.~Stein,
  \emph{{Charge asymmetry in the decay $\eta \to \pi^+ \pi^- \gamma$}},
  \href{https://doi.org/10.1103/PhysRevLett.29.313}{\emph{Phys. Rev. Lett.}
  {\bfseries 29} (1972) 313}.

\bibitem{Gormley:1970qz}
M.~Gormley, E.~Hyman, W.-Y.~Lee, T.~Nash, J.~Peoples, C.~Schultz et~al.,
  \emph{{Experimental determination of the Dalitz-plot distribution of the
  decays $\eta \to \pi^+ \pi^- \pi^0$ and $\eta \to \pi^+ \pi^- \gamma$, and
  the branching ratio $\eta \to \pi^+ \pi^- \gamma/\eta \to \pi^+\pi^-\pi^0$}},
  \href{https://doi.org/10.1103/PhysRevD.2.501}{\emph{Phys. Rev. D} {\bfseries
  2} (1970) 501}.

\bibitem{Jane:1974es}
M.R.~Jane et~al., \emph{{A measurement of the charge asymmetry in the decay
  $\eta \to \pi^+ \pi^- \gamma$}},
  \href{https://doi.org/10.1016/0370-2693(74)90028-8}{\emph{Phys. Lett. B}
  {\bfseries 48} (1974) 265}.

\bibitem{CrystalBall:2005zrs}
{\scshape Crystal Ball} collaboration, \emph{{Test of Charge Conjugation
  Invariance}},
  \href{https://doi.org/10.1103/PhysRevLett.94.041601}{\emph{Phys. Rev. Lett.}
  {\bfseries 94} (2005) 041601}.

\bibitem{McDonough:1988nf}
J.~McDonough et~al., \emph{{New Searches for the C Noninvariant Decay $\pi^0
  \to 3 \gamma$ and the Rare Decay $\pi^0 \to 4 \gamma$}},
  \href{https://doi.org/10.1103/PhysRevD.38.2121}{\emph{Phys. Rev. D}
  {\bfseries 38} (1988) 2121}.

\bibitem{Akdag:2021efj}
H.~Akdag, T.~Isken and B.~Kubis, \emph{{Patterns of C- and CP-violation in
  hadronic \ensuremath{\eta} and \ensuremath{\eta}' three-body decays}},
  \href{https://doi.org/10.1007/JHEP02(2022)137}{\emph{JHEP} {\bfseries 02}
  (2022) 137} [\href{https://arxiv.org/abs/2111.02417}{{\ttfamily
  2111.02417}}].

\bibitem{Gardner:2019nid}
S.~Gardner and J.~Shi, \emph{{Patterns of CP violation from mirror symmetry
  breaking in the $\eta\to\pi^+\pi^-\pi^0$ Dalitz plot}},
  \href{https://doi.org/10.1103/PhysRevD.101.115038}{\emph{Phys. Rev. D}
  {\bfseries 101} (2020) 115038}
  [\href{https://arxiv.org/abs/1903.11617}{{\ttfamily 1903.11617}}].

\bibitem{Manohar:1983md}
A.~Manohar and H.~Georgi, \emph{{Chiral Quarks and the Nonrelativistic Quark
  Model}}, \href{https://doi.org/10.1016/0550-3213(84)90231-1}{\emph{Nucl.
  Phys. B} {\bfseries 234} (1984) 189}.

\bibitem{Weinberg:1989dx}
S.~Weinberg, \emph{{Larger Higgs Exchange Terms in the Neutron Electric Dipole
  Moment}}, \href{https://doi.org/10.1103/PhysRevLett.63.2333}{\emph{Phys. Rev.
  Lett.} {\bfseries 63} (1989) 2333}.

\bibitem{Georgi:1992dw}
H.~Georgi, \emph{{Generalized dimensional analysis}},
  \href{https://doi.org/10.1016/0370-2693(93)91728-6}{\emph{Phys. Lett. B}
  {\bfseries 298} (1993) 187}
  [\href{https://arxiv.org/abs/hep-ph/9207278}{{\ttfamily hep-ph/9207278}}].

\bibitem{Jenkins:2013sda}
E.E.~Jenkins, A.V.~Manohar and M.~Trott, \emph{{Naive Dimensional Analysis
  Counting of Gauge Theory Amplitudes and Anomalous Dimensions}},
  \href{https://doi.org/10.1016/j.physletb.2013.09.020}{\emph{Phys. Lett. B}
  {\bfseries 726} (2013) 697}
  [\href{https://arxiv.org/abs/1309.0819}{{\ttfamily 1309.0819}}].

\bibitem{Gavela:2016bzc}
B.M.~Gavela, E.E.~Jenkins, A.V.~Manohar and L.~Merlo, \emph{{Analysis of
  General Power Counting Rules in Effective Field Theory}},
  \href{https://doi.org/10.1140/epjc/s10052-016-4332-1}{\emph{Eur. Phys. J. C}
  {\bfseries 76} (2016) 485}
  [\href{https://arxiv.org/abs/1601.07551}{{\ttfamily 1601.07551}}].

\bibitem{Anastasi:2016cdz}
{\scshape KLOE-2} collaboration, \emph{{Precision measurement of the
  $\eta\to\pi^{+}\pi^{-}\pi^{0}$ Dalitz plot distribution with the KLOE
  detector}}, \href{https://doi.org/10.1007/JHEP05(2016)019}{\emph{JHEP}
  {\bfseries 05} (2016) 019}
  [\href{https://arxiv.org/abs/1601.06985}{{\ttfamily 1601.06985}}].

\bibitem{Gaspero:2008rs}
M.~Gaspero, B.~Meadows, K.~Mishra and A.~Soffer, \emph{{Isospin analysis of
  $D^0$ decay to three pions}},
  \href{https://doi.org/10.1103/PhysRevD.78.014015}{\emph{Phys. Rev. D}
  {\bfseries 78} (2008) 014015}
  [\href{https://arxiv.org/abs/0805.4050}{{\ttfamily 0805.4050}}].

\bibitem{Prentki:1965tt}
J.~Prentki and M.J.G.~Veltman, \emph{{Possibility of $CP$ violation in
  semistrong interactions}},
  \href{https://doi.org/10.1016/0031-9163(65)91141-8}{\emph{Phys. Lett.}
  {\bfseries 15} (1965) 88}.

\bibitem{BESIII:2016tdb}
{\scshape BESIII} collaboration, \emph{{Amplitude Analysis of the Decays
  $\eta^\prime \rightarrow \pi^+\pi^-\pi^0$ and $\eta^\prime \rightarrow
  \pi^0\pi^0\pi^0$}},
  \href{https://doi.org/10.1103/PhysRevLett.118.012001}{\emph{Phys. Rev. Lett.}
  {\bfseries 118} (2017) 012001}
  [\href{https://arxiv.org/abs/1606.03847}{{\ttfamily 1606.03847}}].

\bibitem{BESIII:2017djm}
{\scshape BESIII} collaboration, \emph{{Measurement of the matrix elements for
  the decays $\eta^{\prime}\rightarrow\eta\pi^+\pi^-$ and
  $\eta^{\prime}\rightarrow\eta\pi^0\pi^0$}},
  \href{https://doi.org/10.1103/PhysRevD.97.012003}{\emph{Phys. Rev. D}
  {\bfseries 97} (2018) 012003}
  [\href{https://arxiv.org/abs/1709.04627}{{\ttfamily 1709.04627}}].

\bibitem{Sakurai:1964}
J.J.~Sakurai, \emph{Invariance principles and elementary particles}, Princeton
  University Press (1964).

\bibitem{Barrett:1965ia}
B.~Barrett, M.~Jacob, M.~Nauenberg and T.N.~Truong, \emph{{Consequences of
  $C$-Violating Interactions in $\eta^{0}$ and $X^{0}$ Decays}},
  \href{https://doi.org/10.1103/PhysRev.141.1342}{\emph{Phys. Rev.} {\bfseries
  141} (1966) 1342}.

\bibitem{Ecker:1987}
G.~Ecker, A.~Pich and E.~{de Rafael}, \emph{{$K\to \pi \ell^+\ell^-$ decays in
  the effective chiral lagrangian of the standard model}},
  \href{https://doi.org/https://doi.org/10.1016/0550-3213(87)90491-3}{\emph{Nucl.
  Phys. B} {\bfseries 291} (1987) 692}.

\bibitem{DAmbrosio:1998gur}
G.~D'Ambrosio, G.~Ecker, G.~Isidori and J.~Portoles, \emph{{The Decays $K \to
  \pi l^+ l^-$ beyond leading order in the chiral expansion}},
  \href{https://doi.org/10.1088/1126-6708/1998/08/004}{\emph{JHEP} {\bfseries
  08} (1998) 004} [\href{https://arxiv.org/abs/hep-ph/9808289}{{\ttfamily
  hep-ph/9808289}}].

\bibitem{Bernstein:1965}
J.~Bernstein, G.~Feinberg and T.D.~Lee, \emph{{Possible $C$, $T$ Noninvariance
  in the Electromagnetic Interaction}},
  \href{https://doi.org/10.1103/PhysRev.139.B1650}{\emph{Phys. Rev.} {\bfseries
  139} (1965) B1650}.

\bibitem{Bazin:1968zz}
M.J.~Bazin, A.T.~Goshaw, A.R.~Zacher and C.R.~Sun, \emph{{An Evaluation of
  Searches for C Nonconservation in eta Decay}},
  \href{https://doi.org/10.1103/PhysRevLett.20.895}{\emph{Phys. Rev. Lett.}
  {\bfseries 20} (1968) 895}.

\bibitem{Cheng:1967zza}
T.P.~Cheng, \emph{{$C$-Conserving Decay $\eta \to \pi^0e^+e^-$ in a
  Vector-Meson-Dominant Model}},
  \href{https://doi.org/10.1103/PhysRev.162.1734}{\emph{Phys. Rev.} {\bfseries
  162} (1967) 1734}.

\bibitem{Ng:1992yg}
J.N.~Ng and D.J.~Peters, \emph{{The Decay of the $\eta$ meson into $\pi \mu^+
  \mu^-$}}, \href{https://doi.org/10.1103/PhysRevD.46.5034}{\emph{Phys. Rev. D}
  {\bfseries 46} (1992) 5034}.

\bibitem{Escribano:2020rfs}
R.~Escribano and E.~Royo, \emph{{A theoretical analysis of the semileptonic
  decays $\eta ^{(\prime )}\rightarrow \pi ^0l^+l^-$ and $\eta ^\prime
  \rightarrow \eta l^+l^-$}},
  \href{https://doi.org/10.1140/epjc/s10052-020-08748-4}{\emph{Eur. Phys. J. C}
  {\bfseries 80} (2020) 1190} [\emph{Erratum}
  \href{https://doi.org/10.1140/epjc/s10052-022-10717-y}{\emph{Eur. Phys. J. C}
  {\bfseries 82} (2022) 743}]
  [\href{https://arxiv.org/abs/2007.12467}{{\ttfamily 2007.12467}}].

\bibitem{Kubis:2010mp}
B.~Kubis and R.~Schmidt, \emph{{Radiative corrections in $K \to \pi \ell^+
  \ell^-$ decays}},
  \href{https://doi.org/10.1140/epjc/s10052-010-1442-z}{\emph{Eur. Phys. J. C}
  {\bfseries 70} (2010) 219} [\href{https://arxiv.org/abs/1007.1887}{{\ttfamily
  1007.1887}}].

\bibitem{Stollenwerk:2011zz}
F.~Stollenwerk, C.~Hanhart, A.~Kup\'s\'c, U.-G.~Mei{\ss}ner and A.~Wirzba,
  \emph{{Model-independent approach to $\eta\to\pi^+\pi^-\gamma$ and
  $\eta'\to\pi^+\pi^-\gamma$}},
  \href{https://doi.org/10.1016/j.physletb.2011.12.008}{\emph{Phys. Lett. B}
  {\bfseries 707} (2012) 184}
  [\href{https://arxiv.org/abs/1108.2419}{{\ttfamily 1108.2419}}].

\bibitem{Hanhart:2013vba}
C.~Hanhart, A.~Kupść, U.-G.~Meißner, F.~Stollenwerk and A.~Wirzba,
  \emph{{Dispersive analysis for $\eta\to \gamma\gamma^*$}},
  \href{https://doi.org/10.1140/epjc/s10052-013-2668-3,
  10.1140/epjc/s10052-015-3429-2}{\emph{Eur. Phys. J. C} {\bfseries 73} (2013)
  2668} [\emph{Erratum}
  \href{https://doi.org/10.1140/epjc/s10052-015-3429-2}{\emph{Eur. Phys. J. C}
  {\bfseries 75} (2015) 242}]
  [\href{https://arxiv.org/abs/1307.5654}{{\ttfamily 1307.5654}}].

\bibitem{Kubis:2015sga}
B.~Kubis and J.~Plenter, \emph{{Anomalous decay and scattering processes of the
  $\eta $ meson}},
  \href{https://doi.org/10.1140/epjc/s10052-015-3495-5}{\emph{Eur. Phys. J. C}
  {\bfseries 75} (2015) 283}
  [\href{https://arxiv.org/abs/1504.02588}{{\ttfamily 1504.02588}}].

\bibitem{Hanhart:2016pcd}
C.~Hanhart, S.~Holz, B.~Kubis, A.~Kup\'s\'c, A.~Wirzba and C.-W.~Xiao,
  \emph{{The branching ratio $\omega \rightarrow \pi ^+\pi ^-$ revisited}},
  \href{https://doi.org/10.1140/epjc/s10052-017-4651-x}{\emph{Eur. Phys. J. C}
  {\bfseries 77} (2017) 98} [\emph{Erratum}
  \href{https://doi.org/10.1140/epjc/s10052-018-5941-7}{\emph{Eur. Phys. J. C}
  {\bfseries 78} (2018) 450}]
  [\href{https://arxiv.org/abs/1611.09359}{{\ttfamily 1611.09359}}].

\bibitem{Holz:2015tcg}
S.~Holz, J.~Plenter, C.-W.~Xiao, T.~Dato, C.~Hanhart, B.~Kubis et~al.,
  \emph{{Towards an improved understanding of $\eta \to \gamma^* \gamma^*$}},
  \href{https://doi.org/10.1140/epjc/s10052-021-09661-0}{\emph{Eur. Phys. J. C}
  {\bfseries 81} (2021) 1002}
  [\href{https://arxiv.org/abs/1509.02194}{{\ttfamily 1509.02194}}].

\bibitem{Holz:2022hwz}
S.~Holz, C.~Hanhart, M.~Hoferichter and B.~Kubis, \emph{{A dispersive analysis
  of $\eta '\rightarrow \pi ^+\pi ^-\gamma$ and $\eta '\rightarrow \ell ^+\ell
  ^-\gamma$}},
  \href{https://doi.org/10.1140/epjc/s10052-022-10247-7}{\emph{Eur. Phys. J. C}
  {\bfseries 82} (2022) 434}
  [\href{https://arxiv.org/abs/2202.05846}{{\ttfamily 2202.05846}}].

\bibitem{Barrett:1966}
B.~Barrett and T.N.~Truong, \emph{{Analysis of $\eta^0$, $X^0 \to \pi^+ \pi^-
  \gamma$ with a Possible C Violation}},
  \href{https://doi.org/10.1103/PhysRev.147.1161}{\emph{Phys. Rev.} {\bfseries
  147} (1966) 1161}.

\bibitem{Wess:1971yu}
J.~Wess and B.~Zumino, \emph{{Consequences of anomalous Ward identities}},
  \href{https://doi.org/10.1016/0370-2693(71)90582-X}{\emph{Phys. Lett. B}
  {\bfseries 37} (1971) 95}.

\bibitem{Witten:1983tw}
E.~Witten, \emph{{Global Aspects of Current Algebra}},
  \href{https://doi.org/10.1016/0550-3213(83)90063-9}{\emph{Nucl. Phys.}
  {\bfseries B223} (1983) 422}.

\bibitem{Jacob:1959at}
M.~Jacob and G.C.~Wick, \emph{{On the General Theory of Collisions for
  Particles with Spin}},
  \href{https://doi.org/10.1016/0003-4916(59)90051-X}{\emph{Annals Phys.}
  {\bfseries 7} (1959) 404}.

\bibitem{Omnes:1958hv}
R.~Omn\`es, \emph{{On the Solution of certain singular integral equations of
  quantum field theory}}, \href{https://doi.org/10.1007/BF02747746}{\emph{Nuovo
  Cim.} {\bfseries 8} (1958) 316}.

\bibitem{Garcia-Martin:2011iqs}
R.~Garc\'ia-Mart\'in, R.~Kami\'nski, J.R.~Pel\'aez, J.~Ruiz~de Elvira and
  F.J.~Yndur\'ain, \emph{{The pion--pion scattering amplitude. IV: Improved
  analysis with once subtracted Roy-like equations up to 1100 MeV}},
  \href{https://doi.org/10.1103/PhysRevD.83.074004}{\emph{Phys. Rev. D}
  {\bfseries 83} (2011) 074004}
  [\href{https://arxiv.org/abs/1102.2183}{{\ttfamily 1102.2183}}].

\bibitem{Akdag:2018}
H.~Akdag, \emph{{Resonanzkopplungen in $\eta^{(\prime)}\to\pi^+\pi^-\gamma$}},
  Bachelor's thesis, Bonn University, 2018.

\bibitem{Nefkens:2002sa}
B.M.K.~Nefkens and J.W.~Price, \emph{{The Neutral decay modes of the eta
  meson}}, \href{https://doi.org/10.1238/Physica.Topical.099a00114}{\emph{Phys.
  Scripta T} {\bfseries 99} (2002) 114}
  [\href{https://arxiv.org/abs/nucl-ex/0202008}{{\ttfamily nucl-ex/0202008}}].

\bibitem{Jarlskog:2002zz}
C.~Jarlskog and E.~Shabalin, \emph{{On searches for CP, $T$, CPT and C
  violation in flavour-changing and flavour-conserving interactions}},
  \href{https://doi.org/10.1238/Physica.Topical.099a00023}{\emph{Phys. Scripta
  T} {\bfseries 99} (2002) 23}.

\bibitem{Guo:2011ir}
F.-K.~Guo, B.~Kubis and A.~Wirzba, \emph{{Anomalous decays of $\eta'$ and
  $\eta$ into four pions}},
  \href{https://doi.org/10.1103/PhysRevD.85.014014}{\emph{Phys. Rev. D}
  {\bfseries 85} (2012) 014014}
  [\href{https://arxiv.org/abs/1111.5949}{{\ttfamily 1111.5949}}].

\bibitem{Cabibbo:1965zzb}
N.~Cabibbo and A.~Maksymowicz, \emph{{Angular Correlations in $K_{e4}$ Decays
  and Determination of Low-Energy $\pi\pi$ Phase Shifts}},
  \href{https://doi.org/10.1103/PhysRev.137.B438,
  10.1103/PhysRev.168.1926}{\emph{Phys. Rev.} {\bfseries 137} (1965) B438}
  [\emph{Erratum} \href{https://doi.org/10.1103/PhysRev.168.1926}{\emph{Phys.
  Rev.} {\bfseries 168} (1968) 1926}].

\bibitem{Barker:2002ib}
A.R.~Barker, H.~Huang, P.A.~Toale and J.~Engle, \emph{{Radiative corrections to
  double Dalitz decays: Effects on invariant mass distributions and angular
  correlations}}, \href{https://doi.org/10.1103/PhysRevD.67.033008}{\emph{Phys.
  Rev. D} {\bfseries 67} (2003) 033008}
  [\href{https://arxiv.org/abs/hep-ph/0210174}{{\ttfamily hep-ph/0210174}}].

\bibitem{Kampf:2018wau}
K.~Kampf, J.~Novotn\'y and P.~S\'anchez-Puertas, \emph{{Radiative corrections
  to double-Dalitz decays revisited}},
  \href{https://doi.org/10.1103/PhysRevD.97.056010}{\emph{Phys. Rev. D}
  {\bfseries 97} (2018) 056010}
  [\href{https://arxiv.org/abs/1801.06067}{{\ttfamily 1801.06067}}].

\bibitem{Berends:1965ftl}
F.A.~Berends, \emph{{The T violating decay of $\pi^0\to3\gamma$}},
  \href{https://doi.org/10.1016/0031-9163(65)90176-9}{\emph{Phys. Lett.}
  {\bfseries 16} (1965) 178}.

\bibitem{Tarasov:1967}
A.V.~Tarasov, \emph{3-photon decay of neutral pions}, {\emph{Sov. J. Nucl.
  Phys.} {\bfseries 5} (1967) 445}.

\bibitem{Dicus:1975cz}
D.A.~Dicus, \emph{{An Estimate of the Rate of the Rare Decay $\pi^0 \to 3
  \gamma$}}, \href{https://doi.org/10.1103/PhysRevD.12.2133}{\emph{Phys. Rev.
  D} {\bfseries 12} (1975) 2133}.

\bibitem{Conzett:1992dn}
H.E.~Conzett, \emph{{Null tests of time reversal invariance}},
  \href{https://doi.org/10.1103/PhysRevC.48.423}{\emph{Phys. Rev. C} {\bfseries
  48} (1993) 423}.

\bibitem{Beyer:1993zw}
M.~Beyer, \emph{{Test of time reversal symmetry in the proton deuteron
  system}}, \href{https://doi.org/10.1016/0375-9474(93)90137-M}{\emph{Nucl.
  Phys. A} {\bfseries 560} (1993) 895}
  [\href{https://arxiv.org/abs/nucl-th/9302002}{{\ttfamily nucl-th/9302002}}].

\bibitem{Uzikov:2015aua}
Y.N.~Uzikov and A.A.~Temerbayev, \emph{{Null-test signal for $T$-invariance
  violation in $pd$ scattering}},
  \href{https://doi.org/10.1103/PhysRevC.92.014002}{\emph{Phys. Rev. C}
  {\bfseries 92} (2015) 014002}
  [\href{https://arxiv.org/abs/1506.08303}{{\ttfamily 1506.08303}}].

\bibitem{Uzikov:2016lsc}
Y.N.~Uzikov and J.~Haidenbauer, \emph{{Polarized proton--deuteron scattering as
  a test of time-reversal invariance}},
  \href{https://doi.org/10.1103/PhysRevC.94.035501}{\emph{Phys. Rev. C}
  {\bfseries 94} (2016) 035501}
  [\href{https://arxiv.org/abs/1607.04409}{{\ttfamily 1607.04409}}].

\bibitem{Eversheim:2017zxl}
D.~Eversheim, Y.~Valdau and B.~Lorentz, \emph{{The Time Reversal Invariance
  Experiment at Cosy (TRIC)}},
  \href{https://doi.org/10.22323/1.281.0177}{\emph{PoS} {\bfseries INPC2016}
  (2017) 177}.

\bibitem{Aksentyev:2017dnk}
{\scshape PAX} collaboration, \emph{{The Test of Time Reversal Invariance at
  Cosy (TRIC)}}, \href{https://doi.org/10.5506/APhysPolB.48.1925}{\emph{Acta
  Phys. Polon. B} {\bfseries 48} (2017) 1925}.

\bibitem{Haxton:1994bq}
W.C.~Haxton, A.~H\"oring and M.J.~Musolf, \emph{{Constraints on T-odd and
  P-even hadronic interactions from nucleon, nuclear, and atomic electric
  dipole moments}}, \href{https://doi.org/10.1103/PhysRevD.50.3422}{\emph{Phys.
  Rev. D} {\bfseries 50} (1994) 3422}.

\end{thebibliography}\endgroup
%---------------------------------------------------------------------------------------------------

\end{document}